\documentclass[aps,preprint]{revtex4}%
\usepackage{amsfonts}
\usepackage{amsmath}
\usepackage{amssymb}
\usepackage{graphicx}%
\begin{document}
\preprint{HEP/123-qed}
\title[ ]{Localized wave solutions of the scalar homogeneous wave equation and their
optical implementation}
\author{Kaido Reivelt}
\affiliation{Institute of Physics University of Tartu}
\author{Peeter Saari}
\affiliation{Institute of Physics University of Tartu}

\begin{abstract}

In recent years the topic of localized wave solutions of the homogeneous
scalar wave equation, i.e., the wave fields that propagate without any
appreciable spread or drop in intensity, has been discussed in many aspects in
numerous publications. In this review the main results of this rather disperse
theoretical material are presented in a single mathematical representation -
the Fourier decomposition by means of angular spectrum of plane waves. This
unified description is shown to lead to a transparent physical understanding
of the phenomenon as such and yield the means of optical generation of such
wave fields.

\end{abstract}
\volumeyear{year}
\volumenumber{number}
\issuenumber{number}
\eid{identifier}
\startpage{1}
\endpage{2}
\maketitle
\tableofcontents

\section{INTRODUCTION}

The birth of the diffraction theory of light dates back to the works of
Francis Maria Grimaldi (1618~--~1663), Robert Hook (1635~--~1703), Christiaan
Huygens (1629~--~1695) and Thomas Young (1773~--~1829) and was mathematically
formulated by Augustin Jean Fresnel (1788~--~1827). Over the two centuries it
has been considered as a very successful theory -- it indeed very precisely
describes the propagation of light in linear media. The foundation of the
diffraction theory is the principle of Huygens which states that (i) all the
points of a wavefront act as the sources of secondary wavelets and (ii) the
field at all the subsequent points is determined by the superposition of those wavelets.

The topic of free-space propagation of wave fields has attracted a renewed
interest in 1983 when James Neill Brittingham claimed \cite{fw1} that he
discovered a family of three-dimensional, nondispersive, source-free,
free-space, classical electromagnetic pulses which propagate in a straight
line in free space at light velocity (in this work he also introduced the term
\textit{focus wave mode} (FWM) for those wave fields). Now, the very idea of
secondary spherical sources in the classical diffraction theory implies that
any optical wave field suffers from lateral and longitudinal spread in the
course of propagation in free space, whereby the diffraction angle of the
spread is the larger the narrower is the field radius. In the view of this
general principle the Brittingham's statement is an astounding one and he
quite rightly used the formula "to convince the scientific community" in its
arguments. However, the original focus wave mode was indeed\textit{\ }the
solutions of the Maxwell's equations and the scientific community had to
resolve this apparent contradiction. As to give the reader an idea of the
initial problem the theoreticians had to tackle with, we reproduce here the
original definition of Brittingham which he deduced by "a very extensive
heuristical fit of various differential equation solutions": given the Maxwell
equations (SI)
\begin{align*}
\mathbf{\nabla\times E}  &  =-\dfrac{\partial\mathbf{B}}{\partial t}\\
\mathbf{\nabla\times H}  &  =\dfrac{\partial\mathbf{D}}{\partial t}\\
\mathbf{\nabla\cdot D}  &  =0\\
\mathbf{\nabla\cdot B}  &  =0\text{,}%
\end{align*}
\emph{\ }where $\mathbf{E}$, $\mathbf{D}$, $\mathbf{H}$, $\mathbf{B}$ and $t$
are electric field, electric flux density, magnetic field, magnetic induction
and time variable, respectively and using cylindrical coordinate system
($\rho,\varphi,z-ct$) the mathematical formulation of the original FWM reads
as
\begin{align*}
D_{\rho}\left(  \rho,\varphi,z,t\right)   &  =\Psi_{1}+\Psi_{1}^{\ast}\\
D_{\varphi}\left(  \rho,\varphi,z,t\right)   &  =\Psi_{2}+\Psi_{2}^{\ast}\\
H_{\rho}\left(  \rho,\varphi,z,t\right)   &  =\Psi_{3}+\Psi_{3}^{\ast}\\
H_{\varphi}\left(  \rho,\varphi,z,t\right)   &  =\Psi_{4}+\Psi_{4}^{\ast}\\
H_{z}\left(  \rho,\varphi,z,t\right)   &  =\Psi_{5}+\Psi_{5}^{\ast}\text{,}%
\end{align*}
where the functions $\Psi_{q}$ ($q=1,2...5$) are written as
\[
\Psi_{q}=A_{q}\left(  \rho,z-ct\right)  G_{1}\left(  \rho,z-ct\right)
G_{2}\left(  z-ct\right)  G_{3}\left(  z-c_{1}t\right)  \Phi^{\prime}\left(
\phi\right)
\]
for $q=1$ and $4$, and
\[
\Psi_{q}=A_{q}\left(  \rho,z-ct\right)  G_{1}\left(  \rho,z-ct\right)
G_{2}\left(  z-ct\right)  G_{3}\left(  z-c_{1}t\right)  \Phi\left(
\phi\right)
\]
for $q=2,3,5$. In those equations
\begin{align*}
G_{1}  &  =\exp\left[  -\frac{\rho^{2}}{4F}\right] \\
G_{2}  &  =\exp\left[  -ik_{1}\left(  z-ct\right)  \right] \\
G_{3}  &  =\exp\left[  ik_{2}\left(  z-c_{1}t\right)  \right]
\end{align*}
and
\[
F=ig\left(  z-ct\right)  +\xi\text{~.}%
\]
The remaining definitions read as%
\[
A_{1}=\frac{D^{TE}}{c^{2}}\left[  \frac{\left(  n+1\right)  cg\rho^{n-1}%
}{F^{n+2}}-\frac{cg\rho^{n+1}}{4F^{n+3}}+\frac{\left(  k_{1}c-k_{2}c\right)
\rho^{n-1}}{F^{n+2}}\right]
\]%
\begin{align*}
A_{2}  &  =-\frac{D^{TE}}{c^{2}}\left[  \frac{n\left(  n+1\right)
cg\rho^{n-1}}{F^{n+2}}-\frac{\left(  3n+4\right)  cg\rho^{n+1}}{4F^{n+3}%
}\right. \\
&  +\frac{n\left(  k_{1}c-k_{2}c_{1}\right)  \rho^{n-1}}{F^{n+1}}\\
&  \left.  +\frac{2cg\rho^{n+3}}{16F^{n+4}}-\frac{2\left(  k_{1}c-k_{2}%
c_{1}\right)  \rho^{n+1}}{4F^{n+2}}\right]
\end{align*}%
\begin{align*}
A_{3}  &  =-\frac{D^{TE}}{c^{2}}\left[  -\frac{n\left(  n+1\right)
g\rho^{n-1}}{F^{n+2}}+\frac{\left(  3n+4\right)  g\rho^{n+1}}{4F^{n+3}}\right.
\\
&  +\frac{n\left(  -k_{1}+k_{2}\right)  \rho^{n-1}}{F^{n+1}}\\
&  \left.  -\frac{2g\rho^{n+3}}{16F^{n+4}}-\frac{2\left(  -k_{1}+k_{2}\right)
\rho^{n+1}}{4F^{n+2}}\right]
\end{align*}%
\begin{align*}
A_{4}  &  =-D^{TE}\left[  -\frac{\left(  n+1\right)  g\rho^{n-1}}{F^{n+2}%
}+\frac{g\rho^{n+1}}{4F^{n+3}}\right. \\
&  \left.  +\frac{\left(  -k_{1}+k_{2}\right)  \rho^{n-1}}{F^{n+1}}\right]
\end{align*}%
\[
A_{5}=-iD^{TE}\left[  \frac{\rho^{n+2}}{4F^{n+3}}-\frac{\left(  n+1\right)
\rho^{n}}{F^{n+2}}\right]  ~\text{,}%
\]
where $D^{TE}$, $g$, $\xi$, $k_{1}$ and $k_{2}$ are constants. The $\Phi$
functions are defined as
\begin{align*}
\Phi\left(  \varphi\right)   &  =\left\{
\begin{array}
[c]{c}%
\sin\left(  n\varphi\right) \\
\cos\left(  n\varphi\right)
\end{array}
\right\} \\
\Phi^{\prime}\left(  \varphi\right)   &  =\left\{
\begin{array}
[c]{c}%
n\cos\left(  n\varphi\right) \\
-n\sin\left(  n\varphi\right)
\end{array}
\right\}
\end{align*}
and the supplemental conditions read%
\begin{align*}
2gk_{2}d_{1}  &  =1\\
k_{2}^{2}d_{2}  &  =\frac{k_{1}}{g}\text{,}%
\end{align*}
where
\begin{align*}
d_{1}  &  =\left(  1-\frac{c_{1}}{c}\right) \\
d_{2}  &  =\left(  1-\frac{c_{1}^{2}}{c^{2}}\right)  \text{.}%
\end{align*}
One has to agree, that the physical idea is very much hidden behind this
mathematical formulation.

Brittingham claimed, that this mathematical formulation (i) satisfy the
homogeneous Maxwell's equations, (ii) is continuous and nonsingular, (iii) has
a three-dimensional pulse structure, (iv) is nondispersive for all time, (v)
move at light velocity in straight lines, and (vi) carry finite
electromagnetic energy. Thus, the formulas above give a mathematical
formulation of a free-space wave field that can be described as a
\textquotedblright light bullet\textquotedblright\ and, though the proof of
the last claim was shown to be faulty by Wu and King \cite{fw2}, the whole
idea was very intricate and rose a considerable scientific interest
\cite{fw3}--\cite{ok6}.

The theoretical work of following years could be divided into the following
topics (see also Ref.~\cite{lw6} for an overview):

In the following publications \cite{fw3,fw5} the original vector field was
reduced to its scalar counterpart and the dominant part of the research work
that followed has been formulated in terms of solutions to homogeneous scalar
wave equation.

The close connection between the FWM's and the solutions of the paraxial wave
equation and Schr\"{o}dinger's equation (which both allow localized solutions)
has been established \cite{fw3,fw4,fw5} -- it has been shown that in terms of
the variables $z+ct$ and $z-ct$, if the solution of the scalar wave equation
is given by the anzatz $\exp\left[  \beta\left(  z+ct\right)  \right]
F\left(  x,y,z-ct\right)  $, the problem can be reduced to one of those equations.

The infinite energy content of the original FWM's has been addressed in
several publications (see
Refs.~\cite{fw3,fw3o1,fw5,fw14,fw16,lw1,g1,g2,g3,g4,g5} and references
therein). First of all, Sezginer \cite{fw3} and Wu and Lehmann \cite{fw3o1}
proved that any finite energy solution of the wave equation irreversibly leads
to dispersion and to spread of the energy. Then Ziolkowski \cite{fw5} pointed
out, that the superpositions of the infinite energy FWM's could result in
finite energy solutions and in following publications a number of finite
energy solutions to the scalar wave equation and Maxwell equations were
deduced -- "electromagnetic directed-energy pulse trains" (EDEPT)
\cite{lw1,g1}, "acoustic directed-energy pulse trains" (ADEPT) \cite{g0b} ,
splash pulses \cite{fw5}, modified power spectrum (MPS) \cite{lw1} pulses,
electromagnetic missiles \cite{mis1,mis2}, various super- and subluminal
pulses \cite{fw15} etc. In correspondence with \cite{fw3,fw3o1} this broader
class of localized waves (LW) have generally extended but finite ranges of
localizations. Also, several alternate infinite energy LW's (Bessel-Gauss
pulses \cite{lw4} for example) were deduced.

In Ref.~\cite{fw14} Besieris \textit{et al }introduced a novel integral
representation for synthesizing those LW's. This bidirectional plane wave
decomposition is based on a decomposition of the solutions of the scalar wave
equations into the forward and backward traveling plane wave solutions and it
has been shown to be a very natural basis for description of LW's (see
Ref.~\cite{lw1} for example).

The FWM's have been interpreted as being related in a special way to the field
of a source, moving on a complex trajectory parallel to the real axis of
propagation \cite{fw5,fw11,fw13,lw1}. This observation linked the FWM's with
the works by Deschamps \cite{o4} and Felsen \cite{o5} where the Gaussian beams
have been described as being paraxially equivalent to spherical waves with
centers at stationary complex locations.

There has been a considerable effort in finding the LW solutions in other
branches of physics, spanning various differential equations like spinor wave
equation \cite{fwz5}, fist-order hyperbolic systems like cold-plasma equation
\cite{fwz10}, Klein-Gordon equation \cite{fw14,fw15}.

In 1988 Durnin \cite{b1} published his paper on so called Bessel beams (see
for example Ref.~\cite{bag5} for an earlier publication on the topic). The
idea attracted much interest and the Bessel beams and their pulsed
counterparts -- X-waves \cite{x1}--\cite{x20} and Bessel-X waves
\cite{bx1,bx2,bx5}~-- became the research field of its own rights. In this
context the issue of the superluminal propagation of a class of LW's has been
considered in Refs.~\cite{x7,x50,x55k,x56k,x57k,x60k,x70k}.

It has been shown that the FWM's can be described as monochromatic Gaussian
beams observed in a moving relativistic inertial reference frame
\cite{fw6,lw6}.

Propagation of optical pulses or beams without any appreciable drop in the
intensity and spread over long distances would be highly desirable in many
applications. The obvious uses could be in fields like optical communication,
monitoring, imaging, and femtosecond laser spectroscopy, also in laser
acceleration of charged particles. Due to this general interest the
experimental generation FWM's and LW's has been discussed in numerous
publications (see Refs.~\cite{g0}--\cite{g25} and references therein). 

The most widely discussed approach has been to use directly the principle of
Huygens and launch the LW's from planar sources \cite{g1o1}. However, it
appears that each point source in such array must (i) have ultra-wide
bandwidth and (ii) be independently drivable as the temporal evolution of the
LW's generally is of the non-separable nature. Due to the present state of the
experiment this approach has not been realized even in radio-frequency domain
(it has been realized in acoustics \cite{g0a,g0b}).

In an another approach it has been shown that the LW's can be launched by the
so-called Gaussian dynamic apertures, that are characterized by an effective
radius that shrinks from an infinite extensions at $t\rightarrow-\infty$ to a
finite value at $t\rightarrow0$, then expands once more to an infinite
dimension as $t\rightarrow\infty$ \cite{g3} or by the spectrally depleted
(finite excitation time) Gaussian apertures \cite{g4,g5,g6,g10}.

It has been shown, that the field from an infinite line source contains a FWM
component \cite{fw17} and the LW's can be generated by a disk source moving
"more slowly than the speed of light" \cite{g15,g25}.

In optics none of those methods is feasible. A practical general idea for
optical generation of LW's was described in \cite{m2,m3} where it was shown
that the angular dispersion of various Bessel beam generators can be used to
produce the necessary coupling between the monochromatic components of the
LW's. In Refs.~\cite{m5,m1,bx5} we also presented the experimental evidence of
the feasibility of this approach. In particular, in Ref.~\cite{m5} we
constructed an optical setup for generation of two-dimensional FWM's and
obtained results from interferometric measurements of the generated wave field
that exhibit all the characteristic properties of the FWM's.

In this review we make an effort to give all the essential results of the
field an unified description in the way that we present them using the Fourier
decomposition methods exclusively. In doing so we unify the notation and
transform the mathematical representation where necessary.

The review is organized as follows:

In the preliminary Chapter~\ref{chIR} we introduce the necessary integral
representations for the solutions of the Maxwell's equations and scalar
homogeneous wave equation. Predominantly we will use the Fourier
representation of the free-space wave fields.

In Chapter~\ref{chPhAp} we deduce what in our opinion is the physically most
comprehensive representation of the FWM's and LW's -- we will show, that the
necessary and sufficient condition for a free-space wave field to be
propagation-invariant is that its support of angular spectrum of plane waves
is of a specific form. Several additional conclusions on the properties of the
LW's will be drawn.

In Chapter~\ref{chOV} we give an outline of the properties of the known
(published) LW's. The material in this section is important, because, to our
best knowledge, this is the first time where the optical feasibility of
certain well-known closed-form LW's is estimated -- we will see that majority
of the known LW's, including the original FWM's, are not realizable in optical domain.

In Chapter~\ref{chOK} we generalize the theory of the propagation-invariant
propagation into the domain of partially coherent wave fields -- we define the
conditions for the propagation-invariance of the mutual coherence function of
the wideband, stochastic, stationary fields. The theory also gives a means of
estimating the effect of spatial and temporal coherence of the source light on
the properties of generated fields and is used in the analysis of the results
of our experiments.

In Chapter~\ref{chG} we present the general idea of the optical generation of
LW's. First of all, the setup for the generation of simplest special case --
optical Bessel-X pulses -- is introduced. Then we show that in Fourier picture
the optical generation of FWM's can be resolved to applying specific angular
dispersion to the Bessel-X pulses and discuss on the finite energy
approximations of the FWM's. Also, the optical generation of partially
coherent propagation-invariant wave fields is discussed.

In Chapter~\ref{chExp} we present the results of the experiments on optical
LW's carried on so far. In particular, we report on experimental measurements
of the whole three-dimensional distribution of the field of optical X waves --
Bessel-X pulses -- and provide the experimental verification of the optical
feasibility of FWM's.

In Chapter~\ref{chSI} we give an outline of our work on self-imaging pulsed
wave fields~-- it appears, that certain discrete superpositions of the FWM's
can be used to compose spatiotemporally self-imaging wave fields that carry
non-trivial three-dimensional images.

\section{\label{chIR}INTEGRAL REPRESENTATIONS OF\ FREE-SPACE ELECTROMAGNETIC
WAVE\ FIELDS}

In this preliminary chapter we introduce the necessary integral
representations for the solutions of the homogeneous Maxwell's equations and
scalar homogeneous wave equation. Only the free-space wave fields are
considered, i.e., the wave fields under investigation do not have any sources
(except perhaps at infinity) and they do not interact with any material
objects. As we will see, such an approach is suitable for our purposes.

\subsection{\label{sMax}Solutions of the Maxwell equations in free space}

In SI units the source-free Maxwell equations can be written as
\begin{subequations}
\begin{align}
\mathbf{\nabla\times E}  &  =-\mu_{0}\dfrac{\partial\mathbf{H}}{\partial t}\\
\mathbf{\nabla\times H}  &  =\varepsilon_{0}\dfrac{\partial\mathbf{E}%
}{\partial t}\\
\mathbf{\nabla\cdot E}  &  =0\\
\mathbf{\nabla\cdot H}  &  =0~\text{,} \label{mx1}%
\end{align}
$\mathbf{E}$ and $\mathbf{H}$ being the electric and magnetic field vectors
respectively, $\mu_{0}$ is the magnetic permittivity of free space,
$\varepsilon_{0}$ is the electric permittivity of free space. As it is well
know, in this special case the components of the electric and magnetic field
vectors satisfy the homogeneous wave equation
\end{subequations}
\begin{subequations}
\begin{align}
\left(  \nabla^{2}-\frac{1}{c^{2}}\frac{\partial^{2}}{\partial t^{2}}\right)
\mathbf{E}\left(  \mathbf{r},t\right)   &  =0\label{mx2}\\
\left(  \nabla^{2}-\frac{1}{c^{2}}\frac{\partial^{2}}{\partial t^{2}}\right)
\mathbf{H}\left(  \mathbf{r},t\right)   &  =0\text{.} \label{mx3}%
\end{align}
In Eqs.~(\ref{mx2}) and (\ref{mx3}) only two of the six field variables are
independent and the Maxwell equations have to used to solve for the other,
dependent field components.

The general solution of the scalar wave equations (\ref{mx3}) can be expressed
as the Fourier decomposition as
\end{subequations}
\begin{subequations}
\begin{align}
\mathbf{E}\left(  \mathbf{r},t\right)   &  =\frac{1}{\left(  2\pi\right)
^{4}}\int_{-\infty}^{\infty}\mathrm{d}\omega\int\int\int_{-\infty}^{\infty
}\mathrm{d}\mathbf{k~}\mathcal{E}\left(  \mathbf{k},\omega\right)  \exp\left[
i\mathbf{kr}-i\omega t\right] \label{mx6}\\
\mathbf{H}\left(  \mathbf{r},t\right)   &  =\frac{1}{\left(  2\pi\right)
^{4}}\int_{-\infty}^{\infty}\mathrm{d}\omega\int\int\int_{-\infty}^{\infty
}\mathrm{d}\mathbf{k~}\mathcal{H}\left(  \mathbf{k},\omega\right)  \exp\left[
i\mathbf{kr}-i\omega t\right]  \text{,} \label{mx7}%
\end{align}
where $\mathcal{E}=\left(  \mathcal{E}_{x},\mathcal{E}_{y},\mathcal{E}%
_{z}\right)  $ and $\mathcal{H}=\left(  \mathcal{H}_{x},\mathcal{H}%
_{y},\mathcal{H}_{z}\right)  $ are plane wave spectrums of the electric and
magnetic field. Specifying, for example $\mathcal{E}_{x}$ $\mathcal{E}_{y}$ as
two solutions of the scalar wave equation we get from $\mathbf{\nabla\cdot
E}=0$ that
\end{subequations}
\begin{equation}
\mathcal{E}_{z}\left(  \mathbf{k},\omega\right)  =-\frac{1}{k_{z}}\left[
k_{x}\mathcal{E}_{x}\left(  \mathbf{k},\omega\right)  +k_{y}\mathcal{E}%
_{y}\left(  \mathbf{k},\omega\right)  \right]  \label{mx11}%
\end{equation}
and from $\mathbf{\nabla\times E}=-\mu_{0}\dfrac{\partial\mathbf{H}}{\partial
t}$%
\begin{subequations}
\begin{align}
\mathcal{H}_{x}\left(  \mathbf{k},\omega\right)   &  =-\frac{1}{\omega
k_{z}\mu_{0}}\left[  k_{x}k_{y}\mathcal{E}_{x}\left(  \mathbf{k}%
,\omega\right)  +\left(  k^{2}-k_{x}^{2}\right)  \mathcal{E}_{y}\left(
\mathbf{k},\omega\right)  \right] \label{mx15}\\
\mathcal{H}_{y}\left(  \mathbf{k},\omega\right)   &  =\frac{1}{\omega k_{z}%
\mu_{0}}\left[  \left(  k^{2}-k_{y}^{2}\right)  \mathcal{E}_{x}\left(
\mathbf{k},\omega\right)  +k_{x}k_{y}\mathcal{E}_{y}\left(  \mathbf{k}%
,\omega\right)  \right] \label{mx16}\\
\mathcal{H}_{z}\left(  \mathbf{k},\omega\right)   &  =\frac{1}{\omega\mu_{0}%
}k_{y}\mathcal{E}_{x}\left(  \mathbf{k},\omega\right)  -k_{x}\mathcal{E}%
_{y}\left(  \mathbf{k},\omega\right)  \text{.} \label{mx17}%
\end{align}
If we substitute the Eqs.~(\ref{mx11}) -- (\ref{mx17}) in (\ref{mx6}) and
(\ref{mx7}) we have a general solution of free-space Maxwell equations as a
superposition of monochromatic plane waves.

The other approach is to determine the vector potential $\mathbf{A}$ as the
solution of the homogeneous wave equation -- if we use the Coulomb gauge and
no sources are present the scalar potential is zero and the fields are given
by \cite{o15}
\end{subequations}
\begin{subequations}
\begin{align}
\mathbf{E}\left(  \mathbf{r},t\right)   &  =-\frac{\partial}{\partial
t}\mathbf{A}\left(  \mathbf{r},t\right) \label{mx22}\\
\mathbf{B}\left(  \mathbf{r},t\right)   &  =\nabla\times\mathbf{A}\left(
\mathbf{r},t\right)  \label{mx23}%
\end{align}
Alternatively, we can determine the Hertz vectors $\mathbf{\Pi}$ from the
homogeneous wave equation, then the fields are given by \cite{o20}
\end{subequations}
\begin{subequations}
\begin{align}
\mathbf{E}\left(  \mathbf{r},t\right)   &  =\nabla\left(  \nabla
\cdot\mathbf{\Pi}^{\left(  e\right)  }\right)  -\mu_{0}\nabla\times
\frac{\partial}{\partial t}\mathbf{\Pi}^{\left(  m\right)  }-\frac{1}{c^{2}%
}\frac{\partial^{2}}{\partial t^{2}}\mathbf{\Pi}^{\left(  e\right)
}\label{mx25}\\
\mathbf{H}\left(  \mathbf{r},t\right)   &  =\nabla\left(  \nabla
\cdot\mathbf{\Pi}^{\left(  m\right)  }\right)  -\varepsilon_{0}\nabla
\times\frac{\partial}{\partial t}\mathbf{\Pi}^{\left(  e\right)  }-\frac
{1}{c^{2}}\frac{\partial^{2}}{\partial t^{2}}\mathbf{\Pi}^{\left(  m\right)
}\text{.} \label{mx26}%
\end{align}
The choice of the vector components of the Hertz vectors and vector potential
generally determine the polarization properties of the resulting vector field.

\subsection{Plane wave expansions of scalar wave fields}

If we assume, that the general solution $\Psi\left(  \mathbf{r},t\right)  $ of
the scalar homogeneous wave equation
\end{subequations}
\begin{equation}
\left(  \nabla^{2}-\frac{1}{c^{2}}\frac{\partial^{2}}{\partial t^{2}}\right)
\Psi^{\prime}\left(  \mathbf{r},t\right)  =0 \label{ang1}%
\end{equation}
can be decomposed into the Fourier superposition of plane waves as%
\begin{equation}
\psi\left(  \mathbf{k},\omega\right)  =\int_{-\infty}^{\infty}\mathrm{d}%
t\int\int\int_{-\infty}^{\infty}\mathrm{d}\mathbf{r~}\Psi^{\prime}\left(
\mathbf{r},t\right)  \exp\left[  -i\mathbf{kr}+i\omega t\right]  \text{,}
\label{ang5}%
\end{equation}
the inverse transform yields%
\begin{equation}
\Psi^{\prime}\left(  \mathbf{r},t\right)  =\frac{1}{\left(  2\pi\right)  ^{4}%
}\int_{-\infty}^{\infty}\mathrm{d}\omega\int\int\int_{-\infty}^{\infty
}\mathrm{d}\mathbf{k~}\psi\left(  \mathbf{k},\omega\right)  \exp\left[
i\mathbf{kr}-i\omega t\right]  \text{.} \label{ang4}%
\end{equation}
The Eq.~(\ref{ang4}) together with the condition%
\begin{equation}
k_{x}^{2}+k_{y}^{2}+k_{z}^{2}=k^{2}=\left(  \frac{\omega}{c}\right)  ^{2}
\label{ang6}%
\end{equation}
which assures, that the Fourier representation satisfies the wave equation
(\ref{ang1}), is the general source-free solution of the scalar homogeneous
wave equation that will be used in this review. The representation
(\ref{ang4}) leads to Whittaker and Weyl type plane wave expansions (for the
discussions on this topic see for example Refs.~\cite{tns5}--\cite{tns18} and
\cite{ok4}).

\subsubsection{\label{ssAW}Whittaker type plane wave expansion}

The dispersion relation (\ref{ang6}) can be inserted into (\ref{ang4}) as a
delta function $\delta(k^{2}-k_{x}^{2}+\allowbreak k_{y}^{2}+k_{z}^{2})$ so
that the integration over $\omega$ yields%
\begin{align}
&  \Psi^{\prime}\left(  \mathbf{r},t\right)  =\frac{1}{\left(  2\pi\right)
^{4}}\int\int\int_{-\infty}^{\infty}\mathrm{d}k_{x}\mathrm{d}k_{y}%
\mathrm{d}k_{z}\mathbf{~}\frac{c}{2k}\label{ang4b}\\
&  \qquad\qquad\times A^{\prime}\left(  k_{x},k_{y},k_{z}\right)  \exp\left[
i\left(  k_{x}x+k_{y}y+k_{z}z-kct\right)  \right]  \text{.}\nonumber
\end{align}
or%
\begin{align}
&  \Psi^{\prime}\left(  \mathbf{r},t\right)  =\frac{1}{\left(  2\pi\right)
^{4}}\int\int\int_{-\infty}^{\infty}\mathrm{d}k_{x}\mathrm{d}k_{y}%
\mathrm{d}k_{z}\mathbf{~}\nonumber\\
&  \qquad\times A\left(  k_{x},k_{y},k_{z}\right)  \exp\left[  i\left(
k_{x}x+k_{y}y+k_{z}z-kct\right)  \right]  \text{~.} \label{ang4a}%
\end{align}
where
\begin{equation}
A\left(  k_{x},k_{y},k_{z}\right)  =\frac{c}{2k}A^{\prime}\left(  k_{x}%
,k_{y},k_{z}\right)  \label{ang4c}%
\end{equation}
If we also introduce the cylindrical coordinate system $\left(  \rho
,z,\varphi\right)  $ in real space and spherical coordinate system $\left(
k,\theta,\phi\right)  $ in $k$-space the Eq.~(\ref{ang4a}) yields%
\begin{align}
&  \Psi^{\prime}\left(  \mathbf{r},t\right)  =\int_{0}^{\infty}\mathrm{d}%
kk^{2}\int_{0}^{\pi}\mathrm{d}\theta\sin\theta\int_{0}^{2\pi}d\phi
\mathbf{~}A\left(  k\sin\theta\cos\phi,k\sin\theta\cos\phi,k\cos\theta\right)
\nonumber\\
&  \qquad\qquad\times\exp\left[  ik\left(  x\sin\theta\cos\phi+y\sin\theta
\sin\phi+z\cos\theta-ct\right)  \right]  \label{ang7}%
\end{align}
(here and hereafter we omit the normalizing constants in front of the
integrals of this type). We can also expand the radial dependence of the
angular spectra as the Fourier series%
\begin{equation}
A\left(  k\sin\theta\cos\phi,k\sin\theta\cos\phi,k\cos\theta\right)
=\sum_{n=-\infty}^{\infty}A_{n}\left(  k,\theta\right)  \exp\left[
in\phi\right]  \label{ang7b}%
\end{equation}
and get another form of (\ref{ang7})%
\begin{align}
&  \Psi\left(  \rho,z,\varphi,t\right)  =\sum_{n=0}^{\infty}\exp\left[  \pm
in\varphi\right]  \int_{0}^{\infty}\mathrm{d}kk^{2}\int_{0}^{\pi}%
\mathrm{d}\theta\sin\theta\mathbf{~}\nonumber\\
&  \qquad\times A_{n}\left(  k,\theta\right)  J_{n}\left(  k\rho\sin
\theta\right)  \exp\left[  ik\left(  z\cos\theta-ct\right)  \right]  \text{,}
\label{ang9}%
\end{align}
where $J_{n}\left(  {}\right)  $ is the $n$-th order Bessel function of the
first kind and we introduced the polar coordinates in real space
($\rho,z,\varphi$), so that $\Psi\left(  \rho,z,\varphi,t\right)
=\allowbreak\Psi^{\prime}\left(  \rho\cos\varphi,\rho\sin\varphi,z,t\right)
$. In the radially symmetric case only the term $n=0$ is taken into account in
Eq.~(\ref{ang9}) and we have%
\begin{align}
\Psi\left(  \rho,z,\varphi,t\right)   &  =\int_{0}^{\infty}\mathrm{d}%
kk^{2}\int_{0}^{\pi}\mathrm{d}\theta\sin\theta~A_{0}\left(  k,\theta\right)
\nonumber\\
&  \times J_{0}\left(  k\rho\sin\theta\right)  \exp\left[  ik\left(
z\cos\theta-ct\right)  \right]  \text{.} \label{ang20}%
\end{align}

If we define $\chi=k\sin\theta$ and again use the Fourier series expansion of
the radial dependence of the angular spectrum, the representation
(\ref{ang4a}) yields
\begin{align}
\Psi\left(  \rho,z,\varphi,t\right)   &  =\sum_{n=0}^{\infty}\int_{-\infty
}^{\infty}\mathrm{d}k_{z}\int_{0}^{\infty}\mathrm{d}\chi\chi A_{n}\left(
\sqrt{k_{z}^{2}+\chi^{2}},\arcsin\tfrac{\chi}{\sqrt{k_{z}^{2}+\chi^{2}}%
}\right) \nonumber\\
&  \times J_{n}\left(  \chi\rho\right)  \exp\left[  \pm in\varphi\right]
\exp\left[  i\left(  k_{z}z-ct\sqrt{\chi^{2}+k_{z}^{2}}\right)  \right]
\text{.} \label{ang8}%
\end{align}
Again, in the radially symmetric case only the term $n=0$ is taken into
account and we have
\begin{align}
\Psi\left(  \rho,z,\varphi,t\right)   &  =\int_{-\infty}^{\infty}%
\mathrm{d}k_{z}\int_{0}^{\infty}\mathrm{d}\chi\chi A_{0}\left(  \sqrt
{k_{z}^{2}+\chi^{2}},\arcsin\tfrac{\chi}{\sqrt{k_{z}^{2}+\chi^{2}}}\right)
\nonumber\\
&  \times J_{0}\left(  \chi\rho\right)  \exp\left[  i\left(  k_{z}%
z-ct\sqrt{\chi^{2}+k_{z}^{2}}\right)  \right]  \text{.} \label{ang17}%
\end{align}

\subsubsection{\label{ssAWe}Weyl type plane wave expansion}

If we use the dispersion relation (\ref{ang6}) to eliminate the variable
$k_{z}$ instead, then Eq.~(\ref{ang4}) can be given the following form
\begin{align}
&  \Psi\left(  \rho,z,\varphi,t\right)  =\sum_{n=0}^{\infty}\exp\left[  \pm
in\varphi\right]  \int_{0}^{\infty}\mathrm{d}k\int_{0}^{\infty}\mathrm{d}%
\chi\chi\nonumber\\
&  \qquad\times A_{n}^{we}\left(  k,\chi\right)  J_{n}\left(  \chi\rho\right)
\exp\left[  i\left(  z\sqrt{k^{2}-\chi^{2}}-kct\right)  \right]  \text{,}
\label{ang26}%
\end{align}
which is the Weyl type superposition over the plane waves (see for example
Ref.~\cite{ok4} for a thorough treatment).

The Weyl type spectrum of plane waves is often derived as the Fourier
transform of the wave field in plane $z=0$. In contrary, the Whittaker type
superposition is calculated as its three-dimensional Fourier transform over
the space. Note however, that the distinction between the two is not clear for
wideband wave fields, as the calculation of Weyl representation requires the
knowledge of the evolution of the wave field on the $z=0$ plane for all times
[see Eq.~(\ref{ang5})].

\subsection{\label{sBd}Bidirectional plane wave decomposition}

The bidirectional plane wave decomposition was introduced by Besieris
\textit{et al }in Ref.~\cite{fw14} and it has been proved to be useful for
description of LW's. It is based on a decomposition of the solutions of the
scalar wave equations into the forward and backward traveling plane wave
solutions, in this representation the general solution to the scalar wave
equation can be written in the form (Eq.~2.22 of Ref.~\cite{fw14})%

\begin{align}
&  \Psi\left(  \rho,\zeta,\eta,\varphi\right)  =\frac{1}{\left(  2\pi\right)
^{2}}\sum_{n=0}^{\infty}\int_{0}^{\infty}\mathrm{d}\tilde{\alpha}\int
_{0}^{\infty}\mathrm{d}\tilde{\beta}\int_{0}^{\infty}\mathrm{d}\chi\chi
C_{n}\left(  \tilde{\alpha},\tilde{\beta},\chi\right) \nonumber\\
&  \qquad\times J_{n}\left(  \chi\rho\right)  \exp\left[  \pm in\varphi
\right]  \exp\left[  -i\tilde{\alpha}\zeta\right]  \exp\left[  i\tilde{\beta
}\eta\right]  \delta\left(  \tilde{\alpha}\tilde{\beta}-\frac{\chi^{2}}%
{4}\right)  \text{,} \label{bidi1}%
\end{align}
where $\eta=z+ct$ and $\zeta=z-ct$. Even though the Eq.~(\ref{bidi1}) differs
noticeably from the Fourier decomposition, there is one to one correspondence
between these two through the change of variables
\begin{subequations}
\begin{align}
k_{z}  &  =\tilde{\alpha}-\tilde{\beta}\label{bidi3}\\
\frac{\omega}{c}  &  =\tilde{\alpha}+\tilde{\beta}\text{,} \label{bidi4}%
\end{align}
or inversely
\end{subequations}
\begin{subequations}
\begin{align}
\tilde{\alpha}  &  =\frac{1}{2}\left(  \frac{\omega}{c}+k_{z}\right)
\label{bidi6}\\
\tilde{\beta}  &  =\frac{1}{2}\left(  \frac{\omega}{c}-k_{z}\right)  \text{.}
\label{bidi7}%
\end{align}
Consequently we can write
\end{subequations}
\begin{align}
\psi\left(  \mathbf{k},\omega\right)   &  \propto\sum_{n=0}^{\infty}%
\exp\left[  \pm in\varphi\right] \nonumber\\
&  \times C_{n}\left[  \frac{1}{2}\left(  \frac{\omega}{c}+k_{z}\right)
,\frac{1}{2}\left(  \frac{\omega}{c}-k_{z}\right)  ,\sqrt{k_{x}^{2}+k_{y}^{2}%
}\right]  \text{.} \label{bidi7a}%
\end{align}
Note that the delta function constraint%
\begin{equation}
4\tilde{\alpha}\tilde{\beta}=\chi^{2} \label{bidi8}%
\end{equation}
in Eq.~(\ref{bidi1}) in the Fourier picture reduces to
\begin{equation}
\left(  \frac{\omega}{c}\right)  ^{2}-k_{z}^{2}-\chi^{2}=0\text{.}
\label{bidi9}%
\end{equation}
For circularly symmetric wave fields the bidirectional expansions yields
\begin{align}
\Psi\left(  \rho,\zeta,\eta,\varphi\right)   &  =\frac{1}{\left(  2\pi\right)
^{3}}\int_{0}^{\infty}\mathrm{d}\tilde{\alpha}\int_{0}^{\infty}\mathrm{d}%
\tilde{\beta}\int_{0}^{\infty}\mathrm{d}\chi\chi C_{0}\left(  \tilde{\alpha
},\tilde{\beta},\chi\right)  J_{0}\left(  \chi\rho\right) \nonumber\\
&  \times\exp\left[  -i\tilde{\alpha}\zeta\right]  \exp\left[  i\tilde{\beta
}\eta\right]  \delta\left(  \tilde{\alpha}\tilde{\beta}-\frac{\chi^{2}}%
{4}\right)  \text{.} \label{bidi11}%
\end{align}

\section{\label{chPhAp}A PRACTICAL APPROACH\ TO SCALAR\newline FWM'S}

\subsection{Propagation invariance of scalar wave fields}

\subsubsection{\label{ssFAng}The angular spectrum of plane waves of the FWM's}

First of all, in literature the term FWM has been used mostly with the
following closed-form solution of the scalar homogeneous wave equation:%
\begin{equation}
\Psi_{f}\left(  \rho,z,\varphi,t\right)  =\exp\left[  i\beta\left(
z+ct\right)  \right]  \frac{a_{1}}{4\pi i\left(  a_{1}+i\zeta\right)  }%
\exp\left[  -\frac{\beta\rho^{2}}{a_{1}+i\zeta}\right]  \label{su0}%
\end{equation}
(Eq.~(2.1) of Ref.~\cite{fw18}). The Weyl and Whittaker type plane wave
spectrums of this wave field have been derived in Refs.~\cite{fw14,fw18} and,
omitting the normalizing constants, the latter reads%
\begin{equation}
A_{0}^{\left(  f\right)  }\left(  k,\theta\right)  =\frac{1}{k}\exp\left[
-\frac{a_{1}k\left(  \cos\theta+1\right)  }{2}\right]  \delta\left(
k-k\cos\theta-2\beta\right)  \text{.} \label{su1}%
\end{equation}
In this respect one can say that the following derivation of the angular
spectrum of plane waves of the FWM's is nothing but the different
interpretation of the results already published. However, the alternate
emphasis in the theory, described in this section (and published in
Ref.~\cite{m2}), have proved to make the difference if the optical generation
of the FWM's is under discussion. Also, the term FWM will be redefined in what follows.

Consider the general solution of the free-space wave equation represented as
the Whittaker type plane wave decomposition Eq.~(\ref{ang9})
\begin{align}
&  \Psi\left(  \rho,z,\varphi,t\right)  =\sum_{n=0}^{\infty}\exp\left[  \pm
in\varphi\right]  \int_{0}^{\infty}\mathrm{d}kk^{2}\int_{0}^{\pi}%
\mathrm{d}\theta\sin\theta~\nonumber\\
&  \qquad\qquad\times A_{n}\left(  k,\theta\right)  J_{n}\left(  k\rho
\sin\theta\right)  \exp\left[  ikz\cos\theta-i\omega t\right]  \text{.}
\label{su4}%
\end{align}
The integral representation of fundamental FWM's can be derived from the
condition that the superposition of Bessel beams in Eq.(\ref{su4}) should form
a nondispersing pulse propagating along the $z$ axis. In terms of group
velocity dispersion of wave packets this condition means that the on-axis
group velocity $v^{g}=\mathrm{d}\omega/\mathrm{d}k_{z}$ should be constant
over the whole spectral range. This restriction allows non-trivial
solutions\ only if we assume that the cone angle in relation $k_{z}%
=k\cos\theta$ is a function of the wave number, i.e., one can write
$\theta\left(  k\right)  $. The corresponding \textit{support} of the angular
spectrum of the plane wave constituents of the pulse, i.e., the volume of the
$k$-space where the angular spectrum of plane waves of the wave field is not
zero, is a cylindrically symmetric surface in the $k$-space and the angular
spectrum can be expressed by means of Dirac delta function as $A_{n}\left(
k,\theta\right)  =B_{n}\left(  k\right)  \delta\left[  \theta-\theta\left(
k\right)  \right]  $.%

%TCIMACRO{\FRAME{ftbFU}{2.6143in}{1.6414in}{0pt}{\Qcb{On the geometrical
%interpretation of the parameters $\beta$ and $\xi$ of the supports of angular
%spectrum of plane waves of FWM's (gray line).}}{\Qlb{fig44}}{fig4_4.jpg}%
%maintain-aspect-ratio TRUE;  display "ICON";  valid_file "F";
%width 2.6143in;  height 1.6414in;  depth 0pt;  original-width 3.9314in;
%original-height 1.9605in;  cropleft "0";  croptop "1";  cropright "1";
%cropbottom "0";  filename 'fig4_4.jpg';file-properties "XNPEU";}}}%
%BeginExpansion
\begin{figure}
[tb]
\begin{center}
\includegraphics[
height=1.6414in,
width=2.6143in
]%
{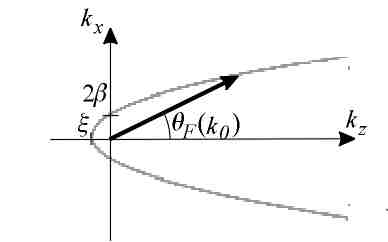}%
\caption{On the geometrical interpretation of the parameters $\beta$ and $\xi$
of the supports of angular spectrum of plane waves of FWM's (gray line).}%
\label{fig44}%
\end{center}
\end{figure}
%EndExpansion
The condition
\begin{equation}
v^{g}=c\frac{dk}{dk_{z}}=\left[  \frac{1}{c}\frac{d}{dk}\left(  k\cos
\theta_{F}\left(  k\right)  \right)  \right]  ^{-1}=\frac{c}{\gamma}\text{,}
\label{su8}%
\end{equation}
where constant $\gamma$ determines the group velocity, yields
\begin{equation}
k_{z}=\gamma k-2\beta\gamma\text{,} \label{su10}%
\end{equation}
where the integration constant $2\beta$ is defined as the wave number of the
plane wave component propagating perpendicularly to the $z$ axis, i.e.,
$\theta_{F}\left(  2\beta\right)  =90%
%TCIMACRO{\U{b0}}%
%BeginExpansion
{{}^\circ}%
%EndExpansion
$ (see Fig.~\ref{fig44} for the geometrical interpretation of the parameter
$\beta$ in $k$-space, the choice is consistent with \cite{fw14,fw18} for
example). Thus, we can write%
\begin{equation}
\cos\theta_{F}\left(  k\right)  =\frac{\gamma\left(  k-2\beta\right)  }{k}
\label{su12}%
\end{equation}
or
\begin{equation}
k_{F}\left(  \theta\right)  =\frac{2\beta\gamma}{\gamma-\cos\theta}\text{.}
\label{su14}%
\end{equation}

It appears in section \ref{ssFClass} that for a subclass of special cases the
above definitions are not appropriate as the corresponding supports of the
angular spectrum of plane waves do not intersect with the $k_{x}$ axis. Then
one should determine an alternate integration constant $\xi$ from the
condition $\theta_{F}\left(  \xi\right)  =180%
%TCIMACRO{\U{b0}}%
%BeginExpansion
{{}^\circ}%
%EndExpansion
$, this choice yields%
\begin{equation}
k_{z}\left(  k\right)  =\left\{
\begin{array}
[c]{c}%
\gamma k-\xi\left(  \gamma+1\right)  ,~\text{if}\mathrm{~}\xi\geq0\\
\gamma k-\xi\left(  \gamma-1\right)  ,~\text{if}\mathrm{~}\xi<0
\end{array}
\right.  \label{su17}%
\end{equation}
(see Fig.~\ref{fig44} for the geometrical interpretation of the parameter
$\xi$ in $k$-space). Thus, we can write
\begin{equation}
\cos\theta_{F}\left(  k\right)  =\frac{\gamma k-\xi\left(  \gamma\pm1\right)
}{k} \label{su19}%
\end{equation}
or
\begin{equation}
k_{F}\left(  \theta\right)  =\frac{\xi\left(  \gamma\pm1\right)  }{\gamma
-\cos\theta} \label{su21}%
\end{equation}
\emph{\ }so that
\begin{equation}
2\beta=\xi\frac{\gamma+1}{\gamma} \label{su22}%
\end{equation}
(as we always have $\beta\geq0$).

The definitions (\ref{su12}) or (\ref{su14}) give the angular spectrum of
plane waves in Eq.~(\ref{su4}) the form
\begin{equation}
A_{n}^{\left(  F\right)  }\left(  k,\theta\right)  =B_{n}\left(  k\right)
\delta\left[  \theta-\theta_{F}\left(  k\right)  \right]  \label{su29}%
\end{equation}
and%
\begin{equation}
A_{n}^{\left(  F\right)  }\left(  k,\theta\right)  =B_{n}\left[  \theta
_{F}\left(  k\right)  \right]  \delta\left[  k-k_{F}\left(  \theta\right)
\right]  \text{,} \label{su31}%
\end{equation}
correspondingly (see Fig.~\ref{fig42} of the section~\ref{ssFClass} for the
set of special cases).

As $k_{z}=k\cos\theta$, our result (\ref{su29}) is consistent with the support
of angular spectrum of plane waves of the original FWM's in Eq.~(\ref{su1}),
the constant $\gamma$ just generalizes to include also FWM's of different
group velocities. Thus, we can conclude that the physically transparent
condition (\ref{su8}) indeed determines the support of the angular spectrum of
plane waves of the FWM's.

\subsubsection{\label{ssFInt}Integral expressions for the field of the scalar
FWM's}

With the angular spectrum of plane waves (\ref{su29}) we can eliminate
variable $\theta$ in integral (\ref{su4}) and get
\begin{align}
&  \Psi_{F}\left(  \rho,z,\varphi,t\right)  =\sum_{n=0}^{\infty}\exp\left[
\pm in\varphi\right]  \int_{0}^{\infty}\mathrm{d}k\,k^{2}\sin\theta_{F}\left(
k\right) \nonumber\\
&  \qquad\times B_{n}\left(  k\right)  J_{n}\left[  k\rho\sin\theta_{F}\left(
k\right)  \right]  \exp\left[  ik\left(  z\cos\theta_{F}\left(  k\right)
-ct\right)  \right]  \text{~,} \label{su40}%
\end{align}
using (\ref{su12}) we can write%
\begin{align}
&  \Psi_{F}\left(  \rho,z,\varphi,t\right)  =\exp\left[  -i2\gamma\beta
z\right]  \sum_{n=0}^{\infty}\exp\left[  \pm in\varphi\right]  \int
_{0}^{\infty}\mathrm{d}k\,k^{2}\sin\theta_{F}\left(  k\right) \nonumber\\
&  \qquad\times B_{n}\left(  k\right)  J_{n}\left[  k\rho\sqrt{1-\left(
\frac{\gamma\left(  k-2\beta\right)  }{k}\right)  ^{2}}\right]  \exp\left[
ik\left(  \gamma z-ct\right)  \right]  \text{.} \label{su43}%
\end{align}
Alternatively, we can eliminate $k$ by means of Eqs.~(\ref{su31}) and get
\begin{align}
&  \Psi_{F}\left(  \rho,z,\varphi,t\right)  =\sum_{n=0}^{\infty}\exp\left[
\pm in\varphi\right]  \int_{0}^{\pi}\mathrm{d}\theta\,\sin\theta~k_{F}%
^{2}\left(  \theta\right) \nonumber\\
&  \qquad\times B_{n}\left[  \theta_{F}\left(  k\right)  \right]  J_{n}\left[
k_{F}\left(  \theta\right)  \rho\sin\theta\right]  \exp\left[  ik_{F}\left(
\theta\right)  \left(  z\cos\theta-ct\right)  \right]  \text{~,} \label{su45}%
\end{align}
again (\ref{su12}) transform the equation to%
\begin{align}
&  \Psi_{F}\left(  \rho,z,\varphi,t\right)  =\sum_{n=0}^{\infty}\exp\left[
\pm in\varphi\right]  \int_{0}^{\pi}\mathrm{d}\theta\,\left(  \frac
{2\beta\gamma}{\gamma-\cos\theta}\right)  ^{2}\sin\theta\nonumber\\
&  \qquad\times B_{n}\left[  \theta_{F}\left(  k\right)  \right]  J_{n}\left[
\frac{2\beta\gamma\rho\sin\theta}{\gamma-\cos\theta}\right]  \exp\left[
i\frac{2\beta\gamma}{\gamma-\cos\theta}\left(  z\cos\theta-ct\right)  \right]
\label{su47}%
\end{align}
[note that analogous expressions can be written using Eqs.~(\ref{su17}) --
(\ref{su21})].

The applied condition (\ref{su8}) implies that the longitudinal shape of the
central peak of the pulsed wave field in Eqs.~(\ref{su40}) -- (\ref{su47}) do
not spread as it propagates in $z$ axis direction. From the integral
expressions it is also obvious that the pulse do not spread in transversal
direction. However, the wave field has what has been called the "local
variations" -- the term $\exp\left[  -i2\gamma\beta z\right]  $ in
(\ref{su43}) implies that only the instantaneous intensity of the wave field
is independent of the propagation distance, in what follows we refer to such
wave fields as \textit{propagation-invariant}.

It is important to note that in Eqs.~(\ref{su29}), (\ref{su31}) and
(\ref{su40}) -- (\ref{su47}) the frequency spectrum is arbitrary. Thus, the
necessary and sufficient condition for the propagation-invariance of the
general pulsed wave field (\ref{su4}) is that its \textit{support }of angular
spectrum of plane waves should be defined by Eq.~(\ref{su12}) or (\ref{su19}).
The statement can also be inverted and one can say that the wave field is
\textit{strictly }propagation-invariant only if its support of angular
spectrum of plane waves is defined by Eq.~(\ref{su29}) or (\ref{su31}) --
indeed, in Eq.~(\ref{su8}) any other choice would lead to the group velocity
dispersion and the pulse would inevitably spread as it propagates. This also
implies, that all the possible solutions of scalar homogeneous wave equation
that have extended depth of propagation as compared to ordinary Gaussian
pulses (see next chapter) should be considered as certain approximations to
the FWM's.

Now, the closed-form expressions like (\ref{su0}) are very convenient in
numerical analysis, however, limiting ourselves to the set of available
closed-form integrals of (\ref{su40}) -- (\ref{su47}) is not reasonable by any
means. In this review we use the term "\textit{focus wave modes}" (FWM) for
all the wave fields that can be represented by the integral expressions
(\ref{su40}) -- (\ref{su47}), whereas the closed-form expression (\ref{su0})
will be called the \textit{original }FWM.

\subsubsection{\label{ssFClass}A physical classification of FWM's}

The recognition, that the spatiotemporal behavior of FWM's is determined only
by the support of their angular spectrum of plane waves enables one to give a
straightforward general classification to the FWM's.%

%TCIMACRO{\FRAME{ftbFU}{4.2644in}{6.0744in}{0pt}{\Qcb{The physical
%classification of the FWM's in terms of sections of the cone $\chi^{2}%
%+k_{z}^{2}-k^{2}=0$ in $\left(  \chi,k_{z},k\right)  $ space. The first two
%columns depict the sections of the cone from two viewpoints, the corresponding
%supports of angular spectrums of plane waves are depicted in third column.}%
%}{\Qlb{fig42}}{fig4_2.jpg}{\special{ language "Scientific Word";
%type "GRAPHIC";  maintain-aspect-ratio TRUE;  display "ICON";
%valid_file "F";  width 4.2644in;  height 6.0744in;  depth 0pt;
%original-width 5.0363in;  original-height 6.8555in;  cropleft "0";
%filename 'fig4_2.jpg';file-properties "XNPEU";}}}%
%BeginExpansion
\begin{figure}
[tb]
\begin{center}
\includegraphics[
height=6.0744in,
width=4.2644in
]%
{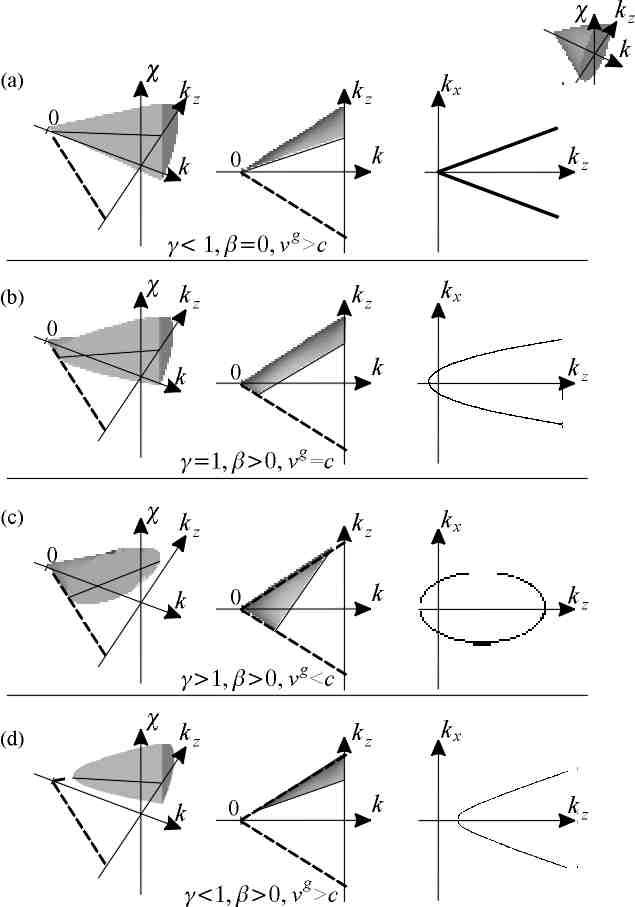}%
\caption{The physical classification of the FWM's in terms of sections of the
cone $\chi^{2}+k_{z}^{2}-k^{2}=0$ in $\left(  \chi,k_{z},k\right)  $ space.
The first two columns depict the sections of the cone from two viewpoints, the
corresponding supports of angular spectrums of plane waves are depicted in
third column.}%
\label{fig42}%
\end{center}
\end{figure}
%EndExpansion
Note, the dispersion relation
\begin{equation}
\chi^{2}+k_{z}^{2}-\left(  \frac{\omega}{c}\right)  ^{2}=0 \label{co1}%
\end{equation}
can be interpreted as a definition of a cone in ($\chi,k_{z},k$) space
\cite{fw15} (see Fig.~\ref{fig42}). In this context the specific supports of
the angular spectrum of plane waves of FWM's in Eqs.~(\ref{su8}) --
(\ref{su21}) have a geometrical interpretation as being the cone sections of
(\ref{co1}) along the planes%
\begin{equation}
k_{z}=\gamma k-2\beta\gamma\label{co3}%
\end{equation}
(\ref{su10}) or%
\begin{equation}
k_{z}\left(  k\right)  =\gamma k-\xi\left(  \gamma\pm1\right)  \label{co4}%
\end{equation}
(\ref{su17}). It can be seen that the possible supports of the angular
spectrums of plane waves can be divided into four explicit special cases (see
Fig.~\ref{fig42}) that can be taken as the natural classification of the FWM's:

\begin{itemize}
\item[1.] $\beta=0$ ($\xi=0$), $\gamma\leq1$, the support is a cone in $k
$-space, typical examples are Bessel-X pulse and X-pulse (the case $\gamma=1 $
corresponds to plane wave pulse);

\item[2.] $\beta\neq0$ ($\xi\neq0$), $\gamma=1$, the support is a paraboloid
in $k$-space, typical example is FWM's, propagating at velocity of light;

\item[3.] $\beta\neq0$ ($\xi\neq0$), $\gamma>1$, the support is an ellipsoid
in $k$-space, the group velocity of the FWM's satisfies $v^{g}<c$;

\item[4.] $\beta\neq0$ ($\xi\neq0$), $\gamma<1$, the support is hyperboloid in
$k$-space, the group velocity of the FWM's satisfies $v^{g}>c$;
\end{itemize}

Thus, there is barely four general types of strictly propagation-invariant
solutions of the scalar wave equation. This point has to be stressed as the
straightforward basic idea we set forward here is often elusive in the general
literature and numerous closed form LW's have been set forward.

\paragraph{Pulsed localized wave fields in dispersive media}

It should be noted at this point that, in principle, the approach can be used
to derive propagation-invariant wave fields for linear dispersive media. In
this case we should replace $k_{z}$ by $k_{z}n\left(  \omega\right)  $ in
Eqs.~(\ref{su8}) -- (\ref{su14}), $n\left(  \omega\right)  $ being the
refractive index of the medium. This modification yields the following
equation for the support of the angular spectrum of plane waves of the FWM in
linear dispersive media:
\begin{equation}
\tilde{\theta}_{F}\left(  \frac{\omega}{c}\right)  =\arccos\left[
\frac{\gamma\left(  \frac{\omega}{c}-2\beta\right)  }{n\left(  \omega\right)
\,\frac{\omega}{c}}\right]  \text{,} \label{su35}%
\end{equation}
$c$ being the velocity of light in vacuum. Eq.~(\ref{su35}) defines the
support of angular spectrum of plane waves to the wave field that propagates
without any longitudinal or transversal spread in linear dispersive media.
This approach -- to use predetermined angular dispersion to suppress the
longitudinal (and transversal) dispersion, though differently formulated, has
been already used in Refs.~\cite{bx2,bx5,ti20} for example.

\subsubsection{\label{ssFRad}The temporal evolution of the FWM's in the radial
direction}

The temporal evolution of the FWM's in the radial direction can be given a
convenient mathematical interpretation. Namely, as Ch\'{a}vez-Cerda \textit{et
al }noted in Ref.~\cite{b16} the monochromatic Bessel beam can be represented
as a superposition of so-called Hankel waves
\begin{subequations}
\begin{align}
\Psi_{m}^{\left(  1\right)  }\left(  \rho,z,t\right)   &  =H_{m}^{\left(
1\right)  }\left(  \chi\rho\right)  \exp\left[  ik_{z}z-i\omega t+im\varphi
\right] \label{r4}\\
\Psi_{m}^{\left(  2\right)  }\left(  \rho,z,t\right)   &  =H_{m}^{\left(
2\right)  }\left(  \chi\rho\right)  \exp\left[  ik_{z}z-i\omega t+im\varphi
\right]  \text{,} \label{r5}%
\end{align}
where
\end{subequations}
\begin{subequations}
\begin{align}
H_{m}^{\left(  1\right)  }\left(  \chi\rho\right)   &  =J_{m}\left(  \chi
\rho\right)  +iN_{m}\left(  \chi\rho\right) \label{r7}\\
H_{m}^{\left(  2\right)  }\left(  \chi\rho\right)   &  =J_{m}\left(  \chi
\rho\right)  -iN_{m}\left(  \chi\rho\right)  \label{r9}%
\end{align}
are the \textit{m}-th order Hankel functions and $N_{m}$ denotes the
\textit{m}-th order Neumann function (the Bessel function of the second kind).
For monochromatic wave fields the two solutions define the diverging and
converging wave in $xy$ plane, in other terms, they form the "sink" and
"source" pair. In those terms the \textit{m}-th order Bessel beam can be
written as
\end{subequations}
\begin{align}
&  J_{m}\left(  \chi\rho\right)  \exp\left[  ik_{z}z-i\omega t+im\varphi
\right]  =\label{r12}\\
&  \qquad\qquad\qquad\left[  H_{m}^{\left(  1\right)  }\left(  \chi
\rho\right)  +H_{m}^{\left(  2\right)  }\left(  \chi\rho\right)  \right]
\exp\left[  ik_{z}z-i\omega t+im\varphi\right] \nonumber
\end{align}
-- this is a standing wave that arise in the superposition of the two Hankel
waves (note how the singularity of the Neumann functions at the origin is eliminated).%

%TCIMACRO{\FRAME{ftbpFU}{3.813in}{5.7977in}{0pt}{\Qcb{(a) Typical
%spatiotemporal field distribution of a FWM; (b) The temporal evolution of the
%FWM in radial direction as the superposition of the pulsed Hankel beams (blue
%solid lines), the amplitude of the corresponding carrier-frequency
%monochromatic Hankel beam is added for comparison (red dotted line); (c) The
%support of the angular spectrum of plane waves of a wave field that is
%propagation-invariant in radial direction (see text).}}{\Qlb{fig415}%
%}{fig4_15.jpg}{\special{ language "Scientific Word";  type "GRAPHIC";
%maintain-aspect-ratio TRUE;  display "ICON";  valid_file "F";  width 3.813in;
%height 5.7977in;  depth 0pt;  original-width 2.3583in;
%original-height 6.3157in;  cropleft "0";  croptop "1";  cropright "1";
%cropbottom "0";  filename 'fig4_15.jpg';file-properties "XNPEU";}} }%
%BeginExpansion
\begin{figure}
[ptb]
\begin{center}
\includegraphics[
height=5.7977in,
width=3.813in
]%
{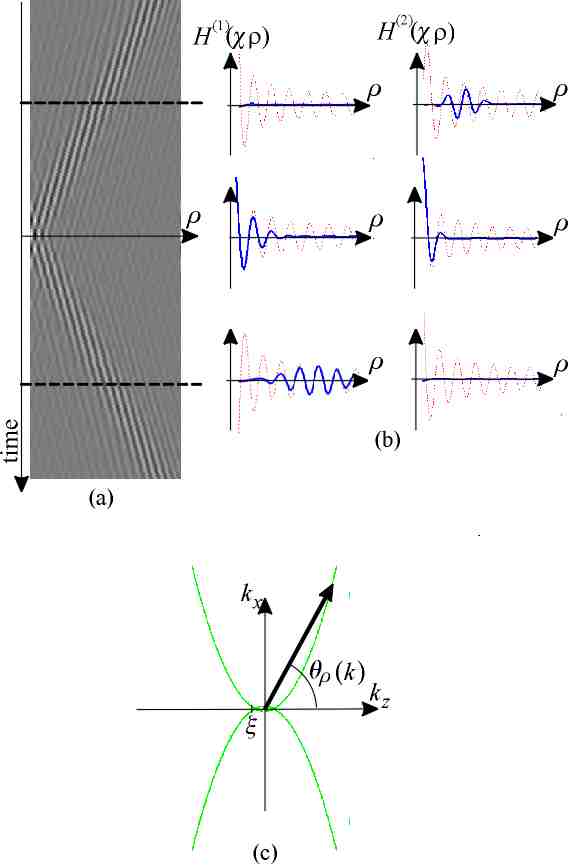}%
\caption{(a) Typical spatiotemporal field distribution of a FWM; (b) The
temporal evolution of the FWM in radial direction as the superposition of the
pulsed Hankel beams (blue solid lines), the amplitude of the corresponding
carrier-frequency monochromatic Hankel beam is added for comparison (red
dotted line); (c) The support of the angular spectrum of plane waves of a wave
field that is propagation-invariant in radial direction (see text).}%
\label{fig415}%
\end{center}
\end{figure}
%EndExpansion
This approach can be easily generalized for the wideband wave fields -- in
this case the superposition of the monochromatic Hankel beams form a
converging or expanding circular pulse in the $xy$ plane. If we also use
condition (\ref{su29}) we get the pulse that corresponds to the radial
evolution of the FWM's. The results of a numerical simulation of its behavior
are depicted in Fig.~\ref{fig415}a and \ref{fig415}b.

Note also, that the radial wave that propagates away from the $z$ axis is
generally not propagation invariant. Indeed, if we follow the arguments of the
section \ref{ssFAng} for \textit{radial }propagation we can write the
condition of propagation-invariance as
\begin{equation}
v^{g}=c\frac{dk}{d\chi}=\left[  \frac{1}{c}\frac{d}{dk}\left(  k\sin\theta
_{F}^{\left(  \rho\right)  }\left(  k\right)  \right)  \right]  ^{-1}=\frac
{c}{\gamma_{\rho}}\text{,} \label{r15}%
\end{equation}
where constant $\gamma_{\rho}$ again determines the group velocity. Specifying
the integration constant $\xi$ again from the condition $\theta_{\rho}\left(
\xi\right)  =180%
%TCIMACRO{\U{b0}}%
%BeginExpansion
{{}^\circ}%
%EndExpansion
$ we can write for the support of angular spectrum of plane waves%
\begin{equation}
k\sin\theta_{F}^{\left(  \rho\right)  }\left(  k\right)  =\gamma_{\rho}%
k-\xi\left(  \gamma_{\rho}+1\right)  \text{.} \label{r17}%
\end{equation}
Thus, we can write
\begin{equation}
\sin\theta_{F}^{\left(  \rho\right)  }\left(  k\right)  =\frac{\gamma_{\rho
}k-\xi\left(  \gamma_{\rho}+1\right)  }{k} \label{r19}%
\end{equation}
(note that in this context $\xi\geq0$).

A typical support of the angular spectrum of plane waves defined by
Eq.~(\ref{r17}) is depicted in Fig.~\ref{fig415}b. So, the FWM is
propagation-invariant in both the $z$ axis direction and radial direction only
in the special case $\xi\equiv0$ where we can write $\gamma_{\rho}%
=\sqrt{1-\gamma^{2}}$. This consequence will be given a further interpretation
in section \ref{ssFTi}.

\subsubsection{\label{ssFLoc}The spatial localization of FWM's}

For most practical cases there is no closed-form integrals to Eq.~(\ref{su40}%
). Consequently, we have to deal with integral transforms and the
straightforward numerical simulation of any realistic situation may be a
tedious task (this is especially true for general LW's where the double
integrals have to be computed). However, for LW's there is a simple method for
qualitative estimate of the resulting wave fields, based on three-dimensional
Fourier transforms (the monochromatic case of the approach was introduced by
McCutchen in Ref.~\cite{mc5} and has been used for example in
Refs.~\cite{mc10,mc15}).%

%TCIMACRO{\FRAME{ftbpFU}{4.6475in}{5.6239in}{0pt}{\Qcb{The Fourier transform
%estimation of the spatial shape of the FWS's (see text).}}{\Qlb{fig41}%
%}{fig4_1.jpg}{\special{ language "Scientific Word";  type "GRAPHIC";
%maintain-aspect-ratio TRUE;  display "ICON";  valid_file "F";
%width 4.6475in;  height 5.6239in;  depth 0pt;  original-width 4.9113in;
%original-height 6.2958in;  cropleft "0";  croptop "1";  cropright "1";
%cropbottom "0";  filename 'fig4_1.jpg';file-properties "XNPEU";}} }%
%BeginExpansion
\begin{figure}
[ptb]
\begin{center}
\includegraphics[
height=5.6239in,
width=4.6475in
]%
{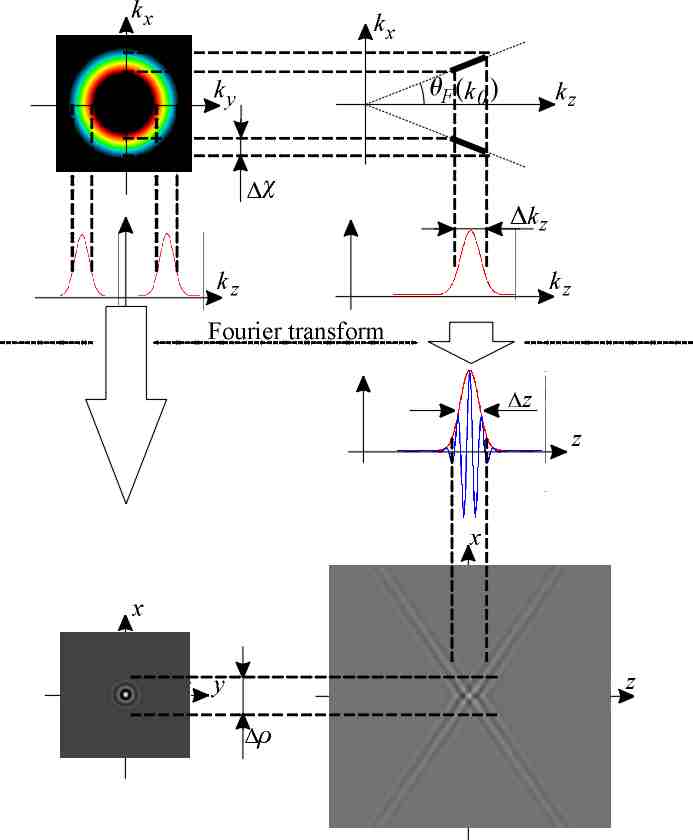}%
\caption{The Fourier transform estimation of the spatial shape of the FWS's
(see text).}%
\label{fig41}%
\end{center}
\end{figure}
%EndExpansion
Let us start with Whittaker type plane wave decomposition in Eq.~(\ref{ang4a})
and set $t=0$:
\begin{align}
&  \Psi^{\prime}\left(  x,y,z,0\right)  =\int\int\int_{-\infty}^{\infty
}\mathrm{d}k_{x}\mathrm{d}k_{y}\mathrm{d}k_{z}\nonumber\\
&  \qquad\qquad\times A\left(  k_{x},k_{y,}k_{z}\right)  \exp\left[  i\left(
k_{x}x+k_{y}y+k_{z}z\right)  \right]  \text{.} \label{lo1}%
\end{align}
Obviously we can write for the field on $z$ axis the relation
\begin{equation}
\Psi^{\prime}\left(  0,0,z,0\right)  =\int_{-\infty}^{\infty}\mathrm{d}%
k_{z}\exp\left[  ik_{z}z\right]  \left\{  \int\int_{-\infty}^{\infty
}\mathrm{d}k_{x}\mathrm{d}k_{y}A\left(  k_{x},k_{y,}k_{z}\right)  \right\}
\text{,} \label{lo2}%
\end{equation}
so that
\begin{equation}
\Psi^{\prime}\left(  0,0,z,0\right)  =\int_{-\infty}^{\infty}\mathrm{d}%
k_{z}\exp\left[  ik_{z}z\right]  A_{xy}\left(  k_{z}\right)  \text{,}
\label{lo3}%
\end{equation}
where
\begin{equation}
A_{xy}\left(  k_{z}\right)  =\int\int_{-\infty}^{\infty}\mathrm{d}%
k_{x}\mathrm{d}k_{y}A\left(  k_{x},k_{y,}k_{z}\right)  \label{lo4}%
\end{equation}
and from the definition of one-dimensional Fourier transform, we can write
\begin{equation}
\Psi^{\prime}\left(  0,0,z,0\right)  =2\pi\mathcal{F}_{z}^{-1}\left[
A_{xy}\left(  k_{z}\right)  \right]  \text{.} \label{lo5}%
\end{equation}
Here $\mathcal{F}_{z}^{-1}\left[  ...\right]  $ denotes the inverse Fourier'
transform in $k_{z}$-direction and the integral (\ref{lo4}) can be thought of
as the projection of the angular spectrum plane waves onto the $z$ axis (see
Fig.~\ref{fig41}). Similarly we can write for the filed in $xy$ plane at $z=0$%
\begin{align}
\Psi^{\prime}\left(  x,y,0,0\right)   &  =\int\int_{-\infty}^{\infty
}\mathrm{d}k_{x}\mathrm{d}k_{y}\exp\left[  ik_{x}x+ik_{y}y\right] \nonumber\\
&  \times\left\{  \int_{-\infty}^{\infty}\mathrm{d}k_{z}A\left(  k_{x}%
,k_{y,}k_{z}\right)  \right\}  \text{,} \label{lo6}%
\end{align}
so that
\begin{equation}
\Psi^{\prime}\left(  x,y,0,0\right)  =\left(  2\pi\right)  ^{2}\mathcal{F}%
_{xy}^{-1}\left[  A_{z}\left(  k_{x},k_{y}\right)  \right]  \text{,}
\label{lo7}%
\end{equation}
where
\begin{equation}
A_{z}\left(  k_{x},k_{y}\right)  =\int_{-\infty}^{\infty}\mathrm{d}%
k_{z}A\left(  k_{x},k_{y,}k_{z}\right)  \label{lo8}%
\end{equation}
and $\mathcal{F}_{xy}^{-1}\left[  ...\right]  $ denotes the two-dimensional
inverse Fourier transform.

Now, having in mind the table of basic one- and two-dimensional Fourier
transforms and the general properties of Fourier transforms, the knowledge of
the defined projections of angular spectrum of plane waves onto the $k_{z}$
axis and $k_{x}k_{y}$ plane allows one immediately estimate the general shape
of the wave field on $z$ axis and $xy$ plane respectively. If we also note
that in studies of the propagation-invariant wave fields the estimates are
valid over the entire $z$ axis (for space-time points $\gamma z-ct$), the
approach can prove to be very useful.

Let us specify the frequency spectrum of the light source $\emph{s}\left(
k\right)  $ as the Gaussian one:%
\begin{equation}
\emph{s}\left(  k\right)  =\exp\left[  -\frac{1}{2}\sigma_{k}^{2}\left(
k-k_{0}\right)  ^{2}\right]  \text{,} \label{lo15}%
\end{equation}
where $k_{0}$ denote the mean wave number of the wave field and $\sigma_{k}$
is determined from the pulse length $\tau_{s}$ of the corresponding plane wave
pulse as
\begin{equation}
\sigma_{k}=\frac{c\tau_{s}}{2\sqrt{2\ln2}}\text{.} \label{lo16}%
\end{equation}
From the known character of the angular spectrum of plane waves of the FWM's
we can approximate for the Gaussian profiles of the $k_{z}$ and $\chi$
projections of the angular spectrum of plane waves
\begin{subequations}
\begin{align}
\sigma_{z}  &  =\frac{\sigma_{k}}{\cos\theta_{F}\left(  k_{0}\right)
}\label{lo17}\\
\sigma_{\rho}  &  =\frac{\sigma_{k}}{\sin\theta_{F}\left(  k_{0}\right)  }
\label{lo19}%
\end{align}
respectively.

The spectral profile of the $k_{z}$-projection of the angular spectrum of
plane waves then reads
\end{subequations}
\begin{equation}
A_{xy}\left(  k_{z}\right)  \propto\exp\left[  -\frac{1}{2}\sigma_{z}%
^{2}\left(  k_{z}-k_{z0}\right)  ^{2}\right]  \label{lo22}%
\end{equation}
with the FWHM (full width at half-maximum)%
\begin{equation}
\Delta k_{z}\approx\Delta k\cos\theta_{F}\left(  k_{0}\right)  =\frac
{2\sqrt{2\ln2}}{\sigma_{z}}\text{,} \label{lo25}%
\end{equation}
where%
\begin{equation}
\Delta k=\frac{2\sqrt{2\ln2}}{\sigma_{k}}\text{.} \label{lo27}%
\end{equation}
The corresponding intensity profile is%
\begin{equation}
\mathcal{F}_{z}^{-1}\left[  A_{xy}\left(  k_{z}\right)  \right]  \propto
\exp\left[  -\frac{z^{2}}{2\sigma_{z}^{2}}\right]  \label{lo29}%
\end{equation}
with FWHM
\begin{equation}
\Delta z\approx\frac{c\tau_{s}}{\cos\theta_{F}\left(  k_{0}\right)  }%
=\sigma_{z}2\sqrt{2\ln2}\text{.} \label{lo31}%
\end{equation}

For the field in transversal direction we can give a good estimate by
recognizing that the intensity profile on $xy$ plane has the Bessel profile
that is multiplied by an envelope. The profile of the latter can be estimated
by the 1D Fourier transform of the projection of the angular spectrum along an
axis and we can write%
\begin{equation}
A_{z}\left(  k_{x},k_{y}\right)  \propto\exp\left[  -\frac{1}{2}\sigma_{\rho
}^{2}\left(  \chi-\chi_{0}\right)  ^{2}\right]  \text{,} \label{lo33}%
\end{equation}
with FWHM%
\begin{equation}
\Delta\chi\approx\Delta k\sin\theta_{F}\left(  k_{0}\right)  =\frac
{2\sqrt{2\ln2}}{\sigma_{\rho}}\text{.} \label{lo39}%
\end{equation}
The corresponding intensity profile reads%
\begin{equation}
\mathcal{F}_{xy}^{-1}\left[  A_{z}\left(  k_{x},k_{y}\right)  \right]  \propto
J_{0}\left(  k\rho\sin\theta_{F}\left(  k_{0}\right)  \right)  \exp\left[
-\frac{\rho^{2}}{2\sigma_{\rho}^{2}}\right]  \label{lo41}%
\end{equation}
with FWHM%
\begin{equation}
\Delta\rho\approx\left\{
\begin{array}
[c]{c}%
\sigma_{\rho}2\sqrt{2\ln2}\text{ -- the envelope\qquad\qquad\qquad\qquad
\qquad}\\
\frac{2\times2.405}{k_{0}\sin\theta_{F}\left(  k_{0}\right)  }\text{ -- the
central peak of the Bessel function}%
\end{array}
\right.  \label{lo44}%
\end{equation}
(see Fig.~\ref{fig41} for an illustration of the description). Note, that as
we can write the ratio
\begin{equation}
\frac{\sigma_{z}}{\sigma_{\rho}}=\frac{\Delta z}{\Delta\rho}=\tan\theta
_{F}\left(  k_{0}\right)  \label{lo21}%
\end{equation}
for the pulse widths in the two directions, we at once can deduce that for
optically feasible FWM's [$\theta_{F}\left(  k_{0}\right)  \ll1$] the central
peak is better localized along the $z$ axis.

\subsection{\label{sFPro}Few remarks on properties of FWM's}

\subsubsection{\label{ssFCa}Causality of FWM's}%

%TCIMACRO{\FRAME{ftbpFU}{2.6039in}{1.5809in}{0pt}{\Qcb{On the causal and
%acausal components of the angular spectrum of plane waves of FWM's. The
%striped region denotes the acausal region of the support of the angular
%spectrum of plane waves.}}{\Qlb{fig45}}{fig4_5.jpg}%
%maintain-aspect-ratio TRUE;  display "ICON";  valid_file "F";
%width 2.6039in;  height 1.5809in;  depth 0pt;  original-width 4.9104in;
%original-height 6.6841in;  cropleft "0";  croptop "1";  cropright "1";
%cropbottom "0";  filename 'fig4_5.jpg';file-properties "XNPEU";}} }%
%BeginExpansion
\begin{figure}
[ptb]
\begin{center}
\includegraphics[
height=1.5809in,
width=2.6039in
]%
{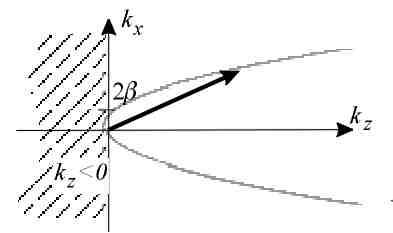}%
\caption{On the causal and acausal components of the angular spectrum of plane
waves of FWM's. The striped region denotes the acausal region of the support
of the angular spectrum of plane waves.}%
\label{fig45}%
\end{center}
\end{figure}
%EndExpansion
In several papers it has been noted, that the original FWM's introduced by
Brittingham and Ziolkowski in Eq.~(\ref{su0}) are not exactly causal as they
include backward propagating plane wave components (see Fig.~\ref{fig45})
\cite{fw13a}. This fact is due to the specific frequency spectrum (\ref{su1})
that leads to the closed-form FWM's (see the overview in following chapter).
In the consequent publications (see Ref.~[\cite{fw18}]) Shaarawi \textit{et.
al.} demonstrated, that the parameters of the spectrum can be chosen so that
the predominant part of the energy of the FWM's is in forward propagating
plane wave components.

In the context of our approach this problem has to be considered as ill-posed
-- as all the wave fields that share the support of the angular spectrum of
plane waves (\ref{su29}) are propagation-invariant regardless of their
frequency spectrum, we can just choose one without the acausal components.

\subsubsection{\label{ssFEv}FWM's and evanescent waves}

The second topic that is closely related to the backward propagating plane
wave components of the original FWM's is the one of evanescent waves
\cite{fw18,fw20}.%

%TCIMACRO{\FRAME{ftbpFU}{3.2647in}{2.6714in}{0pt}{\Qcb{The integration contour
%in Weyl picture of angular spectrum of plane waves. The vertical part of the
%contour where the imaginaty part of the angle $\theta$ is nonzero cancels out
%in integration.}}{\Qlb{fig47}}{fig4_7.jpg}%
%maintain-aspect-ratio TRUE;  display "ICON";  valid_file "F";
%width 3.2647in;  height 2.6714in;  depth 0pt;  original-width 4.9104in;
%original-height 6.6841in;  cropleft "0";  croptop "1";  cropright "1";
%cropbottom "0";  filename 'fig4_7.jpg';file-properties "XNPEU";}} }%
%BeginExpansion
\begin{figure}
[ptb]
\begin{center}
\includegraphics[
height=2.6714in,
width=3.2647in
]%
{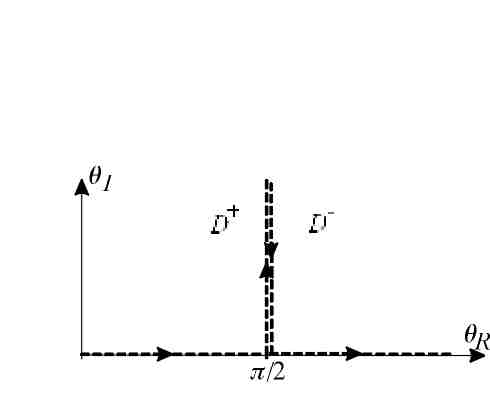}%
\caption{The integration contour in Weyl picture of angular spectrum of plane
waves. The vertical part of the contour where the imaginaty part of the angle
$\theta$ is nonzero cancels out in integration.}%
\label{fig47}%
\end{center}
\end{figure}
%EndExpansion
From the practical point of view it may seem peculiar to introduce the
evanescent waves, the intensity of which decays exponentially, in the context
of the propagation-invariant wave fields where the depth of the propagation
usually extend over several meters. However, the evanescent waves appear
indeed in a Weyl picture of the FWM's. Indeed, from Eqs.~(\ref{ang26}) and
(\ref{su10}) one can write
\begin{equation}
A_{n}^{we}\left(  k,\chi\right)  =B_{n}\left(  k\right)  \delta\left[
k-\sqrt{k^{2}-\chi^{2}}-2\beta\right]  \label{ev5}%
\end{equation}
for the angular spectrum of plane waves of the FWM's so that the field can be
written as%
\begin{align}
&  \Psi\left(  \rho,z,\varphi,t\right)  =\sum_{n=0}^{\infty}\exp\left[  \pm
in\varphi\right]  \int_{0}^{\infty}\mathrm{d}k\int_{0}^{\infty}\mathrm{d}%
\chi\chi~B_{n}\left(  k\right) \label{ev10}\\
&  \qquad\times J_{n}\left(  \chi\rho\right)  \delta\left[  k-\sqrt{k^{2}%
-\chi^{2}}-2\beta\right]  \exp\left[  i\left(  z\sqrt{k^{2}-\chi^{2}%
}-ckt\right)  \right]  \text{.}\nonumber
\end{align}
In Eq.~(\ref{ev10}), for the ranges $\chi<k$ the integration is over
homogeneous plane waves. For $\chi>k$, the wave vector of the plane waves is
purely imaginary and the integration is over the evanescent waves
\cite{fw18,lw6}. The situation may be more apparent if we transform to
variables $\chi,k\rightarrow\theta,k$ and write (for cylindrically symmetric
component only for brevity)
\begin{align}
&  \Psi^{\left(  \pm\right)  }\left(  \rho,z,t\right)  =\int_{0}^{\infty
}\mathrm{d}k~k^{2}\int_{D^{\pm}}\mathrm{d}\theta\cos\theta\sin\theta
\label{ev12}\\
&  \qquad\times B_{0}\left(  k\right)  \delta\left[  k\mp k\cos\theta
-2\beta\right]  J_{0}\left(  k\rho\sin\theta\right)  \exp\left[  \pm
ikz\cos\theta-i\omega t\right] \nonumber
\end{align}
or%
\begin{align}
&  \Psi^{\left(  \pm\right)  }\left(  \rho,z,t\right)  =\int_{D^{\pm}%
}\mathrm{d}\theta\frac{2\beta\sin\theta~B_{0}\left(  \frac{2\beta}%
{\gamma\left(  1-\cos\theta\right)  }\right)  }{\gamma\left(  1-\cos
\theta\right)  }\label{ev13}\\
&  \qquad\times J_{0}\left(  \frac{2\beta\rho\sin\theta}{\gamma\left(
1-\cos\theta\right)  }\right)  \exp\left[  \pm i\frac{2\beta}{\gamma\left(
1-\cos\theta\right)  }\left(  z\cos\theta\mp ct\right)  \right]
\text{.}\nonumber
\end{align}
Here \textquotedblright$+$\textquotedblright\ stands for forward propagating
plane wave components and \textquotedblright$-$\textquotedblright\ stands for
backward propagating plane wave components and the integration is carried out
along the contours $D^{\pm}$ of complex $\theta$ plane, $\chi/k=\sin\left(
\theta_{R}+i\theta_{I}\right)  $ (see Fig.~\ref{fig47}). Also, if the analysis
is carried out for wave fields the angular spectrum of plane waves of which
\textit{has} forward and backward propagating components, the total wave field
can be written as \cite{fw18}
\begin{equation}
\Psi=\left(  \Psi_{h}^{+}+\Psi_{ev}^{+}\right)  +\left(  \Psi_{h}^{-}%
+\Psi_{ev}^{-}\right)  \text{,} \label{ev15}%
\end{equation}
where subscript \textquotedblright$h$\textquotedblright\ denotes homogeneous
component of $\Psi^{+}$ or $\Psi^{-}$, i.e., $0\leq\theta_{R}\leq2\pi$,
$\theta_{I}=0$ and subscript \textquotedblright$ev$\textquotedblright\ denotes
evanescent components, i.e., $\theta_{R}=\pi/2$, $\theta_{I}<0$. It has been
shown \cite{fw18}, that for the evanescent components of a free field one has
\begin{equation}
\Psi_{ev}^{+}=-\Psi_{ev}^{-}\text{,} \label{ev16}%
\end{equation}
so that the Weyl forward and backward propagating components add up resulting
in the source-free solution in Eq.~(\ref{su40}).

Again, in our approach the frequency spectrum is chosen so that the wave
fields do not have any backward propagating components. Consequently, the
integration is only along the real part of the $D^{+}$. Also, it is quite
clear that for the free-space wave fields the presence of the evanescent waves
in the integration (\ref{ev13}) is rather a peculiarity of the Weyl type
angular spectrum of plane waves. For example, if we write the Weyl picture of
a plane wave pulse propagating perpendicularly to $z$ axis, the corresponding
Weyl picture obviously do contain evanescent components. However, there is no
physical content in those components.

\subsubsection{\label{ssFEn}Energy content of scalar FWM's}

As already noted, the total energy content of FWM's is infinite
\cite{fw2,fw5,lw1}. Indeed, as the energy content is calculated as
\begin{equation}
U_{tot}=\int_{-\infty}^{\infty}\mathrm{d}z\int_{0}^{\infty}\mathrm{d}\rho
\rho\int_{0}^{2\pi}\mathrm{d}\varphi\left\vert \Psi_{F}\left(  z,\rho
,\varphi,t\right)  \right\vert ^{2}\text{~.} \label{e5}%
\end{equation}
In the Fourier picture, the Parseval relation and the angular spectrum of
plane waves in Eq.~(\ref{su31}) can be used to yield
\begin{align}
U_{tot}  &  =\sum_{n}\int_{0}^{\infty}\mathrm{d}k\int_{0}^{\pi}\mathrm{d}%
\theta\mathbf{\,}\left\vert \tilde{B}_{n}\left(  k\right)  \delta\left[
\theta-\theta_{F}\left(  k\right)  \right]  \right\vert ^{2}\nonumber\\
&  =\sum_{n}\int_{0}^{\infty}\mathrm{d}k\int_{0}^{\pi}\mathrm{d}%
\theta\mathbf{\,}\left\vert \tilde{B}_{n}\left(  k\right)  \right\vert
^{2}\delta^{2}\left[  \theta-\theta_{F}\left(  k\right)  \right]  \label{e6}%
\end{align}
so that
\begin{equation}
U_{tot}=\infty\label{e9}%
\end{equation}
due to the $\delta^{2}$ in the integrand (here and hereafter the tilde on
angular spectrum indicates that the factor $k^{2}\sin\theta$ is included into
the spectrum).Obviously the relation (\ref{e9}) is valid whenever there is a
delta function in the definition of the angular spectrum of plane waves. Also,
it has been proved that any wave field that is strictly propagation-invariant
has necessarily infinite total energy \cite{fw3,fw3o1}.

The second important energetic parameter of the LW's is their energy flow over
a cross-section per unit time -- obviously, any physically feasible wave field
has to have a finite energy flow. In terms of the previous section and using
the two-dimensional Parseval relation this quantity can be calculated as%
\begin{equation}
\Phi_{xy}=\int\int_{-\infty}^{\infty}\mathrm{d}k_{x}\mathrm{d}k_{y}\left\vert
A_{z}\left(  k_{x},k_{y}\right)  \right\vert ^{2}\text{,} \label{e12}%
\end{equation}
where $A_{z}\left(  k_{x},k_{y}\right)  $ is again the projection\ of the
angular spectrum of plane waves onto $k_{x}k_{y}$ plane. Obviously the
quantity is necessarily infinite, if only the projection of the angular
spectrum can be written in terms of delta function in $k_{x}k_{y}$ plane.
Otherwise the energy flow is finite, provided the function $A_{z}\left(
k_{x},k_{y}\right)  $ is square integrable. The comparison of
Figs.~\ref{fig42} and \ref{fig41} shows that the FWM's generally have finite
total energy flow.

In literature the finite energy LW's have been constructed for example by
means of superpositions of FWM's \cite{fw5,lw1} and by applying finite time
windows \cite{g3,g4,g5,g6,g10}. In section. \ref{ssGFE} we will describe our
approach to this problem as described in Ref.~\cite{m3}.

\subsection{Alternate derivations of scalar FWM's}

\subsubsection{\label{ssFTi}FWM's as cylindrically symmetric superpositions of
tilted pulses}

As to demonstrate the efficiency of the integral transform representations in
describing the properties of FWM's, we give yet another description of FWM's
(Ref.~\cite{m4}).

\begin{figure}
[ptb]
\begin{center}
\includegraphics[
height=6.5743in,
width=2.1837in
]%
{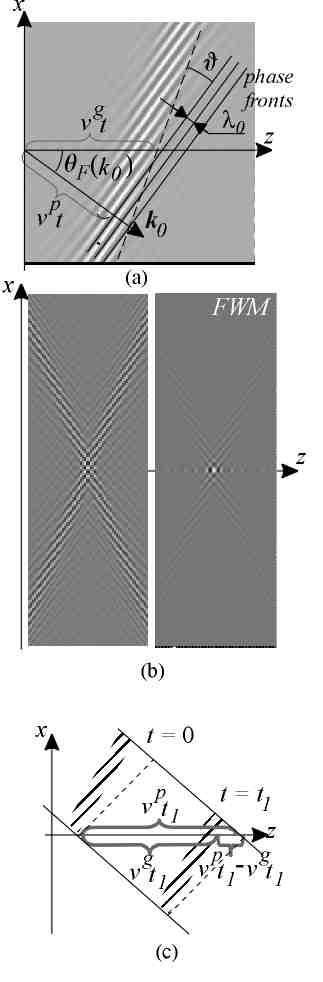}%
\caption{(a) On the field distribution of tilted pulses; (b) comparison of
field of the superposition of a pair of tilted pulses (in left) and of the
corresponding FWM (in right); (c) on the difference of phase and group
velocities of the FWM's (see text).}%
\label{fig411}%
\end{center}
\end{figure}

Let us represent FWM's as the cylindrically symmetric superpositions of the
interfering pairs of certain tilted pulses (see also Ref.~\cite{m1}). In this
representation the field of the FWM's can be expressed as [see
Eqs.~(\ref{ang7}) and (\ref{su29})]%
\begin{align}
\Psi_{F}\left(  \rho,z,t\right)   &  =\int_{0}^{\pi}\mathrm{d}\phi\left[
T\left(  x,y,z,t;\phi\right)  +T\left(  x,y,z,t;\phi+\pi\right)  \right]
\label{t1}\\
&  =\int_{0}^{\pi}\mathrm{d}\phi F^{\prime}\left(  x,y,z,t;\phi\right)
\text{,}\nonumber
\end{align}
where $T\left(  x,y,z,t;\phi\right)  $ denotes the field of the tilted plane
wave pulses, that in the spectral representation are given by
\begin{align}
&  T\left(  x,y,z,t;\phi\right)  =\int_{0}^{\infty}\mathrm{d}k~\tilde
{A}\left(  k,\theta_{F}\left(  k\right)  ,\phi\right) \label{t2}\\
&  \quad\times\exp\left[  ik\left(  x\cos\phi\sin\theta_{F}\left(  k\right)
+y\sin\phi\sin\theta_{F}\left(  k\right)  +z\cos\theta_{F}\left(  k\right)
-ct\right)  \right]  \text{,}\nonumber
\end{align}
where $\tilde{A}\left(  k,\theta_{F}\left(  k\right)  ,\phi\right)  $ is the
angular spectrum of plane waves of the wave field and the angular function
$\theta_{F}\left(  k\right)  $ is defined by Eq.~(\ref{su12}). From
Eqs.~(\ref{t1}) and (\ref{t2}) we get
\begin{align}
&  F^{\prime}\left(  x,y,z,t;\phi\right)  =2\int_{0}^{\infty}\mathrm{d}%
k~\tilde{A}\left(  k,\theta_{F}\left(  k\right)  ,\phi\right) \label{t3}\\
&  \qquad\times\cos\left[  k\sin\theta_{F}\left(  k\right)  \left(  x\cos
\phi+y\sin\phi\right)  \right]  \exp\left[  ik\left(  z\cos\theta_{F}\left(
k\right)  -ct\right)  \right]  \text{.}\nonumber
\end{align}
An example of a tilted pulse with Gaussian frequency spectrum corresponding to
approximately $\tau_{s}\sim4fs$ in Eq.~(\ref{t2}) is depicted in
Fig.~\ref{fig411}a, the corresponding superposition of two tilted pulses in
Eq.~(\ref{t3}) and FWM in Eq.~(\ref{su40}) are depicted in Fig.~\ref{fig411}b).

In this representation the properties of FWM's can be given the following
interpretation:%
%TCIMACRO{\FRAME{ftbpFU}{2.1837in}{6.5743in}{0pt}{\Qcb{(a) On the field
%distribution of tilted pulses; (b) comparison of field of the superposition of
%a pair of tilted pulses (in left) and of the corresponding FWM (in right); (c)
%on the difference of phase and group velocities of the FWM's (see text).}%
%}{\Qlb{fig411}}{fig4_11.jpg}{\special{ language "Scientific Word";
%type "GRAPHIC";  maintain-aspect-ratio TRUE;  display "ICON";
%valid_file "F";  width 2.1837in;  height 6.5743in;  depth 0pt;
%original-width 4.9104in;  original-height 6.6841in;  cropleft "0";
%filename 'fig4_11.jpg';file-properties "XNPEU";}} }%
%BeginExpansion
%EndExpansion

\begin{itemize}
\item[1.] The localized central peak of FWM's is simply the well-known
consequence of taking the axially symmetric superposition of a harmonic
function. Indeed, the interference of the two transform-limited tilted pulses
in Eq.~(\ref{t3}) gives rise to the harmonic interference pattern, the
transversal width of which is proportional to the temporal length of the
tilted pulses (\ref{t2}). The central peak arises due to the constructive
interference of the tilted pulses along the optical axis, formally, the
$\cos\left(  {}\right)  $ function in Eq.~(\ref{t3}) is replaced by
$J_{0}\left(  {}\right)  $ in Eq.~(\ref{su40}) [see Fig.~\ref{fig411}b];

\item[2.] The nondispersing propagation of the optical FWM's wave fields can
be given an alternate wave-optical interpretation. Namely, it can be seen from
Fig.~\ref{fig413}, that in large scale the longitudinal length of the tilted
pulses depends on the distance from the optical axis so that the tilted pulses
have a \textquotedblright waist\textquotedblright\ (this claim is identical to
that given in section \ref{ssFRad} that the radial wave propagating toward the
$z$ axis and back is not propagation-invariant). The relation (\ref{su12})
essentially guarantees, that the waist propagates along the optical axis and
do not spread -- in this case the central peak of the corresponding
cylindrically symmetric superpositions, FWM's (\ref{su40}), also remains
transform-limited;%
%TCIMACRO{\FRAME{ftbpFU}{2.4915in}{1.8645in}{0pt}{\Qcb{The large-scale
%behaviour of the spatial shape of the modulus of the tilted pulses.}%
%}{\Qlb{fig413}}{fig4_13.jpg}{\special{ language "Scientific Word";
%type "GRAPHIC";  maintain-aspect-ratio TRUE;  display "ICON";
%valid_file "F";  width 2.4915in;  height 1.8645in;  depth 0pt;
%original-width 2.7138in;  original-height 1.964in;  cropleft "0";
%filename 'fig4_13.jpg';file-properties "XNPEU";}} }%
%BeginExpansion
\begin{figure}
[ptb]
\begin{center}
\includegraphics[
height=1.8645in,
width=2.4915in
]%
{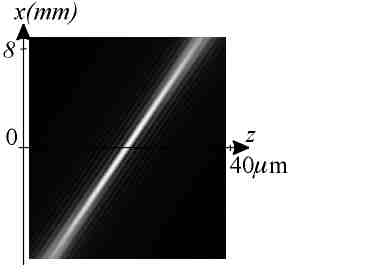}%
\caption{The large-scale behaviour of the spatial shape of the modulus of the
tilted pulses.}%
\label{fig413}%
\end{center}
\end{figure}
%EndExpansion

\item[3.] The local variations of the central peak of the wave field, noted
for example in Ref.~\cite{fw5}, can be explained as the result of the
difference between the phase and group velocities along the optical axis -- as
can be seen from Fig.~\ref{fig411}c the pulse and phase fronts of the tilted
pulses are not parallel;

\item[4.] The group velocity of the wave field can be set by changing the
parameter $\gamma$ in Eq.~(\ref{su12}). The Fig.~\ref{fig411}c gives this
effect a wave optical interpretation -- it can be seen, that the on-axis group
velocity of the wave field directly depends on the angle between the phase
front and pulse front and on the direction of the wave vector of the mean frequency.
\end{itemize}

It is easy to see, that all the presented arguments are equally valid for the
superpositions of tilted pulses in Eq.~(\ref{t3}) and for its cylindrically
symmetric counterparts -- FWM's. Thus, we can state that the defined
interfering pair of tilted pulses possess all the characteristic properties of
FWM's. In fact, the physics behind the two wave fields is similar to the
degree, that we will call the wave field (\ref{t3})
\begin{equation}
F\left(  x,z,t\right)  =\int_{0}^{\infty}\mathrm{d}k~\tilde{B}_{0}\left(
k\right)  \cos\left[  kx\sin\theta_{F}\left(  k\right)  \right]  \exp\left[
ik\left(  z\cos\theta_{F}\left(  k\right)  -ct\right)  \right]  \label{t5}%
\end{equation}
as two-dimensional FWM (2D FWM) in what follows.%

%TCIMACRO{\FRAME{ftbFU}{7.5981cm}{16.0156cm}{0pt}{\Qcb{(a) On the phase and
%group velocity of a plane wave pulse propagating at angle $\theta_{0}$
%relative to $z$-axis; (b) comparison of the field of the superposition of a
%pair of plane wave pulses (in left) and of their corresponding cylindrically
%symmetric superposition -- Bessel-X pulse (in right); (c) on the group
%velocity of Bessel-X pulses.}}{\Qlb{fig412}}{fig4_12.jpg}%
%maintain-aspect-ratio TRUE;  display "ICON";  valid_file "F";
%width 7.5981cm;  height 16.0156cm;  depth 0pt;  original-width 4.9104in;
%original-height 6.6841in;  cropleft "0";  croptop "1";  cropright "1";
%cropbottom "0";  filename 'fig4_12.jpg';file-properties "XNPEU";}} }%
%BeginExpansion
\begin{figure}
[tb]
\begin{center}
\includegraphics[
height=16.0156cm,
width=7.5981cm
]%
{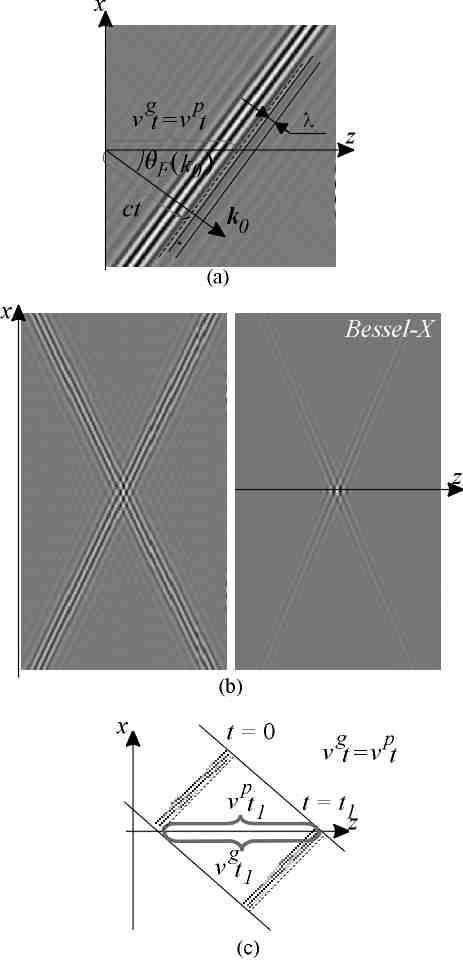}%
\caption{(a) On the phase and group velocity of a plane wave pulse propagating
at angle $\theta_{0}$ relative to $z$-axis; (b) comparison of the field of the
superposition of a pair of plane wave pulses (in left) and of their
corresponding cylindrically symmetric superposition -- Bessel-X pulse (in
right); (c) on the group velocity of Bessel-X pulses.}%
\label{fig412}%
\end{center}
\end{figure}
%EndExpansion
We end this section by noting that the special case of this approach can be
used to discuss the properties of X-type pulses (Ref.~\cite{m1}). In this case
$\theta_{F}\left(  k\right)  =const=\theta_{0}$ and we have the interference
of two plane wave pulses:%
\begin{align}
T\left(  x,y,z,t;\phi\right)   &  =\int_{0}^{\infty}\mathrm{d}k~\tilde
{A}\left(  k,\theta_{F}\left(  k\right)  ,\phi\right) \label{t7}\\
&  \times\exp\left[  ik\left(  x\cos\phi\sin\theta_{0}y\sin\phi\sin\theta
_{0}+z\cos\theta_{0}-ct\right)  \right]  \text{,}\nonumber
\end{align}
so that%
\begin{equation}
F\left(  x,z,t\right)  =\int_{0}^{\infty}\mathrm{d}k~\tilde{B}_{0}\left(
k\right)  \cos\left[  kx\sin\theta_{0}\right]  \exp\left[  ik\left(
z\cos\theta_{0}-ct\right)  \right]  \text{.} \label{t9}%
\end{equation}
(see Fig.~\ref{fig412})

\subsubsection{\label{ssFGau}FWM's as the moving, modulated Gaussian beams}

In literature the closed-form expression (\ref{su0}) for the original FWM's
have been derived with the use of the anzatz \cite{fw3,fw4,fw5}%
\begin{equation}
\Psi\left(  x,y,z,t\right)  =\exp\left[  i\beta\mu\right]  F^{\prime}\left(
x,y,\zeta\right)  \text{~,} \label{gau1}%
\end{equation}
where $\mu=z+ct$ and $\zeta=z-ct$. With (\ref{gau1}) the wave equation
(\ref{ang1}) reduces to the Schr\"{o}dinger equation for $F^{\prime}$%
\begin{equation}
\left(  \Delta_{\bot}+4i\beta\partial_{\zeta}\right)  F^{\prime}\left(
x,y,\zeta\right)  =0 \label{gau2}%
\end{equation}
which, assuming axial symmetry, has a solution of the form \cite{fw5}
\begin{equation}
F^{\prime}\left(  \rho,\zeta\right)  =\frac{1}{4\pi i\left(  a_{1}%
+i\zeta\right)  }\exp\left[  -\frac{\beta\rho^{2}}{a_{1}+i\zeta}\right]
\text{,} \label{gau3}%
\end{equation}
so that one can write the solution similar to the FWM's in Eq.~(\ref{su0})%
\begin{equation}
\Psi_{f}\left(  \rho,\mu,\zeta\right)  =\exp\left[  i\beta\mu\right]
\frac{a_{1}}{4\pi i\left(  a_{1}+i\zeta\right)  }\exp\left[  -\frac{\beta
\rho^{2}}{a_{1}+i\zeta}\right]  \text{~.} \label{gau5}%
\end{equation}

To give the FWM a more convenient form one can use the transform
\begin{equation}
\frac{1}{a_{1}+i\zeta}=\frac{1}{\beta a_{1}^{2}\left(  \zeta\right)  }%
-i\frac{1}{R\left(  \zeta\right)  } \label{gau6}%
\end{equation}
with which the Eq.~(\ref{gau5}) can be shown to yield%
\begin{align}
&  \Psi_{f}\left(  \rho,z,\zeta\right)  =\frac{W_{0}}{4\pi a_{1}\left(
\zeta\right)  }\exp\left[  -i\beta\zeta\right] \label{gau15}\\
&  \qquad\times\exp\left[  -\frac{\rho^{2}}{a_{1}^{2}\left(  \zeta\right)
}+i\frac{\beta\rho^{2}}{R\left(  \zeta\right)  }-i\left(  \arctan\left(
\frac{\zeta}{a_{1}}\right)  -2\beta z\right)  \right]  \text{,}\nonumber
\end{align}
where
\begin{subequations}
\begin{align}
a_{1}\left(  \zeta\right)   &  =W_{0}\left[  1+\left(  \frac{\zeta}{a_{1}%
}\right)  ^{2}\right]  ^{\frac{1}{2}}\label{gau7}\\
R\left(  \zeta\right)   &  =\zeta\left[  1+\left(  \frac{a_{1}}{\zeta}\right)
^{2}\right]  \label{gau8}%
\end{align}
and
\end{subequations}
\begin{equation}
W_{0}=\sqrt{\frac{a_{1}}{\beta}}~\text{.} \label{gau9}%
\end{equation}
If one compares the Eqs.~(\ref{gau15}) - (\ref{gau9}) to those of the
monochromatic Gaussian beam (see Ref.~\cite{o25} for example) one can see
that, the FWM's can be interpreted as moving, modulated Gaussian beams for
which $a_{1}\left(  \zeta\right)  $ and $R\left(  \zeta\right)  $ are the beam
width and radius of curvature respectively and $W_{0}$ is the beam waist at
$\zeta=0$ (see Refs. \cite{fw3,fw4,fw5} for relevant descriptions).%

%TCIMACRO{\FRAME{ftbFU}{2.7908in}{2.2278in}{0pt}{\Qcb{A numerical example of
%the field of the original FWM's.}}{\Qlb{fig410}}{fig4_10.jpg}%
%maintain-aspect-ratio TRUE;  display "ICON";  valid_file "F";
%width 2.7908in;  height 2.2278in;  depth 0pt;  original-width 3.5388in;
%original-height 2.5529in;  cropleft "0";  croptop "1";  cropright "1";
%cropbottom "0";  filename 'fig4_10.jpg';file-properties "XNPEU";}} }%
%BeginExpansion
\begin{figure}
[tb]
\begin{center}
\includegraphics[
height=2.2278in,
width=2.7908in
]%
{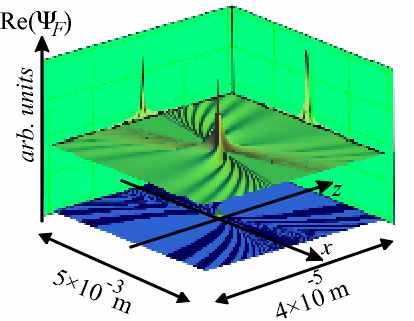}%
\caption{A numerical example of the field of the original FWM's.}%
\label{fig410}%
\end{center}
\end{figure}
%EndExpansion
Now, several interesting consequences can be drawn at this point. Most
importantly, this formal analogy between the FWM's and Gaussian beams is very
conditional and even misleading in some respects. First of all, the constant
$\beta$ is by no means the carrier wave number of the FWM's as one might
expect from the corresponding monochromatic expression -- in the following
Chapter \ref{chOV} we will see that the convenient choice of parameter for
optically feasible FWM's with the carrier wave number $k_{0}\approx
1\times10^{7}\frac{rad}{m}$ the parameter is of the order of magnitude
$\beta\lesssim100\frac{rad}{m}$. Secondly, the requirement of optical
feasibility also implies that $a_{1}\ll1$ (see Sec. \ref{sFWM}) and with this
condition the original FWM's (see Fig.~\ref{fig410}) typically do not resemble
that of the Gaussian beam as they appear in the textbook examples. The reason
for the "abnormal" behaviour is obvious~-- with the above conditions the
direct analogy to the monochromatic case, where $\beta=2\pi/\lambda$, yields
for the beam waist in Eq.$~$(\ref{gau9}) $W_{0}\ll\lambda$. So that we have a
limiting case where the waist of the Gaussian beam is much less than its
wavelength -- clearly here the different physical nature of the FWM's show up.

Next we would like to discuss the claim, often encountered in literature, that
the original FWM's are carrier free wave fields. First of all, in lights of
the general physical considerations in section \ref{ssFLoc} it should be
evident that the non-oscillating shape of the central peak in
Fig.~\ref{fig410} is a direct consequence of the ultra-wide frequency spectrum
of the wave field -- if the pulse length of the corresponding source plane
wave pulse is less than the central wavelength, the resulting FWM is
effectively an half-cycle pulse and in this condition the concept of carrier
wavelength is rather meaningless of course. However, in above sections it was
shown that the general FWM's are not confined to the one particular frequency
spectrum. Correspondingly, we can choose a feasible frequency spectrum and the
carrier free behaviour of the original FWM's should certainly not be mentioned
as the defining property of the original FWM's, this is just a mathematical
peculiarity of a particular integral transform table entry.%

%TCIMACRO{\FRAME{ftbpFU}{3.7974in}{5.6074in}{0pt}{\Qcb{On the character of the
%Gouy phase shift term in the closed form expression of the FWM's: The real
%part of the original FWM (dotted blue line) is depicted for three
%$z$-coordinate values together with the modulus (solid green line) and real
%part (solid black line) of an FWM with narrower bandwidth (see text).}%
%}{\Qlb{fig53}}{fig5_3.jpg}{\special{ language "Scientific Word";
%type "GRAPHIC";  maintain-aspect-ratio TRUE;  display "ICON";
%valid_file "F";  width 3.7974in;  height 5.6074in;  depth 0pt;
%original-width 4.9104in;  original-height 6.6841in;  cropleft "0";
%filename 'fig5_3.jpg';file-properties "XNPEU";}} }%
%BeginExpansion
\begin{figure}
[ptb]
\begin{center}
\includegraphics[
height=5.6074in,
width=3.7974in
]%
{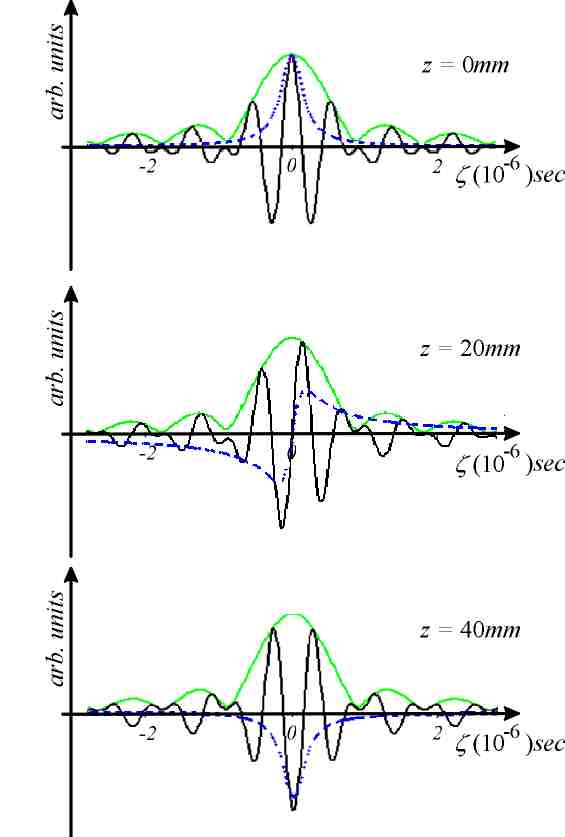}%
\caption{On the character of the Gouy phase shift term in the closed form
expression of the FWM's: The real part of the original FWM (dotted blue line)
is depicted for three $z$-coordinate values together with the modulus (solid
green line) and real part (solid black line) of an FWM with narrower bandwidth
(see text).}%
\label{fig53}%
\end{center}
\end{figure}
%EndExpansion
The issue can be given an alternate description if we note that, using the
analogy to the monochromatic Gaussian beams the term%
\begin{equation}
G\left(  z,\zeta\right)  =i\left(  \arctan\left(  \frac{\zeta}{a_{1}}\right)
+2\beta z\right)  \label{s3}%
\end{equation}
in the expression Eq.~(\ref{gau15}) could be interpreted as the Gouy phase
shift [\cite{o25}] of the FWM's. In previous section \ref{ssFTi} we described
the FWM's as the cylindrically symmetric superpositions of certain tilted
pulses. Now, the original FWM's differ from those, depicted in
Fig.~\ref{fig411}b only by the ultra-wide bandwidths. In Fig.~(\ref{fig53}) we
have depicted the on-axis spatial evolution of an FWM as described by
Eq.~(\ref{gau15}) and of one of reasonable bandwidth, calculated from the
Eq.~(\ref{su40}). The comparison of the two waveforms shows that term
$\exp\left[  i\arctan\left(  \frac{\zeta}{a_{1}}\right)  \right]  $ of the
phase term $G$ can be interpreted as the remnants of the sinusoidal waveform,
lost due to the ultra-wide bandwidth and the term $\exp\left[  i2\beta
z\right]  $ is added as the monothonically growing phase factor that is due to
the difference between the group and phase velocities of FWM's. The latter
term is characteristic to the FWM's only -- instead of having a single focus
with accompanying Gouy phase shift or a "frozen" Gouy phase shift as the
X-type pulses, FWM's have \textit{periodically} evolving phase shift term.

The idea of Gouy phase shift, initially introduced in the Fresnel
approximation of the diffraction theory of monochromatic focused beams, has
attracted a renewed interest recently in the context of propagation of
subcycle Gaussian pulses (see Refs.~\cite{ve35,go15,go20,go25,go30,go35,go40}%
). We believe, that the simple physical interpretation of the term (\ref{s3})
in the context of FWM's, as being the result of the difference between phase
and group velocities of the wave field, might add to the general understanding
of the phenomenon.

\subsubsection{\label{ssFLo}FWM's as the Lorentz transforms of focused
monochromatic beams}

An interesting interpretation to the FWM's can be given in terms of special
theory of relativity. Namely, in Ref.~\cite{fw4} B\'{e}langer demonstrated
that Gaussian monochromatic beams appear as FWM's (Gaussian packetlike beams)
when observed in an inertial system moving at relativistic speeds relative to
the focused wave. In this short note we would like to give an another
mathematical representation to this claim.

\begin{figure}
[ptb]
\begin{center}
\includegraphics[
height=1.9873in,
width=4.9381in
]%
{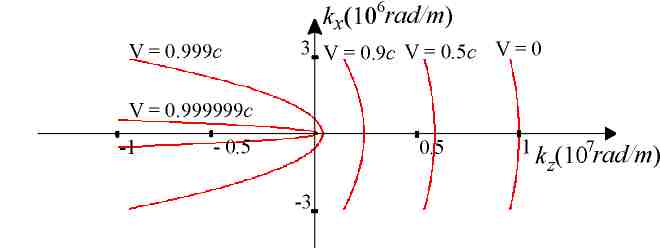}%
\caption{The support of angular spectrum of plane waves of a focused
monochromatic beam as seen from inertial reference systems moving at different
velocities relative to the rest system of the monochromatic beam
($k_{0}=1\times10^{7}\frac{rad}{m}$). Due to the relativistiv Doppler shift
the direction of propagation and the frequency of the monochromatic components
of the focused beam transform so that the beam is seen as the FWM in the
moving reference system.}%
\label{fig518}%
\end{center}
\end{figure}

Suppose we take a focused monochromatic wave of the form
\begin{equation}
\Psi\left(  \rho,z,t\right)  =\int_{0}^{\pi}d\theta K\left(  \theta\right)
\exp\left[  ik_{0}\left(  z\cos\theta-ct\right)  \right]  \label{lor1}%
\end{equation}
with angular spectrum of plane waves%
\begin{equation}
A_{0}\left(  k,\theta\right)  =K\left(  \theta\right)  \delta\left(
k-k_{0}\right)  \text{.} \label{lor2}%
\end{equation}
If we observe the beam from a moving inertial system, the plane wave
components of the field suffer from the relativistic Doppler shift. As the
result, their wave vectors and frequencies transform as described by Lorentz
transformations. Specifically, the wave number of the wave vector and its
longitudinal and transversal components in the inertial frame, moving at speed
$V$ along the $z$ axis, obey equations
\begin{subequations}
\begin{align}
k^{\prime}  &  =\gamma_{l}k_{0}\left(  1-\beta_{l}\cos\theta\right)
\label{lor3a}\\
k_{z}^{\prime}  &  =\gamma_{l}k_{0}\left(  \cos\theta-\beta_{l}\right)
\label{lor3b}\\
k_{x}^{\prime}  &  =k_{0}\sin\theta=k_{x}\label{lor3c}\\
k_{y}^{\prime}  &  =k_{0}\sin\theta=k_{y} \label{lor3d}%
\end{align}
(Eq.~(11.29) of Ref.~\cite{o15}) where $k_{0}$ is the wave number of the wave
field in rest frame and
\end{subequations}
\begin{subequations}
\begin{align}
\beta_{l}\left(  V\right)   &  =\frac{V}{c}\label{lor6}\\
\gamma_{l}\left(  V\right)   &  =\frac{1}{\sqrt{1-\beta^{2}}} \label{lor6a}%
\end{align}
We can use Eq. (\ref{lor3a}) to eliminate $\theta$ from Eq. (\ref{lor3b}) to
get
\end{subequations}
\begin{equation}
k_{z}^{\prime}=-\frac{1}{\beta_{l}\left(  V\right)  }k^{\prime}-\gamma
_{l}\left(  V\right)  k_{0}\left(  \beta_{l}\left(  V\right)  -\frac{1}%
{\beta_{l}\left(  V\right)  }\right)  \label{lor8}%
\end{equation}
and if we define the parameters as
\begin{subequations}
\begin{align}
\gamma\left(  V\right)   &  =\frac{1}{\beta_{l}\left(  V\right)  }%
\label{lor4}\\
\beta\left(  V,k_{0}\right)   &  =\frac{\gamma_{l}\left(  V\right)  k_{0}}%
{2}\text{~,} \label{lor5}%
\end{align}
we can write for the $z$ component of the wave vector
\end{subequations}
\begin{equation}
k_{z}^{\prime}=-\gamma\left(  V\right)  k^{\prime}-2\beta\left(
V,k_{0}\right)  \left(  \frac{1}{\gamma\left(  V\right)  }-\gamma\left(
V\right)  \right)  \text{.} \label{lor7}%
\end{equation}
Thus, if the velocity of the moving frame is close to the speed of light, the
angular spectrum of plane waves of the wave field in the moving frame is the
one of the FWM that moves in negative direction of $z$ axis -- for the FWM's
we have in Eq. (\ref{su10})
\begin{equation}
k_{z}=\gamma k-2\beta\gamma\label{lor9}%
\end{equation}
and using both $k_{0}$ and $V$ as parameters we can model every possible
support of angular spectrum of plane waves of FWM's. In Fig.~\ref{fig518} the
support of angular spectrum of plane waves of the beam as seen from the moving
reference system is depicted for various values of the speed $V$ and fixed
value for $k_{0}$.%
%TCIMACRO{\FRAME{ftbpFU}{4.9381in}{1.9873in}{0pt}{\Qcb{The support of angular
%spectrum of plane waves of a focused monochromatic beam as seen from inertial
%reference systems moving at different velocities relative to the rest system
%of the monochromatic beam ($k_{0}=1\times10^{7}\frac{rad}{m}$). Due to the
%relativistiv Doppler shift the direction of propagation and the frequency of
%the monochromatic components of the focused beam transform so that the beam is
%seen as the FWM in the moving reference system.}}{\Qlb{fig518}}{fig5_18.jpg}%
%maintain-aspect-ratio TRUE;  display "ICON";  valid_file "F";
%width 4.9381in;  height 1.9873in;  depth 0pt;  original-width 4.9104in;
%original-height 1.9588in;  cropleft "0";  croptop "1";  cropright "1";
%cropbottom "0";  filename 'fig5_18.jpg';file-properties "XNPEU";}} }%
%BeginExpansion
%EndExpansion

Note that an alternate approach to describe the LW's in terms of generalized
Lorentz transforms can be found in Ref.~\cite{lw6} -- in this work it was
shown that the superluminal and subluminal Lorentz transformations can be used
to derive LW solutions to the scalar wave equation by boosting known solutions
of the wave equation.

\subsubsection{\label{ssFDo}FWM's as a construction of generalized functions
in the Fourier domain}

In Refs.~\cite{fw15,fw16} Donnelly and Ziolkowski realized, that various
separable and non-separable solutions to the wave equation can be constructed
in spatial and temporal Fourier domain by choosing the Fourier transform of
the solution of the differential equation so that, when multiplied by the
transform of the particular differential operator, it gives zero in the sense
of generalized functions. In the special case of scalar homogeneous wave
equation the corresponding relation reads%
\begin{equation}
\left(  \chi^{2}+k_{z}^{2}-\frac{\omega^{2}}{c^{2}}\right)  \psi\left(
\mathbf{k},\omega\right)  =0\text{,} \label{dn1}%
\end{equation}
where $\psi\left(  \mathbf{k},\omega\right)  $ (\ref{ang5}) is (3+1)
dimensional Fourier transform of the solution of the wave equation
(\ref{ang1}). It can be shown that the function of the general type%
\begin{equation}
\psi\left(  \mathbf{k},\omega\right)  =\Xi\left(  \chi,\beta\right)
\delta\left[  k_{z}-\left(  \beta-\frac{\chi^{2}}{4\beta}\right)  \right]
\delta\left[  \omega-c\left(  \beta+\frac{\chi^{2}}{4\beta}\right)  \right]
\label{dn5}%
\end{equation}
satisfies (\ref{dn1}) and yields all the known FWM's (in the sense defined in
this review). For example the choice \cite{fw15}
\begin{equation}
\Xi\left(  \chi,\beta\right)  =\frac{\pi^{2}}{i\beta}\exp\left[  -\frac
{\chi^{2}a_{1}}{4\beta}\right]  \label{dn6}%
\end{equation}
leads to the original FWM's.

One can notice, that if we eliminate the term $\chi^{2}/4\beta$ from the delta
functions we get the condition (\ref{su10}) and thus the Eq.~(\ref{dn5}) is
yet another transcription of the support of the angular spectrum of plane
waves, derived in section \ref{ssFAng}.

\section{\label{chOV}AN OUTLINE\ OF SCALAR LOCALIZED\ WAVES STUDIED IN
LITERATURE SO\ FAR}

\subsection{Introduction}

Over the years a considerable effort has been made to find closed-form
localized solutions to the homogeneous scalar wave equations. The main aim of
this work is to study the feasibility of LW's in optical domain. Without
debasing the value of those solutions it appears, that this approach often
leads to the source schemes that are difficult to realize even in radio
frequency domain.

Though there has been several publications that provide an unified approach
for the description of LW's \cite{fw15,lw1,lw6}, to our best knowledge, the
optical feasibility of those wave fields has not been estimated in literature.
Moreover, the analysis of the numerical examples that have been published in
literature show, that authors have often choose the parameters of the LW's so
that the frequency spectrum is in the radio frequency domain.

In our opinion in optical domain the best representation for the analysis is
the Whittaker type plane wave decomposition. First of all, the mental picture
of the Fourier lens that produces the two-dimensional Fourier transform of
monochromatic wave field between its focal planes is often very useful in
modeling the optical setups -- we precisely know, how and in what
approximations the elementary components of the Fourier picture, the plane
waves and Bessel beams, can be generated. Secondly, the approach of the
section \ref{ssFLoc} allows us easily estimate the spatial shape of the wave
fields under the discussion.

In the following overview we define the term "optically feasible" by two
rather obvious restrictions:

\begin{itemize}
\item[1.] The frequency spectrum of an optically feasible wave field should be
in optical domain;

\item[2.] The plane wave spectrum of an optically feasible wave field should
not contain plane waves propagating at non-paraxial angles relative to optical axis.
\end{itemize}

The latter requirement can be justified by a very simple geometrical estimate,
described in Fig.~\ref{fig612} -- if the FWM's has to propagate over distances
that exceed the diameter of the source more than, say, five times, the maximum
angle of the plane wave components in the wave field has to be less than $5$ degrees.

Note, that the energy content of most of the wave fields discussed in this
outline is infinite, thus, they are not physically realizable as such.
However, as we will see in chapters that follow, in optical implementations
the finite energy approximations of the LW's follow naturally from the finite
aperture of the setups and \textit{this approximation do not change the
general properties of the LW's}, so that the two conditions for optical
feasibility, posed here, are also valid for LW's with finite energy
content\textit{.}

In our numerical examples we try to optimize the parameters of each LW so that
(i) the frequency spectrum of the wave fields extends from $\sim
0.5\times10^{7}rad/m$ ($\sim1100nm$ of wavelength) to $\sim2\times10^{7}rad/m$
($\sim300nm$ of wavelength), so that $\sigma_{k}\sim3.8\times10^{-7}m$
(\ref{lo16}) for which the length of the corresponding plane wave pulse is
$\sim3fs$ -- the shortest possible pulse length available, (ii) the plane wave
with central wavelength propagate approximately at the angle $\approx0.2%
%TCIMACRO{\U{b0}}%
%BeginExpansion
{{}^\circ}%
%EndExpansion
$ relative to the propagation axis, giving $\sigma_{z}/\sigma_{\rho}%
\approx0.003$ for the approximate ratio of pulse widths in $xy$ and $z$
direction. Note, that specifying the frequency range and cone angle of the
Bessel beam of central wavelength completely determines the parameter $\beta$
-- for $\gamma=1$ we have $\beta\approx40rad/m$ (clarify section \ref{ssFAng}).

\subsection{\label{sFWM}The original FWM's}%

%TCIMACRO{\FRAME{ftbpFU}{3.9972in}{5.9949in}{0pt}{\Qcb{A numerical example of a
%FWM with the parameters $\gamma=1$, $\beta=40\frac{rad}{m}$, $a_{1}%
%=1.4\times10^{-7}m$: (a) The angular spectrum of plane waves in two
%perspectives; (b) The frequency spectrum of the FWM (black line), the
%frequency spectrum of an optically feasible wave field (green line), the angle
%$\theta_{F}\left(  k\right)  $ as the function of the wave number (dashed blue
%line); (c) The field distribution of the FWM.}}{\Qlb{fig52}}{fig5_2.jpg}%
%maintain-aspect-ratio TRUE;  display "ICON";  valid_file "F";
%width 3.9972in;  height 5.9949in;  depth 0pt;  original-width 4.9104in;
%original-height 6.6841in;  cropleft "0";  croptop "1";  cropright "1";
%cropbottom "0";  filename 'fig5_2.jpg';file-properties "XNPEU";}} }%
%BeginExpansion
\begin{figure}
[ptb]
\begin{center}
\includegraphics[
height=5.9949in,
width=3.9972in
]%
{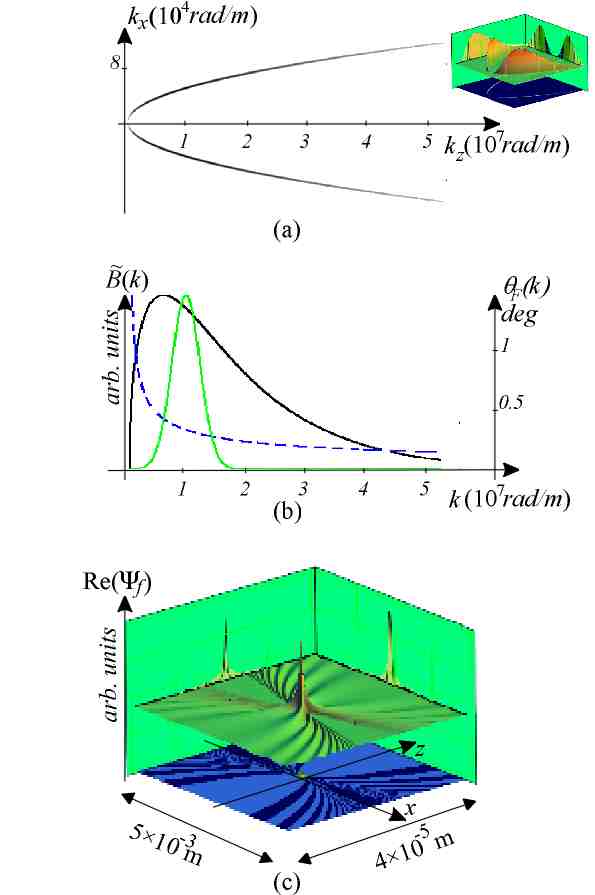}%
\caption{A numerical example of a FWM with the parameters $\gamma=1$,
$\beta=40\frac{rad}{m}$, $a_{1}=1.4\times10^{-7}m$: (a) The angular spectrum
of plane waves in two perspectives; (b) The frequency spectrum of the FWM
(black line), the frequency spectrum of an optically feasible wave field
(green line), the angle $\theta_{F}\left(  k\right)  $ as the function of the
wave number (dashed blue line); (c) The field distribution of the FWM.}%
\label{fig52}%
\end{center}
\end{figure}
%EndExpansion
With Eqs.~(\ref{bidi6}) and (\ref{bidi7}) the angular spectrum of plane waves
of the original scalar FWM's in Eq.~(\ref{su0})%
\begin{equation}
\Psi_{f}\left(  \rho,z,t\right)  =\exp\left[  i\beta\mu\right]  \frac{a_{1}%
}{4\pi i\left(  a_{1}+i\zeta\right)  }\exp\left[  -\frac{\beta\rho^{2}}%
{a_{1}+i\zeta}\right]  \label{fwm11}%
\end{equation}
can be derived from its bidirectional plane wave representation [\cite{fw14}]%
\begin{equation}
C_{0}\left(  \tilde{\alpha},\tilde{\beta},\chi\right)  =\frac{\pi}{2}%
\delta\left(  \tilde{\beta}-2\beta\right)  \exp\left[  -\tilde{\alpha}%
a_{1}\right]  \text{,} \label{fwm1}%
\end{equation}
giving
\begin{equation}
A_{0}\left(  k,\theta\right)  =\frac{\pi}{2}\exp\left[  -\frac{a_{1}k\left(
1+\cos\theta\right)  }{2}\right]  \delta\left(  k-k\cos\theta-2\beta\right)
\label{fwm5}%
\end{equation}
(see Fig.~\ref{fig52}a). Inserting the angular spectrum (\ref{fwm5}) into
integral representation of the type (\ref{ang9}) yields%
\begin{align}
&  \Psi_{f}\left(  \rho,z,t\right)  =\int_{0}^{\infty}\mathrm{d}k~k\exp\left[
-\frac{a_{1}k\left(  1+\cos\theta_{F}\left(  k\right)  \right)  }{2}\right]
\nonumber\\
&  \qquad\times J_{0}\left[  k\rho\sin\theta_{F}\left(  k\right)  \right]
\exp\left[  ik\left(  z\cos\theta_{F}\left(  k\right)  -ct\right)  \right]
\label{fwm9}%
\end{align}
(as compared to (\ref{ang9}) here we have taken into account the $1/k$ term
that appears in (\ref{ang4b}) as to be consistent with [\cite{fw18}] for
example) so that the frequency spectrum of the superposition can be written
as
\begin{equation}
\tilde{B}\left(  k\right)  =k\exp\left[  -\frac{a_{1}k\left(  1+\cos\theta
_{F}\left(  k\right)  \right)  }{2}\right]  \label{fwm13}%
\end{equation}
[the significance of the factor $k\sin\theta_{F}\left(  k\right)  $ will be
discussed in following sections].

The frequency spectrum $\tilde{B}\left(  k\right)  $ in Eq.~(\ref{fwm13}) has
two free parameters, $a_{1}$ and $\beta$, the latter having the same
definition as in Eq.~(\ref{su1}) of section \ref{ssFAng}. As we already noted
in the introduction of this chapter, the choice $\gamma=1$ together with the
frequency range and cone angle of the Bessel beam of central wavelength
determines $\beta=40rad/m$. The single free parameter is $a_{1}$ and a single
parameter does not allow to approximate for any realistic light sources --
Fig.~\ref{fig52}b shows a typical spectrum that can be modeled in terms of
Eq.~(\ref{fwm13}) as compared to the optically feasible frequency spectrum
specified in the introduction of this overview and it can be seen, that the
bandwidth of the wave field is far beyond the reach of any realistic light
source. In fact, due to this extraordinary large bandwidth the original FWM's
in Eq.~(\ref{fwm11}) are essentially half-cycle pulses, as already noted in
section \ref{ssFGau}.

As deduced in section \ref{ssFGau} (and also in terms of the section
\ref{ssFLoc}), the parameter $a_{1}$ determines the waist of the wave field --
in our numerical example $a_{1}=1.4\times10^{-7}$, so that the Eq.~(\ref{gau9}%
) gives%
\[
W_{0}=\sqrt{\frac{a_{1}}{\beta}}\sim6\times10^{-5}m\text{.}%
\]

In literature it has been argued, that the FWM's determined by
Eq.~(\ref{fwm11}) are nonphysical as the wave field contain acausal
components. In the discussion of Ref.~\cite{fw18} it has been shown that the
acausality can be eliminated by proper choice of parameters $a_{1}$ and
$\beta$ -- it has been shown that if $\beta a_{1}<1$, the predominant
contribution to the spectrum comes from the plane waves moving in positive $z$
axis direction. In Fig~\ref{fig52}b it can be seen, that this is indeed the
case, however the field is still far from convenient for any optical
implementation due to ultra-wide bandwidth.

Note, that various closed-form sub- and superluminal FWM's ($\gamma\neq1$)
have been derived for example in Ref.~\cite{fw15}.

\subsection{Bessel-Gauss pulses}%

%TCIMACRO{\FRAME{ftbpFU}{4.9381in}{6.7118in}{0pt}{\Qcb{A numerical example of a
%Bessel-Gauss pulse optimized for optical generation with the parameters
%$\sigma=40000\frac{2\pi}{m}$, $a_{1}=5\times10^{-6}m$, $\beta=40\frac{rad}{m}%
%$, $\gamma=1$: (a) The angular spectrum of plane waves in two perspectives;
%(b) The frequency spectrum of the Bessel-Gauss pulse (black line), the
%frequency spectrum of an optically feasible wave field (green line), the angle
%$\theta_{F}\left(  k\right)  $ as the function of the wave number (dashed blue
%line); (c) The spatial field distribution of the pulse.}}{\Qlb{fig54}%
%}{fig5_4.jpg}{\special{ language "Scientific Word";  type "GRAPHIC";
%maintain-aspect-ratio TRUE;  display "ICON";  valid_file "F";
%width 4.9381in;  height 6.7118in;  depth 0pt;  original-width 4.9104in;
%original-height 6.6841in;  cropleft "0";  croptop "1";  cropright "1";
%cropbottom "0";  filename 'fig5_4.jpg';file-properties "XNPEU";}} }%
%BeginExpansion
\begin{figure}
[ptb]
\begin{center}
\includegraphics[
height=6.7118in,
width=4.9381in
]%
{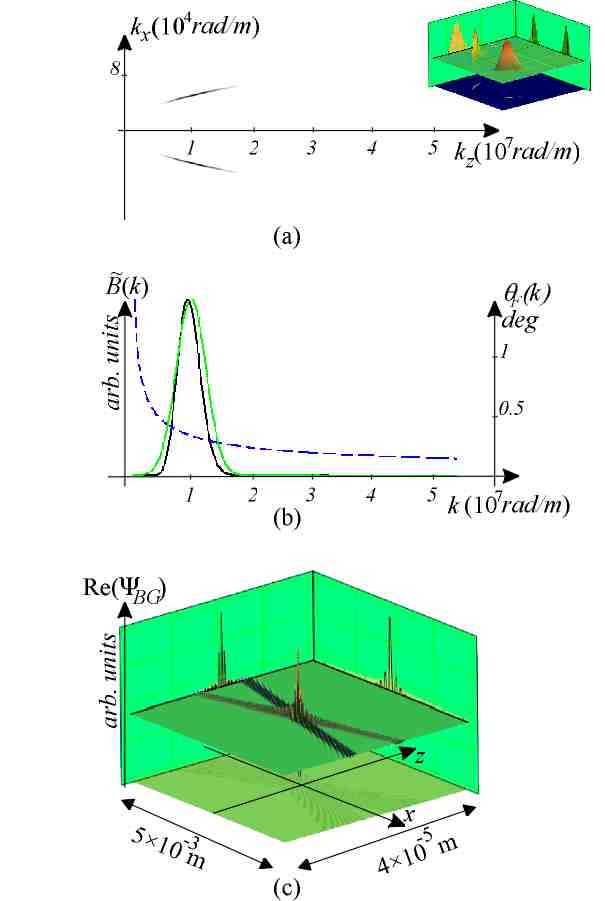}%
\caption{A numerical example of a Bessel-Gauss pulse optimized for optical
generation with the parameters $\sigma=40000\frac{2\pi}{m}$, $a_{1}%
=5\times10^{-6}m$, $\beta=40\frac{rad}{m}$, $\gamma=1$: (a) The angular
spectrum of plane waves in two perspectives; (b) The frequency spectrum of the
Bessel-Gauss pulse (black line), the frequency spectrum of an optically
feasible wave field (green line), the angle $\theta_{F}\left(  k\right)  $ as
the function of the wave number (dashed blue line); (c) The spatial field
distribution of the pulse.}%
\label{fig54}%
\end{center}
\end{figure}
%EndExpansion
The Bessel-Gauss pulses were introduced by Overfelt in Ref.~\cite{lw4} (see
also Refs.~\cite{fw15,g15,g25}). In this publication it was shown, that the
scalar wave field
\begin{align}
\Psi_{BG}\left(  \rho,\mu,\zeta\right)   &  =\frac{a_{1}}{a_{1}+i\zeta}%
J_{0}\left(  \frac{\kappa a_{1}\rho}{a_{1}+i\zeta}\right)  \exp\left[
i\beta\mu\right] \nonumber\\
&  \times\exp\left[  -\frac{\beta\rho^{2}}{a_{1}+i\zeta}\right]  \exp\left[
-i\frac{\kappa^{2}a_{1}\zeta}{4\beta\left[  a_{1}+i\zeta\right]  }\right]
\text{,} \label{bg5}%
\end{align}
where the physical meaning of the parameters $a_{1}$, $\beta$, also $\zeta$
and $\mu$ is consistent with the previous discussion. The expression has an
additional free parameter $\kappa$ as compared to the FWM's in (\ref{fwm11}),
in fact, the latter is the special case of the former in the limiting case
$\kappa\rightarrow0$. The Bessel-Gauss pulses were further investigated in
Ref.~\cite{fw15} where it was shown, that in Fourier picture as in
Eq.~(\ref{ang5}) the spatiotemporal Fourier transform of the field can be
written as
\begin{equation}
\psi_{BG}\left(  \mathbf{k},\omega\right)  =\Xi_{BG}\left(  \chi,\beta\right)
\delta\left[  k_{z}-\left(  \beta-\frac{\chi^{2}}{4\beta}\right)  \right]
\delta\left[  \omega+c\left(  \beta+\frac{\chi^{2}}{4\beta}\right)  \right]
\text{,} \label{bg10}%
\end{equation}
where%
\begin{equation}
\Xi_{BG}\left(  \chi,\beta\right)  =\frac{a_{1}4\pi^{3}}{\beta}I_{0}\left(
\frac{\kappa a_{1}\chi}{2\beta}\right)  \exp\left[  -\frac{\kappa^{2}a_{1}%
}{4\beta}\right]  \exp\left[  -\frac{a_{1}\chi^{2}}{4\beta}\right]
\label{bg12}%
\end{equation}
(see section \ref{ssFDo} for the notation). From Eq.\ (\ref{bg10}) it can be
seen that the support of the plane wave spectrum of the Bessel-Gauss pulses is
the same as described by Eq.~(\ref{su12}) [or Eq.~(\ref{su19})]. The change of
variables in (\ref{bg12}) yields for the frequency spectrum
\begin{align}
B\left(  k\right)   &  =\frac{a_{1}4\pi^{3}}{\beta}I_{0}\left(  \frac{\kappa
a_{1}k\sin\theta_{F}\left(  k\right)  }{2\beta}\right) \nonumber\\
&  \times\exp\left[  -\frac{a_{1}}{4\beta}\left(  \kappa^{2}+k^{2}\sin
^{2}\theta_{F}\left(  k\right)  \right)  \right]  \text{,} \label{bg15}%
\end{align}
where $\kappa>0$, $a_{1}>0$ and $\beta>0$.

In the original paper the Bessel-Gauss pulses were introduced as the wave
fields that are "more highly localized than the fundamental Gaussian solutions
because of its extra spectral degree of freedom". The additional free
parameter is indeed advantageous, however, in our opinion not in the sense
proposed in this publication -- the spatial localization of any wideband
free-space wave field is directly proportional to its bandwidth and the latter
is inappropriately large even for the original FWM's (see Ref.~\cite{fw15} for
a related discussion). It may be the consequence of this general emphasis of
the original paper that it is not generally recognized that the extra
parameter $\kappa$ in Eqs.~(\ref{bg5}) -- (\ref{bg15}) gives one the necessary
degree of freedom to fit an arbitrary bandlimited Gaussian-like spectrum --
from Eq.~(\ref{bg15}) it can be seen, that the central frequency and bandwidth
of the spectra of the pulse are independently adjustable by the parameters
$\kappa$ and $a_{1}$ respectively.

Analogously to the discussion in section \ref{ssFGau} the Bessel-Gauss pulses
can be given the form that, in some respect, resembles that of the
monochromatic Gaussian beam:%
\begin{align}
&  \Psi_{BG}\left(  \rho,z,\zeta\right)  =\exp\left[  -i\beta\zeta\right]
\frac{W_{0}}{a_{1}\left(  \zeta\right)  }J_{0}\left[  \kappa a_{1}\rho\left(
\frac{1}{\beta a_{1}^{2}\left(  \zeta\right)  }-i\frac{1}{R\left(
\zeta\right)  }\right)  \right] \nonumber\\
&  \qquad\times\exp\left[  -\frac{\rho^{2}}{a_{1}^{2}\left(  \zeta\right)
}-\frac{\kappa^{2}a_{1}\zeta}{4\beta R\left(  \zeta\right)  }\right]
\label{bg20}\\
&  \times\exp\left[  -i\left(  \frac{\kappa^{2}a_{1}\zeta}{4\beta^{2}a_{1}%
^{2}\left(  \zeta\right)  }-\frac{\beta\rho^{2}}{R\left(  \zeta\right)
}\right)  -i\left(  \arctan\left(  \frac{\zeta}{a_{1}}\right)  -2\beta
z+\frac{\pi}{2}\right)  \right]  \text{,}\nonumber
\end{align}
here again
\begin{subequations}
\begin{align}
a_{1}\left(  \zeta\right)   &  =W_{0}\left[  1+\left(  \frac{\zeta}{a_{1}%
}\right)  ^{2}\right]  ^{\frac{1}{2}}\label{bg21}\\
R\left(  \zeta\right)   &  =\zeta\left[  1+\left(  \frac{a_{1}}{\zeta}\right)
^{2}\right] \label{bg22}\\
W_{0}  &  =\sqrt{\frac{a_{1}}{\beta}}\text{.} \label{bg23}%
\end{align}
The general form of the Bessel-Gauss pulses (\ref{bg20}) is very advantageous
in the sense that here we can actually write out its carrier wave number~--
often this quantity is elusive for the wideband wave fields. Indeed, around
the point $\zeta=0$, along the optical axis ($\rho=0$) with (\ref{bg21}) we
can write for the $z$ axis component of the carrier wave number
\end{subequations}
\begin{equation}
k_{0z}=\beta+\frac{\sigma^{2}}{4\beta}-\frac{1}{a_{1}} \label{bg30}%
\end{equation}
This result is actually quite significant, if we once more remind that in
literature the FWM's have often been termed as carrier-free wave fields (see
Refs.~\cite{g5,g10} for example). In lights of (\ref{bg30}) we can conclude
that the carrier-free behaviour of the FWM's is indeed caused by the integral
transform table, not by physical arguments.

In the numerical example in Fig.~\ref{fig54} we have optimized the parameters
of the wave field as to match the spectral band specified in the introduction
of this section. Again, the parameter $\beta$ is determined by the bandwidth
and the cone angle of the central frequency as described above. Thus we got:
$\sigma=40000\frac{rad}{m}$, $a_{1}=5\times10^{-6}m$, $\beta=40\frac{rad}{m}$,
$\gamma=1$. The evaluation of the Eq.~(\ref{bg30}) yields $k_{0z}%
=9.80004\times10^{6}\frac{rad}{m}$ and this result is in good correspondence
with the numerical simulations.

In conclusion, the Bessel-Gauss pulses are obviously much more appropriate for
modeling realistic experimental situations.

\subsection{X-type wave fields}

The X-type localized wave fields are characterized by that for their angular
spectrum of plane waves $\beta=0$ in Eq.~(\ref{su12}) [or $\xi=0$ in
Eq.~(\ref{su19})]. This choice implies, that their support of angular spectrum
of plane waves is a cone in $k$-space (see Fig.~\ref{fig42}). Consequently,
the phase and group velocity of X-type pulses are equal (both necessarily
superluminal) and the field propagates without any local changes along the
optical axis.

\subsubsection{Bessel beams}%

%TCIMACRO{\FRAME{ftbpFU}{4.0404in}{6.0009in}{0pt}{\Qcb{A numerical example of a
%monochromatic Bessel beam with the parameters $\theta_{0}=0.223\deg$,
%$k_{0}\sim1\times10^{7}$: (a) The angular spectrum of its plane waves in two
%perspectives; (b) The delta-shaped frequency spectrum; (c) The spatial field
%distribution of the beam.}}{\Qlb{fig59}}{fig5_9.jpg}%
%maintain-aspect-ratio TRUE;  display "ICON";  valid_file "F";
%width 4.0404in;  height 6.0009in;  depth 0pt;  original-width 4.9104in;
%original-height 6.6841in;  cropleft "0";  croptop "1";  cropright "1";
%cropbottom "0";  filename 'fig5_9.jpg';file-properties "XNPEU";}} }%
%BeginExpansion
\begin{figure}
[ptb]
\begin{center}
\includegraphics[
height=6.0009in,
width=4.0404in
]%
{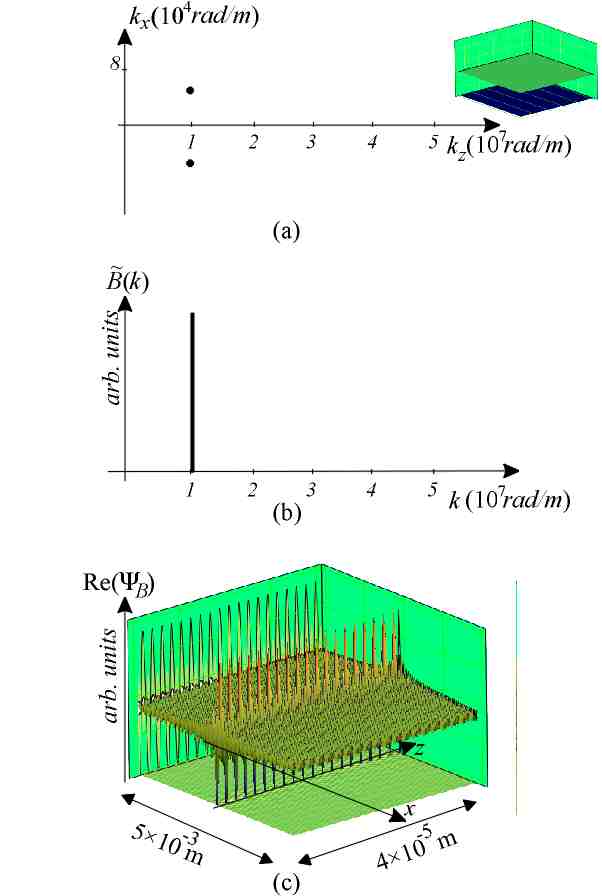}%
\caption{A numerical example of a monochromatic Bessel beam with the
parameters $\theta_{0}=0.223\deg$, $k_{0}\sim1\times10^{7}$: (a) The angular
spectrum of its plane waves in two perspectives; (b) The delta-shaped
frequency spectrum; (c) The spatial field distribution of the beam.}%
\label{fig59}%
\end{center}
\end{figure}
%EndExpansion
The Bessel beams \cite{b1}--\cite{ba82} are the simplest special case of the
propagation-invariant wave fields. Being the exact solutions to the Helmholtz
equation in cylindrical coordinates their field reads as%
\begin{equation}
\Psi_{B}\left(  \rho,z,t\right)  =\sum_{n}c_{n}~J_{n}\left(  k\rho\sin
\theta_{0}\right)  \exp\left[  in\phi\right]  \exp\left[  ik\left(
z\cos\theta_{0}-ct\right)  \right]  \label{bes0}%
\end{equation}
so that for the zeroth order Bessel beam we have%
\begin{equation}
\Psi_{B}\left(  \rho,z,t\right)  =J_{0}\left(  k\rho\sin\theta_{0}\right)
\exp\left[  ik\left(  z\cos\theta_{0}-ct\right)  \right]  \text{.}
\label{bes2}%
\end{equation}
In the Fourier picture, the zeroth--order Bessel beam is the cylindrically
symmetric superposition of the monochromatic plane waves propagating at angles
$\theta_{0}$ relative to $z$ axis, correspondingly, their angular spectrum of
plane waves in Eq.~(\ref{ang20}) reads
\begin{equation}
A_{0}^{\left(  B\right)  }\left(  k,\theta\right)  \sim~\delta\left(
k-k_{0}\right)  \delta\left(  \theta-\theta_{0}\right)  \text{.} \label{bes3}%
\end{equation}
The bidirectional representation of the Bessel beam can be found in
Ref.~\cite{fw14}.

The properties of Bessel beams have been discussed in many publications both
in terms of angular spectrum of plane waves \cite{b1}--\cite{b16} and
diffraction theory \cite{ba5}--\cite{ba82} and their properties are very well
understood today. The interest has been triggered in Refs.~\cite{b1,b2} where
Durnin \textit{et al }presented them as "nondiffracting" solutions of the
homogeneous scalar wave equation -- they demonstrated experimentally, that the
central maximum of the Bessel beams propagates much further than the Rayleigh
range predicts.

Note, that though there has been numerous experiments on Bessel beams, they
are not realizable in experiment in the exact form (\ref{bes2}) -- indeed, the
analysis of section \ref{ssFEn} immediately shows, that this wave field has
both infinite total energy and energy flow over its cross-section. We will
discuss this point in what follows.

In this review the Bessel beams appear as the components of the Fourier
decomposition in Eqs.~(\ref{ang7}) -- (\ref{ang9}) for example. Later in this
review we will refer to their most important properties in some detail. At
this point we just depict its angular spectrum of plane waves with the typical
field distribution (see Fig.~\ref{fig59}).

\subsubsection{X-pulses}%

%TCIMACRO{\FRAME{ftbpFU}{3.8545in}{5.9516in}{0pt}{\Qcb{A numerical example of a
%X-pulse with the parameters $\theta_{0}=0.223\deg$, $\gamma=0.99999$: (a) The
%angular spectrum of plane waves in two perspectives; (b) The frequency
%spectrum of the X-pulse (black line), the frequency spectrum of an optically
%feasible wave field (green line), the angle $\theta_{0}$ as the (constant)
%function of the wave number (dashed blue line); (c) The spatial field
%distribution of the pulse.}}{\Qlb{fig510}}{fig5_10.jpg}%
%maintain-aspect-ratio TRUE;  display "ICON";  valid_file "F";
%width 3.8545in;  height 5.9516in;  depth 0pt;  original-width 4.9104in;
%original-height 6.6841in;  cropleft "0";  croptop "1";  cropright "1";
%cropbottom "0";  filename 'fig5_10.jpg';file-properties "XNPEU";}} }%
%BeginExpansion
\begin{figure}
[ptb]
\begin{center}
\includegraphics[
height=5.9516in,
width=3.8545in
]%
{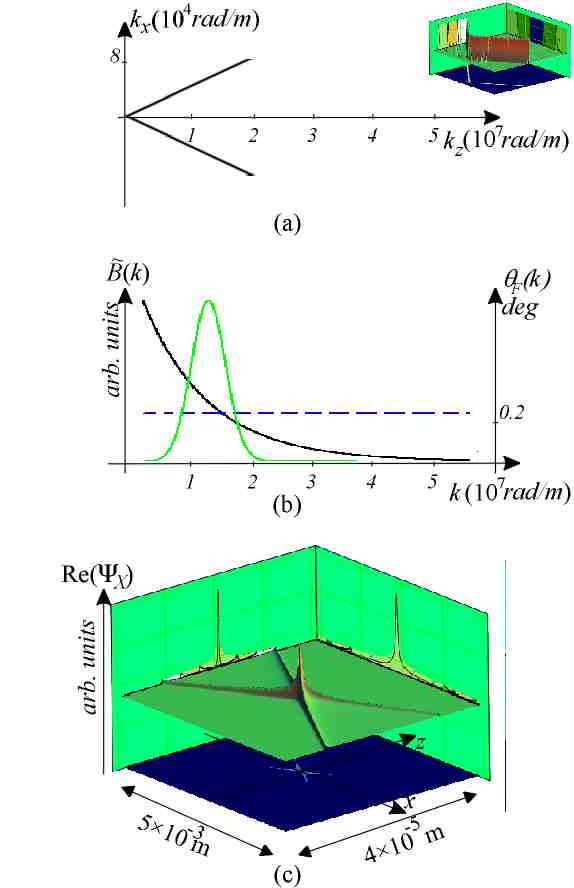}%
\caption{A numerical example of a X-pulse with the parameters $\theta
_{0}=0.223\deg$, $\gamma=0.99999$: (a) The angular spectrum of plane waves in
two perspectives; (b) The frequency spectrum of the X-pulse (black line), the
frequency spectrum of an optically feasible wave field (green line), the angle
$\theta_{0}$ as the (constant) function of the wave number (dashed blue line);
(c) The spatial field distribution of the pulse.}%
\label{fig510}%
\end{center}
\end{figure}
%EndExpansion
In \cite{x1} Lu \textit{et al} demonstrated that the choice
\begin{equation}
A_{0}^{\left(  X\right)  }\left(  k,\theta\right)  =B\left(  k\right)
\delta\left(  \theta-\theta_{0}\right)  \label{x5}%
\end{equation}
in representation (\ref{ang20}) with the frequency spectrum
\begin{equation}
B\left(  k\right)  =\frac{1}{k^{2}}\exp\left[  -ka_{0}\right]  \label{x2}%
\end{equation}
yields the propagation-invariant wave field
\begin{equation}
\Psi_{X}\left(  \rho,z,t\right)  =\frac{a_{0}}{\sqrt{\left(  \rho\sin
\theta_{0}\right)  ^{2}+\left[  a_{0}-i\left(  z\cos\theta_{0}-ct\right)
\right]  ^{2}}} \label{x3}%
\end{equation}
(see Ref.~\cite{x4} for the description of higher order X-pulses). From the
angular spectrum in Eq.~(\ref{x5}) it can be seen, that the support of angular
spectrum of plane waves of the X-pulses is a cone in $k$-space, i.e., all the
plane wave components of the wave field propagate at the equal angle from the
propagation axis. The frequency spectrum of X-pulses in Eq.~(\ref{x5}) is
uniform (see Fig.~\ref{fig510}a) -- the immediate conclusion of the approach
of section \ref{ssFLoc} that the corresponding field should have exponentially
decaying behaviour in both $z$ axis and $xy$ plane is confirmed in
Fig.~\ref{fig510}c.

X-wave fields have been further investigated in Refs.~\cite{x2,x3,x4,x5},
recently the topic have been given an overview and general description in
Ref.~\cite{x15}. We mention here the so called bowtie waves that are generally
introduced as the derivatives of the X-waves:%
\begin{equation}
\Psi_{mX}\left(  \rho,z,t\right)  =\frac{\partial^{m}\Psi^{\prime}\left(
\mathbf{r},t\right)  }{\partial x^{m}}\text{.} \label{x10}%
\end{equation}
The derivatives of X-waves have been shown to possess non-symmetric nature and
have extended localization along a radial direction. In our terms the physical
nature of such wave fields can be interpreted by applying the derivation
operation on the general angular spectrum representation of the free-space
scalar wave fields in Eq.~(\ref{ang7}). We easily get
\begin{align}
&  \frac{\partial^{m}\Psi^{\prime}\left(  \mathbf{r},t\right)  }{\partial
x^{m}}=\frac{1}{\left(  2\pi\right)  ^{4}}\int_{0}^{\infty}\mathrm{d}%
k~k^{2}\int_{0}^{\pi}\mathrm{d}\theta\left(  \sin\theta\right)  ^{m+1}\int
_{0}^{2\pi}\mathrm{d}\phi\cos^{m}\phi\nonumber\\
&  \qquad\times\mathbf{~}A\left(  k\sin\theta\cos\phi,k\sin\theta\cos
\phi,k\cos\theta\right) \nonumber\\
&  \qquad\times\exp\left[  ik\left(  x\sin\theta\cos\phi+y\sin\theta\cos
\phi+z\cos\theta-i\omega t\right)  \right]  \text{~,} \label{x12}%
\end{align}
so that the angular spectrum of plane waves of such wave fields is not
cylindrically symmetric, correspondingly the wave field is a superposition of
higher order monochromatic Bessel beams as described by Eq.~(\ref{ang8}) for example.

Due to the exponential shape of the frequency spectrum the X-waves are not
appropriate for optical implementation.

\subsubsection{\label{ssBX}Bessel-X pulses}%

%TCIMACRO{\FRAME{ftbpFU}{9.9859cm}{15.2182cm}{0pt}{\Qcb{A numerical example of
%a Bessel-X pulse with the parameters $\theta_{0}=0.223\deg$, $\gamma=0.99999$:
%(a) The angular spectrum of plane waves in two perspectives; (b) The frequency
%spectrum of the Bessel-X pulse (black line) as compared to the frequency
%spectrum of an optically feasible wave field (green line), the angle
%$\theta_{0}$ as the (constant) function of the wavel number (dashed blue
%line); (c) The spatial field distribution of the pulse. }}{\Qlb{fig57}%
%}{fig5_7.jpg}{\special{ language "Scientific Word";  type "GRAPHIC";
%maintain-aspect-ratio TRUE;  display "ICON";  valid_file "F";
%width 9.9859cm;  height 15.2182cm;  depth 0pt;  original-width 4.9104in;
%original-height 6.6841in;  cropleft "0";  croptop "1";  cropright "1";
%cropbottom "0";  filename 'fig5_7.jpg';file-properties "XNPEU";}} }%
%BeginExpansion
\begin{figure}
[ptb]
\begin{center}
\includegraphics[
height=15.2182cm,
width=9.9859cm
]%
{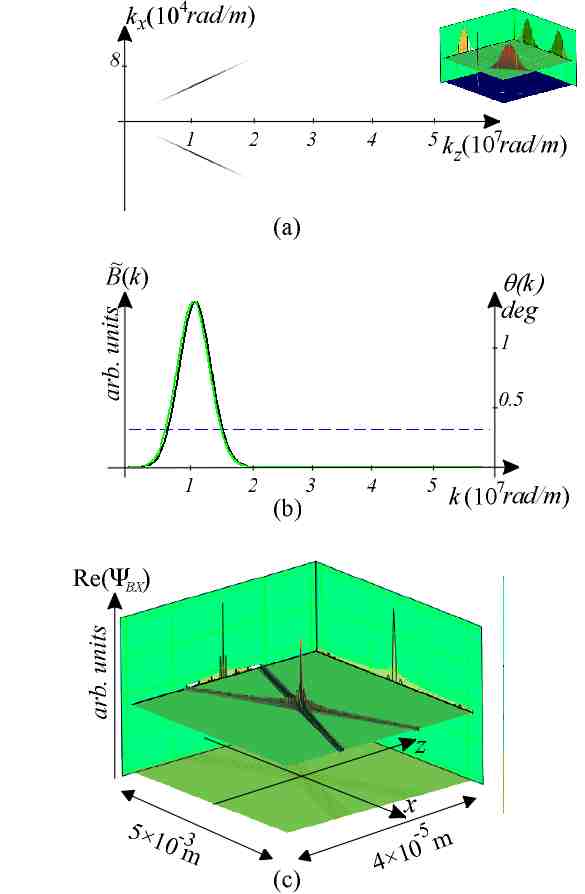}%
\caption{A numerical example of a Bessel-X pulse with the parameters
$\theta_{0}=0.223\deg$, $\gamma=0.99999$: (a) The angular spectrum of plane
waves in two perspectives; (b) The frequency spectrum of the Bessel-X pulse
(black line) as compared to the frequency spectrum of an optically feasible
wave field (green line), the angle $\theta_{0}$ as the (constant) function of
the wavel number (dashed blue line); (c) The spatial field distribution of the
pulse. }%
\label{fig57}%
\end{center}
\end{figure}
%EndExpansion
The Bessel-X pulses were introduced by Saari in Ref.~\cite{bx0,bx1} as the
bandlimited version of X-pulses. Their angular spectrum of plane waves can be
described as%

\begin{equation}
\tilde{A}_{0}^{\left(  BX\right)  }\left(  k,\theta\right)  =\tilde{B}\left(
k\right)  \delta\left(  \theta-\theta_{0}\right)  \text{,} \label{bx1}%
\end{equation}
where
\begin{equation}
\tilde{B}\left(  k\right)  =\frac{\sigma_{k}}{\sqrt{2\pi}}\sqrt{\frac{k}%
{k_{0}}}\exp\left[  -\frac{\sigma_{k}^{2}\left(  k-k_{0}\right)  ^{2}}%
{2}\right]  \text{,} \label{bx2}%
\end{equation}
$\sigma_{k}$ being defined in (\ref{lo16}) and $k_{0}$ being the carrier wave
number, so that for the field one can write
\begin{align}
&  \Psi_{BX}\left(  \rho,z,t\right)  =\int_{0}^{\infty}\mathrm{d}k~\tilde
{B}\left(  k\right) \nonumber\\
&  \qquad\qquad\times J_{0}\left[  k\rho\sin\theta_{0}\right]  \exp\left[
-ik\left(  z\cos\theta_{0}-ct\right)  \right]  \text{.} \label{bx2a}%
\end{align}
The integration in Eq.~(\ref{ang20}) can be carried out to yield \cite{bx1}
\begin{align}
&  \Psi_{BX}\left(  \rho,z,t\right)  =\sqrt{Z\left(  d\right)  }\nonumber\\
&  \qquad\times\exp\left[  -\frac{1}{2\sigma_{k}^{2}}\left(  \rho^{2}\sin
^{2}\theta+d^{2}\right)  \right]  J_{0}\left[  Z\left(  d\right)  \rho
k_{0}\sin\theta\right]  \exp\left[  ik_{0}d\right]  \text{,} \label{bx3}%
\end{align}
where
\begin{equation}
Z\left(  d\right)  =1+\frac{id}{k_{0}\sigma_{k}^{2}} \label{bx4}%
\end{equation}
and
\begin{equation}
d=z\cos\theta-ct\text{.} \label{bx5}%
\end{equation}

From the Eqs.~(\ref{bx1}) and (\ref{bx2}) it can be seen that,again, the
support of angular spectrum of plane waves is a cone in $k$-space (see
Fig.~\ref{fig57}a). However, unlike the X-pulses, the frequency spectrum of
Bessel-X pulse is Gaussian and it can be optimized to approximate that of our
initial conditions. Thus, the Bessel-X pulses are optically feasible in the
sense defined in this chapter.

\subsection{Two limiting cases of the propagation-invariance}

\subsubsection{Pulsed wave fields with infinite group velocity}%

%TCIMACRO{\FRAME{ftbpFU}{3.9539in}{5.7752in}{0pt}{\Qcb{A numerical example of a
%the wave field with infinite group velocity with the parameters $\xi
%=6.7\times10^{6}m$, $\gamma=\infty$: (a) The angular spectrum of plane waves
%in two perspectives; (b) The frequency spectrum of the pulse (black line), the
%frequency spectrum of an optically feasible wave field (green line), the angle
%$\theta_{F}\left(  k\right)  $ as the function of the wave number (dashed blue
%line), (c) The spatial field distribution of the pulse; (d) Three snapshots of
%the temporal evolution of the pulse.}}{\Qlb{fig511}}{fig5_11.jpg}%
%maintain-aspect-ratio TRUE;  display "ICON";  valid_file "F";
%width 3.9539in;  height 5.7752in;  depth 0pt;  original-width 4.9104in;
%original-height 6.6841in;  cropleft "0";  croptop "1";  cropright "1";
%cropbottom "0";  filename 'fig5_11.jpg';file-properties "XNPEU";}} }%
%BeginExpansion
\begin{figure}
[ptb]
\begin{center}
\includegraphics[
height=5.7752in,
width=3.9539in
]%
{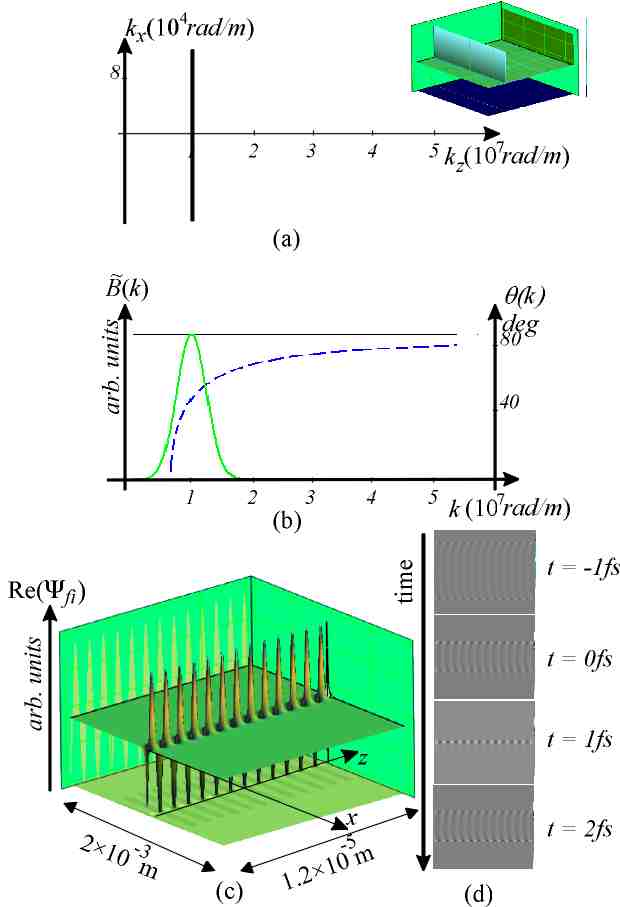}%
\caption{A numerical example of a the wave field with infinite group velocity
with the parameters $\xi=6.7\times10^{6}m$, $\gamma=\infty$: (a) The angular
spectrum of plane waves in two perspectives; (b) The frequency spectrum of the
pulse (black line), the frequency spectrum of an optically feasible wave field
(green line), the angle $\theta_{F}\left(  k\right)  $ as the function of the
wave number (dashed blue line), (c) The spatial field distribution of the
pulse; (d) Three snapshots of the temporal evolution of the pulse.}%
\label{fig511}%
\end{center}
\end{figure}
%EndExpansion
Consider the special case $\gamma=0$ of the support of angular spectrum of
plane waves (\ref{su17}) that reads%
\begin{equation}
k_{z}\left(  k\right)  =\xi=const\text{.} \label{n1}%
\end{equation}
From the general definition of group velocity in Eq.~(\ref{su8}) it is
obvious, that for this particular case we have $v^{g}=\infty$. In what follows
we give a physical description to the wave fields that have such a peculiar property.

A closed-form solution of the homogeneous scalar wave equation that obeys
(\ref{n1}) can be easily found. The angular spectrum of plane waves of the
wave field reads%
\begin{equation}
A_{0}^{\left(  fi\right)  }\left(  k,\theta\right)  =B\left(  k\right)
\delta\left[  \theta-\arccos\left(  \frac{\xi}{k}\right)  \right]  \text{.}
\label{n6}%
\end{equation}
The substitution in Whittaker type superposition (\ref{ang20}) yields
\begin{align}
\Psi_{fi}\left(  \rho,z,t\right)   &  =\int_{0}^{\infty}\mathrm{d}k\tilde
{B}\left(  k\right)  J_{0}\left(  k\rho\sqrt{1-\left(  \frac{\xi}{k}\right)
^{2}}\right) \nonumber\\
&  \times\exp\left[  ik\left(  \frac{\xi}{k}z-ct\right)  \right]  \text{,}
\label{n10}%
\end{align}
where
\begin{equation}
\tilde{B}\left(  k\right)  =k^{2}\sqrt{1-\left(  \frac{\xi}{k}\right)  ^{2}%
}B\left(  k\right)  \text{,} \label{n11}%
\end{equation}
so that%
\begin{equation}
\Psi_{fi}\left(  \rho,z,t\right)  =\exp\left[  i\xi z\right]  \int_{0}%
^{\infty}\mathrm{d}k\tilde{B}\left(  k\right)  J_{0}\left(  \rho\sqrt
{k^{2}-\xi^{2}}\right)  \exp\left[  -ikct\right]  \text{.} \label{n12}%
\end{equation}
If we choose $\tilde{B}\left(  k\right)  =const=1$ and use the integral
transforms \cite{o9}%
\begin{equation}
\int_{0}^{\infty}\mathrm{d}xJ_{0}\left(  b\sqrt{x^{2}-a^{2}}\right)
\cos\left(  xy\right)  =%
%TCIMACRO{\QTATOPD{\{}{.}{\left(  b^{2}-y^{2}\right)  ^{-\frac{1}{2}%
%}e^{-a\left(  b^{2}-y^{2}\right)  ^{\frac{1}{2}}}\text{ \ \ \ \ \ \ \ if
%}0<y<b}{-\left(  y^{2}-b^{2}\right)  ^{-\frac{1}{2}}\sin\left[  a\left(
%y^{2}-b^{2}\right)  ^{\frac{1}{2}}\right]  \text{ if }b<y<\infty} }%
%BeginExpansion
\genfrac{\{}{.}{0pt}{1}{\left(  b^{2}-y^{2}\right)  ^{-\frac{1}{2}%
}e^{-a\left(  b^{2}-y^{2}\right)  ^{\frac{1}{2}}}\text{ \ \ \ \ \ \ \ if
}0<y<b}{-\left(  y^{2}-b^{2}\right)  ^{-\frac{1}{2}}\sin\left[  a\left(
y^{2}-b^{2}\right)  ^{\frac{1}{2}}\right]  \text{ if }b<y<\infty}
%EndExpansion
\label{n13}%
\end{equation}
and
\begin{equation}
\int_{0}^{\infty}\mathrm{d}xJ_{0}\left(  b\sqrt{x^{2}-a^{2}}\right)
\sin\left(  xy\right)  =%
%TCIMACRO{\QTATOPD{\{}{.}%
%{\;\;\;\;\;\;\;\;\;0\;\;\;\;\;\;\;\;\;\;\;\;\;\;\;\;\;\;\;\;\;\;\;\text{,if
%}\;\;\;\;0<y<b}{\left(  y^{2}-b^{2}\right)  ^{-\frac{1}{2}}\cos\left[
%a\left(  y^{2}-b^{2}\right)  ^{\frac{1}{2}}\right]  \text{ if }b<y<\infty} }%
%BeginExpansion
\genfrac{\{}{.}{0pt}{1}{\;\;\;\;\;\;\;\;\;0\;\;\;\;\;\;\;\;\;\;\;\;\;\;\;\;\;\;\;\;\;\;\;\text{,if
}\;\;\;\;0<y<b}{\left(  y^{2}-b^{2}\right)  ^{-\frac{1}{2}}\cos\left[
a\left(  y^{2}-b^{2}\right)  ^{\frac{1}{2}}\right]  \text{ if }b<y<\infty}
%EndExpansion
\label{n14}%
\end{equation}
the integral (\ref{n12}) can be evaluated explicitly to yield
\begin{equation}
\Psi_{fi}\left(  \rho,z,t\right)  =\left\{
\begin{array}
[c]{c}%
\frac{\exp\left[  i\xi z-i\xi\sqrt{\rho^{2}-c^{2}t^{2}}\right]  }{\sqrt
{\rho^{2}-c^{2}t^{2}}}\text{ if }0<tc<\rho\\
i\frac{\exp\left[  i\xi z-\xi\sqrt{c^{2}t^{2}-\rho^{2}}\right]  }{\sqrt
{c^{2}t^{2}-\rho^{2}}}\text{ if }\rho<tc<\infty
\end{array}
\right.  \text{.} \label{n15}%
\end{equation}
Note, that the special case $\xi=0$ yields the cylindrically symmetric
superposition of plane wave pulses propagating perpendicularly to $z$ axis.

The support of the angular spectrum of plane waves of the wave field
(\ref{n15}) is depicted in Fig.~\ref{fig511}a. From the estimates of the
spatial localization of LW's in section \ref{ssFLoc} we can expect the wave
field to be localized in transversal direction at $t=0$, $z=0$ and to be
uniform along the $z$ axis. Indeed, from the Fig.~\ref{fig511}c it can be seen
that at this space-time point the wave field (\ref{n15}) is an approximation
to \textquotedblright light filament\textquotedblright\ along the optical axis.

The temporal evolution of the wave field is depicted in Fig.~\ref{fig511}d.
One can see, that the effect of the infinite group velocity is that the light
filament is focused only at a single time $t=0$ and extends from $-\infty$ to
$\infty$. As to relate to the conventional wave optics, the temporal evolution
of the light filament is a close relative to that of the plane wave pulse in a
plane perpendicular to its wave vector.

As the wave field includes both non-optical frequencies and non-paraxial
angles it is not optically feasible as such. However, it can be shown that a
finite energy approximation to the light filament can in principle be
generated by a cylindrical diffraction grating.

\subsubsection{Pulsed wave fields with frequency-independent beamwidth}%

%TCIMACRO{\FRAME{ftbpFU}{3.9245in}{6.0174in}{0pt}{\Qcb{A numerical example of a
%the pulsed wave field with frequency independent beamwidth with the parameters
%$\alpha_{0}=4\times10^{4}\frac{rad}{m}$, $\gamma\sim0$: (a) The angular
%spectrum of plane waves in two perspectives; (b) The frequency spectrum of the
%pulse (black line), the frequency spectrum of an optically feasible wave field
%(green line), the cone angle of the component Bessel beams as the function of
%the wave number (dashed blue line); (c) The spatial field distribution of the
%pulse.}}{\Qlb{fig512}}{fig5_12.jpg}{\special{ language "Scientific Word";
%type "GRAPHIC";  maintain-aspect-ratio TRUE;  display "ICON";
%valid_file "F";  width 3.9245in;  height 6.0174in;  depth 0pt;
%original-width 4.9104in;  original-height 6.6841in;  cropleft "0";
%filename 'fig5_12.jpg';file-properties "XNPEU";}} }%
%BeginExpansion
\begin{figure}
[ptb]
\begin{center}
\includegraphics[
height=6.0174in,
width=3.9245in
]%
{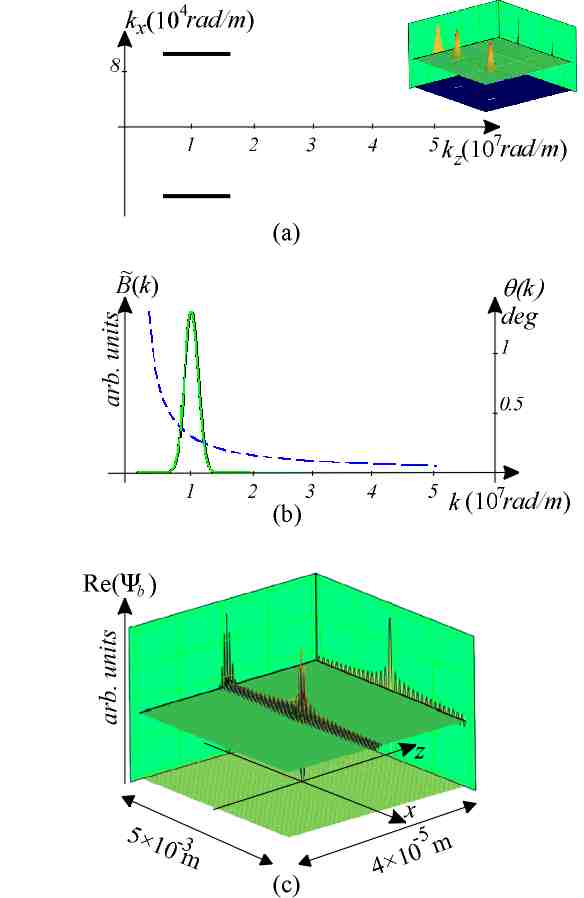}%
\caption{A numerical example of a the pulsed wave field with frequency
independent beamwidth with the parameters $\alpha_{0}=4\times10^{4}\frac
{rad}{m}$, $\gamma\sim0$: (a) The angular spectrum of plane waves in two
perspectives; (b) The frequency spectrum of the pulse (black line), the
frequency spectrum of an optically feasible wave field (green line), the cone
angle of the component Bessel beams as the function of the wave number (dashed
blue line); (c) The spatial field distribution of the pulse.}%
\label{fig512}%
\end{center}
\end{figure}
%EndExpansion
In Ref.~\cite{lw7} Campbell \textit{et al} introduced a wideband wave field
that is a superposition of the Bessel beams the cone angle of which is chosen
so that the condition%
\begin{equation}
k\sin\theta\left(  k\right)  =\alpha_{0}=const \label{fi5}%
\end{equation}
is satisfied for entire bandwidth. The condition (\ref{fi5}) implies, that the
transversal component of the wave vector of every plane wave component of the
wave field is $\alpha_{0}$, the corresponding wave field was called as the
pulsed wave fields with frequency independent beamwidth. The angular spectrum
of plane waves for such choice can be written as
\begin{equation}
\tilde{A}_{0}^{\left(  b\right)  }\left(  k,\theta\right)  =\tilde{B}\left(
k\right)  \delta\left(  \theta-\arcsin\left(  \frac{\alpha_{0}}{k}\right)
\right)  \text{,} \label{fi10}%
\end{equation}
so that the Whittaker superposition in Eq.~(\ref{ang20}) yields
\begin{equation}
\Psi_{b}\left(  \rho,z,t\right)  =J_{0}\left(  \alpha_{0}\rho\right)  \int
_{0}^{\infty}\mathrm{d}k\tilde{B}\left(  k\right)  \exp\left[  ik\left(
z\sqrt{1-\left(  \frac{\alpha_{0}}{k}\right)  ^{2}}-ct\right)  \right]
\text{,} \label{fi12}%
\end{equation}
where
\begin{equation}
\tilde{B}\left(  k\right)  =\alpha_{0}k^{2}~B\left(  k\right)  \text{.}
\label{fi14}%
\end{equation}

The support of the angular spectrum of plane waves of the wave field
Eq.~(\ref{fi12}) is depicted in Fig.~\ref{fig512} (in the numerical example
the frequency spectrum $B\left(  k\right)  $ is Gaussian with the bandwidth
corresponding to $\sim6fs$ pulse). Using the approach of section \ref{ssFLoc}
one can immediately tell the general spatial shape of such wave fields.
Indeed, in this case we have a simple special case, where the projection of
the angular spectrum of plane waves onto the $k_{x}k_{y}$-plane is delta-ring,
correspondingly, the field in transversal direction at $z=0$, $t=0$ should be
of the shape of the Bessel function. As for longitudinal shape, its envelope
is determined by the bandwidth by it Fourier transform, i.e., we should have a
slice of a Bessel beam. The numerical simulation in Fig.~\ref{fig512}d shows
that this estimate is true. Also, one can see that the wave field has
generally infinite energy flow.

Comparing the support in Eq.~(\ref{fi10}) to that of the propagation-invariant
pulsed wave field in Eq.~(\ref{su12}) and (\ref{su29}) one can see, that the
wave field (\ref{fi12}) is not propagation-invariant. Consequently, the
localized part of the wave field spreads as it propagates. We can also suggest
the best condition for limited propagation-invariance~-- the comparison of the
support in Fig.~\ref{fig512}a to those of FWM's in Fig.~\ref{fig42} implies
that for restricted bandwidths the support (\ref{fi5}) could be optimized to
approximate the "horizontal" part of the ellipsoidal supports of the
subluminal FWM's ($\gamma>1$).

\subsection{Physically realizable approximations to FWM's}

As it was explained in section \ref{ssFLoc}, the presence of the delta
function in the support of the angular spectrum of plane waves of the
free-space scalar wave fields necessarily results in infinite total energy
content of the wave field. Consequently, for all the above reviewed wave
fields the total energy content is infinite,
\begin{equation}
U_{tot}=\int_{-\infty}^{\infty}\mathrm{d}z\int_{0}^{\infty}\mathrm{d}\rho
\rho\int_{0}^{2\pi}\mathrm{d}\varphi\left\vert \Psi_{F}\left(  \rho
,z,\varphi,t\right)  \right\vert ^{2}=\infty\text{.} \label{ed1}%
\end{equation}
Here we proceed by reviewing the approaches used in literature to overcome
this difficulty. In later chapters we introduce the approach that is
especially useful for analyzing optical experiments.

\subsubsection{\label{ssED}Electromagnetic directed-energy pulse trains
(EDEPT)}%

%TCIMACRO{\FRAME{ftbpFU}{4.4244in}{2.9542in}{0pt}{\Qcb{On the effect of
%integrating over the parameter $\beta$ on the support of the angular spectrum
%of plane waves of the FWM's: (a) The special case of the "mean" support of the
%angular spectrum of plane waves where $\gamma=1$ ($v^{g}=c$), $\beta\neq0$;
%(b) The special case where $\gamma>0$ ($v^{g}<c$), again, $\beta\neq0$.}%
%}{\Qlb{fig513}}{fig5_13.jpg}{\special{ language "Scientific Word";
%type "GRAPHIC";  maintain-aspect-ratio TRUE;  display "ICON";
%valid_file "F";  width 4.4244in;  height 2.9542in;  depth 0pt;
%original-width 4.9104in;  original-height 6.6841in;  cropleft "0";
%filename 'fig5_13.jpg';file-properties "XNPEU";}} }%
%BeginExpansion
\begin{figure}
[ptb]
\begin{center}
\includegraphics[
height=2.9542in,
width=4.4244in
]%
{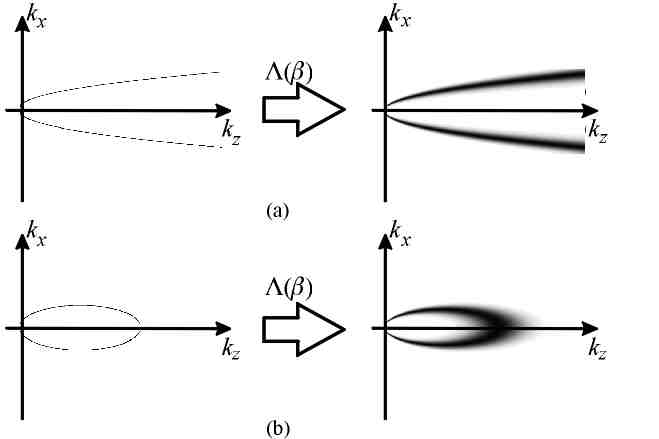}%
\caption{On the effect of integrating over the parameter $\beta$ on the
support of the angular spectrum of plane waves of the FWM's: (a) The special
case of the "mean" support of the angular spectrum of plane waves where
$\gamma=1$ ($v^{g}=c$), $\beta\neq0$; (b) The special case where $\gamma>0$
($v^{g}<c$), again, $\beta\neq0$.}%
\label{fig513}%
\end{center}
\end{figure}
%EndExpansion
One approach has been to construct various continuous superpositions of FWM's
(\ref{fwm11}) over the parameter $\beta$ (see Refs.~\cite{fw5,fw14,fw15,lw1}
and references therein), in this case one writes%
\begin{align}
\Psi_{LW}\left(  z,\rho,t\right)   &  =\int_{0}^{\infty}\mathrm{d}\beta
\Lambda\left(  \beta\right)  \Psi_{F}\left(  z,\rho,t;\beta\right) \nonumber\\
&  =\frac{a_{1}}{4\pi i\left(  a_{1}+i\zeta\right)  }\int_{0}^{\infty
}\mathrm{d}\beta\Lambda\left(  \beta\right)  \exp\left[  s\left(
z,\rho,t\right)  \right]  \text{,} \label{ed5}%
\end{align}
where%
\begin{equation}
s\left(  z,\rho,t\right)  =-\frac{\beta\rho^{2}}{a_{1}+i\zeta}+i\beta\left(
z+ct\right)  \label{ed7}%
\end{equation}
$\Lambda\left(  \beta\right)  $ is a weighting function and the subscript $LW$
means "localized wave". As the supports of the angular spectrum of the FWM's
for different values of parameter $\beta$ generally do not overlap and change
smoothly in $k$-space, the integration indeed eliminates the delta function in
the expression for angular spectrum of plane waves (see Fig.~\ref{fig513}). It
can be shown \cite{lw1} that Eq.~(\ref{ed5}) yields finite total energy wave
field if only the function $\Lambda\left(  \beta\right)  $ satisfies
condition
\begin{equation}
\frac{1}{2a_{1}}\int_{0}^{\infty}\mathrm{d}\beta\left\vert \Lambda\left(
\beta\right)  \right\vert ^{2}\frac{1}{\beta}<\infty\label{ed8}%
\end{equation}
(see Eq.~2.8 of Ref.~\cite{lw1}), i.e., if only $\beta^{-1/2}\Lambda\left(
\beta\right)  $ is square integrable. The LW's of the general form (\ref{ed5})
have been called EDEPT solutions of the scalar wave equation.

From the discussion of previous chapters it is obvious, that the wave field of
the general form (\ref{ed5}) \textit{are not }strictly propagation-invariant.
At first glance it may seem surprising because (i) FWM solutions with
different values of parameter $\beta$ \textit{do }travel without any spread
and (ii) all the FWM's overlap in every space-time point as their group
velocities are equal. However, the effect can be easily understood if we
recollect from section \ref{ssFTi} that the \textit{phase velocities} of the
pulses are different leading to the $z$ axis position dependent interference
and spread of the superposition of the component pulses (see Ref.~\cite{fw3,
fw3o1} for alternate proofs of this claim).

\paragraph{Modified power spectrum pulse (MPS)}%

%TCIMACRO{\FRAME{ftbpFU}{4.0378in}{5.5979in}{0pt}{\Qcb{ A numerical example of
%a MPS with the parameters $a_{1}=1.4\times10^{-7}m$, $a_{2}=4000m$, $q=10$,
%$p=0.0001$, $b=0.002$, ($\gamma=1$, $\beta_{0}=40\frac{rad}{m}$): (a) The
%angular spectrum of plane waves in two perspectives; (b) The $\beta$
%-distribution in bidirectional plane wave spectrum for $\alpha=0$; (c) The
%spatial field distribution of the MPS for $t=0$. }}{\Qlb{fig514}}%
%{fig5_14.jpg}{\special{ language "Scientific Word";  type "GRAPHIC";
%maintain-aspect-ratio TRUE;  display "ICON";  valid_file "F";
%width 4.0378in;  height 5.5979in;  depth 0pt;  original-width 4.9104in;
%original-height 6.6841in;  cropleft "0";  croptop "1";  cropright "1";
%cropbottom "0";  filename 'fig5_14.jpg';file-properties "XNPEU";}} }%
%BeginExpansion
\begin{figure}
[ptb]
\begin{center}
\includegraphics[
height=5.5979in,
width=4.0378in
]%
{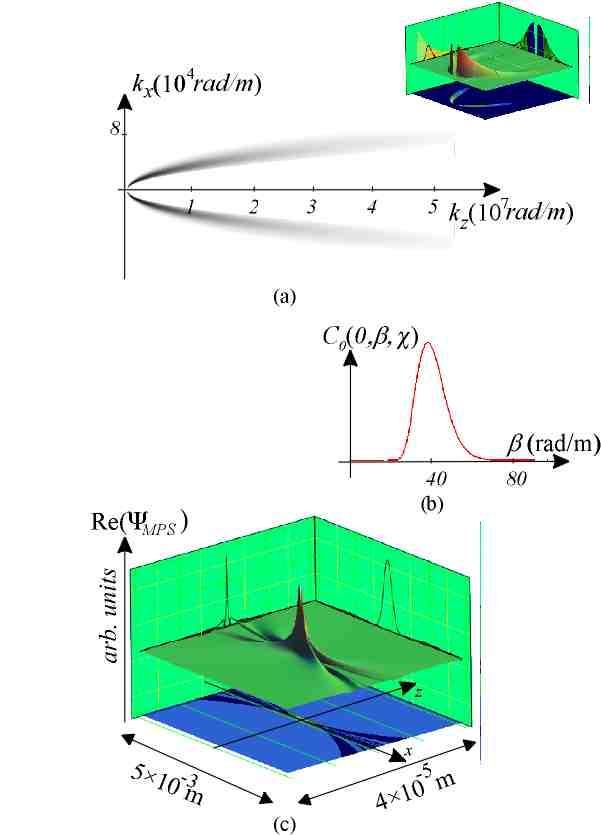}%
\caption{ A numerical example of a MPS with the parameters $a_{1}%
=1.4\times10^{-7}m$, $a_{2}=4000m$, $q=10$, $p=0.0001$, $b=0.002$, ($\gamma
=1$, $\beta_{0}=40\frac{rad}{m}$): (a) The angular spectrum of plane waves in
two perspectives; (b) The $\beta$ -distribution in bidirectional plane wave
spectrum for $\alpha=0$; (c) The spatial field distribution of the MPS for
$t=0$. }%
\label{fig514}%
\end{center}
\end{figure}
%EndExpansion
The modified power spectrum pulses \cite{lw1} have been introduced by the
following bidirectional plane wave spectrum (see Eq.~(3.3) of Ref.~\cite{lw1}
and Eq.~(3.32) of Ref.~\cite{fw14})
\begin{equation}
C_{0}^{\left(  m\right)  }\left(  \tilde{\alpha},\tilde{\beta},\chi\right)
=\left\{
\begin{array}
[c]{c}%
\frac{p\left(  p\tilde{\beta}-b\right)  ^{q-1}}{2\pi\Gamma\left(  q\right)
}\exp\left[  -\left(  \tilde{\alpha}a_{1}+\left(  p\tilde{\beta}-b\right)
a_{2}\right)  \right]  ,\,\text{if}\,\tilde{\beta}>\frac{b}{p}\\
0\quad\quad\quad\quad\quad\quad\quad\quad\quad\quad\quad\text{, if }\frac
{b}{p}>\tilde{\beta}\geq0
\end{array}
\right.  \text{.} \label{mp1}%
\end{equation}
Here $a_{2}$, $b$, $q$ and $p$ are new parameters and $\Gamma$ denotes the
gamma function. Using the relations (\ref{bidi6}) and (\ref{bidi7}) the
corresponding Whittaker type plane wave spectrum can be written as (see also
Eq.~(3.13a) and (3.13b) of Ref.~\cite{lw1})%
\begin{equation}
\tilde{A}_{0}^{\left(  MPS\right)  }\left(  \chi,k_{z}\right)  =\left\{
%TCIMACRO{\QDATOP{%
%\begin{array}
%[c]{c}%
%\frac{p\left[  \frac{p}{2}\left(  k-k_{z}\right)  -b\right]  ^{q-1}}%
%{2\pi\Gamma\left(  q\right)  }\exp\left[  -\frac{\left(  k+k_{z}\right)
%a_{1}}{2}+\left(  b-\frac{\left(  k-k_{z}\right)  p}{2}\right)  a_{2}\right]
%\\
%\quad\quad\quad\quad\quad\quad\quad\quad\text{,~if}\,k_{z}<\frac{p}{b}%
%\frac{\chi^{2}}{4}-\frac{b}{p}%
%\end{array}
%}{%
%\begin{array}
%[c]{c}%
%\quad\\
%\quad0\quad\quad\quad\quad\quad\quad\text{, if }k_{z}>\frac{p}{b}\frac
%{\chi^{2}}{4}-\frac{b}{p}%
%\end{array}
%}}%
%BeginExpansion
\genfrac{}{}{0pt}{0}{%
\begin{array}
[c]{c}%
\frac{p\left[  \frac{p}{2}\left(  k-k_{z}\right)  -b\right]  ^{q-1}}%
{2\pi\Gamma\left(  q\right)  }\exp\left[  -\frac{\left(  k+k_{z}\right)
a_{1}}{2}+\left(  b-\frac{\left(  k-k_{z}\right)  p}{2}\right)  a_{2}\right]
\\
\quad\quad\quad\quad\quad\quad\quad\quad\text{,~if}\,k_{z}<\frac{p}{b}%
\frac{\chi^{2}}{4}-\frac{b}{p}%
\end{array}
}{%
\begin{array}
[c]{c}%
\quad\\
\quad0\quad\quad\quad\quad\quad\quad\text{, if }k_{z}>\frac{p}{b}\frac
{\chi^{2}}{4}-\frac{b}{p}%
\end{array}
}%
%EndExpansion
\right.  \text{,} \label{mp2}%
\end{equation}
where $k=\sqrt{\chi^{2}+k_{z}^{2}}$ and the relation (\ref{bidi8}) has been
used. The field function of the MPS's is described by equation (see
Eqs.~(3.34) and (1.4) of Ref.~\cite{fw14})%
\begin{equation}
\Psi_{MPS}\left(  \rho,\zeta,\eta\right)  =\left[  \frac{1}{4\pi\left(
a_{1}+i\zeta\right)  }\frac{\exp\left(  -\frac{bs}{p}\right)  }{\left(
a_{2}+\frac{s}{p}\right)  ^{q}}\right]  \text{,} \label{mp4}%
\end{equation}
where
\begin{equation}
s=\frac{\rho^{2}}{4\pi\left(  a_{1}+i\zeta\right)  }-i\eta\text{.} \label{mp5}%
\end{equation}

The comparison of the bidirectional plane wave spectra of MPS (\ref{mp1}) with
that of the FWM's (\ref{fwm1}) one can see, that the latter is a special case
$a_{2}=0$, $q=1$ of the former. Consequently, the parameter $a_{1}$ in
(\ref{mp1}) has the same interpretation as in case of FWM's -- it determines
the frequency spectra of the wave field. From (\ref{mp2}) it is also obvious
that the parameter $a_{2}$ determines the width of the $\beta$ distribution
and parameter $b$ determines the central value of $\beta$. As for parameter
$q$, it can be used to optimize the shape of the $\beta$ distribution.

A numerical example of the MPS is depicted in Fig.~\ref{fig514}. In this
example we tried to optimize the parameters so as to satisfy the conditions
for optical feasibility as stated in the introduction of this overview. From
the angular spectrum of plane waves in Fig.~\ref{fig514}a one can see, that
the MPS's generally have the same inconvenience as FWM's -- there no freedom
to choose the frequency spectrum as to optimize for any convenient light
source and they are generally half-cycle pulses.

For an interpretation of MPS's as being the field generated by a combined
point-like source and a sink placed at a complex-number coordinate see
Refs.~\cite{go35,go40}

It is not our aim at this point to study the temporal behaviour of the EDEPT
solutions, thus, the wave field is calculated only for the time $t=0$.

\subsubsection{Splash pulses}%

%TCIMACRO{\FRAME{ftbpFU}{4.0378in}{5.7078in}{0pt}{\Qcb{ A numerical example of
%a Splash mode with the parameters $a_{1}=1.4\times10^{-7}m$, $a_{2}=0.4m$,
%$q=16$, ($\gamma=1$, $\beta_{0}=40\frac{rad}{m}$): (a) The angular spectrum of
%plane waves in two perspectives; (b) The $\beta$ -distribution in
%bidirectional plane wave spectrum for $\tilde{\alpha}=0$; (c) The spatial
%field distribution of the Splash mode for $t=0$. }}{\Qlb{fig515}}%
%{fig5_15.jpg}{\special{ language "Scientific Word";  type "GRAPHIC";
%maintain-aspect-ratio TRUE;  display "ICON";  valid_file "F";
%width 4.0378in;  height 5.7078in;  depth 0pt;  original-width 4.9104in;
%original-height 6.6841in;  cropleft "0";  croptop "1";  cropright "1";
%cropbottom "0";  filename 'fig5_15.jpg';file-properties "XNPEU";}} }%
%BeginExpansion
\begin{figure}
[ptb]
\begin{center}
\includegraphics[
height=5.7078in,
width=4.0378in
]%
{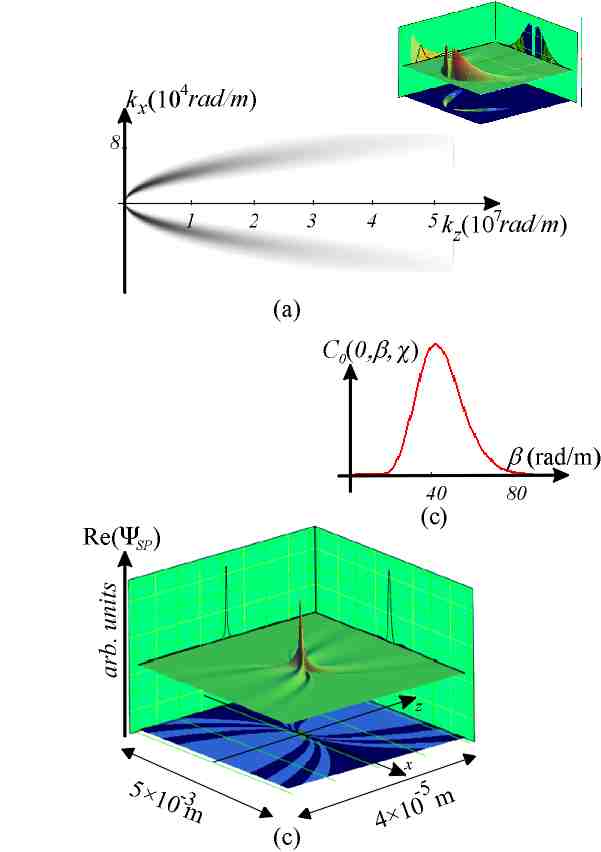}%
\caption{ A numerical example of a Splash mode with the parameters
$a_{1}=1.4\times10^{-7}m$, $a_{2}=0.4m$, $q=16$, ($\gamma=1$, $\beta
_{0}=40\frac{rad}{m}$): (a) The angular spectrum of plane waves in two
perspectives; (b) The $\beta$ -distribution in bidirectional plane wave
spectrum for $\tilde{\alpha}=0$; (c) The spatial field distribution of the
Splash mode for $t=0$. }%
\label{fig515}%
\end{center}
\end{figure}
%EndExpansion
Splash pulses \cite{fw5} appear if one chooses the bidirectional plane wave
spectrum as (Eq.~(3.13) of Ref.~\cite{fw14})%
\begin{equation}
C_{0}^{\left(  SP\right)  }\left(  \tilde{\alpha},\tilde{\beta},\chi\right)
=\frac{\pi}{2}\tilde{\beta}^{q-1}\exp\left[  -\left(  \tilde{\alpha}%
a_{1}+\tilde{\beta}a_{2}\right)  \right]  \text{.} \label{splash1}%
\end{equation}
One can see, that the bidirectional spectrum is similar in the structure as
the one of the MPS (\ref{mp1}). Here again the term $\exp\left[
-\tilde{\alpha}a_{1}\right]  $ can be interpreted as the spectra of the
"central" FWM and the parameters $a_{2}$ and $q$ determine the distribution
function over the parameter $\tilde{\beta}$. The integration in the
bidirectional plane wave decomposition (\ref{bidi1}) can be carried out to
yield (Eq.~(3.19) of Ref.~\cite{fw14}, Eq.~(17) of Ref.~\cite{fw5})
\begin{equation}
\Psi_{SP}\left(  \rho,\zeta,\eta\right)  =\frac{\Gamma\left(  q\right)  }%
{4\pi\left(  a_{1}+i\zeta\right)  }\left[  \left(  a_{2}-i\eta\right)
+\frac{1}{\left(  a_{1}+i\zeta\right)  }\right]  \label{splash2}%
\end{equation}

The wave field has been called as "splash pulse" in Ref.~\cite{fw5} as for its
characteristic spatial shape. However, in our numerical example we tried once
more to find a set of parameters suitable for optical generation. It appeared
(see Fig.~\ref{fig515}), that in this case the angular spectrum of plane waves
is very similar to that of the MPS's as in Fig.~\ref{fig514}.

\subsection{Several more LW's}

To date, the literature on LW's and on propagation of ultrashort
electromagnetic pulses is overwhelming and this overview is by no means
complete. Our aim was to demonstrate the applicability of our approach on most
important special cases.

As already mentioned, in Ref.~\cite{lw6} Besieris \textit{et al }derived
several closed-form superluminal and subluminal LW solutions to the scalar
wave equation by "boosting" known solutions of other Lorentz invariant equations.

In section \ref{ssFDo}, we already reviewed the approach of solving the
homogeneous scalar wave equation and Klein-Gordon equation, introduced in
Refs.~\cite{fw15,fw16} by Donnelly and Ziolkowski. In those works, they also
deduced various closed-form separable and non-separable solutions to the wave equation.

In Ref.~\cite{lw3} Overfelt found a continua of localized wave solutions to
the scalar homogeneous wave, damped wave, and Klein-Gordon equations by means
of a complex similarity transform technic.

The numerous publications that are involved with ultrashort-pulse solutions of
the time-dependent paraxial wave equation (e.g. isodiffracting pulses) should
be mentioned here (see Refs.~\cite{go15}--\cite{go40} and references therein).

\subsection{On the transition to the vector theory}

Even though the scalar theory is often used in description of propagation of
electromagnetic wave fields, generally the solutions to the Maxwell equations
have to be used. However, the latter approach is generally much more involved.

In the context of this review, we investigate the free-space wave fields and
mostly use the angular spectrum representation of the wave fields. In this
context the limitations of the scalar theory can be easily formulated -- the
scalar theory is reasonably accurate if only the plane wave components of the
wave field propagate at small (paraxial) angles relative to optical axis (see
Wolf and Mandel \cite{ok4}, for example). In this review we investigate the
possibilities of optical generation of FWM's (and LW's), correspondingly, the
above formulated restriction is satisfied in all practical cases and we can
restrict ourselves to scalar theory.

(Of course, this is not the case with the original FWM's and LW's published in
literature -- the plane wave components of those wave fields propagate even
perpendicularly to the direction of propagation and in their exact description
the transition to the vector theory is obligatory.)

The vector theory of FWM's and LW's has been formulated and used in several
publications \cite{fw1,fw2,fw3,fw8,lw1,ve20,ve30,ve35,x7}. The preferred
approach has been the use of the Hertz vectors as formulated in
Eqs.~(\ref{mx25}) and (\ref{mx26}). One can refer to the theory expounded by
Ziolkowski \cite{lw1} where he used the Hertz vectors of the form
\begin{subequations}
\begin{align}
\mathbf{\Pi}^{\left(  e\right)  }  &  =\mathbf{z~}\Psi_{F}\label{vv2}\\
\mathbf{\Pi}^{\left(  m\right)  }  &  =\mathbf{z~}\Psi_{F}\text{,} \label{vv3}%
\end{align}
where $\mathbf{z}$ is the unit vector along the propagation axis and $\Psi
_{F}$ is the (localized) solution of the scalar wave equation. With
(\ref{vv2}) and (\ref{vv3}) one get TE or TM field with respect to
$\mathbf{z}$ respectively. A more general treatment can be found in
Ref.~\cite{ve35}, where the Hertz vectors are written as the superpositions of
the solutions of scalar wave equation $\Psi_{i}$ as
\end{subequations}
\begin{subequations}
\begin{align}
\mathbf{\Pi}^{\left(  m\right)  }  &  =\mathbf{x}\sum_{p}a_{p}^{\left(
m\right)  }\Psi_{p}+\mathbf{y}\sum_{q}b_{q}^{\left(  m\right)  }\Psi
_{q}+\mathbf{z}\sum_{s}c_{s}^{\left(  m\right)  }\Psi_{s}\label{vv5}\\
\mathbf{\Pi}^{\left(  e\right)  }  &  =\mathbf{x}\sum_{p}a_{p}^{\left(
e\right)  }\Psi_{p}+\mathbf{y}\sum_{q}b_{q}^{\left(  e\right)  }\Psi
_{q}+\mathbf{z}\sum_{s}c_{s}^{\left(  e\right)  }\Psi_{s}\text{.} \label{vv6}%
\end{align}
The computation of the field components using (\ref{mx25}) and (\ref{mx26}),
although straightforward, results in very complex formulas.

The intuitive analysis of the effect of the transfer to exact vector theory
that is more in the spirit of this review can be carried out in terms of the
results that have been published on vector Bessel beams in
Refs.~\cite{ve10,ve12,ve25,si10}. For example, the result in Ref.~\cite{si10}
reveals, that for TE and TM fields the vector Bessel beams retain their
paraxial-Bessel beam nature up to cone angles $\sim14%
%TCIMACRO{\U{b0}}%
%BeginExpansion
{{}^\circ}%
%EndExpansion
$ and this result indeed amply justifies the use of the scalar theory in this review.

\subsubsection{The derivation of vector FWM's by directly applying the
Maxwell's equations}

To finish this chapter we nevertheless advance in some extent the second
approach mentioned in Sec.~\ref{sMax}, where we gave the general expression
for the plane wave decomposition of the solution of the free-space Maxwell equations.

To find the vector form for the FWM's as described in Eq.~(\ref{su40}) we use
the Eqs.~(\ref{mx6}) -- (\ref{mx17}). In correspondence with Eq.~(\ref{su14})
we choose
\end{subequations}
\begin{align}
\mathcal{E}_{x}\left(  \mathbf{k},\omega\right)   &  =\mathcal{\tilde{E}}%
_{x}\left(  k,\phi\right)  \delta\left[  k_{z}-k\cos\theta_{F}\left(
k\right)  \right] \label{v10}\\
\mathcal{E}_{y}\left(  \mathbf{k},\omega\right)   &  =\mathcal{\tilde{E}}%
_{y}\left(  k,\phi\right)  \delta\left[  k_{z}-k\cos\theta_{F}\left(
k\right)  \right]  \label{v12}%
\end{align}
and this choice yields from Eq.~(\ref{mx6}) and (\ref{mx7})
\begin{align}
E_{i}\left(  \mathbf{r},t\right)   &  =\frac{1}{\left(  2\pi\right)  ^{4}}%
\exp\left[  -2i\beta\gamma z\right] \label{ve4}\\
&  \times\int_{0}^{2\pi}\mathrm{d}\phi\int_{-\infty}^{\infty}\mathrm{d}%
kk^{2}\sin\theta_{F}\left(  k\right)  ~\mathcal{\tilde{E}}_{i}\left(
k,\phi\right)  \times\nonumber\\
&  \times\exp\left[  ik\left(  x\sin\theta_{F}\left(  k\right)  \cos\phi
+y\sin\theta_{F}\left(  k\right)  \sin\phi+\gamma z-ct\right)  \right]
\text{,}\nonumber
\end{align}%
\begin{align}
H_{i}\left(  \mathbf{r},t\right)   &  =\frac{1}{\left(  2\pi\right)  ^{4}}%
\exp\left[  -2i\beta\gamma z\right] \label{ve5}\\
&  \times\int_{0}^{2\pi}\mathrm{d}\phi\int_{-\infty}^{\infty}\mathrm{d}%
kk^{2}\sin\theta_{F}\left(  k\right)  \mathbf{~}\mathcal{\tilde{H}}_{i}\left(
k,\phi\right) \nonumber\\
&  \times\exp\left[  ik\left(  x\sin\theta_{F}\left(  k\right)  \cos\phi
+y\sin\theta_{F}\left(  k\right)  \sin\phi+\gamma z-ct\right)  \right]
\text{,}\nonumber
\end{align}
where from Eqs.~(\ref{mx11}) -- (\ref{mx17})
\begin{equation}
\mathcal{\tilde{E}}_{z}\left(  k,\phi\right)  =-\tan\theta_{F}\left(
k\right)  \left[  \cos\phi\mathcal{\tilde{E}}_{x}\left(  k,\phi\right)
+\sin\phi\mathcal{\tilde{E}}_{y}\left(  k,\phi\right)  \right]  \label{v7}%
\end{equation}
and
\begin{align}
\mathcal{\tilde{H}}_{x}\left(  k,\phi\right)   &  =-\frac{1}{c\mu_{0}%
\cos\theta_{F}\left(  k\right)  }\left[  \sin^{2}\theta_{F}\left(  k\right)
\sin\phi\cos\phi\mathcal{\tilde{E}}_{x}\left(  k,\phi\right)  +\right.
\nonumber\\
&  \left.  +\left(  1-\sin^{2}\theta_{F}\left(  k\right)  \cos^{2}\phi\right)
\mathcal{\tilde{E}}_{y}\left(  \phi,k\right)  \right]  \label{v9}%
\end{align}%
\begin{align}
\mathcal{\tilde{H}}_{y}\left(  k,\phi\right)   &  =\frac{k}{c\mu_{0}\cos
\theta_{F}\left(  k\right)  }\left[  \left(  1-\sin^{2}\theta_{F}\left(
k\right)  \sin^{2}\phi\right)  \mathcal{\tilde{E}}_{x}\left(  k,\phi\right)
+\right. \nonumber\\
&  \left.  +\sin^{2}\theta_{F}\left(  k\right)  \sin\phi\cos\phi
\mathcal{\tilde{E}}_{y}\left(  k,\phi\right)  \right]  \label{v21}%
\end{align}%
\begin{equation}
\mathcal{\tilde{H}}_{z}\left(  k,\phi\right)  =\frac{1}{c\mu_{0}}\sin
\theta_{F}\left(  k\right)  \left[  \sin\phi\mathcal{\tilde{E}}_{x}\left(
k,\phi\right)  -\cos\phi\mathcal{\tilde{E}}_{y}\left(  k,\phi\right)  \right]
\text{.} \label{v23}%
\end{equation}

By expanding
\begin{equation}
\mathcal{\tilde{E}}_{i}\left(  k,\phi\right)  =\sum_{n=-\infty}^{\infty
}\mathcal{\tilde{E}}_{i}\left(  k,n\right)  \exp\left[  in\phi\right]
\text{,} \label{v25}%
\end{equation}
where
\begin{equation}
\mathcal{\tilde{E}}_{i}\left(  k,n\right)  =\frac{1}{2\pi}\int_{0}^{2\pi}%
d\phi\mathcal{\tilde{E}}_{i}\left(  k,\phi\right)  \exp\left[  -in\phi\right]
\text{,} \label{v27}%
\end{equation}
the integration over the $\phi$ can be carried out to yield
\begin{align}
E_{i}\left(  \mathbf{r},t\right)   &  =\frac{1}{\left(  2\pi\right)  ^{2}}%
\exp\left[  -2i\beta\gamma z\right]  \sum_{n}\int_{-\infty}^{\infty}%
\mathrm{d}k\nonumber\\
&  \times\exp\left[  in\phi\right]  L_{i}^{\mathcal{E}}\left(  k,\rho
,n\right)  \exp\left[  ik\left(  \gamma z-ct\right)  \right]  \label{v30}%
\end{align}%
\begin{align}
H_{i}\left(  \mathbf{r},t\right)   &  =\frac{1}{\left(  2\pi\right)  ^{2}}%
\exp\left[  -2i\beta\gamma z\right]  \sum_{n}\int_{-\infty}^{\infty}%
\mathrm{d}k\nonumber\\
&  \times\exp\left[  in\phi\right]  L_{i}^{\mathcal{H}}\left(  k,\rho
,n\right)  \exp\left[  ik\left(  \gamma z-ct\right)  \right]  \text{,}
\label{v32}%
\end{align}
where
\begin{equation}
L_{x}^{\mathcal{E}}\left(  k,\rho,n\right)  =k^{2}\sin\theta_{F}\left(
k\right)  ~\mathcal{\tilde{E}}_{x}\left(  k,n\right)  J_{n}\left(  k\rho
\sin\theta_{F}\left(  k\right)  \right)  \label{v35}%
\end{equation}%
\begin{equation}
L_{y}^{\mathcal{E}}\left(  k,\rho,n\right)  =k^{2}\sin\theta_{F}\left(
k\right)  ~\mathcal{\tilde{E}}_{y}\left(  k,n\right)  J_{n}\left(  k\rho
\sin\theta_{F}\left(  k\right)  \right)  \label{v40}%
\end{equation}
and $L_{z}^{\mathcal{E}}\left(  k,n,\rho\right)  $, $L_{i}^{\mathcal{H}%
}\left(  k,\rho,n\right)  $ can be expressed as the linear combinations of
Bessel functions of different order. (see Refs.~\cite{ve12,si10,si30} for
relevant discussions).

We also note, that in addition to the TM and TE wave fields azimuthally
polarized, radially polarized and circularly polarized vector FWM's can be
derived \cite{si10}.

\subsection{Conclusions.}

The main conclusion of this section are:

\begin{itemize}
\item[1.] At this point it should be clear, that all the possible closed-form
FWM's \textit{can }be analyzed in a single framework where the
\textit{support} of the angular spectrum of plane waves (\ref{su12}) --
(\ref{su29}) is the only definitive property for propagation-invariance. The
question of whether an integration over the support has or has not a
closed-form result is the question of mathematical convenience only.

\item[2.] With a proper choice of parameters some of the closed form FWM's
(Bessel-Gauss pulses, Bessel-X) are well suited for use as the models for
simulating the result of optical experiments. In contrary, the LW's we
reviewed here -- the MPS's, splash pulses and the original FWM's -- are not
feasible in this context. Mostly it is because of the ultra-wide bandwidth and
non-paraxial angular spectrum content of the pulses.

\item[3.] In our opinion, the procedure of modeling finite-thickness supports
for finite energy approximations of FWM's reviewed in this section lacks a
convenient physical interpretation and to estimate its practical value this
topic has to be addressed in the context of a particular launching setup instead.
\end{itemize}

\section{\label{chOK}LOCALIZED WAVES IN THE THEORY OF
PARTIALLY\ COHERENT\ WAVE\ FIELDS}

Every electromagnetic field in nature has some fluctuations associated with
it-- even the purest laser light is not exactly coherent. However, in optical
region the fluctuations are too rapid for direct measurement. Their existence
can be deduced from suitable experiments where the correlation between these
fluctuations of field variables at two or more space-time points are measured.
The second-order coherence theory gives a precise measure of those
correlations for any two space-time points and formulates the dynamic laws
which the corresponding correlation functions obey. It provides a unified
treatment of all well-known interference and polarization phenomena of
traditional optics.

Our interest in introduction of the coherence theory is two-fold. First of
all, we can generalize the concept of propagation-invariance into the more
general class of optical fields where the (idealistic) fully coherent laser
light is but a special case. As the consequence we not only obtain a more
general view of the subject but also a practical approach towards the
experimental evidence of validity of the theory presented above.

In what follows we again confine ourselves to free fields only, i.e., within
the spatiotemporal domain the fields under investigation do not contain
sources (except perhaps at infinity) and they do not interact with any
material objects.

\subsection{Propagation-invariance in domain of partially coherent fields in
second order coherence theory}

\subsubsection{General definitions}

Let us start with some general notion on the subject (see Ref.~\cite{ok4}).
The definitive characteristics of a (generally) stochastic wave field in
second order coherence theory is the ensemble cross-correlation function,
often named as mutual coherence function for two points that can be defined
as
\begin{equation}
\Gamma\left(  \mathbf{r}_{1},\mathbf{r}_{2},t_{1},t_{2}\right)  =\left\langle
V^{\ast}(\mathbf{r}_{1},t_{1})V(\mathbf{r}_{2},t_{2})\right\rangle
_{e}\text{.} \label{ok0}%
\end{equation}
In this equation $V(\mathbf{r},t)$ is the (complex) field of a particular
realization of a source-free scalar field and angle brackets denote the
averaging over the ensemble of realizations, in essence the function "measure"
the correlation that exists between the light vibrations of field $V\left(
\mathbf{r},t\right)  $ at the space-time points $\left(  \mathbf{r}_{1}%
,t_{1}\right)  $ and $\left(  \mathbf{r}_{2},t_{2}\right)  $ (in terms of
statistics the two could be called processes). The direct calculation of
ensemble averages for such two stochastic processes generally requires
determination of their joint (two-fold) probability densities $p_{2}$, with
which one can write
\begin{equation}
\Gamma\left(  \mathbf{r}_{1},\mathbf{r}_{2},t_{1},t_{2}\right)  =\int\int
V_{1}^{\ast}V_{2}p_{2}\left[  V_{1}^{\ast},\mathbf{r}_{1},t_{1};V_{2}%
\mathbf{r}_{2},t_{2}\right]  \mathrm{d}V_{1}^{\ast}\mathrm{d}V_{2}\text{,}
\label{ok1}%
\end{equation}
where we have denoted the continuous set of field values in the two space-time
points as
\begin{equation}
V_{i}=V(\mathbf{r}_{i},t_{i})\text{, \quad}\,i=1,2\text{.} \label{ok2}%
\end{equation}

The joint probability density is generally unknown, however, there are several
special cases for which the mutual coherence function can be directly
calculated. For example, the two processes, $V_{1}$ and $V_{2}$ can be
statistically independent of each other, in this case the joint probability
density is completely separable:
\begin{equation}
p_{2}\left[  V_{1}^{\ast},\mathbf{r}_{1},t_{1};V_{2}\mathbf{r}_{2}%
,t_{2}\right]  =p_{1}\left(  V_{1}^{\ast},\mathbf{r}_{1},t_{1}\right)
p_{1}\left(  V_{2},\mathbf{r}_{2},t_{2}\right)  \text{,} \label{ok3}%
\end{equation}
so that the mutual coherence function is also separable, giving
\begin{align}
\Gamma\left(  \mathbf{r}_{1},\mathbf{r}_{2},t_{1},t_{2}\right)   &  =\int
V_{1}^{\ast}p_{1}\left(  V_{1}^{\ast},\mathbf{r}_{1},t_{1}\right)
\mathrm{d}V_{1}^{\ast}\int V_{2}p_{1}\left(  V_{2},\mathbf{r}_{2}%
,t_{2}\right)  \mathrm{d}V_{2}\nonumber\\
&  =\left\langle V^{\ast}(\mathbf{r}_{1},t_{1})\right\rangle _{e}\left\langle
V(\mathbf{r}_{2},t_{2})\right\rangle _{e}\text{,} \label{ok4}%
\end{align}
where $p_{1}$ is the first order probability density. The second well-known
limiting case is the complete correlation (or complete mutual coherence)
between the two processes. In this case the knowledge of one process
completely determines the second process, so that the two processes as well as
the field $V(\mathbf{r},t)$ are non-stochastic in nature. The first order
probability density then have to have the form
\begin{equation}
p_{1}\left(  V,\mathbf{r,}t\right)  =\delta\left(  V-V_{c}\left(
\mathbf{r,}t\right)  \right)  \text{,} \label{ok5}%
\end{equation}
where $V_{c}\left(  \mathbf{r,}t\right)  $ is the deterministic function of
the space-time point. For the second order probability density we have
\begin{equation}
p_{2}\left[  V_{1}^{\ast},\mathbf{r}_{1},t_{1};V_{2},\mathbf{r}_{2}%
,t_{2}\right]  =\delta\left(  V_{1}^{\ast}-V_{c}^{\ast}\left(  \mathbf{r}%
_{1}\mathbf{,}t_{1}\right)  \right)  \delta\left(  V_{2}-V_{c}\left(
\mathbf{r}_{2}\mathbf{,}t_{2}\right)  \right)  \label{ok6}%
\end{equation}
and for the mutual coherence function of the coherent wave field
\begin{equation}
\Gamma\left(  \mathbf{r}_{1},\mathbf{r}_{2},t_{1},t_{2}\right)  =V_{c}^{\ast
}\left(  \mathbf{r}_{1}\mathbf{,}t_{1}\right)  V_{c}\left(  \mathbf{r}%
_{2}\mathbf{,}t_{2}\right)  \text{.} \label{ok7}%
\end{equation}

The third special case is the stationary stochastic field. The random process
is said to be stationary if all the probability densities governing the
fluctuations of the field are invariant under an arbitrary translation of the
origin of time, i.e., if
\begin{align}
p_{1}\left[  V_{1}^{\ast},\mathbf{r}_{1},t\right]   &  =p_{1}\left[
V_{1}^{\ast},\mathbf{r}_{1},t_{1}+T\right] \label{ok8}\\
p_{2}\left[  V_{1}^{\ast},\mathbf{r}_{1},t_{1};V_{2},\mathbf{r}_{2}%
,t_{2}\right]   &  =p_{2}\left[  V_{1}^{\ast},\mathbf{r}_{1},t_{1}%
+T;V_{2},\mathbf{r}_{2},t_{2}+T\right]  \text{.} \label{ok9}%
\end{align}
In this case the mutual coherence function depends only on the difference of
time, so that it is often written as
\begin{equation}
\Gamma\left(  \mathbf{r}_{1},\mathbf{r}_{2},t_{2},t_{1}\right)  =\Gamma\left(
\mathbf{r}_{1},\mathbf{r}_{2},t_{2}-t_{1}\right)  \text{.} \label{ok10}%
\end{equation}
Note, that for the statistically independent processes the conditions
(\ref{ok3}) and (\ref{ok8}) imply that $\Gamma\left(  \mathbf{r}%
_{1},\mathbf{r}_{2},t_{1},t_{2}\right)  \equiv0$.

The above definitions are of general nature. Let us now constraint ourselves
to the stochastic superpositions of solutions of scalar wave fields. In our
context we could proceed as follows. For scalar wave fields the general
stochastic process $V(\mathbf{r},t)$ should satisfy the homogeneous scalar
wave equation and this obviously implies certain restrictions to the form of
the corresponding mutual coherence function. In the context of present study
the most convenient way to introduce them is to represent the field as the
Whittaker-type plane wave expansion:
\begin{equation}
V(\mathbf{r},t)=\int_{0}^{\infty}dk\int_{0}^{\pi}d\theta\int_{0}^{2\pi}%
d\phi\ a(k,\theta,\phi)\ \exp\left[  ik\left(  \mathbf{rn}-ct\right)  \right]
\text{,} \label{ok15}%
\end{equation}
where $a(k,\theta,\phi)$ is the stochastic, properly normalized realization of
the angular spectrum of the wave field and $\mathbf{n=}\left[  \sin\theta
\cos\phi,\ \sin\theta\sin\phi,\ \cos\theta\right]  $ is the directional unit
vector of the plane wave. In this representation the mutual coherence function
can be expanded to
\begin{align}
&  \Gamma\left(  \mathbf{r}_{1},\mathbf{r}_{2},t_{1},t_{2}\right)
=\label{ok16}\\
&  \qquad\times\iint\nolimits_{0}^{\infty}\mathrm{d}k_{1}\mathrm{d}k_{2}%
\iint\nolimits_{0}^{\pi}\mathrm{d}\theta_{1}\mathrm{d}\theta_{2}%
\iint\nolimits_{0}^{2\pi}\mathrm{d}\phi_{1}\mathrm{d}\phi_{2}\mathcal{A}%
\left(  k_{1},\phi_{1},\theta_{1},k_{2},\phi_{2},\theta_{2}\right)
\ \nonumber\\
&  \qquad\qquad\times\exp\left[  -ik_{1}\left(  \mathbf{r}_{1}\mathbf{n}%
_{1}-ct_{1}\right)  \right]  \exp\left[  ik_{2}\left(  \mathbf{r}%
_{2}\mathbf{n}_{2}-ct_{2}\right)  \right]  \text{,}\nonumber
\end{align}
where the $\mathcal{A}$ is the (Whittaker type) angular correlation function
(cross-angular spectrum density), defined as
\begin{equation}
\mathcal{A}\left(  k_{1},\phi_{1},\theta_{1},k_{2},\phi_{2},\theta_{2}\right)
\equiv\left\langle a^{\ast}(k_{1},\theta_{1},\phi_{1})a(k_{2},\theta_{2}%
,\phi_{2})\right\rangle \label{ok17}%
\end{equation}
and the mutual coherence function obeys the coupled wave equations
\begin{align}
\left(  \nabla_{1}^{2}-\frac{1}{c^{2}}\frac{\partial^{2}}{\partial t_{1}^{2}%
}\right)  \Gamma\left(  \mathbf{r}_{1},\mathbf{r}_{2},t_{1},t_{2}\right)   &
=0\nonumber\\
\left(  \nabla_{2}^{2}-\frac{1}{c^{2}}\frac{\partial^{2}}{\partial t_{2}^{2}%
}\right)  \Gamma\left(  \mathbf{r}_{1},\mathbf{r}_{2},t_{1},t_{2}\right)   &
=0\text{.} \label{ok18}%
\end{align}

The angular correlation function $\mathcal{A}\left(  {}\right)  $ is the
ensemble average of the product of the complex amplitudes $a\left(  {}\right)
$ of corresponding plane waves, i.e., it is the measure of correlation between
two plane wave components of the field. Note, that the stochasticity in this
representation lies in the correlation of amplitudes and phases of the
strictly monochromatic plane wave components in different realizations of the
field. In the coherent limit the phases and amplitudes depend on the
parameters in non-stochastic manner and do not depend on the specific realization.

In practice the ensemble averages of stationary wave fields are often replaced
by the corresponding time averages~-- it can be shown that for certain wave
fields the two averages are equal, the property is referred to as ergodicity.
To this point an apparent contradiction can be seen in representation
(\ref{ok15})~-- the stochastic part of the definition, the angular correlation
function $a\left(  {}\right)  $, do not change in time. Of course, this is
primarily the consequences of the fact that the temporal evolution of a
source-free, forward-propagating, scalar wave field is fully determined by its
values on a plane. However, to understand how the time averaging appears in
this representation one has to notice that if the field is stationary and the
correlations die out sufficiently rapidly as $t_{2}-t_{1}\rightarrow\infty$
and $\left\vert \mathbf{r}_{2}-\mathbf{r}_{1}\right\vert \rightarrow\infty$,
i.e., if the field is also ergodic, a single realization of the field can be
divided up into sections of shorter lengths that are uncorrelated and contain
all the statistical information about a realization. The time average of the
field then is the average over those shorter realizations. Note, that in this
picture the stochastic angular spectrum (and angular correlation function),
being the Fourier representation of those shorter sections, \textit{depend} on
time and the time average has to be interpreted as an average over the angular
spectrums of the shorter sections of the realization.

\subsubsection{Propagation-invariance in second order coherence theory}

The very nature of electromagnetic wave fields implies that their mutual
coherence function, the correlation of fluctuations of the field at two
space-time points, should be a propagating quantity. Hence, we can also define
the spatial localization of the mutual coherence function and discuss its
spread, either transversal or longitudinal. And this is the point where one
can use the ideas of LW's in coherent theory-- we should find the conditions
when the mutual coherence has some localization quality and when the
localization is preserved during the propagation in free space.

The mutual coherence function obeys the coupled wave equations (\ref{ok18}).
Thus, the localization and propagation of mutual coherence function obeys the
same laws as the field and in complete analogy to the coherent theory in
section \ref{ssFAng} it can be shown that the instantaneous intensity of a
particular stochastic wideband field (essentially a trail of pulses) propagate
without change along the $z$ axis only if the monochromatic components of the
field are coupled as
\begin{equation}
a_{F}(k,\theta,\phi)=a(k,\theta,\phi)\delta\left[  \theta-\theta_{F}\left(
k\right)  \right]  \text{,} \label{ok19}%
\end{equation}
where the function $\theta_{F}\left(  k\right)  $ is defined in Eq.~\ref{su12}%
, correspondingly, for the longitudinal component of the wave vectors of plane
waves we have
\begin{equation}
k_{z}=k\cos\theta_{F}\left(  k\right)  =\gamma k-2\beta\gamma\label{ok20}%
\end{equation}
and the field can be expressed as
\begin{align}
&  F^{\prime}(\mathbf{r},t)=\exp\left[  -i2\beta\gamma z\right]  \int
_{0}^{2\pi}\mathrm{d}\phi\int_{0}^{\infty}\mathrm{d}k\ \tilde{a}(k,\theta
_{F}\left(  k\right)  ,\phi)\label{ok21}\\
&  \qquad\times\exp\left[  ik\left(  x\cos\phi\sin\theta_{F}\left(  k\right)
+y\sin\phi\sin\theta_{F}\left(  k\right)  +\gamma z-ct\right)  \right]
\text{,}\nonumber
\end{align}
As already noted above, the cylindrically symmetric case of the field reads
\begin{align}
&  F\left(  \rho,z,t\right)  =\exp\left[  -i2\gamma\beta z\right]  \int
_{0}^{\infty}\mathrm{d}k\,\nonumber\\
&  \qquad\times\tilde{b}\left(  k\right)  J_{0}\left[  k\rho\sin\theta
_{F}\left(  k\right)  \right]  \exp\left[  ik\left(  \gamma z-ct\right)
\right]  \text{.} \label{ok21a}%
\end{align}
The Eqs.~(\ref{ok21}) and (\ref{ok21a}) are regarded the definition of FWM's
in second order coherence theory in what follows. One can see the intimate
relevance between the mathematical definitions of coherent and partially
coherent theory, the only difference being the stochastic nature of the
angular spectrum of plane waves for the latter. The direct correspondence
between the two also allows us to transfer the results on finite energy
content LW's described above, so that the FWM (\ref{ok21}) or (\ref{ok21a}) is
essentially the infinite energy content limit of the broader class of fields,
LW's. Here we proceed by studying the consequences of the stochastic nature of
the angular spectrum of plane waves of the partially coherent FWM's.

The angular correlation function for the partially coherent FWM's reads
\begin{align}
&  \left\langle a_{F}^{\ast}\left(  k_{1},\theta_{1},\phi_{1}\right)
a_{F}\left(  k_{2},\theta_{2},\phi_{2}\right)  \right\rangle =\label{ok22}\\
&  \qquad\qquad\qquad\mathcal{A}_{F}\left(  k_{1},k_{2},\phi_{1},\phi
_{2}\right)  \delta\left[  \theta_{1}-\theta_{F}\left(  k_{1}\right)  \right]
\delta\left[  \theta_{2}-\theta_{F}\left(  k_{2}\right)  \right]
\text{,}\nonumber
\end{align}
i.e., the cross-angular spectrum vanishes unless $\theta_{1}=\theta_{2}%
=\theta_{F}\left(  k\right)  $. The general expression for the mutual
coherence function of the propagation-invariant partially coherent FWM's
reads
\begin{align}
&  \Gamma_{F}\left(  \mathbf{r}_{1},\mathbf{r}_{2},t_{1},t_{2}\right)
=\exp\left[  -i2\beta\gamma\left(  z_{2}-z_{1}\right)  \right]  \int
_{0}^{\infty}\int_{0}^{\infty}\mathrm{d}k_{1}\mathrm{d}k_{2}\int_{0}^{2\pi
}\int_{0}^{2\pi}\mathrm{d}\phi_{1}\mathrm{d}\phi_{2}\nonumber\\
&  \qquad\times\mathcal{A}_{F}\left(  k_{1},k_{2},\phi_{1},\phi_{2}\right)
\exp\left\{  ik_{2}\left[  \mathbf{r}_{\bot2}\mathbf{n}_{\bot2}\left(
k_{2}\right)  +\left(  \gamma z_{2}-ct_{2}\right)  \right]  \right\}
\nonumber\\
&  \qquad\qquad\times\exp\left\{  -ik_{1}\left[  \mathbf{r}_{\bot1}%
\mathbf{n}_{\bot1}\left(  k_{1}\right)  +\left(  \gamma z_{1}-ct_{1}\right)
\right]  \right\}  \text{,} \label{ok23}%
\end{align}
where $\mathbf{r}_{\bot j}=\left(  x_{j},y_{j}\right)  $, and $\mathbf{n}%
_{\bot j}\left(  k_{j}\right)  =\left(  \cos\phi_{j}\sin\theta_{F}\left(
k\right)  ,\sin\phi_{j}\sin\theta_{F}\left(  k\right)  \right)  $, $j=1,2$.
The mutual coherence function (\ref{ok23}) depends on the longitudinal
coordinate and time instant through expression $z\gamma-ct$ (or, equivalently,
$z-v^{g}t$) and through the $z$ dependent overall phase and can be expressed
as
\begin{align}
\Gamma_{F}\left(  \mathbf{r}_{1},\mathbf{r}_{2},t_{1},t_{2}\right)   &
=\exp\left[  -i2\beta\gamma\left(  z_{2}-z_{1}\right)  \right] \nonumber\\
&  \times G\left(  x_{1},y_{1},z_{1}\gamma-ct_{1},x_{2},y_{2},z_{2}%
\gamma-ct_{2}\right)  \text{.} \label{ok24}%
\end{align}
This is the direct analog of the corresponding expression in the coherent
theory (\ref{gau1}).

Without the loss of generality we can introduce a partitioning of the angular
correlation function as
\begin{equation}
\mathcal{A}_{F}\left(  k_{1},k_{2},\phi_{1},\phi_{2}\right)  \equiv
\mathcal{V}^{\ast}\left(  k_{1},\phi_{1}\right)  \mathcal{V}\left(  k_{2}%
,\phi_{2}\right)  \mathcal{C}\left(  k_{1},k_{2},\phi_{1},\phi_{2}\right)
\text{.} \label{ok25}%
\end{equation}
Since the modified angular correlation function $\mathcal{C}$ depends on all
variables in non-factored manner, this expression still describes the most
general case of the non-stationary non-homogeneous partially coherent FWM's.
The factors $\mathcal{V}\left(  ..\right)  $ in Eq.~(\ref{ok25}) can be given
the interpretation as being the square root of the angular spectrum density
$S\left(  k,\phi\right)  $ multiplied by a phase constant, i.e.,%
\begin{align}
\mathcal{V}\left(  k,\phi\right)   &  =\left\langle a^{\ast}(k,\theta
_{F}\left(  k\right)  ,\phi)a(k,\theta_{F}\left(  k\right)  ,\phi
)\right\rangle ^{\frac{1}{2}}\exp\left[  i\upsilon\left(  k,\phi\right)
\right] \nonumber\\
&  =\left[  S\left(  k,\phi\right)  \right]  ^{\frac{1}{2}}\exp\left[
i\upsilon\left(  k,\phi\right)  \right]  \text{,} \label{ok27}%
\end{align}
where the phase factor $\upsilon\left(  k,\phi\right)  $ is a real quantity
and is essentially the extracted phase of the factor $\mathcal{V}\left(
..\right)  $.

As the factors $\mathcal{V}\left(  ..\right)  $ do not depend on specific
realization, the statistical, correlation properties of the field have to be
determined by the function $\mathcal{C}$ , for example, the special case where
the function is constant corresponds to full correlation-- obviously the
integrand in Eq.~(\ref{ok23}) is nonzero only for the pairs of plane waves
that correlate in different realizations for the ensemble.

To summarize this section we note, that the defining property of the partially
coherent FWM's is the same as for their fully coherent counterparts: the
support of their angular spectrum of plane waves obeys the linearity condition
in Eq.~(\ref{ok20}). In following section we discuss several limiting special
cases of the angular correlation function in Eq.~(\ref{ok25}).

Note, that the following approach can be considered as the generalization of
the discussion on monochromatic propagation-invariant wave fields in
Refs.~\cite{ok1,ok2,ok6}.

\subsection{\label{sOF}Special cases of partially coherent FWM's}

\subsubsection{\label{ssOF0}Coherent limit}

For fully coherent wave fields the phases and amplitudes of the plane waves in
Eqs.~(\ref{ok24}) and (\ref{ok25}) are invariant of the specific realization
of the field and all the plane wave components are fully correlated, hence,
the non-factored part of the angular correlation function can be expressed as
\begin{equation}
\mathcal{C}\left(  k_{1},k_{2},\phi_{1},\phi_{2}\right)  \mathcal{\propto
\,}const=1 \label{o1}%
\end{equation}
the angular correlation function $\mathcal{A}$ factorizes to
\begin{equation}
\mathcal{A}_{F}\left(  k_{1},k_{2},\phi_{1},\phi_{2}\right)  \mathcal{=V}%
^{\ast}\left(  k_{1},\phi_{1}\right)  \mathcal{V}\left(  k_{2},\phi
_{2}\right)  \label{o7}%
\end{equation}
and the mutual coherence function factorizes as
\begin{equation}
\Gamma_{F}\left(  \mathbf{r}_{1},\mathbf{r}_{2},t_{1},t_{2}\right)  =F^{\ast
}\left(  \mathbf{r}_{1},t_{1}\right)  F\left(  \mathbf{r}_{2},t_{2}\right)
\text{,} \label{o10}%
\end{equation}
where
\begin{align}
F\left(  \mathbf{r},t\right)   &  =\exp\left[  -i2\beta\gamma z\right]
\int_{0}^{2\pi}d\phi\int_{0}^{\infty}dk\ \nonumber\\
&  \times\mathcal{V}\left(  k,\phi\right)  \exp\left[  ik\left(
\mathbf{r}_{\bot}\mathbf{n}_{\bot}\left(  k\right)  +\left(  \gamma
z-ct\right)  \right)  \right]  \text{.} \label{o12}%
\end{align}
Thus, the factor $\mathcal{V}\left(  k,\phi\right)  $ as defined in
Eq.~(\ref{ok25}) appears as the plane wave spectrum of the coherent wave field
the multiplier $\exp\left[  ib\left(  k,\phi\right)  \right]  $ defining the
phase of the monochromatic components. The field (\ref{o12}) could be
generated from mode-locked femtosecond laser pulses in a stable linear-optical setup.

To conclude this section we note, that the Eq.~(\ref{ok23}) implies that the
mutual coherence function of a nonstationary field is generally time-dependent
quantity. However in optical domain the function of the field are far too
rapid for direct measurement, so, in optical experiments a time-averaged
mutual coherence function appears
\begin{equation}
\left\langle \Gamma\left(  \mathbf{r}_{1},\mathbf{r}_{2},t_{1},t_{2}\right)
\right\rangle =\frac{1}{\Delta T}\int_{\Delta T}\mathrm{d}t~V^{\ast}\left(
\mathbf{r}_{1},t-t_{1}\right)  V\left(  \mathbf{r}_{2},t-t_{2}\right)
\text{.} \label{o14}%
\end{equation}
If we confine ourselves to cylindrically symmetric wave fields the
time-averaging can be carried out to yield%
\begin{align}
&  \Gamma_{F}\left(  \mathbf{r}_{1\bot},\mathbf{r}_{2\bot},\Delta
z,\gamma\Delta z-c\tau\right)  =\exp\left[  -i\beta\gamma\Delta z\right]
\int_{0}^{\infty}\mathrm{d}k\left\vert \mathcal{V}\left(  k\right)
\right\vert ^{2}\label{o16}\\
&  \;\;\;\;\;\;\;\times J_{0}\left(  k\rho_{1}\sin\theta_{F}\left(  k\right)
\right)  J_{0}\left(  k\rho_{2}\sin\theta_{F}\left(  k\right)  \right)
\exp\left[  ik\left(  \gamma\Delta z-c\tau\right)  \right]  \text{.}\nonumber
\end{align}

\subsubsection{\label{ssOF1}FWM's with frequency noncorrelation}

Consider the special case where the plane waves of different wave number are
not correlated, however, the field is spatially fully coherent at some
particular frequency throughout a volume, i.e., the field is completely
coherent in space-fequency domain. In this case the nonfactored part of the
angular correlation function $\mathcal{A}_{F}$ has the form
\begin{equation}
\mathcal{C}\left(  k_{1},k_{2},\phi_{1},\phi_{2}\right)  \mathcal{\propto
}\delta\left(  k_{1}-k_{2}\right)  \text{,} \label{f1}%
\end{equation}
so that
\begin{equation}
\mathcal{A}_{F}\left(  k_{1},k_{2},\phi_{1},\phi_{2}\right)  \mathcal{=V}%
^{\ast}\left(  k_{1},\phi_{1}\right)  \mathcal{V}\left(  k_{2},\phi
_{2}\right)  \delta\left(  k_{1}-k_{2}\right)  \label{f3}%
\end{equation}
and the field is stationary, i.e., its statistical properties do not depend on
time origin. The cross-spectral density of the field also factorized
(\cite{ok4}, Eq.~(4.5.73))
\begin{equation}
W\left(  \mathbf{r}_{1},\mathbf{r}_{2},k\right)  =\mathcal{U}^{\ast}\left(
\mathbf{r}_{1},k\right)  \mathcal{U}\left(  \mathbf{r}_{2},k\right)  \text{,}
\label{f5}%
\end{equation}
where
\begin{align}
\mathcal{U}\left(  \mathbf{r},k\right)   &  \equiv\exp\left[  -i\beta\gamma
z\right]  \int_{0}^{2\pi}\mathrm{d}\phi\mathcal{V}\left(  k,\phi\right)
\ \exp\left[  ik\left(  \mathbf{r}_{\bot}\mathbf{n}_{\bot}\left(  k\right)
+\left(  \gamma z-ct\right)  \right)  \right] \nonumber\\
&  =\exp\left[  -i\beta\gamma z\right]  \sum_{n=0}^{\infty}\exp\left[  \pm
in\varphi\right]  \mathcal{V}_{n}\left(  k\right)  J_{n}\left[  k~\mathbf{r}%
_{\bot}\mathbf{n}_{\bot}\left(  k\right)  \right]  \exp\left[  ik\left(
\gamma z-ct\right)  \right]  \label{f7}%
\end{align}
is the temporal Fourier transform of the field and the quantity $\mathcal{V}%
\left(  k\right)  $ can be found from the relation
\begin{equation}
\int_{0}^{2\pi}\mathrm{d}\phi\mathcal{V}\left(  k,\phi\right)  \ \exp\left[
ik\mathbf{r}_{\bot}\mathbf{n}_{\bot}\right]  =\sum_{n=0}^{\infty}%
\mathcal{V}_{n}\left(  k\right)  \exp\left[  \pm in\varphi\right]
J_{n}\left(  k~\mathbf{r}_{\bot}\mathbf{n}_{\bot}\right)  \text{.} \label{f11}%
\end{equation}
So that the cross-spectral density function can be expressed as
\begin{align}
&  W(x_{1},z_{1},x_{2},z_{2},k)=\exp\left[  -i2\beta\gamma\left(  z_{2}%
-z_{1}\right)  \right]  \sum_{n=0}^{\infty}\exp\left[  \pm i\phi\left(
m-n\right)  \right] \label{f12a}\\
&  \quad\quad\quad\times\left\vert \mathcal{V}_{n}\left(  k\right)
\right\vert ^{2}J_{n}\left(  k\rho_{1}\sin\theta_{F}\left(  k\right)  \right)
J_{m}\left(  k\rho_{2}\sin\theta_{F}\left(  k\right)  \right)  \exp\left[
ik\gamma\left(  z_{2}-z_{1}\right)  \right]  \text{.}\nonumber
\end{align}

The mutual coherence function can be written either by taking the Fourier'
transform of the cross-spectral density (\ref{f12a})
\begin{align}
\Gamma_{F}\left(  \mathbf{r}_{1},\mathbf{r}_{2},\tau\right)   &  =\left\langle
V^{\ast}\left(  \mathbf{r}_{1},t\right)  V\left(  \mathbf{r}_{2}%
,t+\tau\right)  \right\rangle \nonumber\\
&  =\int_{0}^{\infty}\mathrm{d}k\ \mathcal{U}^{\ast}\left(  \mathbf{r}%
_{1},k\right)  \mathcal{U}\left(  \mathbf{r}_{2},k\right)  e^{-ikc\tau
}\text{,} \label{f13}%
\end{align}
or by inserting the angular correlation function into the general
Eq.~(\ref{ok23}). As the result we get
\begin{align}
&  \Gamma_{F}\left(  \mathbf{r}_{1\bot},\mathbf{r}_{2\bot},\Delta
z,\gamma\Delta z-c\tau\right)  =\exp\left[  -i\beta\gamma\Delta z\right]
\nonumber\\
&  \quad\quad\quad\quad\quad\times\int_{0}^{2\pi}\int_{0}^{2\pi}\mathrm{d}%
\phi_{1}\mathrm{d}\phi_{2}\int_{0}^{\infty}\mathrm{d}k\mathcal{V}^{\ast
}\left(  \phi_{1},k\right)  \mathcal{V}\left(  \phi_{2},k\right) \label{f17}\\
&  \quad\quad\quad\quad\times\exp\left[  ik\left(  \mathbf{r}_{\bot
1}\mathbf{n}_{\bot1}\left(  k_{1}\right)  -\mathbf{r}_{\bot2}\mathbf{n}%
_{\bot2}\left(  k_{2}\right)  +\gamma\Delta z-c\tau\right)  \right]
\text{.}\nonumber
\end{align}
The integration over the azimuthal angle gives:
\begin{align}
&  \Gamma_{F}\left(  \mathbf{r}_{1\bot},\mathbf{r}_{2\bot},\Delta
z,\gamma\Delta z-c\tau\right)  =\exp\left[  -i\beta\gamma\Delta z\right]
\sum_{n=0}^{\infty}\sum_{m=0}^{\infty}\label{f19}\\
&  \quad\quad\quad\quad\times\exp\left[  \pm i\left(  n\varphi_{1}%
-m\varphi_{2}\right)  \right]  \int_{0}^{\infty}\mathrm{d}k\mathcal{V}%
_{n}^{\ast}\left(  k\right)  \mathcal{V}_{m}\left(  k\right) \nonumber\\
&  \quad\quad\quad\quad\times J_{n}\left(  k\rho_{1}\sin\theta_{F}\left(
k\right)  \right)  J_{m}\left(  k\rho_{2}\sin\theta_{F}\left(  k\right)
\right)  \exp\left[  ik\left(  \gamma\Delta z-c\tau\right)  \right]
\text{,}\nonumber
\end{align}
so that for cylindrically symmetric fields we have%
\begin{align}
&  \Gamma_{F}\left(  \mathbf{r}_{1\bot},\mathbf{r}_{2\bot},\Delta
z,\gamma\Delta z-c\tau\right)  =\exp\left[  -i\beta\gamma\Delta z\right]
\int_{0}^{\infty}\mathrm{d}k\left\vert \mathcal{V}_{0}\left(  k\right)
\right\vert ^{2}\nonumber\\
&  \quad\quad\quad\times J_{0}\left(  k\rho_{1}\sin\theta_{F}\left(  k\right)
\right)  J_{0}\left(  k\rho_{2}\sin\theta_{F}\left(  k\right)  \right)
\exp\left[  ik\left(  \gamma\Delta z-c\tau\right)  \right]  \text{,}
\label{f20}%
\end{align}
this result is identical to the time-averaged mutual coherence function in
(\ref{o16}).

In general the full coherence of the light in space-frequency domain can be
achieved by filtering the incoherent light from a nearly blackbody source by a
small (delta) pinhole. The corresponding partially coherent FWM's, described
by the Eq.~(\ref{ok23}) are characterized by the two properties:

\begin{itemize}
\item[1.] Their mutual coherence function depends only on the difference
$\Delta z$, i.e., its is invariant of the origin on the $z$ axis. This also
means, that the mutual coherence function propagates without change.

\item[2.] In general the intensity of the stationary field can be expressed
by
\begin{equation}
I\left(  \mathbf{r}\right)  =\Gamma\left(  \mathbf{r},\mathbf{r},0\right)
\label{f21}%
\end{equation}
then from Eq.~(\ref{f19}) it can be seen that the averaged intensity of the
FWM's is described by
\begin{align}
I\left(  \mathbf{r}\right)   &  =\sum_{n=0}^{\infty}\sum_{m=0}^{\infty}%
\exp\left[  \pm i\left(  n\varphi_{1}-m\varphi_{2}\right)  \right]
\label{f22}\\
&  \times\int_{0}^{\infty}\mathrm{d}k\mathcal{V}_{n}^{\ast}\left(  k\right)
\mathcal{V}_{m}\left(  k\right)  J_{n}\left(  k\rho\sin\theta_{F}\left(
k\right)  \right)  J_{m}\left(  k\rho\sin\theta_{F}\left(  k\right)  \right)
\nonumber
\end{align}
and this expression do not depend on location on $z$ axis. In particular, if
the field is cylindrically symmetric we have
\begin{equation}
I\left(  \mathbf{r}\right)  =\int_{0}^{\infty}\mathrm{d}k\left\vert
\mathcal{V}_{0}\left(  k\right)  \right\vert ^{2}J_{0}^{2}\left(  k\rho
\sin\theta_{F}\left(  k\right)  \right)  \text{.} \label{f25}%
\end{equation}
For wideband fields the integration yields a localized on-axis spot, for
quasi-\allowbreak monochro\allowbreak matic wave fields the intensity in near
axis volume is a close approximation to that of the zeroth-order partially
coherent Bessel beam $I\left(  \mathbf{r}\right)  \sim J_{0}^{2}\left(
k\rho\sin\theta_{F}\left(  k\right)  \right)  $.
\end{itemize}

\subsubsection{\label{ssOF2}FWM's with angular noncorrelation}

Consider the special case where the plane waves of different azimuthal angle
are not correlated (are directionally $\delta$ correlated) but for every
particular direction the components add up coherently. In this case the
nonfactored part of the angular correlation function $\mathcal{A}_{F}$ of the
partially coherent FWM's has the form
\begin{equation}
\mathcal{C}\left(  k_{1},k_{2},\phi_{1},\phi_{2}\right)  \mathcal{\propto
}\delta\left(  \phi_{1}-\phi_{2}\right)  \text{,} \label{a1}%
\end{equation}
so that
\begin{equation}
\mathcal{A}_{F}\left(  k_{1},k_{2},\phi_{1},\phi_{2}\right)  \mathcal{=V}%
^{\ast}\left(  k_{1},\phi_{1}\right)  \mathcal{V}\left(  k_{2},\phi
_{2}\right)  \delta\left(  \phi_{1}-\phi_{2}\right)  \label{a3}%
\end{equation}
and for the mutual coherence function we have
\begin{align}
&  \Gamma_{F}\left(  \mathbf{r}_{1},\mathbf{r}_{2},t_{1},t_{2}\right)
=\exp\left[  -i2\beta\gamma\left(  z_{2}-z_{1}\right)  \right] \nonumber\\
&  \qquad\qquad\times\int_{0}^{2\pi}\mathrm{d}\phi\int_{0}^{\infty}\int
_{0}^{\infty}\mathrm{d}k_{1}\mathrm{d}k_{2}\mathcal{V}^{\ast}\left(
k_{1},\phi\right)  \mathcal{V}\left(  k_{2},\phi\right) \nonumber\\
&  \qquad\qquad\times\exp\left[  -ik_{1}\left[  \mathbf{r}_{\bot1}%
\mathbf{n}_{\bot}\left(  k_{1}\right)  -\left(  \gamma z_{2}-ct_{2}\right)
\right]  \right] \nonumber\\
&  \qquad\qquad\times\exp\left[  ik_{2}\left[  \mathbf{r}_{\bot2}%
\mathbf{n}_{\bot}\left(  k_{2}\right)  +\left(  \gamma z_{2}-ct_{2}\right)
\right]  \right]  ~\text{,} \label{a5}%
\end{align}
The integration over the azimuthal angle gives:
\begin{align}
&  \Gamma_{F}\left(  \mathbf{r}_{1},\mathbf{r}_{2},t_{1},t_{2}\right)
=\exp\left[  -i2\beta\gamma\left(  z_{2}-z_{1}\right)  \right] \nonumber\\
&  \qquad\qquad\times\sum_{n=0}^{\infty}\exp\left[  \pm in\phi\right]
\int_{0}^{\infty}\int_{0}^{\infty}\mathrm{d}k_{1}\mathrm{d}k_{2}%
\mathcal{V}_{n}^{\ast}\left(  k_{1}\right)  \mathcal{V}_{n}\left(
k_{2}\right) \nonumber\\
&  \qquad\qquad\times J_{n}\left(  k_{1}\rho_{2}\sin\theta_{F}\left(
k_{1}\right)  \right)  J_{n}\left(  k_{2}\rho_{2}\sin\theta_{F}\left(
k_{2}\right)  \right) \nonumber\\
&  \qquad\qquad\times\exp\left[  ik_{1}\left(  \gamma z_{1}-ct_{1}\right)
\right]  \exp\left[  -ik_{2}\left(  \gamma z_{2}-ct_{2}\right)  \right]
\text{,} \label{a7}%
\end{align}
The expression (\ref{a5}) can be rewritten as
\begin{equation}
\Gamma_{F}\left(  \mathbf{r}_{1},\mathbf{r}_{2},t_{1},t_{2}\right)  =\int
_{0}^{2\pi}\mathrm{d}\phi\Psi^{\ast}\left(  \phi,\mathbf{r}_{1},t_{1}\right)
\Psi\left(  \phi,\mathbf{r}_{2},t_{2}\right)  \text{,} \label{a9}%
\end{equation}
where
\begin{align}
\Psi &  \left(  \phi,\mathbf{r}_{j},t_{j}\right)  \equiv\exp\left[
-i2\beta\gamma z_{j}\right]  \int_{0}^{\infty}\mathrm{d}k\mathcal{V}\left(
k,\phi\right) \label{a11a}\\
&  \qquad\times\exp\left[  -ik\left(  \mathbf{r}_{\bot j}\mathbf{n}_{\bot
}\left(  k\right)  +\left(  \gamma z-ct_{j}\right)  \right)  \right]
~\text{.}\nonumber
\end{align}

In this special case the plane wave constituents for every particular
direction add up coherently resulting in a pulsed tilted plane wave
constituents $\Psi\left(  \phi,\mathbf{r},t\right)  $ that have a (generally)
specific profile for every direction $\phi$. The integration in Eq.~(\ref{a5})
generally yields an incoherent angular mixture of fully coherent plane wave
pulses and $\Gamma$ factorizes only asymptotically in the far field. Such
fields might be formed from mode-locked femtosecond laser pulses in an optical
set-up containing elements or parameters which fluctuate (slowly as compared
to the pulse repetition rate) -- moving scattering media, etc.

The FWM's of this type have the following properties:

\begin{itemize}
\item[1.] Again, the mutual coherence function propagates without any change.
However, the time origin and the origin on the longitudinal axis is important
now as the field is not stationary.

\item[2.] The time averaged intensity can be found from the mutual coherence
functions as
\begin{align}
&  \Gamma_{F}\left(  \mathbf{r},\mathbf{r},t,t\right)  =\int_{0}^{2\pi
}\mathrm{d}\phi\int_{0}^{\infty}\int_{0}^{\infty}\mathrm{d}k_{1}%
\mathrm{d}k_{2}\mathcal{V}^{\ast}\left(  k_{1},\phi_{1}\right)  \mathcal{V}%
\left(  k_{2},\phi_{2}\right)  \times\label{a13}\\
&  \qquad\times\exp\left[  -i\left(  k_{2}\mathbf{n}_{\bot}\left(
k_{2}\right)  -k_{1}\mathbf{n}_{\bot}\left(  k_{1}\right)  \right)
\mathbf{r}_{\bot}+\left(  k_{2}-k_{1}\right)  \left(  \gamma z-ct\right)
\right]  \text{,}\nonumber
\end{align}
so that
\begin{equation}
I\left(  \mathbf{r}\right)  =\int_{0}^{2\pi}\mathrm{d}\phi\int_{0}^{\infty
}\mathrm{d}k\left\vert \mathcal{V}\left(  k,\phi\right)  \right\vert ^{2}
\label{a15}%
\end{equation}
and due to the noncorrelation of the tilted pulses the recording system see
uniform intensity distribution. However, the instantaneous intensity $V^{\ast
}V$ in this case strongly depends on location on $z$ axis and time origin.
\end{itemize}

\subsubsection{\label{ssOF3}FWM's with angular and frequency noncorrelation}

We conclude this discussion with the special case where all the plane waves
are uncorrelated. The non-factored part of the angular correlation function
then can be expressed as
\begin{equation}
\mathcal{C}\left(  k_{1},k_{2},\phi_{1},\phi_{2}\right)  \mathcal{\propto
}\delta\left(  \phi_{1}-\phi_{2}\right)  \delta\left(  k_{1}-k_{2}\right)
\text{,} \label{af1}%
\end{equation}
so that the angular correlation function $\mathcal{A}_{F}$ does not factorize
neither with respect of the angular variables nor of the frequency:
\begin{equation}
\mathcal{A}_{F}\left(  k_{1},k_{2},\phi_{1},\phi_{2}\right)  \mathcal{=V}%
^{\ast}\left(  k_{1},\phi_{1}\right)  \mathcal{V}\left(  k_{2},\phi
_{2}\right)  \delta\left(  \phi_{1}-\phi_{2}\right)  \delta\left(  k_{1}%
-k_{2}\right)  \text{.} \label{af2}%
\end{equation}
The corresponding mutual coherence function can be expressed as
\begin{align}
\Gamma_{F}\left(  \mathbf{r}_{1},\mathbf{r}_{2},t_{1},t_{2}\right)   &
\equiv\Gamma_{F}\left(  \mathbf{r}_{1}-\mathbf{r}_{2},t_{1}-t_{2}\right)
\label{af4}\\
&  =\exp\left[  -i2\beta\gamma\left(  z_{2}-z_{1}\right)  \right]  \int
_{0}^{\infty}\mathrm{d}k\int_{0}^{2\pi}\mathrm{d}\phi\left\vert \mathcal{V}%
\left(  k,\phi\right)  \right\vert ^{2}\nonumber\\
&  \times\exp\left[  -ik\left[  \mathbf{n}_{\bot}\left(  k\right)  \left(
\mathbf{r}_{\bot1}-\mathbf{r}_{\bot1}\right)  -\left(  \gamma\Delta z-c\Delta
t\right)  \right]  \right]  \text{,}\nonumber
\end{align}
The integration over the azimuthal angle gives:
\begin{align}
&  \Gamma_{F}\left(  \mathbf{r}_{1},\mathbf{r}_{2},t_{1},t_{2}\right)
=\exp\left[  -i2\beta\gamma\left(  z_{2}-z_{1}\right)  \right] \label{af6}\\
&  \qquad\times\sum_{n=0}^{\infty}\exp\left[  \pm in\phi\right]  \int
_{0}^{\infty}\mathrm{d}k\left\vert \mathcal{V}_{n}\left(  k\right)
\right\vert ^{2}\nonumber\\
&  \qquad\times J_{n}\left[  k\left(  \rho_{2}-\rho_{1}\right)  \sin\theta
_{F}\left(  k\right)  \right]  \exp\left\{  ik\left[  \gamma\left(
z_{2}-z_{1}\right)  -c\left(  t_{2}-t_{1}\right)  \right]  \right\}
\text{.}\nonumber
\end{align}

The above equations show, that in case of superposition of completely
uncorrelated plane waves the mutual coherence function $\Gamma$ is homogeneous
and stationary, i.e., it depends only on differences of its arguments.

The FWM's of this type have the following properties:

\begin{itemize}
\item[1.] Despite of the complete noncorrelation the mutual coherence function
of this type of partially coherent FWM's can be localized and it propagates
without any change. Particularly, the field is propagation-invariant in the
sense that $\Gamma$ depends on the longitudinal coordinate $z$ through the
difference $z_{2}-z_{1}$ only.

\item[2.] The time-averaged instantaneous intensity is described by
\begin{equation}
I\left(  \mathbf{r}\right)  =\int_{0}^{2\pi}\mathrm{d}\phi\int_{0}^{\infty
}\mathrm{d}k~\left\vert \mathcal{V}\left(  \phi,k\right)  \right\vert
^{2}=const\text{,} \label{af8}%
\end{equation}
i.e., it is uniform in space.
\end{itemize}

This is the field, that may be viewed as the opposite of an coherent FWM in
variable-spatial-coherence optics: its intensity is uniform along all spatial
directions and in time, whereas the sharply peaked behavior, characteristic to
FWM's, reveals itself in the mutual coherence functions $\Gamma_{F}$.

\subsection{Conclusions}

In this chapter, we have generalized the concept of propagation-invariance
into the domain of partially coherent wave fields. In the case of partially
coherent LW's the propagation-invariant, spatially localized field variable is
its mutual coherence function. Nevertheless the mathematical description is
similar to that used for coherent wave fields -- the angular correlation
function of the partially coherent LW's, being the counterpart of the angular
spectrum of plane waves in the second order coherence theory, is defined using
principles known from coherent theory.

\section{\label{chG}OPTICAL GENERATION OF LW'S}

\subsection{Introduction}

Despite the extensive theoretical work carried out on FWM's and LW's, for the
long time there was very few experimental verification of the concept of
propagation-invariance of pulsed wave fields. The only experimental
verification on LW's was the launching the "acoustic directed energy pulse
trains" for ultrasonic waves in water \cite{g0a,g0b}. Theoretically the
problem has been addressed in numerous publications \cite{g0}] -- \cite{g25}
very seriously -- in a letter to the author of this review of 17.12.2002
Pierre Hillion wrote: "Please do accept my apology but it makes so a long time
that I am working with focus wave modes that I am skeptical on the possibility
of man-made focus wave modes; perhaps only Nature in cosmic events?". In the
lights of the discussion of the overview in Chapter~\ref{chOV} one has to
agree with this opinion, but with the following concretization. In optical
domain this opinion is true because (i) today we do not have a coherent light
source that generates half-cycle pulses in visible region and (ii) due to the
non-paraxial angular spectrum of plane waves the original FWM's are more like
the modes of a cylindrical resonator. However, in the preceding chapters we
have shown, that the two requirements are nothing but the peculiarities of a
closed-form integral of the corresponding general expressions that describe
propagation-invariant pulsed wave fields. In this chapter, we demonstrate that
the wave fields that are essentially the limited-bandwidth modifications of
the original FWM's \textit{can }be generated in optical domain.

Majority of the publications on generation of FWM's and LW's (except of course
those that are the part of this review) have discussed possibilities of
launching them from an array (matrix) of (point) sources. This approach has
proved to be very involved -- as it has been noted in literature \cite{g1},
the LW solutions generally cannot be written as the product of a function only
of time and a function only of space, i.e., the LW solutions are
mathematically nonseparable in the space-time coordinates. As a consequence,
if one tries to use the principle of Huygens and launch the LW's from a planar
source, it appears that each point of this source have to be driven
independently. In other words, the nonseparability of the LW's implies that
the frequency spectrum of every point-source is different and one has to drive
a separately addressable array of wideband Huygens sources with a function
identical to a LW. The analysis presented in literature \cite{g10}
demonstrates that if we could build that kind of matrix of sources, the
launched wave field would be a perfectly causal FWM. The finite aperture of
the source or the time-limiting of the excitation do not destroy the localized
propagation of the generated pulses.

In optical domain this approach has a fundamental drawback -- the frequencies
of the wave fields are of the order of magnitude $10^{15}Hz$ and the idea of
driving a matrix of independent ultra-wideband sources in this frequency range
cannot be realized in experiment. The minor deficiency is that in the Huygens
representation the propagation properties of the LW's are not physically
transparent in the degree they are in angular spectrum representation.%

%TCIMACRO{\FRAME{ftbpFU}{2.9611in}{2.0609in}{0pt}{\Qcb{The principal scheme of
%a setup that generates optical FWM's. Here L is a lens and AS is a chromatic
%angular slit. The angular dispersion arises as the consequence of the
%frequency dependence of diameter of the chromatic angular slit.}}{\Qlb{fig64}%
%}{fig6_4.jpg}{\special{ language "Scientific Word";  type "GRAPHIC";
%maintain-aspect-ratio TRUE;  display "ICON";  valid_file "F";
%width 2.9611in;  height 2.0609in;  depth 0pt;  original-width 4.9104in;
%original-height 3.1419in;  cropleft "0";  croptop "1";  cropright "1";
%cropbottom "0";  filename 'fig6_4.jpg';file-properties "XNPEU";}} }%
%BeginExpansion
\begin{figure}
[ptb]
\begin{center}
\includegraphics[
height=2.0609in,
width=2.9611in
]%
{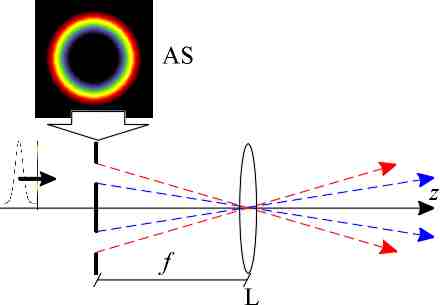}%
\caption{The principal scheme of a setup that generates optical FWM's. Here L
is a lens and AS is a chromatic angular slit. The angular dispersion arises as
the consequence of the frequency dependence of diameter of the chromatic
angular slit.}%
\label{fig64}%
\end{center}
\end{figure}
%EndExpansion
As an example of an alternative approach, characteristic in optics, consider a
setup consisting of a Fourier lens of focal length $f$ and of a circularly
symmetric mask with transfer function
\begin{equation}
t\left(  R,k\right)  =\emph{s}\left(  k\right)  \delta\left[  R-f\tan
\theta_{F}\left(  k\right)  \right]  \text{,} \label{gg2}%
\end{equation}
where $R$ is the radial coordinate on the ring and $\theta_{F}\left(
k\right)  $ is the angular function of a FWM defined by Eq.~(\ref{su12}) --
this is essentially an annular ring mask, the diameter of which depends on the
wave number of the light as $R\left(  k\right)  =f\tan\theta_{F}\left(
k\right)  $ (see Fig.~\ref{fig64}). It is well known, that each monochromatic
point-source of wave number $k$ at some radial distance $\rho$ on the focal
plane of an ideal the lens results in an apertured plane wave that subtends
the angle $\theta_{F}\left(  k\right)  $ relative to $z$ axis behind the lens.
The angular spectrum of plane waves of the total field then can be described
as (here the infinite aperture is assumed)%
\begin{equation}
A_{0}\left(  k,\theta\right)  =\emph{s}\left(  k\right)  \exp\left[
i\upsilon\left(  k\right)  \right]  \delta\left[  \theta-\theta_{F}\left(
k\right)  \right]  \text{,} \label{gg3}%
\end{equation}
where $\upsilon\left(  k\right)  $ is some phase factor specific to the setup.
The support of the angular spectrum of plane waves of the wave field is the
one of the FWM's (\ref{su29}). If we also compensate for the phase factor in
Eq.~(\ref{gg3}) by applying the conjugated phase chirp to the input pulse the
Fourier representation directly suggest that a transform-limited FWM can be
generated by illuminating a "chromatic" annular ring mask by a specifically
chirped plane wave pulse. Though such chromatic mask is not very practical
either, the advantage of this approach over the Gaussian aperture \cite{g10}
in optical domain is obvious -- this model is essentially static.

\subsection{\label{sGE}Feasible approach to optical generation FWM's}

\begin{figure}
[ptb]
\begin{center}
\includegraphics[
height=6.0909in,
width=2.0176in
]%
{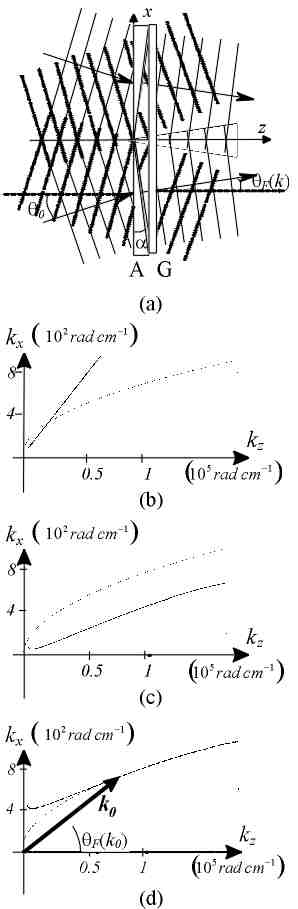}%
\caption{The general idea for optical generation of the 3D (2D) FWM's: (a) The
FWM generator consist of an axicon A and of circular diffractional grating G
(or two wedges and linear diffractional grating respectively if we generate 2D
FWM's), the FWM can be observed in the conical volume (striped region) behind
the diffraction grating; (b) The support of the angular spectrum of plane
waves of the initial wave field on the FWM generator (solid line). Here and
hereafter the dotted line denotes the support of angular spectrum of plane
waves of the (2D) FWM under discussion, the dashed lines denote the bands of
the frequency spectrum of the light used in our experiments (note the
difference in scales between $k_{x}$ and $k_{z}$ axis); (c) The support of the
angular spectrum of plane waves behind the axicon; (d) The support of the
angular spectrum of plane waves at the exit of the FWM generator as compared
with the theory.}%
\label{fig65}%
\end{center}
\end{figure}

Let us introduce the general idea in terms of 2D FWM's as defined in section
\ref{ssFTi}. Consider the pair of plane wave pulses propagating at angles
$\theta_{0}$ and $-\theta_{0}$ relative to the optical axis (see
Fig.~\ref{fig65}a). Obviously we can introduce a tilt into their angular
spectrum of plane waves by means of the angularly dispersive elements like
diffraction gratings or prisms (wedges) and provided that the resulting tilted
pulses overlap, we can observe the interference of two tilted pulses in
near-axis conical volume [see the striped region in Fig.~\ref{fig65}a]. If the
introduced tilt of the plane wave components is such that the condition in the
Eq.~(\ref{su12}), i.e.,
\begin{equation}
\cos\theta_{F}\left(  k\right)  =\frac{\gamma\left(  k-2\beta\right)  }{k}
\label{gg5}%
\end{equation}
is satisfied for every $k$ and if we ignore for the while the diffractional
edge effect that appear due to the finite aperture of the optical elements,
the interference pattern in the near-axis volume is that of the 2D FWM.%
%TCIMACRO{\FRAME{ftbpFU}{2.0176in}{6.0909in}{0pt}{\Qcb{The general idea for
%optical generation of the 3D (2D) FWM's: (a) The FWM generator consist of an
%axicon A and of circular diffractional grating G (or two wedges and linear
%diffractional grating respectively if we generate 2D FWM's), the FWM can be
%observed in the conical volume (striped region) behind the diffraction
%grating; (b) The support of the angular spectrum of plane waves of the initial
%wave field on the FWM generator (solid line). Here and hereafter the dotted
%line denotes the support of angular spectrum of plane waves of the (2D) FWM
%under discussion, the dashed lines denote the bands of the frequency spectrum
%of the light used in our experiments (note the difference in scales between
%$k_{x}$ and $k_{z}$ axis); (c) The support of the angular spectrum of plane
%waves behind the axicon; (d) The support of the angular spectrum of plane
%waves at the exit of the FWM generator as compared with the theory.}%
%}{\Qlb{fig65}}{fig6_5.jpg}{\special{ language "Scientific Word";
%type "GRAPHIC";  maintain-aspect-ratio TRUE;  display "ICON";
%valid_file "F";  width 2.0176in;  height 6.0909in;  depth 0pt;
%original-width 4.9104in;  original-height 6.6841in;  cropleft "0";
%filename 'fig6_5.jpg';file-properties "XNPEU";}} }%
%BeginExpansion
%EndExpansion

The cylindrically symmetric case (FWM) can be considered similarly -- one has
to find the means to generate the superposition of Bessel beams, the support
of the angular spectrum of which satisfies condition (\ref{gg5}). Also, the
use of the angular dispersion of the known generators of Bessel beams, i.e.,
the devices that transform a monochromatic plane wave into a Bessel beam,
could be an appropriate idea.%

%TCIMACRO{\FRAME{ftbpFU}{3.0874in}{1.4408in}{0pt}{\Qcb{The Bessel--X pulse
%generator. Here L stands for lens and AS for angular slit.}}{\Qlb{fig63}%
%}{fig6_3.jpg}{\special{ language "Scientific Word";  type "GRAPHIC";
%maintain-aspect-ratio TRUE;  display "ICON";  valid_file "F";
%width 3.0874in;  height 1.4408in;  depth 0pt;  original-width 4.9104in;
%original-height 6.6841in;  cropleft "0";  croptop "1";  cropright "1";
%cropbottom "0";  filename 'fig6_3.jpg';file-properties "XNPEU";}} }%
%BeginExpansion
\begin{figure}
[ptb]
\begin{center}
\includegraphics[
height=1.4408in,
width=3.0874in
]%
{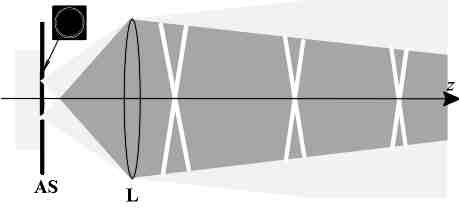}%
\caption{The Bessel--X pulse generator. Here L stands for lens and AS for
angular slit.}%
\label{fig63}%
\end{center}
\end{figure}
%EndExpansion
As the first step we should generate the cylindrically symmetric counterpart
of the pair of plane wave pulses -- the Bessel--X pulse (see section
\ref{ssBX}). The general approach for this is quite straightforward --
according to (\ref{bx1}) we have to use the Bessel beam generator without any
angular dispersion and illuminate it with a wideband pulse. The obvious choice
is the annular slit in the back focal plane of a Fourier lens (see
Fig.~\ref{fig63}). The simple geometrical considerations show that in this
case the wave field behind the lens is the cylindrically symmetric
superposition of plane wave pulses -- for the cone angle of the Bessel beam of
the wave number $k$ we find $\sin\theta\left(  k\right)  =\sin\left(
\arctan\left(  D/f\right)  \right)  \equiv\sin\theta_{0}$, where $D$ is the
diameter of the angular slit and $f$ is the focal length of the lens, so that
the generated wave field can be approximated by (\ref{su40})%
\begin{equation}
\Psi_{BX}\left(  \rho,z,t\right)  =\int_{0}^{\infty}\mathrm{d}k\,\emph{s}%
\left(  k\right)  J_{0}\left(  k\rho\sin\theta_{0}\right)  \exp\left[
ik\left(  z\cos\theta_{0}-ct\right)  \right]  \text{,} \label{gg10}%
\end{equation}
where $\emph{s}\left(  k\right)  $ is the frequency spectrum of the source field.

Using the Bessel-X pulses as input, the problem of optical generation of FWM's
reduces to the modeling of a set of diffractive elements that transform the
conical support of the angular spectrum of plane waves so that the angular
dispersion of the output pulse approximates the one described by the condition
in Eq.~(\ref{gg5}) and to the \textquotedblright compression\textquotedblright%
\ of the resulting wave field by compensating for the relative phases between
its monochromatic components so that a transform-limited pulse appears on the
optical axis.

There are various Bessel beam generators described in literature such as
axicons, circular diffraction gratings, etc. (see, e.g., Ref.~\cite{b1} --
\cite{b16} and references therein). By illuminating those elements with a
plane-wave pulse, we obviously get a superposition of Bessel beams, the
support of the angular spectrum of which is determined by the wavelength
dispersion of cone angle of the optical element $\theta\left(  k\right)  $.
For example, an axicon is characterized by the complex transmission function
$\exp\left[  ik\tan\alpha\left(  1-n\left(  k\right)  \right)  \right]  $,
where $\alpha$ is the angle formed by the conical surface with a flat surface
and $n\left(  k\right)  $ is the refractive index of the axicon (see, e.g.,
Ref.~\cite{ba22}). The method of stationary phase \cite{ok4}, applied as in
Ref.~\cite{ba15}, yields $\sin\theta\left(  k\right)  =\tan\alpha\left[
1-n\left(  k\right)  \right]  $. Hence, the angular spectrum of plane waves of
the corresponding polychromatic wave field can be approximately described by
\begin{equation}
A_{0}\left(  k,\theta\right)  =\,\emph{s}\left(  k\right)  \delta\left[
\theta-\arcsin\left(  \tan\alpha\left(  1-n\left(  k\right)  \right)  \right)
\right]  \text{.} \label{gg15}%
\end{equation}
Substitution into Eq.\thinspace(\ref{ang9}) yields
\begin{align}
\Psi\left(  \rho,z,t\right)   &  =\int_{0}^{\infty}\mathrm{d}k~\emph{s}\left(
k\right)  J_{0}\left[  k\rho\left(  \tan\alpha\left(  1-n\left(  k\right)
\right)  \right)  \right] \label{gg18}\\
&  \times\exp\left[  ik\left[  z\cos\left(  \arcsin\left(  \tan\alpha\left(
1-n\left(  k\right)  \right)  \right)  \right)  -ct\right]  \right] \nonumber
\end{align}
and we can conclude that ideal Bessel beam generators can be used to design
wave fields with cylindrically symmetric, two--dimensional supports of angular
spectrum. Likewise, a circular grating yields for the cone angle $\sin
\theta\left(  k\right)  =2\pi/kd$, where $d$ is the grating constant
(first--order diffraction is assumed). The polychromatic wave field can be
written as
\begin{align}
\Psi\left(  \rho,z,t\right)   &  =\int_{0}^{\infty}\mathrm{d}k~\emph{s}\left(
k\right)  J_{0}\left[  k\rho\left(  \frac{2\pi}{kd}\right)  \right]
\label{gg19}\\
&  \times\exp\left[  ik\left(  z\cos\left(  \arcsin\left(  \frac{2\pi}%
{kd}\right)  \right)  -ct\right)  \right] \nonumber
\end{align}
(note, that the Eqs. (\ref{gg18}) and (\ref{gg19}) do not count correctly for
the longitudinal intensity change of Bessel beams behind axicons and circular
diffraction gratings \cite{ba15,ba57}, they are brought about just to
exemplify the use of the angular dispersion in our problem).

The support\thinspace of angular spectrum\thinspace of a FWM (\ref{su12})
cannot be approximated by a single diffractive element. However, we will show
below that aside from the conventional configuration, where axicons and
circular gratings are used to transform a plane wave into a Bessel beam, those
elements can also be used to \textit{change} the cone angle of Bessel beams.
This property allows us to use a set of those Bessel beam generators as to
design more complex supports of the angular spectrum.%

%TCIMACRO{\FRAME{ftbpFU}{2.0583in}{5.6178in}{0pt}{\Qcb{A qualitative
%description of a Bessel beam as a superposition of diverging (dotted lines)
%and converging (solid lines) conical waves in plane $z=0$ : (a) free--space
%evolution of the conical components; (b) diffraction on a circular opaque
%mask; (c) refraction on an axicon. }}{\Qlb{fig67}}{fig6_7.jpg}%
%maintain-aspect-ratio TRUE;  display "ICON";  valid_file "F";
%width 2.0583in;  height 5.6178in;  depth 0pt;  original-width 4.9104in;
%original-height 6.6841in;  cropleft "0";  croptop "1";  cropright "1";
%cropbottom "0";  filename 'fig6_7.jpg';file-properties "XNPEU";}} }%
%BeginExpansion
\begin{figure}
[ptb]
\begin{center}
\includegraphics[
height=5.6178in,
width=2.0583in
]%
{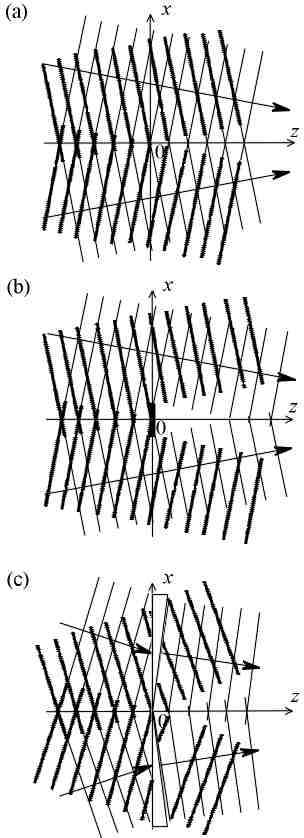}%
\caption{A qualitative description of a Bessel beam as a superposition of
diverging (dotted lines) and converging (solid lines) conical waves in plane
$z=0$ : (a) free--space evolution of the conical components; (b) diffraction
on a circular opaque mask; (c) refraction on an axicon. }%
\label{fig67}%
\end{center}
\end{figure}
%EndExpansion
The generic features of the diffraction of Bessel beams on circularly
symmetric optical elements can be understood by means of the following model.
A zeroth--order Bessel function in the off--axis region can be well
approximated by \cite{o2}
\begin{equation}
J_{0}\left(  a\right)  \approx\frac{1}{2}\sqrt{\frac{2}{\pi a}}\left\{
\exp\left[  i\left(  a-\frac{\pi}{4}\right)  \right]  +\exp\left[  -i\left(
a-\frac{\pi}{4}\right)  \right]  \right\}  \text{.} \label{gg21}%
\end{equation}
Substitution in (\ref{bes2}) yields
\begin{align}
\Psi_{B}\left(  \rho,0,t\right)   &  \approx\frac{1}{2}\sqrt{\frac{2}{\pi
k\rho\sin\theta}}\left\{  \exp\left[  ik\rho\sin\theta-i\frac{\pi}{4}\right]
\right. \nonumber\\
&  +\left.  \exp\left[  -ik\rho\sin\theta+i\frac{\pi}{4}\right]  \right\}
\exp\left[  -ikct\right]  \label{gg22}%
\end{align}
for the Bessel beam\thinspace\ in plane $z=0$. In this expansion a Bessel beam
in this plane\thinspace is a superposition of two conical waves,\thinspace as
shown in Fig.~\ref{fig67}a (see also Ref.~\cite{b14}). The evolution of those
components can be qualitatively analyzed by means of ray--tracing, if we also
recognize that the characteristic field distribution of a Bessel beam arises
in the volume where the conical waves interfere. Typical examples, the
free-space evolution and diffraction on circularly symmetric mask, are
depicted in Fig.~\ref{fig67}a and Fig.~\ref{fig67}b. As one can see, such
well-known features of Bessel beams as finite propagation length \cite{b1} and
regeneration behind an opaque screen appear immediately.

Let us place an axicon\thinspace(or circular grating) onto plane $z=0$ (see
Fig.~\ref{fig67}c). A qualitative analysis by means of ray-tracing immediately
reveals, that the element changes the cone angle of both conical waves. The
\textquotedblright converging\textquotedblright\ conical wave forms a Bessel
beam in a conical near axis volume, but the cone angle is now different. The
\textquotedblright diverging\textquotedblright\ conical component leaves the
near axis region. Hence, axicons and diffraction gratings change the cone
angle of a Bessel beam in the sense that the regenerated Bessel beam after the
element is of a different cone angle.%

%TCIMACRO{\FRAME{ftbpFU}{4.0612in}{1.3534in}{0pt}{\Qcb{(a) A circular
%diffraction grating on the surface of an axicon -- a composite optical
%element, which can be used for generation of FWM's; (b) An optical setup for
%generation of FWM's. A plane wave pulse is incident upon an annular slit (AS).
%Bessel--X pulse with cone angle $\theta$ behind a Fourier lens (L) is incident
%upon the composite optical element (AG).}}{\Qlb{fig68}}{fig6_8.jpg}%
%maintain-aspect-ratio TRUE;  display "ICON";  valid_file "F";
%width 4.0612in;  height 1.3534in;  depth 0pt;  original-width 4.9104in;
%original-height 1.9588in;  cropleft "0";  croptop "1";  cropright "1";
%cropbottom "0";  filename 'fig6_8.jpg';file-properties "XNPEU";}} }%
%BeginExpansion
\begin{figure}
[ptb]
\begin{center}
\includegraphics[
height=1.3534in,
width=4.0612in
]%
{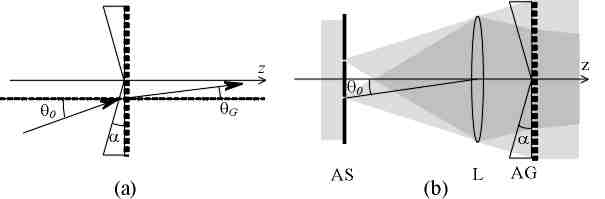}%
\caption{(a) A circular diffraction grating on the surface of an axicon -- a
composite optical element, which can be used for generation of FWM's; (b) An
optical setup for generation of FWM's. A plane wave pulse is incident upon an
annular slit (AS). Bessel--X pulse with cone angle $\theta$ behind a Fourier
lens (L) is incident upon the composite optical element (AG).}%
\label{fig68}%
\end{center}
\end{figure}
%EndExpansion
Let us find the cone angle of the resulting Bessel beam for a composite
optical element-- a circular diffraction grating on the surface of an axicon
(see Fig.~\ref{fig68}a). We assume that a Bessel--X pulse with cone angle
$\theta_{0}$, generated by means of an annular slit and Fourier lens, is
incident upon it (see Fig.~\ref{fig68}b). Snell's law and the grating equation
yield the following equation for the cone angle of the resulting Bessel beam:
\begin{equation}
\sin\theta_{G}\left(  k\right)  =\frac{2\pi}{kd}+n\left(  k\right)
\sin\left\{  -\alpha+\arcsin\left[  \frac{1}{n\left(  k\right)  }\sin\left(
\theta_{0}+\alpha\right)  \right]  \right\}  \text{.} \label{gg23}%
\end{equation}
Here $d$ is the grating constant and $n\left(  k\right)  $ is the refractive
index of the axicon material. Sign conventions are chosen so that the angles
$\alpha$, $\theta_{0}$, $\theta_{G}\left(  k\right)  $ are positive in
Fig.~\ref{fig68}a. First--order diffraction is assumed. If the angles $\alpha
$, $\theta_{0}$ are small, so that $\sin x\sim\arcsin x\sim x$,
Eq.~(\ref{gg23}) yields
\begin{equation}
\theta_{G}\left(  k\right)  =\frac{2\pi}{kd}+\alpha\left[  1-n\left(
k\right)  \right]  +\theta_{0}\text{.} \label{gg24}%
\end{equation}

Similar results can be obtained by means of the method of stationary phase.
Below we use the approach in Ref.~\cite{ba15}. Let the initial field on the
diffractive element be the converging conical component of a Bessel beam
$\frac{1}{2}\sqrt{2/\left(  \pi k\rho\sin\theta_{0}\right)  }\exp\left[
ik\rho\sin\theta_{0}\right]  $. The amplitude transmission function of the
element is $\exp\left[  ik\rho\left(  2\pi/kd+\tan\alpha\left(  1-n\left(
k\right)  \right)  \right)  \right]  $. The corresponding Fresnel diffraction
integral can be given as
\begin{equation}
\Psi\left(  \rho,z\right)  =\frac{1}{i\lambda z}e^{ik\left(  z+\frac{\rho^{2}%
}{2z}\right)  }\int\nolimits_{0}^{D}\mathrm{d}\rho^{\prime}f\left(
\rho^{\prime}\right)  \exp\left[  ik\mu\left(  \rho^{\prime}\right)  \right]
\text{,} \label{gg25}%
\end{equation}
where
\begin{equation}
f\left(  \rho^{\prime}\right)  =\rho^{\prime}\frac{1}{2}\sqrt{\frac{2}{\pi
k\rho^{\prime}\sin\theta_{0}}}2\pi J_{0}\left(  \frac{k\rho\rho^{\prime}}%
{z}\right)  \label{gg26}%
\end{equation}
and
\begin{equation}
\mu\left(  \rho^{\prime}\right)  =\frac{\rho^{\prime2}}{2z}-\rho^{\prime
}\left(  \frac{2\pi}{kd}+\tan\alpha\left(  1-n\left(  k\right)  \right)
+\sin\theta_{0}\right)  \text{.} \label{gg27}%
\end{equation}
The first derivative of $\mu\left(  \rho^{\prime}\right)  $ yields exactly one
critical point $\rho_{c}^{\prime}$, namely
\begin{equation}
\rho_{c}^{\prime}=z\left(  \frac{2\pi}{kd}+\tan\alpha\left(  1-n\left(
k\right)  \right)  +\sin\theta_{0}\right)  \equiv z\sin\theta_{SP}\left(
k\right)  \label{gg28}%
\end{equation}
and, according to Ref.~\cite{ba15}, the leading contribution to (\ref{gg25})
behaves as
\begin{equation}
\frac{1}{i\lambda z}\exp\left[  ik\left(  z+\frac{\rho^{2}}{2z}\right)
\right]  \int\nolimits_{0}^{D}\mathrm{d}\rho^{\prime}f\left(  \rho^{\prime
}\right)  \exp\left[  ik\mu\left(  \rho^{\prime}\right)  \right]  \quad
\quad\quad\label{gg29}%
\end{equation}%
\[
\quad\quad\quad\quad\quad\quad\quad\quad\quad\quad\quad\quad\quad\approx
\frac{1}{i\lambda z}\exp\left[  ik\left(  z+\frac{\rho^{2}}{2z}\right)
\right]  \frac{f\left(  \rho_{c}^{\prime}\right)  \exp\left[  ik\mu\left(
\rho_{c}^{\prime}\right)  \right]  }{\sqrt{k\mu^{\left(  2\right)  }\left(
\rho_{c}^{\prime}\right)  }}\text{,}%
\]
where $\mu^{\left(  2\right)  }\left(  \rho_{c}^{\prime}\right)  $ denotes the
value of the second derivative $\mu\left(  \rho_{c}^{\prime}\right)  $ at the
critical point. Substituting\thinspace(\ref{gg28}) into (\ref{gg29}) and
omitting some position-independent factors we get
\begin{align}
\Psi\left(  \rho,z\right)   &  \approx\frac{1}{i\lambda z}\exp\left[
ik\left(  z+\frac{\rho^{2}}{2z}\right)  \right]  \frac{1}{\sqrt{k\frac{1}{z}}%
}z\sin\theta_{SP}\left(  k\right) \nonumber\\
&  \times\frac{1}{2}\sqrt{\frac{2}{\pi kz\sin\theta_{SP}\left(  k\right)
\sin\theta_{0}}}\left[  2\pi J_{0}\left(  \frac{k\rho z\sin\theta_{SP}\left(
k\right)  }{z}\right)  \right] \nonumber\\
&  \times\exp\left[  ik\left(  \frac{\left(  z\sin\theta_{SP}\left(  k\right)
\right)  ^{2}}{2z}-z\sin^{2}\theta_{SP}\left(  k\right)  \right)  \right]
\text{,} \label{gg30}%
\end{align}
so that%
\begin{equation}
\Psi\left(  \rho,z\right)  \sim J_{0}\left(  k\rho\sin\theta_{SP}\left(
k\right)  \right)  \exp\left[  ik\frac{\rho^{2}}{2z}\right]  \exp\left[
ikz\cos\theta_{SP}\left(  k\right)  \right]  \text{,} \label{gg31}%
\end{equation}
where relation $1-\sin^{2}\theta/2\sim\sqrt{1-\sin^{2}\theta}=\cos\theta$ is
used. We also have $k\left(  \rho^{2}/2z\right)  \ll\pi/2$ for the far--field
in near--axis region and Eq.~(\ref{gg30}) reads
\begin{equation}
\Psi\left(  \rho,z,t\right)  \approx J_{0}\left(  k\rho\sin\theta_{SP}\left(
k\right)  \right)  \exp\left[  ikz\cos\theta_{SP}\left(  k\right)  \right]
\text{.} \label{gg32}%
\end{equation}%
%TCIMACRO{\FRAME{ftbpFU}{4.3578in}{3.4679in}{0pt}{\Qcb{The experimental
%demonstration of the use of the angularly dispersive Bessel beam generators
%for change of the cone angle of monochromatic Bessel beams. In the setup: AS,
%annular slit; L, lens; A, "hollow" axicon. In the left and right pane the CCD
%image of the Bessel beam at the plane A and B are depicted respectively. }%
%}{\Qlb{fig89}}{fig8_9.jpg}{\special{ language "Scientific Word";
%type "GRAPHIC";  maintain-aspect-ratio TRUE;  display "ICON";
%valid_file "F";  width 4.3578in;  height 3.4679in;  depth 0pt;
%original-width 4.7176in;  original-height 2.3531in;  cropleft "0";
%filename 'fig8_9.jpg';file-properties "XNPEU";}} }%
%BeginExpansion
\begin{figure}
[ptb]
\begin{center}
\includegraphics[
height=3.4679in,
width=4.3578in
]%
{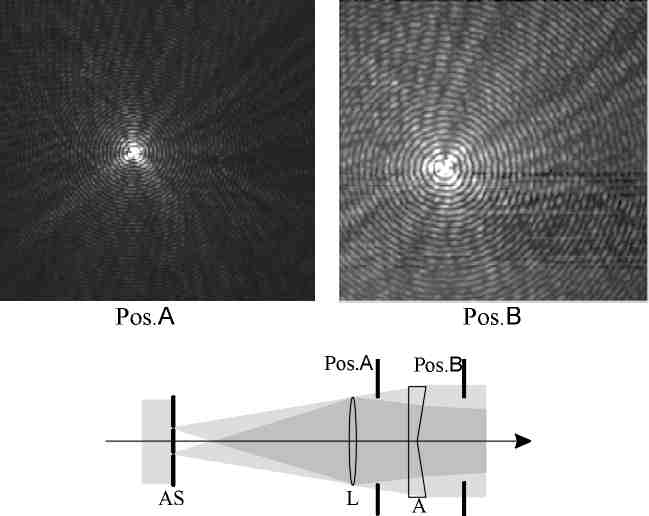}%
\caption{The experimental demonstration of the use of the angularly dispersive
Bessel beam generators for change of the cone angle of monochromatic Bessel
beams. In the setup: AS, annular slit; L, lens; A, "hollow" axicon. In the
left and right pane the CCD image of the Bessel beam at the plane A and B are
depicted respectively. }%
\label{fig89}%
\end{center}
\end{figure}
%EndExpansion
Thus, the method of stationary phase yields a similar result as the
qualitative ray tracing analysis\thinspace-- the output field of an axicon (or
circular grating), illuminated by a Bessel beam, is a Bessel beam. If we
assume a small angle limit i.e. $\sin x\sim\tan x\sim x$ in Eq. (\ref{gg28}),
the generated cone angle $\theta_{SP}$ is also exactly the same as predicted
by the ray tracing analysis in Eq.(\ref{gg24}) and we have shown that the
Bessel beam generators can be used to change the cone angle of Bessel beams.
We also carried out experimental proof of this principle, the typical results
are depicted in Fig.~\ref{fig89}.

The wavelength dispersion of the cone angle of the Bessel beam constituents of
a FWM $\theta_{F}\left(  k\right)  $ is determined by Eq.\thinspace
(\ref{su14}). Combining this condition with the dispersion of the cone angle
of the discussed setup\thinspace$\theta_{G}\left(  k\right)  $ (\ref{gg24}),
we get
\begin{equation}
\arccos\left(  \frac{\gamma\left(  k-2\beta\right)  }{k}\right)  =\frac{2\pi
}{kd}+\alpha\left(  1-n\left(  k\right)  \right)  +\theta_{0}\text{.}
\label{gg33}%
\end{equation}
In following chapters we will see that the three free parameters $\left(
\alpha,\theta_{0},d\right)  $ can be adjusted so that the relation
(\ref{gg33}) is satisfied in a very good approximation in a limited spectral range.

As to finish this section we note that care must be taken when specifying the
frequency spectrum $B\left(  k\right)  $ or $\tilde{B}\left(  k\right)
=\allowbreak k^{2}\sin\theta\left(  k\right)  $ in Eqs.~(\ref{gg3}),
(\ref{gg10}), (\ref{gg15}) and (\ref{gg19}). In terms of the frequency
spectrum of the light source $\emph{s}\left(  k\right)  $ -- generally the
choice the transform depends on the particular setup under consideration. If
we set $\emph{s}\left(  k\right)  =\tilde{B}\left(  k\right)  $, we integrate
over the monochromatic Bessel beams with amplitudes $\emph{s}\left(  k\right)
$. However, for example in the principal setup in Fig.~\ref{fig64} it is quite
obvious that the amplitudes of the Bessel beams depend on the diameter of the
annular slit -- the larger diameter means larger area of the slit and more
transmitted energy. Consequently, the "chromatic" annular ring mask has a
transfer function that is not constant but can be approximated by $\sim
k\sin\theta\left(  k\right)  =\allowbreak\chi$ and one should use the
weighting function $\tilde{B}\left(  k\right)  =\allowbreak k\sin\theta\left(
k\right)  \emph{s}\left(  k\right)  $ in the superposition over the Bessel
beams or
\begin{equation}
\tilde{A}_{0}\left(  k,\theta\right)  =k\sin\theta\left(  k\right)
\emph{s}\left(  k\right)  \exp\left[  i\upsilon\left(  k\right)  \right]
\delta\left[  \theta-\theta_{F}\left(  k\right)  \right]  \label{g55}%
\end{equation}
in place of (\ref{gg3}) for the angular spectrum of plane waves (here
$\upsilon\left(  k\right)  $ stands for the phase distortions of the setup).
As for the special cases of annular slit with constant diameter, axicon and
circular diffraction grating the analogous arguments show that the replacement
$\emph{s}\left(  k\right)  =\tilde{B}\left(  k\right)  $ applies. (note also
how the otherwise exponential frequency spectrum of the FWM's in
Eq.~(\ref{fwm5}) behaves as a Gaussian-like in (\ref{fwm13}) due to the
$k\sin\theta_{F}\left(  k\right)  $ term in the expression of the Whittaker
type plane wave expansion (\ref{ang9})). Of course, the generated wave field
is propagation-invariant regardless of the weighting function in the
superposition and generally in any optical experiment only the bandwidth of
the source have to be concerned about.

\subsection{\label{ssGFE}Finite energy approximations to FWM's}

We already noted in the overview in Chapter \ref{chOV} that the total energy
content of FWM's is infinite and as such they are not realizable in any
physical experiment. We also referenced some of the most commonly used
approaches to derive finite energy approximations to FWM's. Obviously, in
experimental situation a natural choice is to consider the approximations that
correspond to the specific launching mechanism. In other words, the
approximations of the type (\ref{ed5}) should be given a physical content, in
terms of limitations of a real experimental setup. In what follows we derive
the finite energy approximation of FWM's that is due to the finite aperture of
the optical system introduced in previous chapter (Ref.~\cite{m3}).

In Fourier picture the starting point for this discussion is obvious. In this
picture any realistic (finite-aperture) optical system generates a
superposition of \textit{apertured} monochromatic Bessel beams so that the
field in the exit plane the FWM generator can be described by (for brevity we
restrict ourselves to cylindrically symmetric case)%
\begin{equation}
\Psi_{B}\left(  \rho,0,0;k\right)  =t\left(  \rho\right)  J_{0}\left[
k\rho\sin\theta_{F}\left(  k\right)  \right]  \text{,} \label{en1}%
\end{equation}
where $t\left(  \rho\right)  $ is the complex-amplitude transmission function
of the aperture of the setup. One just has to show that (i) the resulting
superposition of the apertured Bessel beams still represents a wave field that
propagate as a FWM, and (ii) the superposition has finite energy content.

\subsubsection{Apertured (finite energy flow approximations to) Bessel beams}

The simplest mathematical model for the monochromatic Bessel beams is the
exact solution of the scalar homogeneous wave equation in Eq.~(\ref{bes2})
\begin{equation}
\Psi_{B}\left(  \rho,z,t\right)  =J_{0}\left(  k\rho\sin\theta_{0}\right)
\exp\left[  ik\left(  z\cos\theta_{0}-ct\right)  \right]  \text{.} \label{en0}%
\end{equation}
In this case the beam can be described in Fourier' picture by the single
delta-ring in $k$-space and the transformation of the beam in free space and
in optical elements can be easily estimated by simple geometrical
constructions. However, such closed form Bessel beams are not square
integrable. The most apparent approach to deduce physically realizable
approximations to Bessel beams has been to apply a finite aperture to the
system. The problem has been discussed in many works, mostly in terms of
Fresnel approximation of scalar diffraction theory (see Refs.~\cite{ba5}%
--\cite{ba82} and references therein).

As an alternative, we could use the Gaussian windowing profile and consider
the paraxial wave equation -- this combination yields a closed mathematical
form of the beams, that have been called the Bessel-Gauss beams (see
Refs.~\cite{ba10,ba25,ba45, ba59,ba65} and references therein). However, the
paraxial approximation lacks the physical transparency of the angular spectrum
representation. Still, it should be noted, that the Bessel-Gauss beams
\textit{can} be used to construct wave fields that behave like LW's
propagating in free space.%

%TCIMACRO{\FRAME{ftbpFU}{3.9608in}{1.561in}{0pt}{\Qcb{On the finite aperture
%approximations to the Bessel beams. The optical setup is the simple Bessel
%beam generator consisting of annular ring AS of diameter D, and the lens of
%aperture $D$ and focal length $f$. The striped region denotes the near axis
%volume where the generated wave field approximates closely the behaviour of
%the finite aperture Bessel beam.}}{\Qlb{fig612}}{fig6_12.jpg}%
%maintain-aspect-ratio TRUE;  display "ICON";  valid_file "F";
%width 3.9608in;  height 1.561in;  depth 0pt;  original-width 3.9314in;
%original-height 1.9605in;  cropleft "0";  croptop "1";  cropright "1";
%cropbottom "0";  filename 'fig6_12.jpg';file-properties "XNPEU";}} }%
%BeginExpansion
\begin{figure}
[ptb]
\begin{center}
\includegraphics[
height=1.561in,
width=3.9608in
]%
{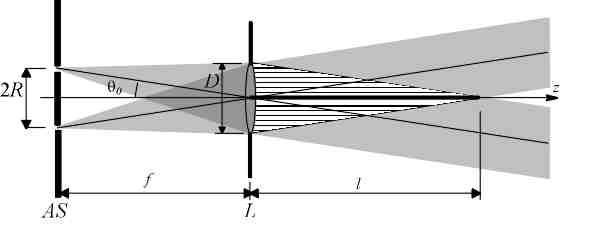}%
\caption{On the finite aperture approximations to the Bessel beams. The
optical setup is the simple Bessel beam generator consisting of annular ring
AS of diameter D, and the lens of aperture $D$ and focal length $f$. The
striped region denotes the near axis volume where the generated wave field
approximates closely the behaviour of the finite aperture Bessel beam.}%
\label{fig612}%
\end{center}
\end{figure}
%EndExpansion
We will use the generally accepted facts that (i) applying finite aperture to
a Bessel beam provides us with a finite energy flow wave field, that is a very
good approximation of the infinite-aperture Bessel beams (\ref{en1}) in a
certain finite-depth, near axis volume (see, e.g., Ref.~\cite{b2}) and (ii)
that the polychromatic superpositions of those apertured Bessel beams
approximate very closely the superpositions of \textquotedblright
non-apertured\textquotedblright\ Bessel beams in this volume \cite{b14}. Such
a behavior can be easily explained in the Fourier picture where a
monochromatic Bessel beam is a cylindrically symmetric superposition of plane
waves, that propagate at angle $\theta$ relative to $z$ axis. Indeed, as the
apertured plane waves approximate their infinite aperture counterparts very
closely in their central parts, one can also observe a very good approximation
to the infinite-aperture Bessel beam in this near-axis volume (see striped
region in Fig.~\ref{fig612} and Ref.~\cite{ba62} for a more detailed
description in terms of diffraction theory). If the cone angle of a Bessel
beam is small, as it is always the case in paraxial optical systems, the
apertured Bessel beam would behave as its infinite-aperture counterpart
(\ref{en1}) for several meters of propagation.

The situation in Fig.~\ref{fig612} can be modeled by applying the transmission
function $t\left(  \rho\right)  $ of the aperture to the Bessel beams. In Weyl
picture this operation is equivalent to calculating the two-dimensional
Fourier transform of the transversal amplitude distribution $t\left(
\rho\right)  J_{0}\left(  \chi_{0}\rho\right)  $, where $\chi_{0}$ stands for
radial projection of the wave vector of the Bessel beam. Given the Weyl type
angular spectrum of plane waves of the infinite-aperture Bessel beam
\begin{equation}
A\left(  \chi\right)  =K\delta\left(  \chi-\chi_{0}\right)  \text{,}
\label{en2}%
\end{equation}
where $K$ is a constant, the Fourier transform of (\ref{en1}) in point $z=0$,
$t=0$ can be found to be
\begin{align}
A_{ApB}\left(  \chi\right)   &  =\frac{K}{\left(  2\pi\right)  ^{2}}\,T\left(
\chi\right)  \ast\delta\left(  \chi-\chi_{0}\right) \nonumber\\
&  =\frac{\chi_{0}K}{\left(  2\pi\right)  ^{2}}\int_{0}^{2\pi}\mathrm{d}\phi
T\left(  \sqrt{\chi^{2}+\chi_{0}^{2}-2\chi\chi_{0}\cos\left(  \phi-\phi
_{0}\right)  }\right)  \text{,} \label{en3}%
\end{align}
where $T\left(  \chi\right)  $ is the two-dimensional Fourier transform of the
transmission function $t\left(  \rho\right)  $ and $\ast$ denotes convolution
operation (see also Ref.~\cite{si25}). The argument of the function $T\left(
{}\right)  $ in (\ref{en3}) has an interpretation as being the distance
between the points $\left(  \chi,\phi\right)  $ and $\left(  \chi_{0},\phi
_{0}\right)  $. As for all convenient apertures the function $T\left(
\chi\right)  $ is well-localized around zero, the major contribution to the
integral (\ref{en3}) comes from small values of $\phi$ and one can write in
good approximation
\begin{equation}
A_{ApB}\left(  \chi\right)  \approx\frac{\chi_{0}K}{2\pi}T\left(  \chi
-\chi_{0}\right)  \text{.} \label{en4}%
\end{equation}
The interpretation of the expression (\ref{en4}) is straightforward: the
finite aperture gives the support of angular spectrum of a monochromatic
Bessel beam a finite \textquotedblright width\textquotedblright. Exact form of
the support is determined by the complex-amplitude transmission function,
however, the well-known set of fundamental Fourier transform pairs gives a
good idea of what the support of angular spectrum looks like, without any calculations.

\subsubsection{\label{ssApFWM}Apertured FWM's}

First of all, it has been demonstrated both numerically and experimentally
that the superposition of apertured Bessel beam behaves like the superposition
of their infinite-aperture counterpart in near-axis volume as defined in
Fig.~\ref{fig612}. Thus we can claim that the substitution of
infinite-aperture Bessel beams in (\ref{su29}) by their apertured counterparts
should generate a finite energy flow wave field that is a good approximation
of the FWM (\ref{su40}) in this finite volume. In other words, in this
near-axis volume the apertured wave field can be well approximated by the
formulas of infinite-aperture wave fields.

Obviously, the finite aperture has a similar effect on the angular spectrum
support of a FWM -- the delta function in Eq.~(\ref{su29}) is substituted by a
weighting function and the angular spectrum of plane waves of apertured FWM's
can be written as
\begin{align}
A_{ApF}\left(  k,\chi\right)   &  =\frac{\chi_{F}\left(  k\right)  B\left(
k\right)  }{\left(  2\pi\right)  ^{2}}\int_{0}^{2\pi}\mathrm{d}\varphi
\nonumber\\
&  \times T\left(  \sqrt{\chi^{2}+\chi_{F}\left(  k\right)  ^{2}-2\chi\chi
_{F}\left(  k\right)  \cos\varphi}\right) \nonumber\\
&  \approx\frac{\chi_{F}\left(  k\right)  B\left(  k\right)  }{2\pi}T\left[
\chi-\chi_{F}\left(  k\right)  \right]  \text{,} \label{en5}%
\end{align}
where $\chi_{F}\left(  k\right)  =k\sin\theta_{F}\left(  k\right)  $.
Consequently, the Weyl type plane wave expansion of the wave field behind the
aperture reads
\begin{align}
\Psi_{ApF}\left(  \rho,z,t\right)   &  =\int_{0}^{\infty}\mathrm{d}k\int
_{0}^{\infty}\mathrm{d}\chi\,\chi\,A_{ApF}\left(  k,\chi\right) \nonumber\\
&  \times J_{0}\left(  \rho\chi\right)  \exp\left[  ik\left(  z\sqrt
{k^{2}-\chi^{2}}-ct\right)  \right]  \text{.} \label{en8}%
\end{align}
Alternatively, the transformation $\chi=k\sin\theta$ gives the expression
(\ref{en8}) the following form
\begin{align}
\Psi_{ApF}\left(  \rho,z,t\right)   &  =\int_{0}^{\infty}\mathrm{d}kk^{2}%
\int_{0}^{2\pi}\mathrm{d}\theta\sin\theta\cos\theta\,A_{ApF}\left(
k,k\sin\theta\right) \nonumber\\
&  \times J_{0}\left(  k\rho\sin\theta\right)  \exp\left[  ik\left(
z\cos\theta-ct\right)  \right]  \text{.} \label{en9}%
\end{align}
%

%TCIMACRO{\FRAME{ftbpFU}{4.2505in}{1.3941in}{0pt}{\Qcb{The comparison of the
%supports of angular spectrums of plane waves of (a) an apertured FWM and (b)
%of an EDEPT type superposition of FWM's as described in above sections.}%
%}{\Qlb{fig613}}{fig6_13.jpg}{\special{ language "Scientific Word";
%type "GRAPHIC";  maintain-aspect-ratio TRUE;  display "ICON";
%valid_file "F";  width 4.2505in;  height 1.3941in;  depth 0pt;
%original-width 4.9104in;  original-height 1.9588in;  cropleft "0";
%filename 'fig6_13.jpg';file-properties "XNPEU";}} }%
%BeginExpansion
\begin{figure}
[ptb]
\begin{center}
\includegraphics[
height=1.3941in,
width=4.2505in
]%
{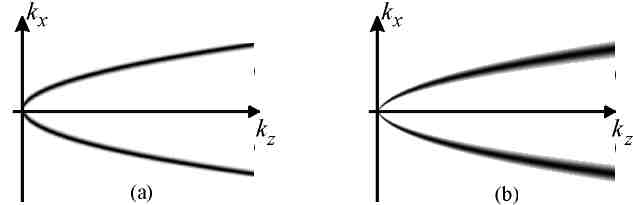}%
\caption{The comparison of the supports of angular spectrums of plane waves of
(a) an apertured FWM and (b) of an EDEPT type superposition of FWM's as
described in above sections.}%
\label{fig613}%
\end{center}
\end{figure}
%EndExpansion
The support of the angular spectrum of plane waves (\ref{en5}) of the derived
wave field is depicted on Fig. \ref{fig613}a. In correspondence with
Eq.~(\ref{en4}) it has a finite \textquotedblright thickness\textquotedblright%
. We can outline the main difference between the support of angular spectrum
of LW's, proposed in this section (\ref{en9}) and that of EDEPT's (\ref{ed5})
discussed in section \ref{ssED}. The comparison of their supports of angular
spectrum of plane waves on Fig. \ref{fig613}a and \ref{fig613}b (respectively)
shows, that the transversal \textquotedblright width\textquotedblright\ in
$k_{x}k_{y}$ plane of angular spectrum of plane waves of our apertured FWM's
is a constant-- such a result is a consequence of applying aperture with a
wavelength-independent complex-amplitude transmission function. On the other
hand, the transversal width is generally not constant for the superpositions
of FWM's in Eq.(\ref{ed5}). The description in this section gives the property
a straightforward interpretation: the corresponding aperture has a wavelength
dependent complex-amplitude transmission function. In the context of our
discussion, where the main goal is optical feasibility, such an approach
should be regarded as an impractical one.

The spatial localization of the result (\ref{en8}) can be estimated by the
approach in section \ref{ssFLoc} -- obviously the resulting wave field still
has the characteristic narrow central peak. Also, in correspondence with the
note in the beginning of current section, the field of apertured and
non-apertured FWM's do not differ noticeably in the near-axis volume at $t=0$.
Thus, if the aperture of the system is reasonably large (several millimeters)
the only qualitative effect of the finite aperture is the reduced propagation
length of the wave field.

There is several ways to estimate the propagation length of the apertured
FWM's. The simplest possibility is still use the geometrical construction in
Fig.~\ref{fig612} from which we can write
\begin{equation}
l_{1}\left(  k_{\min}\right)  =\frac{D}{2\tan\left[  \theta_{F}\left(
k_{\min}\right)  \right]  }\text{,} \label{en15}%
\end{equation}
where $k_{\min}$ denotes the minimal wave vector of the spectrum of the wave
field (note, that for the supports of FWM's this generally corresponds to
maximum angle $\max\left[  \theta_{F}\left(  k\right)  \right]  $, i.e., to
the minimal propagation length of the corresponding Bessel beams [clarify
Fig.~\ref{fig612})]. The second possibility is to estimate the propagation
length of the apertured monochromatic Bessel beams by means of the approach in
the section \ref{ssFLoc} (in present case where the wave field is essentially
monochromatic, this is actually the McCutchen theory as introduced in
Ref.~\cite{mc5}). From the Eq.~(\ref{lo21}) we at once get
\begin{equation}
l_{2}\left(  k_{\min}\right)  =\Delta z=\frac{\Delta\rho}{\tan\theta
_{F}\left(  k_{\min}\right)  }\text{,} \label{en17}%
\end{equation}
i.e., we get identical result as in Eq.~(\ref{en15}). The third possibility is
to estimate the maximum group- and phase velocity uncertainty of the finite
width support of the angular spectrum of plane waves (\ref{en3}).

In conclusion, in our setup the finite energy (physically realizable)
approximations to FWM's are introduced quite plainly by the finite aperture of
the optical system. In Fourier domain this corresponds to smoothing of the
delta function in the angular spectrum of plane waves of the wave fields.
Also, the double integrals in Eq.~(\ref{en8}) can in some detail be handled
without excessive numerical calculations.

\subsubsection{\label{ssOnFin}On finite time window excitation of the FWM's}

In literature, there has been several works on generation of FWM's where the
finite total energy has been achieved by limiting the excitation time of the
source array \cite{g4,g5,g6,g10}. It has been shown, that, indeed, such
approach do not destroy the localized propagation of the generated pulses. It
is intuitively clear, that the finite time excitation results in a
superposition of the longitudinal "fragments" of the monochromatic Bessel
beams. In this picture the reasonable time window indeed do not corrupt the
behaviour of the central part of the wave field, and it is still a good
approximation that of the exact monochromatic Bessel beam. What the time
window \textit{do} is that it broadens the frequency spectrum of the points of
the source matrix. Correspondingly, the support of the angular spectrum of the
generated wave field would look much like the one in Fig. \ref{fig613}a.
Actually, the excitation time of our setup is also finite and the two
broadening effects~-- due the aperture and the finite excitation time~--
appear simultaneously, however, the one that is caused by the finite aperture
is several orders of magnitude stronger.

\subsection{Optical generation of partially coherent LW's}

The direct comparison of the equations describing the angular spectrum of
plane waves of coherent and partially coherent LW's in Eqs.~(\ref{su10}) --
(\ref{su47}) and (\ref{ok19}) -- (\ref{ok21a}) respectively implies, that the
main part of the generating setup should be the same for both cases. Indeed,
the partially coherent quasi-monochromatic plane waves transform in linear
optical systems similarly as the coherent monochromatic plane waves, at least
the description of the previous chapter concerning refraction and diffraction
in grating hold for both cases. The qualitative difference lies in the
correlations\textit{\ }between the plane wave components.

\subsubsection{\label{ssLS}The light source}

According to our general idea of the optical generation of FWM's in
Sec.~\ref{sGE}, we need a well-directed wideband partially coherent plane wave
as the initial field in our setup. In mathematical limit the angular spectrum
of plane waves of such initial field is described by
\begin{equation}
a(\phi,\theta,k)=s(k)\delta\left[  \theta\right]  \text{,} \label{ls5}%
\end{equation}
giving for the field
\begin{equation}
V(\mathbf{r},t)=\frac{1}{2\pi}\int_{0}^{\infty}\mathrm{d}k\ s(k)\exp\left[
ik\left(  z-ct\right)  \right]  \text{,} \label{ls9}%
\end{equation}
where $s\left(  k\right)  $ is generally stochastic function. Such field is
fully coherent in transversal direction. In longitudinal direction the
coherence time $\tau_{c}$ is determined by the bandwidth of the light $\Delta
k$ and the reciprocity inequality \cite{ok4}%
\begin{equation}
\tau_{c}\Delta k\geq\frac{1}{2c}\text{.} \label{ls10}%
\end{equation}
For the ensemble average, we can write%
\begin{equation}
\left\langle s^{\ast}(k_{1})s(k_{2})\right\rangle =\mathcal{S}\left(
k\right)  \delta\left(  k_{2}-k_{1}\right)  \text{,} \label{ls10a}%
\end{equation}
where $\mathcal{S}\left(  k\right)  $ denotes the spectral density (power
spectrum) of the light source so that the mutual coherence function reads as%
\begin{equation}
\Gamma\left(  \Delta z,\tau\right)  =\int_{0}^{\infty}\mathrm{d}%
k\mathcal{S}\left(  k\right)  \exp\left[  ik\left(  \Delta z-c\tau\right)
\right]  \text{.} \label{ls11}%
\end{equation}
In other words, the mutual coherence function of light field behaves like a
plane wave pulse of the duration $\tau_{s}$ in coherent optics.

The traditional approach to generate such field is to use a thermal light
source, for example the superhigh-pressure Xe-arc lamp, and spatially filter
the light by means of a pinhole. The well-known drawback of the choice is the
hugely reduced signal level as compared to laser sources. The alternative is
to use a directional white-light continuum source.

\subsubsection{FWM's with frequency noncorrelation}

As it was explained in section \ref{ssOF1} (clarify Eqs.~(\ref{f5}) and
(\ref{f7})), the FWM's with frequency noncorrelation differ from their fully
coherent counterparts only by the lack of correlation between the fluctuations
of their Fourier components of different frequency. Obviously such fields can
be generated by illuminating the setup in Fig.~\ref{fig68}b with a wideband
partially coherent plane wave as described in Eqs.~(\ref{ls5}) -- (\ref{ls11}).

\subsubsection{FWM's with angular noncorrelation}%

%TCIMACRO{\FRAME{ftbpFU}{4.1874in}{1.535in}{0pt}{\Qcb{The principal setup for
%generation of FWM's with angular noncorrelation: AS, annular slit; L1, L2, L3,
%lenses; AG, annular grism; D, diffuser;}}{\Qlb{fig69}}{fig6_9.jpg}%
%maintain-aspect-ratio TRUE;  display "ICON";  valid_file "F";
%width 4.1874in;  height 1.535in;  depth 0pt;  original-width 4.9104in;
%original-height 1.9588in;  cropleft "0";  croptop "1";  cropright "1";
%cropbottom "0";  filename 'fig6_9.jpg';file-properties "XNPEU";}} }%
%BeginExpansion
\begin{figure}
[ptb]
\begin{center}
\includegraphics[
height=1.535in,
width=4.1874in
]%
{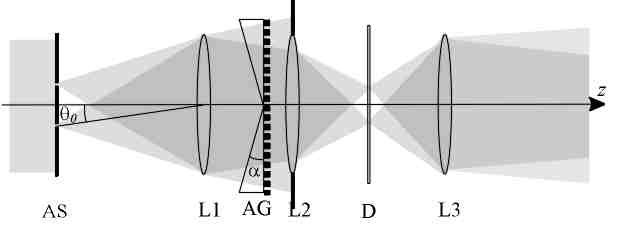}%
\caption{The principal setup for generation of FWM's with angular
noncorrelation: AS, annular slit; L1, L2, L3, lenses; AG, annular grism; D,
diffuser;}%
\label{fig69}%
\end{center}
\end{figure}
%EndExpansion
According to section \ref{ssOF2} the only difference between the fully
coherent FWM's and this special case is the lack of correlation between the
tilted pulses that compose the coherent FWM's. Thus, to generate a FWM with
angular noncorrelation one has to illuminate the setup by coherent light and
somehow break the correlation between the tilted pulses propagating at
different polar angles. This can be done by means of the modified setup
depicted in Fig.~\ref{fig69} (see Ref.~\cite{ok6} for the description of
corresponding quasi-monochromatic case). In this setup the pair of lenses L2,
L3 is inserted into the path of FWM's so that in the focal plane of L2 the
two-dimensional Fourier transform of the FWM's -- the characteristic
concentric rings -- appear. In this plane we can insert a weak diffuser as to
modify the amplitude and phase of the tilted pulses.

\subsubsection{FWM's with frequency and angular noncorrelation}

In this special case one has to illuminate the setup in Fig~\ref{fig69} with
the white-light source as described in Eqs.~(\ref{ls5}) -- (\ref{ls11}).

\subsection{Conclusions. Optical generation of general LW's}

As to conclude this chapter we note that except for the ultra-wide bandwidths,
required to generate the wave fields described in Chapter \ref{chOV} the
concept of propagation-invariance is well realizable in optical domain -- we
have shown, that the angular dispersion of various Bessel beam generators can
be used to transform the support of the angular spectrum of plane waves so
that to approximate that of the FWM's in some limited near-axis volume. The
choice of the combination of the elements may be a problem in some cases,
however, the chances are good for finding satisfactory combination.
Consequently, one can launch the wave fields with the characteristic central
peak that propagates over reasonable distances.%

%TCIMACRO{\FRAME{ftbpFU}{2.3644in}{1.9847in}{0pt}{\Qcb{Optical generation of
%LW's that contain plane waves that subtend non-paraxial angles with respect to
%the optical axis.}}{\Qlb{fig611}}{fig6_11.jpg}%
%maintain-aspect-ratio TRUE;  display "ICON";  valid_file "F";
%width 2.3644in;  height 1.9847in;  depth 0pt;  original-width 4.9104in;
%original-height 1.9588in;  cropleft "0";  croptop "1";  cropright "1";
%cropbottom "0";  filename 'fig6_11.jpg';file-properties "XNPEU";}} }%
%BeginExpansion
\begin{figure}
[ptb]
\begin{center}
\includegraphics[
height=1.9847in,
width=2.3644in
]%
{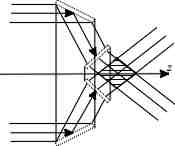}%
\caption{Optical generation of LW's that contain plane waves that subtend
non-paraxial angles with respect to the optical axis.}%
\label{fig611}%
\end{center}
\end{figure}
%EndExpansion
Note, that the LW's containing plane waves that propagate at nonparaxial
relative to $z$-axis can in principle be generated within the framework of
this approach. Nevertheless, this requires very non-conventional optical
elements like conical mirrors and diffraction gratings as sketched in
Fig.~\ref{fig611}. Also, in this case the propagation distance of the
generated LW's is quite short as can be seen from Eq.~(\ref{en15}).

\subsection{On the physical nature of propagation-invariance of pulsed wave
fields}

As to conclude the theoretical part of this review we interpret some of the
properties of the LW's in the context of classical diffraction theory (see
Ref.~\cite{x49k} for a relevant discussion).

The first note has to be made on the definition of the propagation length of
LW's. Namely, it has been claimed in several publications (see Ref.~\cite{g0a}
for example) that the LW's propagate over extended distances as compared to
that, defined by the Rayleigh range $Z_{R}$, the well known estimate for the
scale length of the falloff in intensity behind of a Gaussian aperture in the
diffraction theory, defined as (see Ref.~\cite{o25} for example)%
\begin{equation}
Z_{R}=\frac{\pi W_{0}^{2}}{\lambda}\text{,} \label{pn10}%
\end{equation}
$W_{0}$ being the minimum spot size (radius) of the beam. Indeed, for the
apertured FWM's the radial diameter of its central spot is approximately
(\ref{lo44})
\[
d\approx\frac{4.81}{k_{0}\sin\theta_{F}\left(  k_{0}\right)  }=1.5\times
10^{-4}m
\]
(again, $\gamma=1$, $\beta=40\frac{rad}{m}$) and if we consider a planar
source with diameter $D=1cm$, this spot travels (\ref{en15})%
\begin{equation}
l_{1}\left(  k_{\min}\right)  =\frac{D}{2\tan\left[  \theta_{F}\left(
k_{\min}\right)  \right]  }=0.9m\text{,} \label{pn15}%
\end{equation}
The Rayleigh range (\ref{pn10}) of the Gaussian pulse with radius $W_{0}=d$ is
$Z_{R}\approx8.8cm\allowbreak\ll l_{1}\left(  k_{\min}\right)  $. However, in
our opinion such estimates are misleading and should not be used without the
following additional details.%

%TCIMACRO{\FRAME{ftbpFU}{4.6838in}{2.5408in}{0pt}{\Qcb{The comparison of the
%radial field distribution of a FWM (left pane) with the field that is
%integrated over the polar angle (right pane). }}{\Qlb{fig516}}{fig5_16.jpg}%
%maintain-aspect-ratio TRUE;  display "ICON";  valid_file "F";
%width 4.6838in;  height 2.5408in;  depth 0pt;  original-width 4.9104in;
%original-height 6.6841in;  cropleft "0";  croptop "1";  cropright "1";
%cropbottom "0";  filename 'fig5_16.jpg';file-properties "XNPEU";}} }%
%BeginExpansion
\begin{figure}
[ptb]
\begin{center}
\includegraphics[
height=2.5408in,
width=4.6838in
]%
{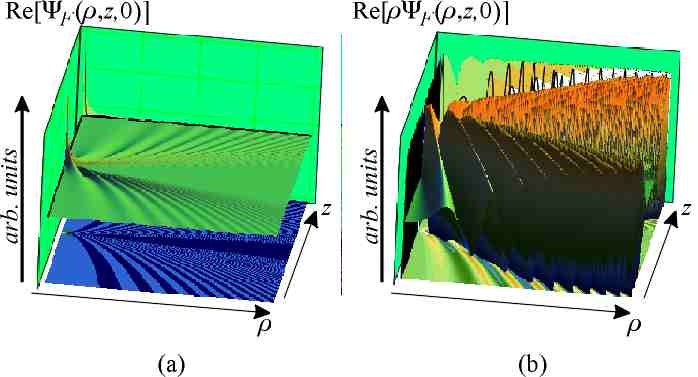}%
\caption{The comparison of the radial field distribution of a FWM (left pane)
with the field that is integrated over the polar angle (right pane). }%
\label{fig516}%
\end{center}
\end{figure}
%EndExpansion
The effect of diffraction is typically manifested when an obstacle is placed
in the path of a light field. For (pulsed) beams the definition (\ref{pn10})
determines minimal spread corresponding to the waist radius $W_{0}$. In other
words, this parameter determines the minimum radius of a circular aperture
that can be placed in the path of the beam without significantly distorting
its behaviour behind the aperture. Now, as to compare the two estimates we
have to ask, what is the waist radius of an apertured FWM's?

It appears, that the situation is similar to that with the monochromatic
Bessel beams for which the energy content in every transverse lobe is
approximately constant and equal to the energy content in the central maximum
(see Ref.~\cite{x49k} for example). In the case of apertured FWM's the
amplitude of the branches of the characteristic X-shape fall off approximately
as $1/\rho$. Thus, the integrated energy density as the function of radial
coordinate $\rho$ is approximately constant, as the integration over the polar
angle adds the factor $\rho$ to the amplitude of the wave field (see
Fig.~\ref{fig516}). Thus, only a minor part of the energy of the apertured
FWM's is contained in its central lobe and if one has to compare the
propagation length of the apertured FWM's with the Rayleigh range of the
Gaussian pulses, one should take $W_{0}=D$ instead of $W_{0}=d$ in
(\ref{pn10}).%
%TCIMACRO{\FRAME{ftbpFU}{4.8248in}{2.7415in}{0pt}{\Qcb{The comparison of the
%focusing of (a) FWM's and (b) Gaussian pulses (see text).}}{\Qlb{fig517}%
%}{fig5_17.jpg}{\special{ language "Scientific Word";  type "GRAPHIC";
%maintain-aspect-ratio TRUE;  display "ICON";  valid_file "F";
%width 4.8248in;  height 2.7415in;  depth 0pt;  original-width 4.9104in;
%original-height 1.9588in;  cropleft "0";  croptop "1";  cropright "1";
%cropbottom "0";  filename 'fig5_17.jpg';file-properties "XNPEU";}} }%
%BeginExpansion
\begin{figure}
[ptb]
\begin{center}
\includegraphics[
height=2.7415in,
width=4.8248in
]%
{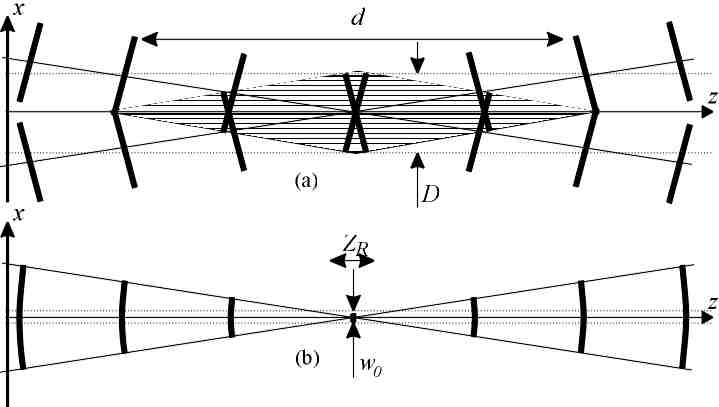}%
\caption{The comparison of the focusing of (a) FWM's and (b) Gaussian pulses
(see text).}%
\label{fig517}%
\end{center}
\end{figure}
%EndExpansion

As the matter of fact, such comparison of the Gaussian pulse and apertured
FWM's is not appropriate as\textit{\ focusing} of the two wave fields are of
qualitatively different physical nature. Namely, the Gaussian pulses are
composed of monochromatic components with \textit{curved} phase fronts, the
phase fronts of the monochromatic components of the FWM's are conical. In
lights of this difference, one can say that the FWM's\textit{\ are never
focused} in the conventional sense of the term and the term, focus wave mode,
is rather misleading.

In focusing Gaussian pulses most of its energy content can be concentrated
into a single spot for a time moment. A cunning mental picture of the
energetic propertied of focusing in FWM's can be acquired if we suppose that
we have an ideal (apertured) FWM generator and suppose that we illuminate it
with a plane wave pulse. Then central peak of each monochromatic Bessel beam
component of the generated FWM's has the amplitude equal to the amplitude of
the corresponding monochromatic plane wave component of the initial plane wave
pulse. Indeed, the effect of the FWM generator to this monochromatic plane
wave component is that its initial energy in the $k$-space (a finite width
spot on the $k_{z}$-axis) is smeared over the finite width toroid in the
$k$-space. In real space each point in this toroid will compose an apertured
plane wave in the Bessel beam and the net on-axis amplitude of the Bessel beam
is the integrated amplitude over the toroid in $k$-space, i.e., again the
amplitude of the initial monochromatic plane wave component. Now, for the
superposition of the Bessel beams we can say, that the only place where the
constructive interference occurs is the central spot of the apertured FWM.
Thus, we can say, that given the initial plane wave pulse with the amplitude
$A$\ and aperture $D$,\textit{\ the amplitude of the central spot of an
apertured FWM behind an ideal FWM generator is that of the initial plane wave
pulse }$A$\textit{, the rest of the energy is in the sidelobes of the
generated FWM's. }This consequence is a good illustration of the qualitative
difference between the FWM' and Gaussian pulse -- due to the curved phase
fronts of its monochromatic components the latter can indeed effectively
transfer most of its energy to a single spot for a time moment.

\section{\label{chExp}THE EXPERIMENTS}

In our experiments we realized the optical setups for two LW's -- the
apertured Bessel-X pulses and apertured FWM's and used an interferometric
cross-correlation method with time-integrated intensity recording to study the
generated wave fields. The material of this chapter is published in
Refs.~I\ and V.

\subsection{FWM's in interferometric experiments}

A straightforward method for recording the complicated field shape of a
coherent FWM or Bessel-X pulses would use a CCD camera with a gate in front of
it, which should possess a temporal resolution and a variable firing delay
both in subfemtosecond range. As such a gate is not realizable, any workable
idea of experiment has to resort to a field cross-correlation technique (see
Refs. \cite{ok7}--\cite{ok17} and references therein).

In our experiments the wave field under investigation $F\left(  \mathbf{r}%
,t\right)  $ interfere with a reference wave $V_{P}$:%
\begin{equation}
V_{\Sigma}\left(  \mathbf{r},t\right)  =F\left(  \mathbf{r},t\right)
+V_{P}\left(  \mathbf{r},t\right)  \text{.} \label{eo1}%
\end{equation}
For the reference wave we can write
\begin{equation}
V_{P}\left(  \mathbf{r},t\right)  =\int_{0}^{\infty}\mathrm{d}k\,\emph{s}%
\left(  k\right)  \upsilon_{P}\left(  k\right)  \exp\left[  ik\left(
z-c\left(  t+\Delta t\right)  \right)  \right]  \text{,} \label{eo3}%
\end{equation}
where $\emph{s}\left(  k\right)  $ is the (generally stochastic) frequency
spectrum of the light source, $\upsilon_{P}\left(  k\right)  $ is the spectral
phase shift introduced by the optics in the reference arm of the
interferometer, $\left\vert \upsilon_{P}\left(  k\right)  \right\vert \equiv1$
and $\Delta t$ denote the variable time delay between the signal and reference
wave fields. For the wave field under investigation we have
\begin{equation}
F\left(  \mathbf{r},t\right)  =\int_{0}^{\infty}\mathrm{d}k\,\emph{s}\left(
k\right)  \,\upsilon_{F}\left(  k\right)  J_{0}\left[  kx\sin\theta_{F}\left(
k\right)  \right]  \exp\left[  ik\left(  z\cos\theta_{F}\left(  k\right)
-ct\right)  \right]  \text{,} \label{eo5}%
\end{equation}
where $\upsilon_{F}\left(  k\right)  $ is again the undesirable spectral phase
shift from the setup, $\left\vert \upsilon_{F}\left(  k\right)  \right\vert
\equiv1$ (we have assumed here that the FWM generator transform the input
light so that for every spectral component $\emph{s}\left(  k\right)  $ the
amplitude of the central spot of the corresponding Bessel beam is also
$\emph{s}\left(  k\right)  $). The averaged intensity of the resulting wave
field can be expressed as
\begin{equation}
\left\langle V_{\Sigma}^{\ast}V_{\Sigma}\right\rangle =\left\langle
V_{P}^{\ast}V_{P}\right\rangle +\left\langle F^{\ast}F\right\rangle
+2\operatorname{Re}\left\langle F^{\ast}V_{P}\right\rangle \label{oe10}%
\end{equation}
(here the exact meaning of the angle brackets depends on the statistical
properties of the light source of the experiment).

The quantities $\left\langle V_{P}^{\ast}V_{P}\right\rangle $ and
$\left\langle F^{\ast}F\right\rangle $ denote time-independent intensity of
the wave field. Specifically, the first term in the sum is the uniform
intensity of the plane wave pulse:
\begin{equation}
\left\langle V_{P}^{\ast}V_{P}\right\rangle =\int_{0}^{\infty}\mathrm{d}%
k\mathcal{S}\left(  k\right)  \text{,} \label{In1}%
\end{equation}
where again $\mathcal{S}\left(  k\right)  =\left\langle \emph{s}^{\ast}\left(
k\right)  \emph{s}\left(  k\right)  \right\rangle $ is the spectral density.
The second term is the time-averaged intensity of $F$:
\begin{equation}
\left\langle F^{\ast}F\right\rangle =\int_{0}^{\infty}\mathrm{d}%
k\mathcal{S}\left(  k\right)  J_{0}^{2}\left[  kx\sin\theta_{F}\left(
k\right)  \right]  \text{.} \label{In2}%
\end{equation}
In principle the two components can be eliminated from the results by
recording them separately and by numerically subtracting them from the interferograms.

From Eqs.~(\ref{eo3}) and (\ref{eo5}) we can write%
\begin{align}
&  2\operatorname{Re}\left\langle F^{\ast}V_{P}\right\rangle
=2\operatorname{Re}\left\langle \int_{0}^{\infty}\mathrm{d}k_{1}\emph{s}%
^{\ast}\left(  k_{1}\right)  \upsilon_{F}^{\ast}\left(  k_{1}\right)  \right.
\nonumber\\
&  \qquad\qquad\times J_{0}\left(  k_{1}\rho\sin\theta_{F}\left(  k\right)
\right)  \exp\left[  -ik_{1}\left(  z\cos\theta_{F}\left(  k\right)
-ct\right)  \right] \nonumber\\
&  \qquad\qquad\times\left.  \int_{0}^{\infty}\mathrm{d}k_{2}\emph{s}\left(
k_{2}\right)  \upsilon_{P}\left(  k_{2}\right)  \exp\left[  ik_{2}\left(
z-c\left(  t-\Delta t\right)  \right)  \right]  \right\rangle \text{.}
\label{in5}%
\end{align}
In our discussion we concentrated on two limiting special cases of the
classification in Sec.~\ref{sOF} -- the fully coherent LW's (section
\ref{ssOF0}) and the LW's with frequency noncorrelation (section \ref{ssOF1}).
In both cases the averaging in (\ref{in5}) yields%
\begin{align}
&  \left\langle F^{\ast}V_{P}\right\rangle =\int_{0}^{\infty}\mathrm{d}%
k\,\mathcal{S}\left(  k\right)  \upsilon_{F}^{\ast}\left(  k\right)
\upsilon_{P}\left(  k\right) \nonumber\\
&  \qquad\times J_{0}\left(  kx\sin\theta_{F}\left(  k\right)  \right)
\exp\left[  ikz\left(  \cos\theta_{F}\left(  k\right)  -1\right)  +ikc\Delta
t\right]  \text{.} \label{in19}%
\end{align}
Here for the partially coherent field we have used Eq.~(\ref{ls10a}):%
\begin{equation}
\left\langle \emph{s}^{\ast}\left(  k_{1}\right)  \emph{s}\left(
k_{2}\right)  \right\rangle =\mathcal{S}\left(  k_{1}\right)  \delta\left(
k_{1}-k_{2}\right)  \text{,} \label{in11}%
\end{equation}
for coherent fields the $\delta\left(  k_{2}-k_{1}\right)  $ appears as the
time-averaging over the term $\exp\left[  ict\left(  k_{2}-k_{1}\right)
\right]  $. Equation (\ref{in19}) can be given the form
\begin{align}
&  \left\langle F^{\ast}V_{P}\right\rangle =\exp\left[  i2\beta\gamma
z\right]  \int_{0}^{\infty}\mathrm{d}k\,\mathcal{S}\left(  k\right)
\upsilon_{F}^{\ast}\left(  k_{1}\right)  \upsilon_{P}\left(  k\right)
\nonumber\\
&  \qquad\quad\times J_{0}\left(  k\rho\sin\theta_{F}\left(  k\right)
\right)  \exp\left\{  ik\left[  z\left(  1-\gamma\right)  -c\Delta t\right]
\right\}  \text{.} \label{in20}%
\end{align}

The above mathematical description yields identical results for the coherent
and partially coherent fields, i.e., the results of such experiments generally
do not depend on the correlations between the Fourier components of different
frequency of the wave fields -- it is well known that in any interferometric
experiment the phase information of wave fields is necessarily lost. In other
words, the results of the experiments do not depend on whether we use the
transform-limited femtosecond pulses or a source of a stationary white noise.

The latter consequence is of great practical significance. In our overview in
Chapter \ref{chOV} we used a spectrum that corresponds to a $3fs$ laser pulse
and showed that the corresponding FWM's and Bessel-X pulse had a good spatial
localization (see Fig.~\ref{fig52} and \ref{fig57}). However, computer
simulations, or even simple geometrical estimations show that if the
autocorrelation time of the source field $\tau$ exceeds $\sim10$ femtoseconds,
the characteristic X branching occurs too far from the axis $z$ and in this
narrow-band limit the resulting wave field would be nothing but a trivial
interference of quasi-monochromatic plane waves. Thus, the bandwidth of the
light source is a very challenging part of the setup. In what follows we add
to the reputation of incoherent sources as being the poor mans femtosecond
source and confine ourselves to the special case of frequency non-correlating fields.

\subsection{Experiment on optical Bessel-X pulses}

\subsubsection{Setup}%

\begin{figure}
[ptb]
\begin{center}
\includegraphics[
height=4.7746in,
width=3.4177in
]%
{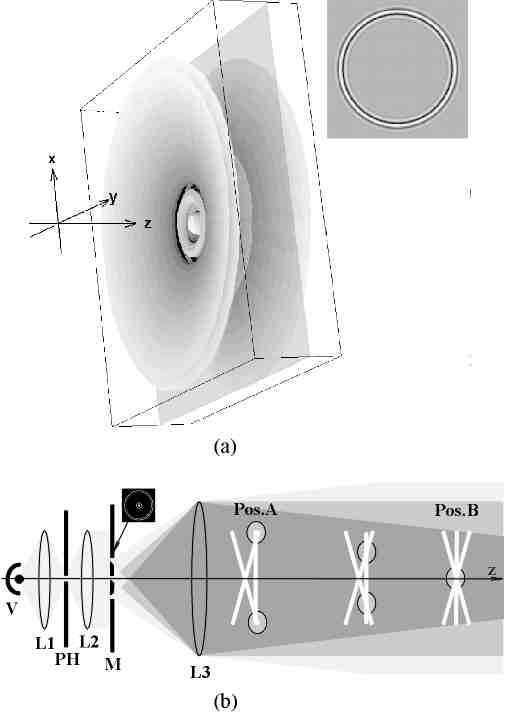}%
\caption{(a) Intensity profile of a computer-simulated Bessel-X pulse flying
in space, shown as surfaces on which the field intensity is equal to a
fraction $0.13\left(  1/e^{2}\right)  $ of its maximum value in the central
point. The field intensity outside the central bright spot has been multiplyed
by the radial distance in order to reveal the weak off-axis side-lobes. Inset:
amplitude distribution in the plane shown as intersecting the pulse. The plots
have been computed for 3-fsec near-Gaussian-spectrum source pulse with carrier
wavelenght $\lambda=0.6\mu m$ and the angle $\theta_{0}=0.223\deg$. For these
parameters the dimensions of the plot $xyz$ box are 20 times 20 times $\mu m$;
(b) Optical scheme of the experiment. Mutual instantaneous placement of the
Bessel-X pulse and the plane wave pulse is shown for three recording positions
(two of which labeled in accordance with Fig.~\ref{fig82}). The ovals indicate
toroid-like correlation volumes where co-propagating Bessel-X and plane wave
pulses interfere at different propagation distances along the $z$-axis. L's,
lenses; M, mask with Durnin's annular slit and an additional central pinhole
for creating the plane wave; PH, cooled pinhole $10\mu m$ in diameter to
assure the transversal coherence of the light from the source V. In case
source V generates a non-transform-limited pulses or a white cw noise, the
bright shapes depict propagation of the correlation functions instead of the
pulses.}%
\label{fig81}%
\end{center}
\end{figure}
%EndExpansion
Our setup for the interferometric experiments on optical Bessel-X pulses is
depicted in Fig.~\ref{fig81}b -- as compared to the Bessel-X pulse generator
in Fig.~\ref{fig63} the pinhole is made in the centre of the annular ring mask
as to form the reference field (plane wave pulse) behind the lens L. For the
mathematical description of the situation we have to choose $\beta=0$ in
Eq.~(\ref{in20}), i.e., $\theta\left(  k\right)  \equiv\theta_{0}%
=\allowbreak\arccos\gamma=\arccos c/v$. Setting also $\Delta t=0$ we get%
\begin{align}
\left\langle F^{\ast}V_{P}\right\rangle  &  =\int_{0}^{\infty}\mathrm{d}%
k\mathcal{S}\left(  k\right)  \upsilon_{F}^{\ast}\left(  k_{1}\right)
\upsilon_{P}\left(  k\right) \nonumber\\
&  \times J_{0}\left(  k\rho\sin\theta_{0}\right)  \exp\left[  ikz_{m}\left(
1-\gamma\right)  \right]  \label{bex1}%
\end{align}
where $\mathcal{S}\left(  k\right)  $ is the spectral density of the light
source (note that the annular ring mask has the uniform spectral response
function as explained in the end of the Sec.~\ref{sGE}). In Eq.~(\ref{bex1})
$z_{m}$ denotes the distance along the optical axis of the setup. To
understand the significance of this parameter we have to remember the
superluminal group velocity of the Bessel-X pulse -- during the flight the
latter catch-up with the reference plane wave pulse. In this context the
coordinate $z_{m}$ is the distance of the recording device from this catch-up
point (see Fig.~\ref{fig81}b).

If we define in the general expression for the mutual coherence function of
the wave fields in Eq.~(\ref{f20}) the origin of the $z$ axis as being in the
point $z=ct$ so that $\Delta z$ can be replaced by the distance from the pulse
centre $z_{\Delta}$ and set $\tau=0$, $\mathbf{r}_{\bot1}\equiv\mathbf{r}%
_{\bot}$, $\mathbf{r}_{\bot2}=0$ the result reads
\begin{equation}
\Gamma_{F}\left(  \mathbf{r}_{\bot},0,z_{\Delta}\right)  =\exp\left[
i2\beta\gamma z_{\Delta}\right]  \int_{0}^{\infty}\mathrm{d}k\left\vert
\mathcal{V}\left(  k\right)  \right\vert ^{2}J_{0}\left(  k\rho\sin\theta
_{F}\left(  k\right)  \right)  \exp\left[  ik\gamma z_{\Delta}\right]
\text{.} \label{in25}%
\end{equation}
If we also set $\beta=0$ we get%
\begin{equation}
\Gamma_{BX}\left(  \mathbf{r}_{\bot},0,z_{\Delta}\right)  =\int_{0}^{\infty
}\mathrm{d}k\left\vert \mathcal{V}\left(  k\right)  \right\vert ^{2}%
J_{0}\left(  k\rho\sin\theta_{0}\right)  \exp\left[  ik\gamma z_{\Delta
}\right]  \text{.} \label{bex3}%
\end{equation}
Comparing the Eqs.~(\ref{bex1}) and (\ref{bex3}) we see that if $\left\vert
\mathcal{V}\left(  k\right)  \right\vert ^{2}=\allowbreak\mathcal{S}\left(
k\right)  $ and $\upsilon_{F}^{\ast}\left(  k_{1}\right)  \upsilon_{P}\left(
k\right)  \allowbreak\equiv1$, we have
\begin{equation}
\left\langle F^{\ast}V_{P}\right\rangle =\Gamma_{BX}\left(  \mathbf{r}_{\bot
},0,z_{m}\frac{1-\gamma}{\gamma}\right)  \text{,} \label{bex4}%
\end{equation}
so that
\begin{equation}
z_{\Delta}=z_{m}\frac{1-\gamma}{\gamma}\text{.} \label{bex5}%
\end{equation}
The interpretation of the small factor $\left(  1-\gamma\right)  /\gamma$ in
(\ref{bex5}) is that the setup serves as a \textquotedblright$z$ axis
microscope\textquotedblright\ for recording the mutual coherence function
(\ref{bex1}) along the $z$-axis which scales the micrometer-range $z$
dependence of the field into a centimeter range.

If we compare the expression (\ref{bex3}) with the hypothetically measurable
field distribution given as the real part of the Bessel-X pulse in
Eq.~(\ref{bx2a}),
\begin{equation}
\Psi_{BX}\left(  \rho,z,t\right)  =\int_{0}^{\infty}\mathrm{d}k~\emph{s}%
\left(  k\right)  J_{0}\left[  k\rho\sin\theta_{0}\right]  \exp\left[
ik\left(  z\gamma-ct\right)  \right]  \label{bex7}%
\end{equation}
we conclude that the experiment reveals the whole spatiotemporal structure of
the Bessel-X field. The natural price we have to pay for resorting to the
correlation measurements is replacing the spectrum $\emph{s}\left(  k\right)
$ with its autocorrelation, which is a minor issue in the case of
transform-limited source pulses. Nevertheless, we cannot claim, that we
actually detect the field under investigation (see also Ref.~\cite{m1} for a
relevant discussion). Indeed, as the absolute phases of the plane wave
components are inevitably lost in any linear interferometric experiments only
the spatial amplitude distribution of the wave field can be detected.

In the reasons described above we took advantage of the insensitivity of
Eq.~(\ref{bex1}) to the source field phase and used a white light noise from a
superhigh pressure Xe-arc lamp instead of a laser as the field source to
achieve the $\sim3$ femtosecond correlation time in our experiment (\textit{V}
in Fig.~\ref{fig81}b). Thus we implemented the third special case of
Sec.~\ref{sOF} -- the Bessel-X field with frequency noncorrelation.

\subsubsection{Results of the experiment}%

%TCIMACRO{\FRAME{ftbpFU}{3.5777in}{5.348in}{0pt}{\Qcb{(a) Samples of the
%experimental recordings and processing of the intensity distributions measured
%at the positions Pos.A and Pos.B along the $z$-axis as shown in
%Fig.~\ref{fig81}b. Left column, total interference pattern of the
%cross-correlated fields; middle column, lateral intensity distribution of the
%Bessel-X field alone. In the right column gray (about 50 per cent) corresponds
%to zero level and dark to negative values; (b) Comparison of the result of the
%experiment (right panel) with a computer-simulated Bessel-X pulse field.}%
%}{\Qlb{fig82}}{fig8_2.jpg}{\special{ language "Scientific Word";
%type "GRAPHIC";  maintain-aspect-ratio TRUE;  display "ICON";
%valid_file "F";  width 3.5777in;  height 5.348in;  depth 0pt;
%original-width 4.9104in;  original-height 6.6841in;  cropleft "0";
%filename 'fig8_2.jpg';file-properties "XNPEU";}} }%
%BeginExpansion
\begin{figure}
[ptb]
\begin{center}
\includegraphics[
height=5.348in,
width=3.5777in
]%
{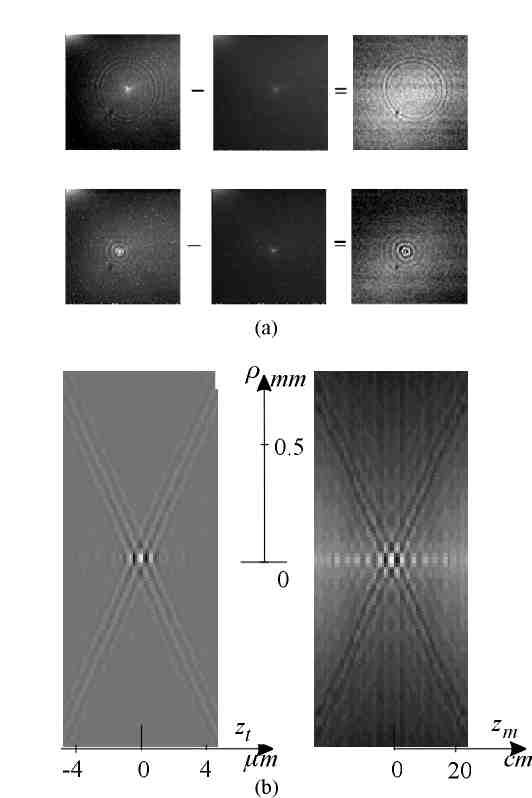}%
\caption{(a) Samples of the experimental recordings and processing of the
intensity distributions measured at the positions Pos.A and Pos.B along the
$z$-axis as shown in Fig.~\ref{fig81}b. Left column, total interference
pattern of the cross-correlated fields; middle column, lateral intensity
distribution of the Bessel-X field alone. In the right column gray (about 50
per cent) corresponds to zero level and dark to negative values; (b)
Comparison of the result of the experiment (right panel) with a
computer-simulated Bessel-X pulse field.}%
\label{fig82}%
\end{center}
\end{figure}
%EndExpansion

The recordings at 70 points on the $z$ axis (from behind the L3 lens up to a
point a few centimeters beyond the origin) with a 0.5-cm step were performed
with a cooled CCD-camera \emph{EDC--1000TE}, which has $2.64\times2.64~mm$
working area containing $192\times165$ pixels, and processed by a PC as follows.

First, the subtraction of the Bessel-X field intensity was performed (see
Fig.~\ref{fig82}a), whereas the same procedure with the plane wave field
intensity, due to its practically even distribution, was found to be
unnecessary. In order to reduce noise and the dimensionality of the data
array, an averaging over the polar angle in every recording was carried out by
taking advantage of the axial symmetry of the field. Thus we got a 1D array
containing up to hundred significant elements from every $192\times165 $
matrix recorded. Seventy such arrays formed a matrix, which, having in mind
the known symmetry of the real part of Eq.~(\ref{bex1}), was mirrored in the
lateral and the axial plane. The result is compared in Fig.~\ref{fig82}b with
the Bessel-X field distribution in an axial plane, computed from
Eq.~(\ref{bex7}) for a model spectrum. The latter was taken as a convex curve
covering the whole visible region from blue to near infrared (up to $0.9~\mu
m$) in order to simulate the effective light spectrum in the experiment, which
is a product of the Xe-arc spectrum and the sensitivity curve of the camera.
The central (carrier) frequency was chosen corresponding to wavelength
$0.6~\mu m$ which had been determined from the fringe spacing of an
autocorrelation pattern recorded for the light source with the same CCD
camera. The left- and right-hand tails of the central X-like structure are
more conspicuous in the experimental pattern due to unevenness of the real
Xe-arc spectrum. The different scaling of the horizontal axes of the two
panels is in accordance with the\textit{\ }$z$ axis \textit{\textquotedblright%
}magnification\textquotedblright\ factor of the experimental setup.

Observing the obvious agreement between theoretical and experimental patterns,
we arrive at a conclusion that we have really recorded the characteristic
spatiotemporal profile of an optical realization of the nonspreading
axisymmetric Bessel-X field.

\subsection{Experiment on optical FWM's}

\subsubsection{3D FWM's and 2D FWM's, the mathematical description of the
experiment}

To prove the feasibility of the approach for optical generation of apertured
FWM's described in Chapter \ref{chG} one has to implement the setup depicted
in Fig.~\ref{fig68}b and show that the generated wave field indeed behaves as
a FWM in interferometric experiments. However, without the loss of generality
the task can be simplified as follows.

In section \ref{ssFTi} we demonstrated that all the defining properties of the
FWM's -- its propagation-invariance, the characteristic field distribution and
pre-determined group velocity -- can be studied in terms of the specific pair
of interfering tilted pulses, the 2D FWM's in Eq.~(\ref{t3}):%
\begin{align}
&  F_{2D}\left(  x,y,z,t\right)  =\int_{0}^{\infty}\mathrm{d}k\,\tilde
{B}\left(  k\right) \nonumber\\
&  \qquad\qquad\times\cos\left[  kx\sin\theta_{F}\left(  k\right)  \right]
\exp\left[  ik\left(  z\cos\theta_{F}\left(  k\right)  -ct\right)  \right]
\label{j1}%
\end{align}
(here we have set $\phi=0$). In other words, we have shown that the peculiar
propagation of FWM's is assured exclusively by the specific coupling between
the wave number and the direction of propagation of plane wave components of
the FWM's as defined by the function $\theta_{F}\left(  k\right)  $ in
Eq.~(\ref{su12}).

In \cite{m5} we experimentally demonstrated the feasibility of 2D FWM's in
Eq.~(\ref{j1}) as the physical concept which is much more transparent in this
sort of experiments. However, it also appeared, that the fabrication and
polishing of a high-quality, large aperture concave conical surfaces is still
a complicated task.

The mutual coherence function for the 2D FWM's can be easily deduced from the
one for the (3D) FWM's (\ref{f20}). In complete analogy with the discussion in
section \ref{ssFTi} we replace the Bessel function $J_{0}\left(  {}\right)  $
by $\cos\left(  {}\right)  $, choose appropriate coordinates and get%
\begin{align}
&  \Gamma_{2D}\left(  x_{1},x_{2},\Delta z,\gamma\Delta z-c\tau\right)
=\exp\left[  -i\beta\gamma\Delta z\right]  \int_{0}^{\infty}\mathrm{d}%
k~\left\vert \mathcal{V}\left(  k\right)  \right\vert ^{2}\nonumber\\
&  \quad\qquad\times\cos\left(  kx_{1}\sin\theta_{F}\left(  k\right)  \right)
\cos\left(  kx_{2}\sin\theta_{F}\left(  k\right)  \right)  \exp\left[
ik\left(  \gamma\Delta z-c\tau\right)  \right]  \text{.} \label{j5}%
\end{align}
Also, the mathematical description of the experiments with 2D FWM's is
analogous to that of the three-dimensional one. In the spatial intensity
distribution of the interferometric experiment%
\begin{equation}
\left\langle V_{\Sigma}^{\ast}V_{\Sigma}\right\rangle =\left\langle
V_{P}^{\ast}V_{P}\right\rangle +\left\langle F_{2D}^{\ast}F_{2D}\right\rangle
+2\operatorname{Re}\left\langle F_{2D}^{\ast}V_{P}\right\rangle \text{,}
\label{j6}%
\end{equation}
we have for the intensity of the 2D FWM
\begin{equation}
\left\langle F_{2D}^{\ast}F_{2D}\right\rangle =\int_{0}^{\infty}%
\mathrm{d}k\mathcal{S}\left(  k\right)  \cos^{2}\left(  kx\sin\theta
_{G}\left(  k\right)  \right)  \label{j7}%
\end{equation}
where the angular function $\theta_{G}\left(  k\right)  $ [see Eq.~(\ref{gg23}%
)] is determined by the specific setup and $\mathcal{S}\left(  k\right)  $ is
the spectral density of the light source (see notes in the end of the
Sec.~\ref{sGE}). For the third term instead of Eq.~(\ref{in20}), we have%
\begin{align}
\left\langle F_{2D}^{\ast}V_{P}\right\rangle  &  =\exp\left[  2\beta\gamma
z\right]  \int_{0}^{\infty}\mathrm{d}k\,\mathcal{S}\left(  k\right)
\upsilon_{F}^{\ast}\left(  k_{1}\right)  \upsilon_{P}\left(  k\right)
\nonumber\\
&  \times\cos\left(  k\rho\sin\theta_{G}\left(  k\right)  \right)
\exp\left\{  ik\left[  z\left(  1-\gamma\right)  -c\Delta t\right]  \right\}
\text{.} \label{j7a}%
\end{align}
In (\ref{j7a}) for FWM's we have to set $\gamma=1$, so that%
\begin{align}
\left\langle F_{2D}^{\ast}V_{P}\right\rangle  &  =\exp\left[  2\beta z\right]
\int_{0}^{\infty}\mathrm{d}k\mathcal{S}\left(  k\right)  \upsilon_{F}^{\ast
}\left(  k_{1}\right)  \upsilon_{P}\left(  k\right) \nonumber\\
&  \times\cos\left(  k\rho\sin\theta_{G}\left(  k\right)  \right)  \exp\left[
-ikc\Delta t\right]  \text{.} \label{j8}%
\end{align}

In analogy with the case of Bessel-X fields we can define in (\ref{j5})
$\Delta z=0$, $\mathbf{r}_{\bot2}=0$, so that the mutual coherence functions
of the 2D FWM reads%
\begin{equation}
\Gamma_{2D}\left(  x,0,0,-c\tau\right)  =\int_{0}^{\infty}\mathrm{d}%
k~\left\vert \mathcal{V}\left(  k\right)  \right\vert ^{2}\cos\left(
kx\sin\theta_{G}\left(  k\right)  \right)  \exp\left[  -ikc\tau\right]
\text{.} \label{j9}%
\end{equation}
Thus, if we record the interference pattern at $z=0$, the comparison of
Eqs.~(\ref{j9}) and (\ref{j7a}) yields
\begin{equation}
\left\langle F_{2D}^{\ast}V_{P}\right\rangle _{\Delta t}=\Gamma_{2D}\left(
x,0,0,-c\Delta t\right)  \text{,} \label{j19}%
\end{equation}
and we can conclude that the mutual coherence function of the 2D FWM's can be
studied by recording the intensity of the interference picture as the function
of the delay $\Delta t$ between the signal and reference fields.

Note, that the equation (\ref{j8}) can be given the form
\begin{align}
&  \left\langle F_{2D}^{\ast}V_{P}\right\rangle =\int_{0}^{\infty}%
\mathrm{d}k\mathcal{S}\left(  k\right) \nonumber\\
&  \qquad\times\cos\left(  kx\sin\theta_{G}\left(  k\right)  \right)
\exp\left[  ikz\cos\theta_{G}\left(  k\right)  -ikc\left(  t_{0}-\Delta
t\right)  \right]  \text{,} \label{j15}%
\end{align}
where $t_{0}=z/c$ and the constant has the interpretation of being the time
that a wave field propagating at group velocity $c$ travels the distance $z$
to the plane of measurement. The integral expression (\ref{j15}) is very
similar to the one describing the field of the 2D FWM's (\ref{j1}), the only
difference being that the theoretical angular function $\theta_{F}\left(
k\right)  $ is replaced by the $\theta_{G}\left(  k\right)  $ and frequency
spectrum is replaced by the power spectrum $\mathcal{S}\left(  k\right)  $ in
(\ref{j15}). Again, as the absolute phases of the plane wave components are
inevitably lost in any linear interferometric experiments we can detect only
the amplitude distribution of the wave field. However, the interferograms
\textit{do} carry information about the defining, most essential
characteristic of the FWM's -- their \textit{support} of the angular spectrum
of plane waves. Indeed, the general structure of the interference patterns in
Eq.~(\ref{j15}) is primarily determined by the angular function $\theta
_{G}\left(  k\right)  $, and it resembles the corresponding transform-limited
wave field only if $\theta_{G}\left(  k\right)  =\allowbreak\theta_{F}\left(
k\right)  $ in good approximation over the entire bandwidth of the field.

\subsubsection{Setup}%

%TCIMACRO{\FRAME{ftbpFU}{4.497in}{5.6879in}{0pt}{\Qcb{Experimental setup for
%generating 2D FWM's and recording its interference with plane wave pulses. The
%FWM generator can be seen in grayed area; M's, mirrors; L's, lenses; BS's,
%beam splitters; W's, wedges; G, diffractional grating; AL, Xe-arc lamp; PH,
%pinhole; GP's, compensating glass plates.}}{\Qlb{fig83}}{fig8_3.jpg}%
%maintain-aspect-ratio TRUE;  display "ICON";  valid_file "F";  width 4.497in;
%height 5.6879in;  depth 0pt;  original-width 4.9104in;
%original-height 6.6841in;  cropleft "0";  croptop "1";  cropright "1";
%cropbottom "0";  filename 'fig8_3.jpg';file-properties "XNPEU";}} }%
%BeginExpansion
\begin{figure}
[ptb]
\begin{center}
\includegraphics[
height=5.6879in,
width=4.497in
]%
{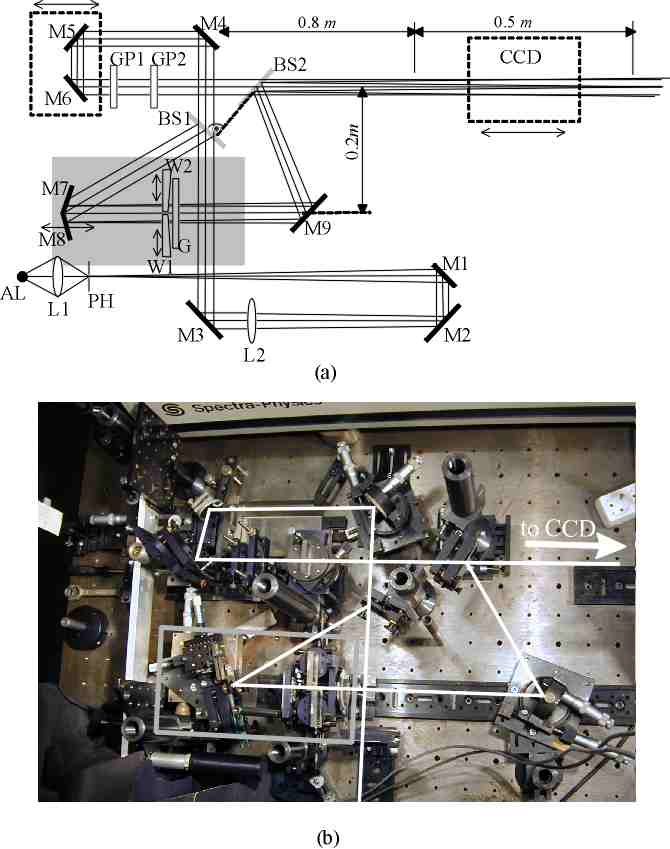}%
\caption{Experimental setup for generating 2D FWM's and recording its
interference with plane wave pulses. The FWM generator can be seen in grayed
area; M's, mirrors; L's, lenses; BS's, beam splitters; W's, wedges; G,
diffractional grating; AL, Xe-arc lamp; PH, pinhole; GP's, compensating glass
plates.}%
\label{fig83}%
\end{center}
\end{figure}
%EndExpansion
%TCIMACRO{\FRAME{ftbpFU}{3.0208in}{5.4008in}{0pt}{\Qcb{(a) The power spectrum
%of the light used in our setup; (b) The angular spectrum of the plane waves
%generated in the setup (solid black line) as compared to the theoretical one
%(dotted cyan line); (c) The deviation of generated support, here
%$k_{1}=7.4\times10^{6}\frac{rad}{m}$ ($\lambda=849nm$), $k_{2}=10.6\times
%10^{7}\frac{rad}{m}$ ($\lambda=593nm$), $k_{3}=1.6\times10^{7}\frac{rad}{m}$
%($\lambda=381nm$).}}{\Qlb{fig84}}{fig8_4.jpg}%
%maintain-aspect-ratio TRUE;  display "ICON";  valid_file "F";
%width 3.0208in;  height 5.4008in;  depth 0pt;  original-width 3.5388in;
%original-height 2.3566in;  cropleft "0";  croptop "1";  cropright "1";
%cropbottom "0";  filename 'fig8_4.jpg';file-properties "XNPEU";}} }%
%BeginExpansion
\begin{figure}
[ptbptb]
\begin{center}
\includegraphics[
height=5.4008in,
width=3.0208in
]%
{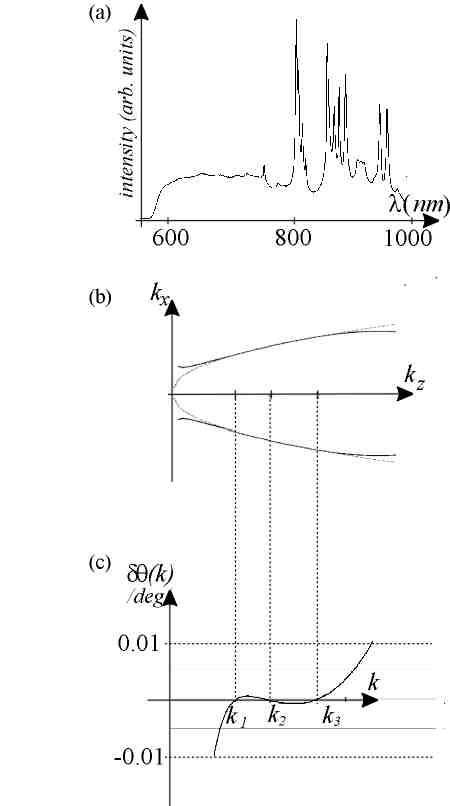}%
\caption{(a) The power spectrum of the light used in our setup; (b) The
angular spectrum of the plane waves generated in the setup (solid black line)
as compared to the theoretical one (dotted cyan line); (c) The deviation of
generated support, here $k_{1}=7.4\times10^{6}\frac{rad}{m}$ ($\lambda
=849nm$), $k_{2}=10.6\times10^{7}\frac{rad}{m}$ ($\lambda=593nm$),
$k_{3}=1.6\times10^{7}\frac{rad}{m}$ ($\lambda=381nm$).}%
\label{fig84}%
\end{center}
\end{figure}
%EndExpansion
The conversion of the FWM generator in Fig.~\ref{fig68}b to the
two-dimensional case is straightforward -- we just replace the axicon and
circular diffraction by their one-dimensional counterparts -- prisms (wedges)
and a diffraction grating. The initial field on the elements is an interfering
pair of pulsed plane waves.

The setup of our experiment is depicted in Fig.~\ref{fig83}. The main part of
it is the (2D) FWM generator that consist of the mirrors M7 and M8, of the two
wedges W1 and W2 and of a blazed diffraction grating G.(see grayed area in
Fig.~\ref{fig83}). The FWM generator is placed into an arm of an
interferometer as will be explained below.

In our experiment we implemented a 2D FWM with following parameters:
$\beta=\allowbreak40rad/m$, $\gamma=1$ ($v_{g}=c$) giving $\theta_{F}\left(
k_{0}\right)  \approx\allowbreak0.23%
%TCIMACRO{\U{b0}}%
%BeginExpansion
{{}^\circ}%
%EndExpansion
$ if $k_{0}=\allowbreak7.8\times10^{6}rad/m$ ($\lambda_{0}\approx800nm$) (see
Fig.~\ref{fig84}a for the spectral density of the light source and
Fig.~\ref{fig84}b for the support of the angular spectrum of plane waves of
the specified 2D FWM).

The FWM generator has three free parameters: $\alpha$ -- the apex angle of the
wedges, $\theta_{0}$ -- the angle that the initial pulsed plane waves subtend
with the optical axis of the setup and $d$ -- the groove spacing of the
diffraction grating. As to find the values for the parameters that give the
best fit between the generated support of the angular spectrum of plane waves
and the support of angular spectrum of plane waves of the theoretical FWM's in
Eq.~(\ref{su12})%
\begin{equation}
\theta_{F}\left(  k\right)  =\arccos\left(  \frac{k-2\beta}{k}\right)
\label{j20}%
\end{equation}
($\gamma=1$) we combine the (\ref{j20}) with the Eq.~(\ref{gg24}) and write
the following system of equations:%
\begin{equation}
\arccos\left(  \frac{\gamma\left(  k_{m}-2\beta\right)  }{k_{m}}\right)
=\frac{2\pi}{k_{m}d}+\alpha\left(  1-n\left(  k_{m}\right)  \right)
+\theta_{0}\;,m=1,2,3\text{.} \label{j21}%
\end{equation}
We specify the three wave numbers as $k_{1}=7.4\times10^{6}\frac{rad}{m}$
($\lambda=849nm$), $k_{2}=10.6\times10^{7}\frac{rad}{m}$ ($\lambda=593nm$),
$k_{3}=1.6\times10^{7}\frac{rad}{m}$ ($\lambda=381nm$) and assumed that the
wedges are made of the optical glass BK\thinspace7 for which the refractive
index $n\left(  k\right)  $ is known to an accuracy better than $10^{-5}$. The
system\thinspace(\ref{j21}) then yields
\begin{align}
\alpha &  =1.2044\times10^{-2}rad\nonumber\\
d  &  =3.7494\times10^{-4}m\label{j22}\\
\theta_{0}  &  =9.4683\times10^{-3}rad~\text{.}\nonumber
\end{align}
As inserted into Eq.~(\ref{gg23}), the maximum deviation $\delta\theta\left(
k\right)  =\theta_{F}\left(  k\right)  -\theta_{G}\left(  k\right)  $ for the
wavelength dispersion of the cone angle in the selected spectral range is as
small as $5\times10^{-4}\deg$, i.e., $<0.2\%$. A comparison of the
corresponding supports of angular spectrum and the exact form of deviation
$\delta\theta\left(  k\right)  $ are depicted in Figs.~\ref{fig84}b and
\ref{fig84}c (see also Fig.~\ref{fig65}). As for rough estimation of the
spread of the pulse due to the $\delta\theta\left(  k\right)  $ one can
estimate the corresponding maximum group velocity dispersion $\Delta v^{g}$
and compare this with the mean wavelength of the pulse. The numerical
simulations show that the approximation is indeed good enough for propagating
the central peak of the FWM's over several meters.

The FWM generator has been placed in what is basically a specially designed,
modified Mach-Zehnder interferometer. The interferometer consists of two
beamsplitters and of identical broadband non-dispersive mirrors. The input
field from the light source is split by the beamsplitter BS1 into the fields
that travel through the two arms of the interferometer, the one with the FWM
generator and the arm for the reference beam. The mirrors M5 and M6 form a
delay line, they were translated by the \textit{Burleigh Inchworm} linear step
motor, the $1\mu m$ translation step of which was reduced to $65nm$ by a
transmission mechanism. The mirror M7 was continuously translatable as to
correct for the time-shift between the two tilted pulses. The wedges W1 and W2
were transversally translatable as to balance the material dispersion they
introduce to the plane wave pulses (see text below). We used \textit{Kodak
Megaplus~1.6i} CCD camera with the $1534\times1024$ matrix resolution and
10~bit pixel depth. The linear dimensions of the matrix are $13.8mm$%
(H)$\times9.2mm$(V), the pixel size is $9\mu m\times9\mu m$.

Again, in our experiment we used the filtered light form a superhigh pressure
Xe-arc lamp, giving $\approx5\,fsec$ correlation time for the input field [see
Fig.~\ref{fig84}a for the power spectrum of the light]. To ensure good
transversal coherence over the clear aperture of the setup the required
maximum diameter $\approx15\mu m$ of the pinhole and focal length $2m$\ of the
collimating Fourier lens L1 was estimated from the van Cittert-Zernike theorem
\cite{ok4} for the mean wavelength of the light $\lambda_{0}=800nm$. As the
result of filtering, the total power of the signal on the $\approx1.5cm^{2}$
CCD chip was very low, approximately $0.03\mu W$.

Due to the short coherence time of the source field, the experiment is highly
sensitive to the phase distortions (spectral phase shift) introduced by the
dispersive optical elements of the system -- the beamsplitters and the FWM
generator. In the FWM generator there is three possible sources of undesirable
dispersion: (1) the propagation in the glass substrate of the diffraction
grating, (2) the propagation between the grating and the axicon where the
support of angular spectrum of the wave field is not appropriate for the free
space propagation, i.e., it does not obey the Eq.~(\ref{su12}) and (3) the
propagation in the wedges. The beamsplitters in our setup are identical and if
we set them perpendicularly and orient the coated sides so that each beam
passes the glass substrate of the beamsplitter twice, the arms of the
interferometer remain balanced. The influence of the propagation between the
elements can be made neglible by placing them close to each other. The
character of the undesirable dispersion in the wedges can be estimated from
the following considerations. The entrance wave field on the wedges is the
transform-limited Bessel-X pulse, so, the on-axis part of the pulse is also
transform-limited and should pass through the wedges unchanged, i.e., without
any additional spectral phase shift. Consequently, the wedges should be
produced and aligned so that their thickness is zero on the optical axis. As
the apex angle of the wedges is very small in our setup ($\approx0.7%
%TCIMACRO{\U{b0}}%
%BeginExpansion
{{}^\circ}%
%EndExpansion
$), this is not very practical approach and we consider the finite thickness
on the axis as the source of additional spectral phase shift instead. Thus,
the composite spectral phase shift of the FWM generator can be described as
the phase distortion introduced by the substrate of the diffraction grating
and by a glass plate of the material of the wedges, the thickness of which is
equal to the thickness of the wedges on the optical axis. We balanced the arms
of the interferometer by inserting material dispersion into the reference arm
of the setup by means of two appropriate glass plates (GP1 and GP2 in
Fig.~\ref{fig83}a).

\subsubsection{Results of the experiment}%

%TCIMACRO{\FRAME{ftbpFU}{4.0612in}{5.0609in}{0pt}{\Qcb{(a) A typical
%interferogram in the setup as recorded by the CCD camera; (b) The interference
%pattern in the setup as the function of the delay between the signal and
%reference wave fields in three positions of the CCD camera (see text for more
%detailed description).}}{\Qlb{fig85}}{fig8_5.jpg}%
%maintain-aspect-ratio TRUE;  display "ICON";  valid_file "F";
%width 4.0612in;  height 5.0609in;  depth 0pt;  original-width 4.875in;
%original-height 7.4806in;  cropleft "0";  croptop "1";  cropright "1";
%cropbottom "0";  filename 'fig8_5.jpg';file-properties "XNPEU";}} }%
%BeginExpansion
\begin{figure}
[ptb]
\begin{center}
\includegraphics[
height=5.0609in,
width=4.0612in
]%
{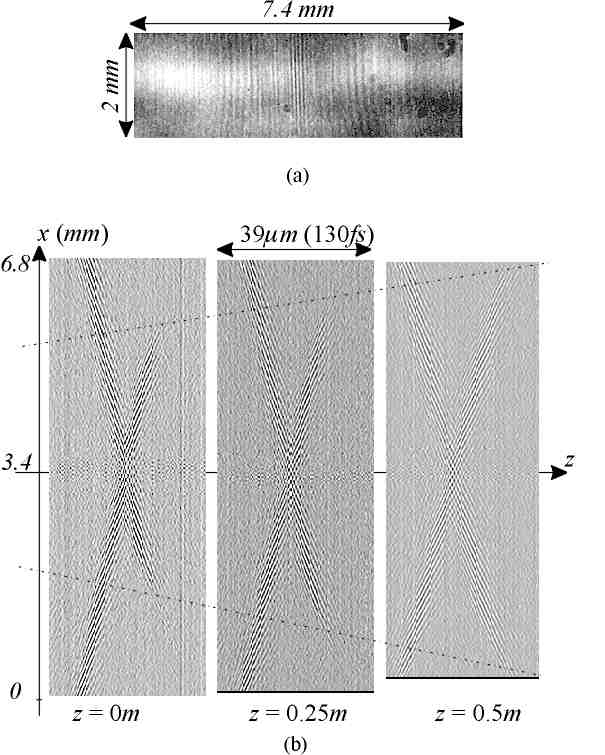}%
\caption{(a) A typical interferogram in the setup as recorded by the CCD
camera; (b) The interference pattern in the setup as the function of the delay
between the signal and reference wave fields in three positions of the CCD
camera (see text for more detailed description).}%
\label{fig85}%
\end{center}
\end{figure}
%EndExpansion
In first experiment we recorded the time-averaged interference pattern of the
2D FWM and the reference wave field as the function of the time delay between
the two. The experiment can be mathematically modeled by varying parameter
$\Delta t$ in Eq.~(\ref{j8}). We scanned the time-delay at three $z$-axis
positions, $z=0cm$, $z=25cm$, $z=50cm$ (the origin of the $z$-axis is about
$30cm$ away from the beamsplitter BS2 in Fig.~\ref{fig83}a). In each
experiment we recorded $300$ interferograms, the time-delay step was
$0.43\,fsec$ ($0.13\,\mu m$).

In a typical interference pattern in our experiment [see Fig.~\ref{fig85}a]
the sharp vertical interference fringes in the center correspond to the second
term in the interference sum (\ref{j6}) -- this is the time-averaged,
\textquotedblright propagation-invariant\textquotedblright\ time-averaged
intensity of the 2D FWM. The fringes can also be interpreted as the
autocorrelation function of the interfering tilted pulses [see
Fig.~\ref{fig84}a for the corresponding power spectrum]. In this experiment
the intensity of the wave field under study do not carry any important
information, so we subtracted it numerically from the results in
Fig.~\ref{fig85}b. The interference fringes that are symmetrical at both sides
of the central part correspond to the third, most important term in this sum.
It can be seen from Fig.~\ref{fig85}a, that due to the low signal level the
recorded interferograms are quite noisy. To get the better signal-to-noise
ratio, we averaged the data in the interferograms over the rows and used the
resulting one-dimensional data arrays instead.

The results of the experiment are depicted in Fig.~\ref{fig85}b. We can see,
that there is a good qualitative resemblance between the measured $x\Delta t$
plot of the interference pattern and the theoretical field distribution of the
2D FWM's in Fig.~\ref{fig411}b as it was predicted by the Eq.~(\ref{j15}) --
one can clearly recognize the two interfering tilted pulses forming the
characteristic X-branching, also the phase fronts in the tilted pulses can be
seen. The wave field is definitely transform-limited, so we have managed to
compensate for the spectral phase shift in the 2D FWM generator.

We can also see, that the interference pattern does not show any spread over
the $0.5\,m$ distance, consequently, the wave field does not spread in the
course of propagation.

An additional detail can be found in Fig.~\ref{fig85}b: the tilted pulses do
not extend across the whole picture but are cut out (see dashed lines in
Fig.~\ref{fig85}b). Also, the \textquotedblright edges\textquotedblright\ of
the tilted pulses move away from the optical axis. This effect can be clearly
interpreted as the consequence of the finite extent of the tilted pulses, as
illustrated in Figs.~\ref{fig65}a and \ref{fig67} -- the dashed lines simply
mark the borders of the volume, where the two tilted pulses intersect, i.e.,
the borders of the volume, where the 2D FWM exist [see the striped area in
Fig.~\ref{fig65}a].%

%TCIMACRO{\FRAME{ftbpFU}{2.7579in}{6.0943in}{0pt}{\Qcb{The interference pattern
%as the function of the CCD camera position (see text for detailed
%description).}}{\Qlb{fig86}}{fig8_6.jpg}%
%maintain-aspect-ratio TRUE;  display "ICON";  valid_file "F";
%width 2.7579in;  height 6.0943in;  depth 0pt;  original-width 4.9104in;
%original-height 6.6841in;  cropleft "0";  croptop "1";  cropright "1";
%cropbottom "0";  filename 'fig8_6.jpg';file-properties "XNPEU";}} }%
%BeginExpansion
\begin{figure}
[ptb]
\begin{center}
\includegraphics[
height=6.0943in,
width=2.7579in
]%
{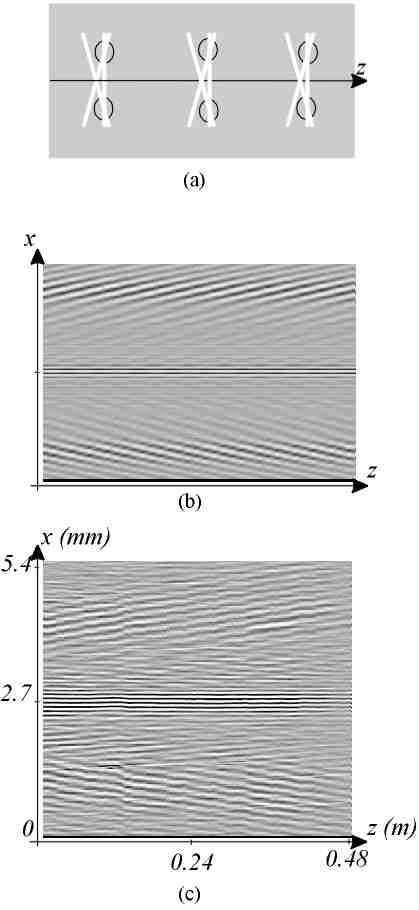}%
\caption{The interference pattern as the function of the CCD camera position
(see text for detailed description).}%
\label{fig86}%
\end{center}
\end{figure}
%EndExpansion
In the second experiment we recorded the interference pattern as the function
of the propagation distance $z$. The experiment can be simulated by varying
$z$ coordinate in Eq.~(\ref{j8}). We recorded $240$ interferograms, the step
of the CCD camera position was $3.1\,mm$. The numerical simulation of the
experiment and the results of the experiment are depicted in Figs.~\ref{fig86}%
b and \ref{fig86}c respectively.

The experiment can be easily interpreted -- the position-invariant envelope of
the interference pattern is the consequence of the fact, that the group
velocities of the propagation-invariant 2D FWM and the reference field are
equal, $c$, so that the overlapping volume of the two fields do not change in
the course of propagation (see Fig.~\ref{fig86}a). The $z$ dependent finer
structure of the interferograms is the consequence of the fact, that the phase
velocities of the plane wave pulse and 2D FWM are not equal, i.e., we have
also $v^{g}\neq v^{p}$ for the phase and group velocities of the 2D FWM. The
result of the experiment in Fig.~\ref{fig86}c show good qualitative agreement
with the theory.

We can also determine the parameter $\beta$ from our experiment -- the
exponent multiplier in Eq.~(\ref{j8}) reads $\exp\left[  i2\beta z\right]  $,
thus $\beta=\pi/z_{0}$ where $z_{0}$ is the period of the variations along the
$z$-axis. From the result in Fig.~\ref{fig86}a we estimated $z_{0}%
\approx7.5cm$, so that $\beta\approx42\,rad/m$, which result is in good
agreement with the theory.

Thus, we have shown, that the generated wave field has all the characteristic
properties described in previous theoretical sections and the validity of the
general idea has been given an experimental proof.

\section{\label{chSI}SELF-IMAGING OF PULSED WAVE\newline FIELDS}

Self-imaging, also known as Talbot effect is, in its original sense, a
well-known phenomenon in classical wave optics where certain wave fields
reproduce their transversal intensity distribution at periodic spatial
intervals in the course of propagation (see Refs.~\cite{si1} -- \cite{si30},
and references therein). The effect has been studied extensively by means of
Fresnel diffraction theory and the angular spectrum representation of scalar
wave fields. As a result, the general, physically transparent conditions have
been formulated the transversal intensity distribution of a wave field have to
obey to be self-imaging. It is also well known that the monochromatic
propagation-invariant wave fields (Bessel beams) constitute a special class of
self-imaging wave fields, the mathematical description of the two being much
the same.

In recent years the effect has attracted a renewed interest, as the concept
has been generalized into the domain of pulsed wave fields by several authors
(see Refs. \cite{si20,si28,si29,si37,si40}, and references therein). The
phenomenon has been discussed in the context of fiber optics \cite{si20} and
also as a property of spatial, wideband wave fields \cite{si28,si29,si37,si40}.

In what follows we show, that the discrete superpositions of FWM's over the
parameter $\beta$
\begin{equation}
\Psi^{\left(  SI\right)  }\left(  \rho,z,\varphi,t\right)  =\sum_{q}C_{q}%
\Psi_{F}\left(  z,\rho,\varphi,t;\beta_{q}\right)  \label{ss5}%
\end{equation}
can be used as the self-imaging \textquotedblleft pixels\textquotedblright\ of
a spatiotemporally self-imaging three-dimensional spatial image -- the wave
fields of this type can reproduce spatial separated copies of its initial
three-dimensional intensity distribution at specific time intervals. The
results in this chapter are published in Ref.~\cite{m4}.

(It is important to note, that our discussion is closely related to those in
Refs. [\cite{si28}] and \cite{si40} -- those publications consider essentially
the same problem, however the analysis is different in each occasion)

\subsection{Monochromatic self-imaging}

The self-imaging condition for the monochromatic wave field has been defined
as (see, e.g., Ref. \cite{si10})%
\begin{equation}
\Psi\left(  \rho,z,\varphi,t\right)  =\Psi\left(  \rho,z+d,\varphi,t\right)
\text{.} \label{ss10}%
\end{equation}
With the condition (\ref{ss10}) and the Whittaker type angular spectrum of
plane waves of monochromatic wave fields (see Eq.~(\ref{ang9})%
\begin{align}
\Psi\left(  \rho,z,\varphi,t;k\right)   &  =\sum_{n=0}^{\infty}\exp\left[  \pm
in\varphi\right]  \int_{0}^{\pi}\mathrm{d}\theta\mathbf{~}\tilde{A}_{n}\left(
k,\theta\right) \nonumber\\
&  \times J_{n}\left(  k\rho\sin\theta\right)  \exp\left[  ik\left(
z\cos\theta-ct\right)  \right]  \label{ss11}%
\end{align}
one can easily deduce the condition for self-imaging for the monochromatic
scalar wave fields that reads \cite{si10}%
\begin{equation}
kd\cos\theta=\psi+2\pi q\text{,} \label{ss12}%
\end{equation}
where $q$ is an integer and $\psi$ is an arbitrary phase factor. The relation
(\ref{ss12}) implies, that a monochromatic wave field (\ref{ang7})
periodically reconstructs its initial transversal field distribution if only
its angular spectrum of plane waves is sampled so that the condition%
\begin{equation}
k_{z}=\frac{\psi+2\pi q}{d}\text{,} \label{ss15}%
\end{equation}
where $k_{z}$ is again the $z$ component of the wave vector, is satisfied for
every plane wave component of the wave field. Applying the condition
(\ref{ss15}) and including only the axially symmetric terms in the summation,
we get the following expression for the general cylindrically symmetric,
monochromatic, self-imaging wave field:%
\begin{equation}
\Psi\left(  \rho,z,t;k\right)  =\exp\left[  -i\omega t\right]  \sum_{q}%
a_{q}J_{0}\left[  k\rho\sqrt{1-\left(  \frac{2\pi q}{d}\right)  ^{2}}\right]
\exp\left[  i\frac{2\pi q}{d}z\right]  \text{,} \label{ss20}%
\end{equation}
where we have denoted
\begin{equation}
a_{q}=\tilde{A}_{0}\left(  k,\arccos\frac{2\pi q}{kd}\right)  \label{ss25}%
\end{equation}
and $\psi=0$ is assumed. Thus, we have a discrete superposition of Bessel
beams the cone angles are specified by Eq.~(\ref{ss15}). The physical content
of Eqs.~(\ref{ss15}) and (\ref{ss20}) can be summarized by saying that the
monochromatic self-imaging is essentially the phenomenon where the wave field
is a discrete superposition of wave fields with different phase velocities so
that the total field, being the superposition of the composite fields, depends
periodically on the distance.

The on-axis superposition in (\ref{ss20}),%
\begin{equation}
\Psi\left(  \rho,z,t;k\right)  =\exp\left[  -i\omega t\right]  \sum_{q}%
a_{q}\exp\left[  i\frac{2\pi q}{d}z\right]  \label{ss30}%
\end{equation}
can be recognized as the Fourier series representation of a periodic function
along the optical axis. However, in (\ref{ss30}) the condition (\ref{ss12})
together with the causality requirement $k_{z}>0$ imply that%
\begin{equation}
0<q<\frac{kd}{2\pi}\text{.} \label{ss31}%
\end{equation}
An example of the on axis intensity distribution of a superposition of Bessel
beams in (\ref{ss30}) is depicted in Fig.~\ref{fig107}.

\subsection{Self-imaging of pulsed wave fields}%

%TCIMACRO{\FRAME{ftbpFU}{4.4209in}{5.4483in}{0pt}{\Qcb{The snapshots of the
%temporal evolution of the spatial intensity distributions of the pulsed
%self-imaging wave field, the three-dimensional self-imaging \textquotedblleft
%pixel\textquotedblright, at 26 femtosecond time intervals. Due to the axial
%symmetry, the distribution is shown in one meridional (say, $xz$) plane.}%
%}{\Qlb{fig91}}{fig9_1.jpg}{\special{ language "Scientific Word";
%type "GRAPHIC";  maintain-aspect-ratio TRUE;  display "ICON";
%valid_file "F";  width 4.4209in;  height 5.4483in;  depth 0pt;
%original-width 4.9104in;  original-height 6.6841in;  cropleft "0";
%filename 'fig9_1.jpg';file-properties "XNPEU";}} }%
%BeginExpansion
\begin{figure}
[ptb]
\begin{center}
\includegraphics[
height=5.4483in,
width=4.4209in
]%
{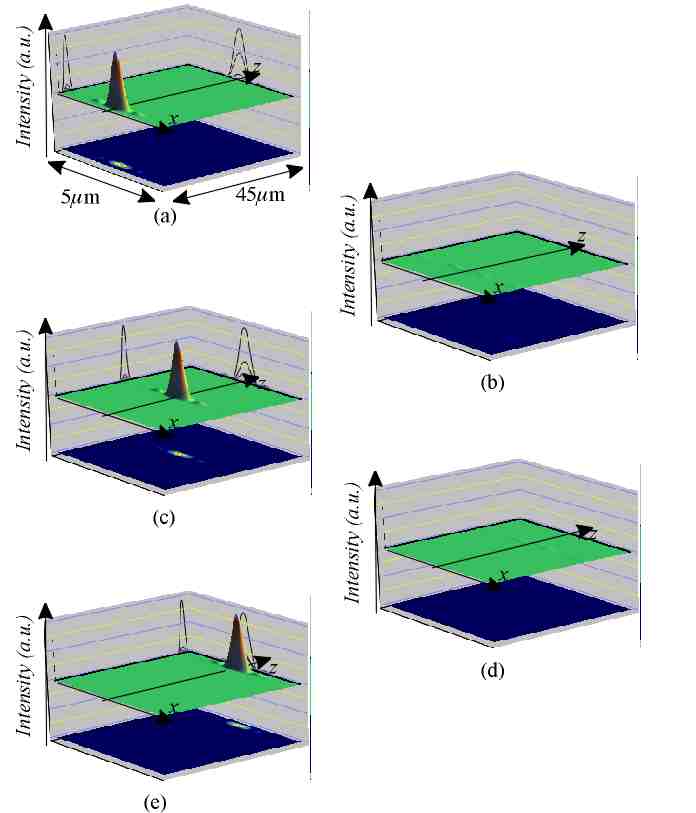}%
\caption{The snapshots of the temporal evolution of the spatial intensity
distributions of the pulsed self-imaging wave field, the three-dimensional
self-imaging \textquotedblleft pixel\textquotedblright, at 26 femtosecond time
intervals. Due to the axial symmetry, the distribution is shown in one
meridional (say, $xz$) plane.}%
\label{fig91}%
\end{center}
\end{figure}
%EndExpansion
From the material in Chapter \ref{chPhAp} we know that the phase velocities
$v^{p}$ of the FWM's with different parameter $\beta$ along the optical axis
are generally different and we can write%
\begin{equation}
v^{p}\left(  \beta\right)  =\frac{c}{\cos\theta_{F}\left(  k_{0},\beta\right)
}=\frac{ck_{0}}{\gamma\left(  k_{0}-2\beta\right)  }\text{.} \label{ss32}%
\end{equation}
Thus, the superposition of FWM's over the parameter $\beta$ in (\ref{ss5})%
\begin{equation}
\Psi^{\left(  SI\right)  }\left(  \rho,z,\varphi,t\right)  =\sum_{q}C_{q}%
\Psi_{F}\left(  z,\rho,\varphi,t;\beta_{q}\right)  \label{ss37}%
\end{equation}
is taken over a set of overlapping, non-spreading optical pulses that (1) are
transversally localized, i.e., their transversal intensity distribution have a
single narrow intense peak, (2) have equal carrier frequency, (3) propagate at
equal group velocities, but (4) have different phase velocities. In complete
analogy with the monochromatic self-imaging we could suggest that if the
component pulses satisfy certain conditions, the single narrow peak of the
wave field could periodically vanish and reconstruct its initial (localized)
transversal intensity distribution. As a result we could get a wave field the
temporal evolution of which can be perceived as a spatial arrow of
sequentially visible light spots (see Fig.~\ref{fig91}). In this sense, such
superposition could be considered self-imaging and, due to its spatial
localization, it could be used as a pulsed \textquotedblleft self-imaging
pixel\textquotedblright\ of a transversal or even spatial image.

Consider a discrete superposition of a set of the tilted pulses in
Eq.~(\ref{t2})
\begin{equation}
T^{\left(  SI\right)  }\left(  x,z,t\right)  =\sum_{q}a_{q}\int_{0}^{\infty
}\mathrm{d}k\,\tilde{A}\left(  k\right)  \exp\left[  ik\left(  x\sin\theta
_{F}\left(  k\right)  +z\cos\theta_{F}\left(  k\right)  -ct\right)  \right]
\text{,} \label{ss36}%
\end{equation}
the function $\theta_{F}\left(  k,\beta_{q}\right)  $ being determined by the
Eq. (\ref{su12}). The expression can be given a readily interpretable form if
we approximate the $x$ and $z$ components of the wave vector by%
\begin{equation}
k_{x}\left(  k,\beta\right)  =k_{x}\left(  k_{0},\beta\right)  +\left.
\frac{d}{dk}k_{x}\left(  k,\beta\right)  \right\vert _{k=k_{0}}\left(
k-k_{0}\right)  \label{ss38}%
\end{equation}%
\begin{align}
k_{z}\left(  k,\beta\right)   &  =k_{z}\left(  k_{0},\beta\right)  +\left.
\frac{d}{dk}k_{z}\left(  k,\beta\right)  \right\vert _{k=k_{0}}\left(
k-k_{0}\right) \nonumber\\
&  =k_{z}\left(  k_{0},\beta\right)  +\gamma\left(  k-k_{0}\right)  \text{,}
\label{ss40}%
\end{align}
where we have used Eq.~(\ref{su10}). Substitution of relations (\ref{ss38})
and (\ref{ss40}) in Eq. (\ref{ss36}) yields
\begin{align}
&  T^{\left(  SI\right)  }\left(  x,z,t\right)  \cong\sum\limits_{q}a_{q}%
L_{q}\left(  x,z\gamma-ct\right) \nonumber\\
&  \qquad\times\exp\left[  ik_{0}\left(  x\sin\theta_{F}\left(  k_{0}%
,\beta_{q}\right)  +z\cos\theta_{F}\left(  k_{0},\beta_{q}\right)  -ct\right)
\right]  \text{,} \label{ss46}%
\end{align}
where $\exp\left[  ik_{0}\left(  x\sin\theta_{F}\left(  k_{0},\beta
_{q}\right)  +z\cos\theta_{F}\left(  k_{0},\beta_{q}\right)  -ct\right)
\right]  $ is the carrier wavelength plane wave component of the tilted pulse
(\ref{ss36}) and%
\begin{align}
&  L_{q}\left(  x,z\gamma-ct\right)  =\int\nolimits_{\,-k_{0}}^{\,\infty
}\mathrm{d}k{}A\left(  k+k_{0}\right) \nonumber\\
&  \qquad\times\exp\left[  ik\left(  x\left.  \frac{d}{dk}k_{x}\left(
k,\beta_{q}\right)  \right\vert _{k_{0}}+z\gamma-ct\right)  \right]
\label{ss51}%
\end{align}
is an approximation to the non-spreading traveling envelope of the pulse. In
near axis volume we can write%
\begin{align}
&  T^{\left(  SI\right)  }\left(  x,z,t\right)  \cong L\left(  x,z\gamma
-ct\right)  \sum\limits_{q}a_{q}\label{ss55}\\
&  \qquad\qquad\times\exp\left[  ik_{0}\left(  x\sin\theta_{F}\left(
k_{0},\beta_{q}\right)  +z\cos\theta_{F}\left(  k_{0},\beta_{q}\right)
-ct\right)  \right]  \text{.}\nonumber
\end{align}
The Eq.{~}(\ref{ss55}) is essentially a product of a propagating pulse
$L(x,z\gamma-ct$) and of a term, that is a mathematical equivalent of the
superposition of the monochromatic carrier-wavelength plane waves that
propagate at angles $\theta_{F}\left(  k_{0},\beta_{q}\right)  $ to the
optical axis. According to our general idea, the latter term should be
self-imaging in the conventional, monochromatic sense of the term in
Eq.~(\ref{ss20}) -- in this case the product (\ref{ss55}) behaves as a pulse
that vanishes and reconstructs itself periodically.

\begin{figure}
[ptb]
\begin{center}
\includegraphics[
height=2.015in,
width=2.8305in
]%
{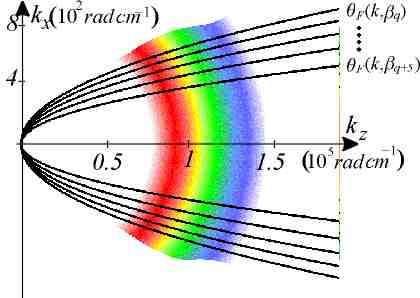}%
\caption{An example of the set of supports of the angular spectrums of plane
waves of a self-imaging superposition of the tilted pulses (see text). }%
\label{fig105}%
\end{center}
\end{figure}

In complete analogy with the Eq.{~}(\ref{ss12}) the condition for such
behavior can be written as%
\begin{equation}
k={}d\cos\theta_{F}\left(  k_{0},\beta_{q}\right)  =\psi+2\pi q\text{.}
\label{ss57}%
\end{equation}
and with the Eq.{~}(\ref{su10}) we can write%
\begin{equation}
\frac{2\pi q}{d}=\gamma k_{0}-2\gamma\beta\text{~,} \label{ss59}%
\end{equation}
so that we get a discrete set of constants $\beta$ for the superposition
(\ref{ss36}):%
\begin{equation}
\beta_{q}=\frac{k_{0}}{2}-\frac{\pi q}{\gamma d} \label{ss61}%
\end{equation}
(see Fig.{~}\ref{fig105} for an example). The direction of propagation of the
carrier wavelength plane wave component of the tilted pulse (\ref{ss36}) can
be found by combining the Eqs.{~}(\ref{su10}) and (\ref{ss61}): we can write%
\begin{equation}
\theta_{F}\left(  k,\beta_{q}\right)  =\arccos\left(  \frac{\gamma\left(
k-k_{0}\right)  }{k}+\frac{2\pi}{kd}q\right)  \text{,} \label{ss63}%
\end{equation}
so that%
\begin{equation}
\theta_{F}\left(  k_{0},\beta_{q}\right)  =\arccos\left(  \frac{2\pi}{k_{0}%
d}q\right)  \text{.} \label{ss66}%
\end{equation}
Note, that again, in (\ref{ss66}) the angle $\theta_{F}\left(  k,\beta
_{q}\right)  $ have to satisfy the condition%
\begin{equation}
0<q<\frac{k_{0}d}{2\pi}\text{~.} \label{ss69}%
\end{equation}

With those conditions we can write the cylindrical superposition of the tilted
pulses in Eq.~(\ref{ss36})
\begin{equation}
\Psi^{\left(  SI\right)  }\left(  \rho,z,t\right)  =\int_{0}^{2\pi}%
\mathrm{d}\varphi~T^{\left(  SI\right)  }\left(  x,z,t;\varphi\right)
\label{ss77}%
\end{equation}
as
\begin{align}
&  \Psi^{\left(  SI\right)  }\left(  \rho,z,t\right)  =\sum\limits_{q}%
a_{q}\exp\left[  -i2\gamma\beta_{q}z\right]  \int\nolimits_{0}^{\infty
}\mathrm{d}k{}\tilde{A}\left(  k\right) \label{ss80}\\
&  \qquad\times J_{0}\left[  k\rho\sqrt{1-\left(  \frac{\gamma\left(
k-2\beta_{q}\right)  }{k}\right)  ^{2}}\right]  \exp\left[  ik\left(
z\gamma-ct\right)  \right]  \text{.}\nonumber
\end{align}
and this is the discrete superposition of FWM's we were looking for in
Eq.~(\ref{ss37}).%

%TCIMACRO{\FRAME{ftbpFU}{2.8305in}{2.015in}{0pt}{\Qcb{An example of the set of
%supports of the angular spectrums of plane waves of a self-imaging
%superposition of the tilted pulses (see text). }}{\Qlb{fig105}}{fig10_5.jpg}%
%maintain-aspect-ratio TRUE;  display "ICON";  valid_file "F";
%width 2.8305in;  height 2.015in;  depth 0pt;  original-width 8.4864in;
%original-height 10.99in;  cropleft "0";  croptop "1";  cropright "1";
%cropbottom "0";  filename 'fig10_5.jpg';file-properties "XNPEU";}} }%
%BeginExpansion
%EndExpansion
%TCIMACRO{\FRAME{ftbpFU}{4.7141in}{2.0349in}{0pt}{\Qcb{(a) The Fourier spectrum
%and (b) the spatial amplitude of a train of sinusoidal waves}}{\Qlb{fig107}%
%}{fig10_7.jpg}{\special{ language "Scientific Word";  type "GRAPHIC";
%maintain-aspect-ratio TRUE;  display "ICON";  valid_file "F";
%width 4.7141in;  height 2.0349in;  depth 0pt;  original-width 4.7176in;
%original-height 1.9597in;  cropleft "0";  croptop "1";  cropright "1";
%cropbottom "0";  filename 'fig10_7.jpg';file-properties "XNPEU";}} }%
%BeginExpansion
\begin{figure}
[ptbptb]
\begin{center}
\includegraphics[
height=2.0349in,
width=4.7141in
]%
{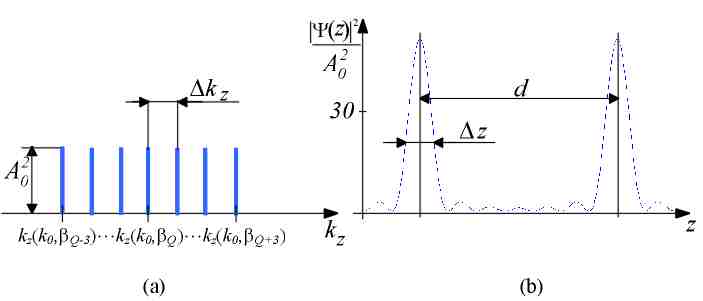}%
\caption{(a) The Fourier spectrum and (b) the spatial amplitude of a train of
sinusoidal waves}%
\label{fig107}%
\end{center}
\end{figure}
%EndExpansion
The on-axis longitudinal shape of the superposition in Eq.{~}(\ref{ss80}) can
be easily evaluated for the most practical case of a uniform superposition of
$(2n+1)$ tilted pulses, centered around some carrier spatial frequency
$k_{z}\left(  k_{0},\beta_{Q}\right)  $ (see Fig.{~}\ref{fig107}).
Superposition can be expressed as%
\begin{equation}
\Psi\left(  z\right)  =A_{0}\frac{\sin\left[  \frac{1}{2}\left(  2n+1\right)
\Delta k_{z}z\right]  }{\sin\left(  \frac{1}{2}\Delta k_{z}z\right)  }\text{,}
\label{ss75}%
\end{equation}
where $\Delta k_{z}$ is the interval between the spatial frequencies (see
Ref.{~}\cite{o28} for example). For this case the self-imaging distance is
$d=2\pi/\Delta k_{z}$ and the width of the peaks of the resulting function is
$\Delta z\approx2\pi/(2n+1)\Delta k_{z}$. The result of the evaluation of
Eq.~(\ref{ss66}) for a superposition of seven tilted pulses is shown on
Fig.{~}\ref{fig107}b.

A numerical example of the self-imaging behavior of a superposition of five
FWM's in (\ref{ss80}) is depicted in Fig.~\ref{fig91}. In this example the
self-imaging distance $d=2\times10^{-5}m$, $\gamma=1$, $k_{0}=1\times
10^{7}rad{/}m$, $\sigma_{k}=3.8\times10^{-7}m$ ($\approx3fs$) $q=27,28,...,31$
with $\beta_{q}$ being determined by Eq.~(\ref{ss61}) ($\beta_{q}%
=7.6\times10^{5}rad/m...1.3\times10^{5}rad/m$) and $\theta_{F}\left(
k_{0},\beta_{q}\right)  $ by Eq.{~}(\ref{ss66}).

\begin{figure}
[ptb]
\begin{center}
\includegraphics[
height=4.6276in,
width=4.9372in
]%
{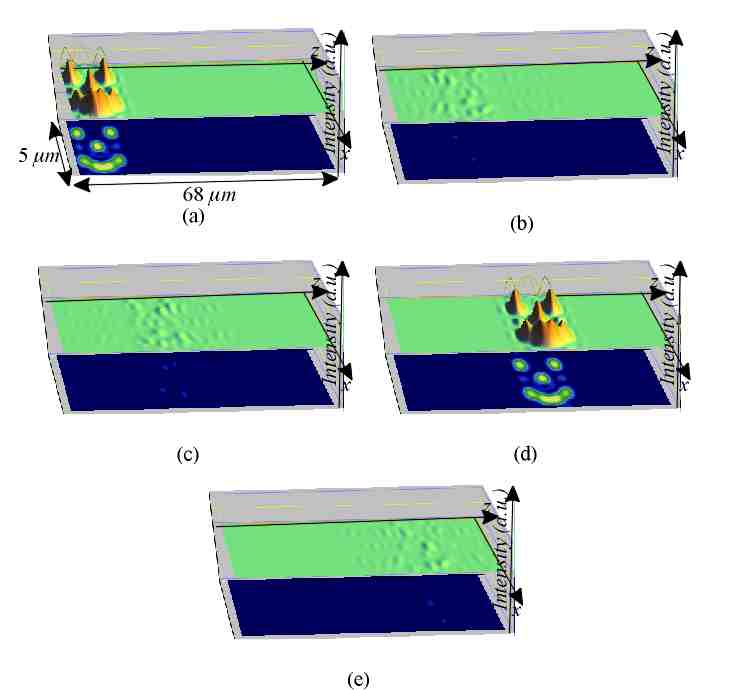}%
\caption{A numerical example of the evolution of a self-imaging spatial image
(smiling human face) consisting of eight self-imaging pixels. The snapshots
are taken at 19 femtosecond time intervals.}%
\label{fig92}%
\end{center}
\end{figure}

As the second example we demonstrate the self-imaging transmission of a
non-trivial spatial image, depicted on Fig.{~\ref{fig92}}a. The image consists
of eight \textquotedblleft pixels\textquotedblright-- the self-imaging
superpositions of FWM's{~}-- specified in previous example. The numerical
examples clearly show that the concept, in principle, is applicable for
constructing wave fields that self-image three-dimensional images. Still, the
experimental realization of such wave fields is not trivial.%
%TCIMACRO{\FRAME{ftbpFU}{4.9372in}{4.6276in}{0pt}{\Qcb{A numerical example of
%the evolution of a self-imaging spatial image (smiling human face) consisting
%of eight self-imaging pixels. The snapshots are taken at 19 femtosecond time
%intervals.}}{\Qlb{fig92}}{fig9_2.jpg}{\special{ language "Scientific Word";
%type "GRAPHIC";  maintain-aspect-ratio TRUE;  display "ICON";
%valid_file "F";  width 4.9372in;  height 4.6276in;  depth 0pt;
%original-width 4.9104in;  original-height 6.6841in;  cropleft "0";
%filename 'fig9_2.jpg';file-properties "XNPEU";}} }%
%BeginExpansion
%EndExpansion

\section{CONCLUSIONS}

In this review we developed a physically transparent, comprehensive theory for
the description of propagation-invariance of the scalar wideband wave fields
in terms of Whittaker type angular spectrum of plane waves. This
representation was demonstrated to be very useful in discussions on the
physical nature of the localized wave transmission phenomenon and propose a
general idea for optical generation of the localized waves.

\section{NOTATIONS USED\ IN\ THIS\ REVIEW}%

\begin{tabular}
[c]{ll}%
$\rho,z,\varphi$ & -- cylindrical coordinates system\\
$k_{x},k_{y},k_{z}$ & -- the Cartesian components of the wave vector
$\mathbf{k}$\\
$k_{0}$ & -- carrier wave number\\
$k,\theta,\phi$ & -- spherical coordinates in $k$-space;\\
$\chi=k_{\rho}=k\sin\theta$ & \\
$\gamma$, $\gamma_{\rho}$ & -- see Eq.~(\ref{su8}), (\ref{r15})\\
$\beta,\xi$ & -- see Eqs. (\ref{su10}) and (\ref{su17})\\
$\mu=z+ct$ & \\
$\zeta=z-ct$ & \\
$v^{g}$, $v^{p}$ & -- group velocity (\ref{su8}) and phase velocity\\
$a_{1},a_{2},\kappa,b,p,q$ & -- parameters, see Eqs.~(\ref{fwm5}),
(\ref{bg5}), (\ref{mp1})\\
$\theta_{F}\left(  k\right)  \equiv\theta_{F}\left(  k,\beta\right)  $,
$\theta_{F}^{\left(  \rho\right)  }\left(  k\right)  $ & -- see
Eqs.~(\ref{su12}), (\ref{r19})\\
$\theta_{G}\left(  k\right)  $ & -- see Eq.~(\ref{gg23})\\
$k_{F}\left(  \theta\right)  $ & -- see Eq.~(\ref{su14})\\
$\tilde{\alpha}=\frac{1}{2}\left(  \frac{\omega}{c}+k_{z}\right)  $ & \\
$\tilde{\beta}=\frac{1}{2}\left(  \frac{\omega}{c}-k_{z}\right)  $ & \\
$\tau=t_{2}-t_{1}$, $\Delta z$ & -- the time- and $z$ coordinate difference\\
$\tau_{s}$, $\sigma_{k}$, $\sigma_{z}$, $\sigma_{\rho}$ & -- see
Eqs.~(\ref{lo16}), (\ref{lo17}), (\ref{lo19})\\
$\Delta t$, $z_{\Delta}$, $z_{m}$ & -- see Eqs.~(\ref{eo3}), (\ref{in25}),
(\ref{bex1})
\end{tabular}
\vspace*{0.5cm}\newline%
\begin{tabular}
[c]{ll}%
$\psi\left(  \mathbf{k},\omega\right)  =\mathcal{F}\left[  \Psi\left(
\mathbf{r},t\right)  \right]  $ & -- see Eq.~(\ref{ang5})\\
$A\left(  k_{x},k_{y,}k_{z}\right)  $ & -- see Eq.~(\ref{ang4a})\\
$A_{n}\left(  k,\theta\right)  $ & -- see Eq.~(\ref{ang9})\\
$A_{z}\left(  k_{x},k_{y}\right)  $, $A_{xy}\left(  k_{z}\right)  $ & -- see
Eq.~(\ref{lo4}), (\ref{lo8})\\
$A_{n}^{we}\left(  k,\chi\right)  $ & -- see Eqs.~(\ref{ang26})\\
$B_{n}\left(  k\right)  $, $B_{0}\left(  k\right)  \equiv B\left(  k\right)  $
& -- see Eq.~(\ref{su29})\\
$\tilde{A}\left(  k_{x},k_{y,}k_{z}\right)  $, $\tilde{A}_{n}\left(
k,\theta\right)  $, $\tilde{B}_{n}\left(  k\right)  $ & -- see note after
Eq.~(\ref{e9})\\
$C_{n}\left(  \tilde{\alpha},\tilde{\beta},\chi\right)  $ & -- see
Eq.~(\ref{bidi1})\\
$\Xi\left(  \chi,\beta\right)  $ & -- see Eq.~(\ref{dn5})\\
$a(k,\theta,\phi)$ & -- stochastic angular spectrum of plane waves
\end{tabular}
\vspace*{0.5cm}\newline%
\begin{tabular}
[c]{ll}%
$\Psi^{\prime}\left(  x,y,z,t\right)  $, $\Psi\left(  \rho,z,\varphi,t\right)
$ & -- see Eqs.~(\ref{ang4}), (\ref{ang8})\\
$\mathbf{E}\left(  \mathbf{r},t\right)  $, $\mathbf{H}\left(  \mathbf{r}%
,t\right)  $,$\mathbf{A}\left(  \mathbf{r},t\right)  $, $\mathbf{\Pi}^{\left(
e\right)  }$, $\mathbf{\Pi}^{\left(  m\right)  }$ & -- see Sec.~\ref{sMax}\\
$\Psi_{F}\left(  \rho,z,\varphi,t\right)  $, $\Psi_{f}\left(  \rho
,z,\varphi,t\right)  $ & -- see Eqs.~(\ref{su40}), (\ref{fwm11})\\
$T\left(  x,y,z,t;\phi\right)  $, $F\left(  x,z,t\right)  $ & -- see
Eqs.~(\ref{t2}), (\ref{t5})\\
$\Gamma\left(  \mathbf{r}_{1},\mathbf{r}_{2},t_{1},t_{2}\right)  $ & -- mutual
coherence function, Eq.~(\ref{ok0})\\
$W\left(  \mathbf{r}_{1},\mathbf{r}_{2},k\right)  $ & -- cross-spectral
density\\
$\mathcal{A}\left(  k_{1},k_{2},\theta_{1},\theta_{2},\phi_{1},\phi
_{2},\right)  $ & -- see Eq.~(\ref{ok17}), angular correlation function\\
$\mathcal{C}\left(  k_{1},k_{2},\phi_{1},\phi_{2}\right)  $ & -- see
Eq.~(\ref{ok25})\\
$\mathcal{V}\left(  k,\phi\right)  $ & -- see Eq.~(\ref{ok25})\\
$\emph{s}\left(  k\right)  $ & -- see Eq.~(\ref{lo15}), frequency spectrum of
light source\\
$\mathcal{S}\left(  k\right)  =\left\vert \emph{s}\left(  k\right)
\right\vert ^{2}$ & -- spectral density (power spectrum) of light source
\end{tabular}

\end{document}